%% file: dissertation.tex
\newcommand{\squishlist}{
   \begin{list}{$\bullet$}
    { \setlength{\itemsep}{0pt}      \setlength{\parsep}{0pt}
      \setlength{\topsep}{0pt}       \setlength{\partopsep}{0pt}
      \setlength{\leftmargin}{1em} \setlength{\labelwidth}{1em}
      \setlength{\labelsep}{0.5em} } }

\newcommand{\squishend}{
    \end{list}  }

\documentclass[12pt]{report}%
\usepackage[T1]{fontenc}
\usepackage{times}
\usepackage{fullpage}
\usepackage{amsmath}
\usepackage{amsfonts}%
\usepackage{amssymb}
\usepackage{url}
\usepackage{hyperref}
\usepackage[top=1in,bottom=1in,left=1in,right=1in]{geometry}
\usepackage{setspace}
\usepackage{graphicx}
\usepackage{subcaption}
\usepackage{wrapfig}
\usepackage{titlesec}
\usepackage{xspace}
\usepackage{multirow}
\usepackage{booktabs}
\usepackage{slashbox}
\usepackage{algorithm2e}
\usepackage{xfrac}
\usepackage{setspace}
\def\httilde{\mbox{\tt\raisebox{-.5ex}{\symbol{126}}}}

\def\title{\textbf{Providing High and Controllable Performance in Multicore Systems Through Shared Resource Management}}

\def\author{Lavanya Subramanian}

\begin{document}
\setcounter{page}{1}
\include{ThesisCover}
\newpage
\include{dedication}

\newpage
\doublespacing
\pagenumbering{roman}
\input{chapters/abstract}

\newpage
\input{acknowledgements}
\newpage

\tableofcontents
\newpage
\listoftables
\newpage
\listoffigures
\newpage
%\linespread{2}
\setcounter{page}{1}
\pagenumbering{arabic}
\input{chapters/introduction}
\newpage
\input{chapters/background}

\newpage
\input{chapters/blacklisting}

\newpage
\input{chapters/mise-model}

\newpage
\input{chapters/applications-mise-model}

\newpage
\input{chapters/asm-model}

\newpage
\input{chapters/applications-asm}

\input{chapters/conclusions}

\bibliographystyle{hplain}
\bibliography{references}
\end{document}

%% file: ThesisCover.tex
%% Title page
%%Courtesy of Reshma Shetty
%%modified by Huaiwei Liao

\def\name{Lavanya Subramanian}
\def\addrone{5000 Forbes Ave}
\def\addrtwo{Pittsburgh, PA 15213}

\def\degree{Doctor of Philosophy}
\def\pastbsdegree{B.E., Electronics and Communication, Madras Institute of Technology}
\def\pastmsdegree{M.S., Electrical and Computer Engineering, Carnegie Mellon University}
\def\deptname{Electrical and Computer Engineering}

\def\submissiondate{{August, 2015}}

\def\advisor{Prof. Onur Mutlu}
\def\supervisoronedept{Department of Electrical and Computer Engineering}

\def\membertwo{Prof. James Hoe}
\def\membertwodepta{Department of Electrical and Computer Engineering}

\def\memberone{Prof. Greg Ganger}
\def\memberonedepta{Department of Electrical and Computer Engineering}

\def\memberthree{Dr. Ravi Iyer, Intel}
\def\memberthreedepta{Intel}

%\def\academicoffice{Dalia Gabour}
%\def\academicofficetitle{Academic Administrator}
%\def\academicofficedept{Biological Engineering Division}

%%%%%%%%%%%%%%%%%%%%%%%%%%%%%%%%%
%%% Formatting of Title Page
%%%%%%%%%%%%%%%%%%%%%%%%%%%%%%%%%
\thispagestyle{empty}
\vspace*{0.8in}
\begin{center}
{\Large \bf
   \title\\
}
\vspace{0.6in}
{\large \bf \it Submitted in partial fulfillment of the requirements for\\
the degree of\\
\degree\\
in\\
\deptname\\}
\vspace{.3in}
{\large \name\\
\pastbsdegree\\
\pastmsdegree\\
}
\vspace{1in}
{\large \bf Thesis Committee:\\}
{\large Advisor: \advisor\\
\memberone\\
\membertwo\\
\memberthree\\}
\vspace{.4in}
{\large Carnegie Mellon University\\
Pittsburgh, PA\\}
\vspace{0.25in}
{\large July, 2015\\}
\vspace{0.25in}
{\large Copyright \copyright~2015 Lavanya Subramanian}

%{\Large \bf 
%   Carnegie Mellon University
%\\ \deptname \\}
%\vspace{.25in}
%{\Large \bf
% Dissertation
%\\ \degree \\}
%	\vspace{.25in}
%   {\Large \bf \underline{Title:}}  \\                    
%   \vspace{.15in}
%   
%{\Large \bf 
%   \title\\}
%%    \\ \titletwo \\}                                            
%\vspace{.25in}
%Date of Submission: \submissiondate
%%\vspace{.15in}
%%\\ \submissiondate
%
\end{center}

\vspace{.25in}

%\def\authorsig{{\small \sc (Signature of Author)}}
%\def\supervisorsig{{\small \sc (Signature of Supervisor)}}
%\def\academicofficesig{{\small \sc (Signature of Academic Office)}}

%\begin{tabular}{rlc}
%\\ {\small \sc Submitted by:}
%                            & \author  & \\
%%                            & \addrone & \\
%%                            & \addrtwo & \\
%%\\ {\small \sc Signed:}
%%                            & \cline{1-1} \\
%
%\\ {\small \sc Advisor:}
%                            & \supervisorone  & \\
%%                            & \supervisoronetitle & \\
%                            & \supervisoronedept & \\
%%\\ {\small \sc Signed:}
%%                            & \cline{1-1} \\
%                            			    
%%\\ {\small \sc Academic Office:}
%%                            & \academicoffice  & \\
%%                            & \academicofficetitle & \\
%%                            & \academicofficedept & \\
%%\\ {\small \sc Signed:}
%%                            & \cline{1-1} \\
%
%\\ {\small \sc Committee Members:}
%                            & \memberone  & \\
%                            & \memberonedepta & \\
%%                            & \memberonedeptb & \\
%                            \\
%                            & \membertwo  & \\
%                            & \membertwodepta & \\
%%                            & \membertwodepta & \\
%                            \\
%                            & \memberthree  & \\
%                            & \memberthreedepta & \\
%
%			    
%\end{tabular}

%% file: dedication.tex
\thispagestyle{empty}
\vspace*{\fill}
\begin{center}
{\large \it In memory of my beloved granddad Rajamony (1917 - 2005)\\
who was so full of life until his last breath}
\end{center}
\vfill
%\vspace*{-100mm}

%% file: chapters/abstract.tex
\chapter*{Abstract}
\vspace*{-10mm}

Multiple applications executing concurrently on a multicore system
interfere with each other at different shared resources such
as main memory and shared caches. Such inter-application
interference, if uncontrolled, results in high system performance
degradation and unpredictable application slowdowns. While
previous work has proposed application-aware memory scheduling as
a solution to mitigate inter-application interference and improve
system performance, previously proposed memory scheduling
techniques incur high hardware complexity and unfairly slowdown
some applications. Furthermore, previously proposed
memory-interference mitigation techniques are not designed to
precisely control application performance. 

This dissertation seeks to achieve high and controllable
performance in multicore systems by mitigating and quantifying the
impact of shared resource interference. First, towards mitigating
memory interference and achieving high performance, we propose the
Blacklisting memory scheduler. We observe that ranking
applications individually with a total order based on memory
access characteristics, like previous schedulers do, leads to high
hardware cost, while also causing unfair application slowdowns.
The Blacklisting memory scheduler overcomes these shortcomings
based on two key observations. First, we observe that, to mitigate
interference, it is sufficient to separate applications into only
two groups, one containing applications that are vulnerable to
interference and another containing applications that cause
interference, instead of ranking individual applications with a
total order. Vulnerable-to-interference group is prioritized over
the interference-causing group. Second, we show that this grouping
can be efficiently performed by simply counting the number of
consecutive requests served from each application -- an
application that has a large number of consecutive requests served
is dynamically classified as interference-causing. The
Blacklisting memory scheduler, designed based on these insights,
achieves high system performance and fairness, while incurring
significantly lower complexity than state-of-the-art
application-aware schedulers.

Next, towards quantifying the impact of memory interference and
achieving controllable performance in the presence of memory
bandwidth interference, we propose the Memory Interference induced
Slowdown Estimation (MISE) model. The MISE model estimates
application slowdowns due to memory interference based on two
observations. First, the performance of a memory-bound application
is roughly proportional to the rate at which its memory requests
are served, suggesting that request-service-rate can be used as a
proxy for performance. Second, when an application's requests are
prioritized over all other applications' requests, the application
experiences very little interference from other applications. This
provides a means for estimating the uninterfered
request-service-rate of an application while it is run alongside
other applications. Using the above observations, MISE estimates
the slowdown of an application as the ratio of its uninterfered
and interfered request service rates. We propose simple changes to
the above model to estimate the slowdown of non-memory-bound
applications. We propose and demonstrate two use cases that can
leverage MISE to provide soft performance guarantees and high
overall performance/fairness.

Finally, we seek to quantify the impact of shared cache
interference on application slowdowns, in addition to memory
bandwidth interference. Towards this end, we propose the
Application Slowdown Model (ASM). ASM builds on MISE and observes
that the performance of an application is strongly correlated
with the rate at which the application accesses the shared cache.
This is a more general observation than that of MISE and
holds for all applications, thereby enabling the estimation of
slowdown for any application as the ratio of the uninterfered to
the interfered shared cache access rate. This reduces the problem
of estimating slowdown to estimating the shared cache access rate
of the application had it been run alone on the system. ASM
periodically estimates each application's cache-access-rate-alone
by minimizing interference at the main memory and quantifying
interference at the shared cache. We propose and demonstrate
several use cases of ASM that leverage it to provide soft
performance guarantees and improve performance and fairness.

%% file: acknowledgements.tex
\chapter*{Acknowledgments}
My educational journey until this point has been fruitful and very
memorable, thanks to all the wonderful people who have been a part
of my journey. First and foremost, I am grateful to my advisor,
Prof. Onur Mutlu, who was willing to take me on as his student
despite my lack of background in computer architecture then and
give me the mentorship, time and encouragement to build
background and grow as a researcher. Onur's emphasis on clarity in
thinking, speaking and writing has been a major influence in
shaping me. I am also very thankful to Onur for providing me with
the resources and freedom to carry out research and for always finding
the right opportunities by way of collaborations and internships
to further my research.

I would like to thank my committee members, Prof. Greg Ganger,
Prof. James Hoe and Dr. Ravi Iyer for their time, effort and
inputs in bringing this dissertation to completion. Special thanks
to James for his encouragement and feedback even since my early
years at Carnegie Mellon University. Thanks to Greg for his incisive insights on various
aspects of my work, from his unique perspective as the storage QoS
expert on my committee. Thanks to Ravi for his many inputs and
insights on QoS and for giving me the opportunity to intern with
his spirited and warm group in Intel Labs, Hillsboro. 

The SAFARI group has been a great source of critical feedback,
ideas and fun. I am incredibly thankful for everything I've
learned from this group of smart, enthusiastic and hard-working
graduate students over the years. Vivek Seshadri has been a great
friend and lab mate. He has been an amazing sounding board for new
ideas. Several of the ideas in this thesis have evolved a lot
through numerous discussions with Vivek. Thanks a lot to Yoongu
Kim for all his feedback on writing and presentation. Yoongu's
high standards for research, presentation and writing are an
inspiration. I am thankful to Chris Fallin for his critical
feedback on research and for all I learned from him during the
many times we worked together on TAing, course work, quals. Thanks
a lot to Samira Khan for the many discussions on research and
life, in general. Donghyuk Lee's DRAM expertise and drive to keep
learning are admirable. Kevin Chang's methodical approach to
research and problem solving have been very useful in many of the
projects we have worked together on. Thanks to Rachata
Ausavarungnirun for his helpful nature. Many thanks to visiting
researchers Hui Wang, Hiroyuki Usui and interns Harsha Rastogi,
Arnab Ghosh for working with me on different research projects.
Thanks to Gennady Pekhimenko for being a great cube mate. His work
ethic and discipline are admirable. Thanks to Justin Meza,
Hongyi Xin, Nandita Vijayakumar, Yang Li, Yixin Luo, Kevin Hsieh
and Amirali Bouramand for the many discussions and dinners. 

Besides members of the SAFARI group, several graduate students
have been a great source of advice and encouragement at several
points. Thanks to Siddharth Garg for getting me inducted into the
workings of graduate school and the sound advice, when I was still
learning the ropes as a first year graduate student. Thanks to
Michael Papamichael for all the inputs, advice and discussions
over the years at Carnegie Mellon University. Michael's genuine
passion for research and his ability to explain concepts so
clearly are inspiring. Thanks to Karthik Lakshmanan for the many
discussions and inputs during my early years. Thanks to Anagha
Kulkarni for her encouragement and the many discussions we have
had about graduate school and life in general when at MSR,
Redmond. Thanks to Michelle Goodstein for her company and her
perspectives on life. Many thanks to Elaine Lawrence, Samantha
Goldstein, Karen Lindenfelser, Nathan Snizaski, Marilyn Patete,
Debbie Scappatura and Olivia Vadnais for helping me navigate
through administrative aspects. Thanks also to the CMU shuttle and
escort drivers who have safely ferried me home several late
nights.

I enjoyed my internships at Intel Labs, Hillsboro and Microsoft
Research, Redmond. Thanks to Li Zhao for being a very hands-on and
involved mentor during my internship at Intel Labs. Li was always
available to brainstorm and discuss. Thanks to Trishul Chilimbi,
Sriram Sankar and Kushagra Vaid for being great mentors during my
internship at MSR. Thanks to Thomas Moscibroda for the weekly
brainstorming sessions when I was at MSR. Many thanks to Gabriel
Loh for his mentorship. I would like to thank National Science
Foundation (NSF), Semiconductor Research Corporation (SRC),
Gigascale Systems Research Center (GSRC) and Intel for generously
supporting my research over the years and Carnegie Mellon
University for supporting me with the John and Claire Bertucci
fellowship

I am very thankful to all my teachers from school and professors
from undergrad for instilling in me a basic sense of curiosity and
an urge to learn. Special thanks to Ms. Jennifer, Ms. Nargis, Ms.
Bhuvaneswari, Prof. Mala John and Prof. Kannan. Thanks also to
several seniors from my undergraduate institution who served as
role models and a source of inspiration. I am grateful to Dr.
Sasikanth Avancha from Intel for mentoring me on my undergraduate
project and my manager at SanDisk, Radhakrishnan Nair, for his
support when applying to graduate school.

Graduate school is a long and intense journey, with many highs and
lows. I am very thankful for the support systems I had through
grad school, in terms of friends, room mates and family. Thanks to
Anusha Venkatramani, Ashwati Krishnan, Aishwarya Sukumar, Lavanya
Iyer, Swati Sarraf and Manali Bhutiyani for being great and very
understanding room mates and friends. Thanks to Abhay Ramachandra,
Aditi Pandya, Arun Kannawadi, Arvind Muralimohan, Athula
Balachandran, Bhavana Dalvi, Divya Hariharan, Divya Sharma, Janani
Mukundan, Madhumitha Ramesh, Mahesh Joshi, Natasha Kholgade,
Niranjini Rajagopal, Ramkumar Krishnan, Siddharth Komini Babu,
Siddharth Gopal, Suchita Ramachandran, Swaminathan Ramesh, Varoon
Shankar for their companionship through different points of grad
school. Thanks to Aishwarya R, Anusha Radhakrishnan, Ramya
Guptha, Gayatri Singaravelu, Mukund Kalyanaraman, Sudharsan
Seshadri and Arunachalam Annamalai for their friendship and
support from afar. 

My family has been a big source of support, encouragement and
comfort all through my journey. Thanks to my mother, Bhuvaneswari
and my father, Subramanian for instilling in me the value of
education. My mother, a math teacher, inculcated in me an interest
in math from a young age and has been a big source of
encouragement. My father's work ethic has been an inspiration. My
grand parents have been extremely supportive of all my endeavors,
right from a very young age and I am grateful to them for all the
warmth, support and care over the years. My grand father
Rajamony's enthusiasm until his very last breath was infectious.
My grand mother Jambakam's appreciation of and strong belief in
the need for a solid education, despite her not having had access
to it is admirable. I am also very grateful to my paternal grand
mother Visalakshi for her encouragement - she, for some reason,
thought I was going to be an engineer since I was ten years old. I
wish she had seen me finish my PhD.

My uncle, Kumar, has been a great source of encouragement, knowledge and
fun. He introduced me to science through fun experiments at home.
He introduced me to the wonderful world of books and libraries
from a very young age. He has had an immense role in shaping me
during my formative years. My sister, Lakshmi, has been a pillar of support.
Through school and undergrad, she was always willing and eager to
get me any books and resources I needed. She has been extremely
supportive and involved in all my major academic/career decisions.
My parents-in-law and sister-in-law, Devi have been a big source of
encouragement over the past couple of years.

Finally, Kaushik Vaidyanathan has been a rock solid pair of
dependable shoulders, over the past several years as my best
friend and more recently, my husband. Right from pushing me to
apply to grad school to listening to my long rants during times of
self doubt to putting up with my erratic schedules to what not,
his support has been immense and has made this whole journey
possible and that much more fun and enjoyable. I will not even
attempt to say thank you for that will trivialize all that he has
done for me.

%% file: chapters/introduction.tex
\chapter{Introduction}
\label{chap:introduction}

\section{Problem}
Applications executing concurrently on a multicore chip contend
with each other to access shared resources such as main memory and
shared caches. The main memory has limited bandwidth, driven by
constraints on pin counts. If the available shared cache capacity
and memory bandwidth are not managed well, different applications
can harmfully interfere with each other, resulting in significant
degradation in both system performance and individual application
performance. Furthermore, the slowdown experienced by an
application due to inter-application interference at these shared
resources depends on the other concurrently running applications
and the available memory bandwidth and shared cache capacity.
Hence, different applications experience different and
unpredictable slowdowns.

\begin{figure*} [ht!]
  \centering
  \begin{subfigure}{0.47\textwidth}
    \centering
    \includegraphics[scale=0.3, angle=270]{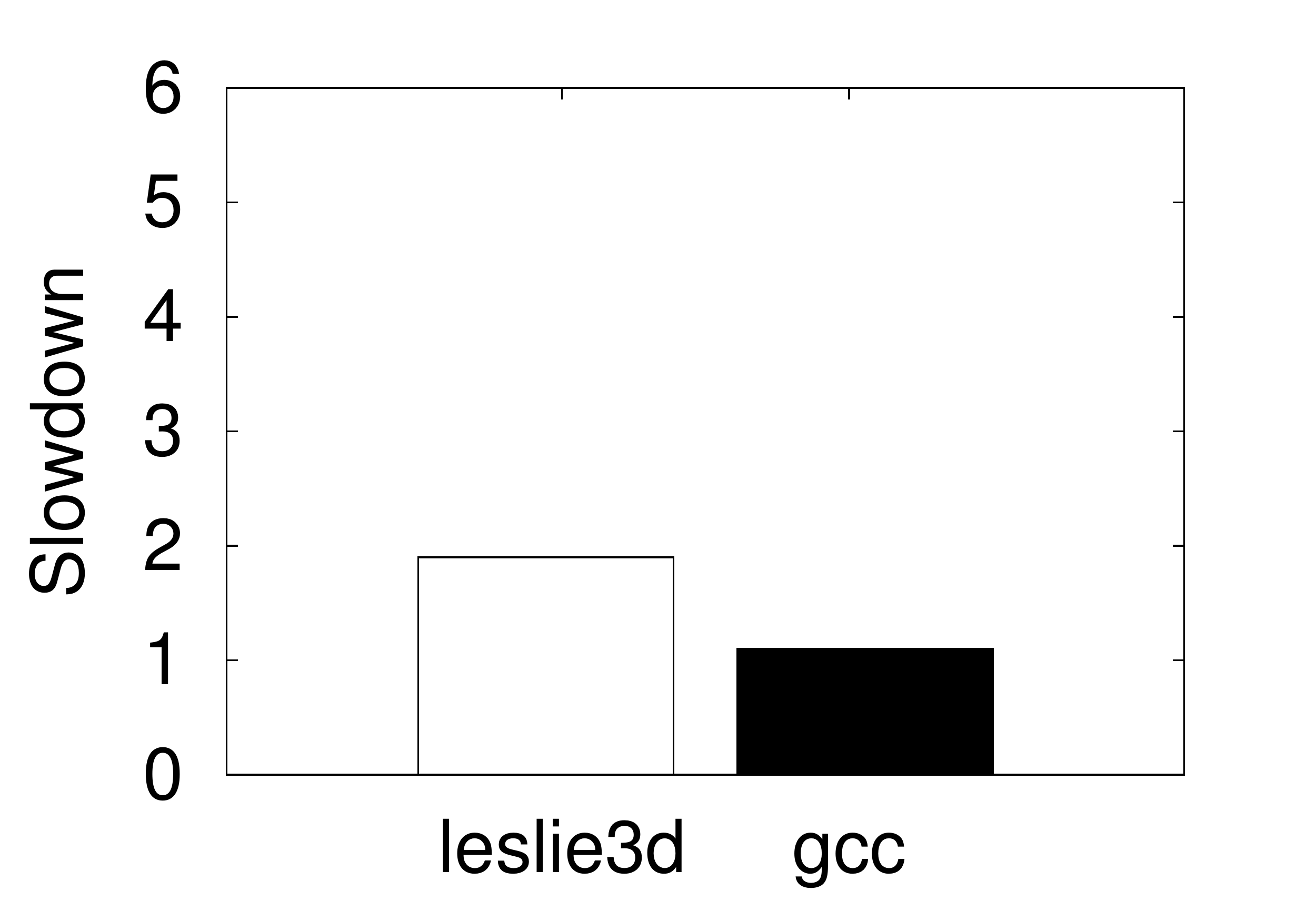}
    \caption{leslie3d co-running with gcc}
  \end{subfigure}
  \begin{subfigure}{0.47\textwidth}
    \centering
    \includegraphics[scale=0.3, angle=270]{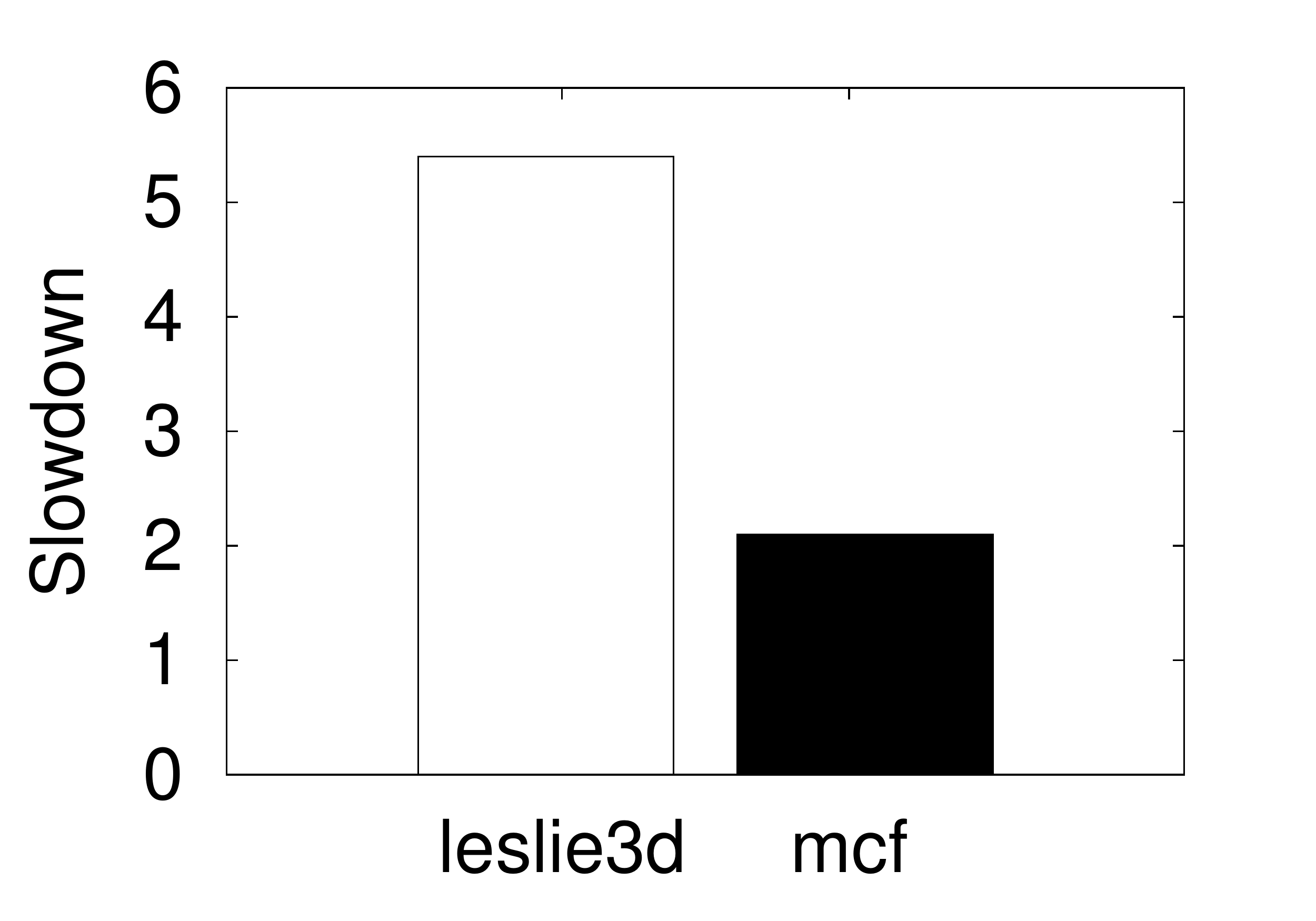}
    \caption{leslie3d co-running with mcf}
  \end{subfigure}
  \caption{leslie3d's slowdown compared to when run alone}
  \label{fig:slowdown-plot}
\end{figure*}

Figure~\ref{fig:slowdown-plot} shows leslie3d, an application from
the SPEC CPU 2006 suite when it is run with two different
applications, gcc and mcf, on a simulated two-core system where
the cores share a main memory channel. As can be seen, leslie3d
and mcf slow down significantly due to shared resource
interference (gcc does not slow down since it is largely
compute-bound and does not access the main memory much).
Furthermore, leslie3d slows down by 1.9x when it is run with gcc,
an application that rarely accesses the main memory. However,
leslie3d slows down by 5.4x when it is run with mcf, which
frequently accesses the main memory. This is one representative
example demonstrating that an application experiences different
slowdowns when run with applications that have different shared
resource access characteristics. We observe this behavior across a
variety of applications, as also observed by previous
works~\cite{mph, stfm, parbs, fqm}, resulting in high and
unpredictable applications slowdowns.

Inter-application interference and the resultant high application
slowdowns are a major problem in most multicore systems, where
multiple applications share resources. Furthermore, the
unpredictable nature of the slowdowns is particularly undesirable
in several scenarios where some applications are critical and need
to meet requirements on their performance. For instance, in a data
center/virtualized environment where multiple applications, each
potentially from a different user, are consolidated on the same
machine, it is common for each user to require a certain
guaranteed performance for their application. Another example is
in mobile systems where interactive and non-interactive jobs share
resources and the interactive jobs need to meet deadlines/frame rate
requirements. \emph{Our main research objective is to mitigate and
quantify shared resource interference, towards the end of achieving
high and controllable application performance, through simple and
implementable slowdown estimation and shared resource management
techniques.}

\section{Our Solutions}

\subsection{The Blacklisting Memory Scheduler}
Towards achieving our goal of high system performance and
fairness, we propose the Blacklisting memory scheduler, a simple
memory scheduler design that is able to achieve high performance
and fairness at low cost by mitigating interference at the main
memory. Although the problem of memory interference mitigation has
been much explored, with memory request scheduling being the
prevalent solution direction, we observe that previously proposed
memory schedulers are both complex and unfair. The main source of
this complexity and unfairness is the notion of ranking
applications individually, with a total rank order, based on
applications' memory access characteristics. Computing and
enforcing ranks incurs high hardware complexity, both in terms of
logic and storage overhead.  As a result, the critical path
latency and chip area of ranking-based application-aware memory
schedulers is significantly higher compared to application-unaware
schedulers. For example, Thread Cluster Memory Scheduler
(TCM)~\cite{tcm}, a state-of-the-art application-aware scheduler
is 8x slower and 1.8x larger than a commonly-employed
application-unaware scheduler, FRFCFS~\cite{frfcfs}. Furthermore,
when a total order based ranking is employed, applications that
are at the bottom of the ranking stack get heavily deprioritized
and unfairly slowed down.  This greatly degrades system fairness.

In order to overcome these shortcomings of previous ranking-based
schedulers, we propose the Blacklisting memory scheduler (BLISS)~\cite{bliss,bliss-arxiv}
based on two new observations. First, in contrast to forming a total rank order
of all applications (as done in prior works), we find that, to
mitigate interference, it is sufficient to i) separate
applications into \emph{only two} groups, one group containing
applications that are vulnerable to interference and another
containing applications that cause interference, and ii)
prioritize the requests of the {\em vulnerable-to-interference}
group over the requests of the {\em interference-causing} group.
Second, we observe that applications can be
efficiently classified as either {\em vulnerable-to-interference}
or {\em interference-causing} by simply counting the number of
consecutive requests served from an application in a short time
interval. 

BLISS achieves better system performance and fairness than the
best-performing previous schedulers, while incurring significantly
low complexity. However, BLISS does not tackle the problem of
unpredictable application slowdowns.

\subsection{The Memory Interference induced Slowdown Estimation
(MISE) Model} 

Towards tackling the problem of unpredictable application
slowdowns, we first propose to estimate/quantify and control
application slowdowns in the presence of interference at the main
memory. First, we estimate application slowdowns using the Memory
Interference induced Slowdown Estimation (MISE) model~\cite{mise}.
The MISE model accurately estimates application slowdowns based on
two key observations. First, the performance of a memory-bound
application is roughly proportional to the rate at which its
memory requests are served. This observation suggests that we can
use request-service-rate as a proxy for performance, for
memory-bound applications. As a result, slowdown of such an
application can be computed as the ratio of the
request-service-rate when the application is run alone on a system
to that when it is run alongside other interfering applications.
Second, the alone-request-service-rate of an application can be
estimated by giving the application's requests the highest
priority in accessing memory. Giving an application's requests the
highest priority in accessing memory results in very little
interference from other applications' requests. As a result, most
of the application's requests are served as though the application
has all the memory bandwidth for itself, allowing the system to
gather a good estimate for the alone-request-service-rate of the
application. We adapt these observations and extend the model to
estimate slowdowns of applications that are not bound at memory
too. 

Accurate slowdown estimates from the MISE model can enable several
mechanisms to achieve both high and controllable performance. We build two such
mechanisms on top of our proposed model to demonstrate its
effectiveness.

\subsection{The Application Slowdown Model (ASM)}

The MISE model estimates slowdowns due to interference at the main
memory. However, it does not take into account interference at the
shared caches and assumes caches are private. The Application
Slowdown Model (ASM)~\cite{asm-tech-report} estimates slowdowns due to both shared cache
and main memory interference. ASM does so by exploiting the
observation that the performance of each application is roughly
proportional to the rate at which it accesses the shared cache.
This observation builds on MISE's observation on correlation
between memory request service rate and performance. However, it
is more general and applies to all applications, unlike MISE's
observation that applies only to memory-bound applications. ASM
estimates alone-cache-access-rate in two steps. First, ASM
\emph{minimizes interference for an application at the main
memory} by giving the application's requests the highest priority
at the memory controller, similar to MISE. Doing so also enables
ASM to get an accurate estimate of the average cache miss service
time of the application had it been run alone (to be used in the
next step). Second, \emph{ASM quantifies the effect of
interference at the shared cache} by using an auxiliary tag store
to determine the number of shared cache misses that would have
been hits if the application did not share the cache with other
applications (contention misses). This aggregate contention miss
count is used along with the average miss service time (from the
previous step) to estimate the actual time it would have taken to
serve the application's requests had it been run alone.

We present and evaluate several mechanisms that can leverage ASM's
slowdown estimates towards achieving different goals such as high
performance, fairness, bounded application slowdowns and fair billing,
thereby demonstrating the model's effectiveness.

\section{Thesis Statement}

\emph{High and controllable performance} can be achieved in multicore
systems through simple and implementable mechanisms to \emph{mitigate
and quantify shared resource interference}.

\section{Contributions}

This dissertation makes the following major contributions:

\begin{itemize}
\item
This dissertation makes the observation that it is not necessary
to rank individual applications with a total rank order, like most
previous ranking-based application-aware memory schedulers do, in
order to mitigate interference between applications. This
observation enables the design of the Blacklisting memory
scheduler, a low-complexity memory scheduling technique that is
able to achieve high performance and fairness, by simply
categorizing applications as interference-causing or vulnerable. 
\item
This dissertation makes the observation that an application's
performance is roughly proportional to the rate at which requests
are generated to/served at a shared resource. This observation can
serve as a general principle enabling the estimation of
progress/slowdowns at different shared resources.
\item
This dissertation presents the Memory Interference induced
Slowdown Estimation (MISE) model that accurately estimates
application slowdowns in the presence of memory interference as
the ratio of uninterfered to interfered request service rates,
based on the correlation between request service rate and
performance.
\item
This dissertation presents the Application Slowdown Model (ASM)
that accurately estimates application slowdowns due to both shared
cache and main memory interference, by minimizing interference at
the main memory and quantifying interference at the shared cache.
\item
This dissertation builds several resource management mechanisms on
top of MISE and ASM that leverage their slowdown estimates to
provide high performance, fairness and bounded slowdowns,
demonstrating MISE/ASM's effectiveness in estimating slowdowns.
\end{itemize}

%Several previous works have tackled the problem of memory
%interference mitigation with the goal of improving performance.
%However, there have been few previous proposals that attempt to
%provide performance predictability, in the presence of shared
%resource interference. We will provide a review of previously
%proposed techniques in Section~\ref{sec:related-work}. 

\section{Dissertation Outline}
This dissertation is organized into eight chapters.
Chapter~\ref{chap:background} presents background on memory system
organization and discusses related prior work on shared resource
management and providing Quality of Service (QoS).
Chapter~\ref{chap:blacklisting} presents the design of the
Blacklisting memory scheduler (BLISS) and evaluates it against
state-of-the-art memory request schedulers.
Chapter~\ref{chap:mise} presents the Memory Interference induced
Slowdown Estimation (MISE) model and its evaluation against
previous slowdown estimation techniques.
Chapter~\ref{chap:mise-applications} presents memory bandwidth
management schemes that leverage the MISE model to provide bounded
application slowdowns and fairness. Chapter~\ref{chap:asm}
presents the Application Slowdown Model (ASM) and compares it
against previous schemes that estimate slowdown due to both shared
cache and main memory interference.
Chapter~\ref{chap:asm-applications} presents several use cases
that leverage slowdown estimates from ASM to provide high
performance, fairness and bounded slowdowns. Finally,
Chapter~\ref{chap:conclusions} presents conclusions and future
research directions that are enabled by this dissertation.

%% file: chapters/background.tex
\chapter{Background and Related Prior Work}
\label{chap:background}

The problem of shared resource interference has been a significant
deterrent to achieving high and controllable system performance.
Not surprisingly, several previous works have attempted to
mitigate interference at both the shared caches and main memory,
with the goal of improving system performance. However, few
previous works have tackled the problem of unpredictable
application slowdowns in the presence of shared resource
interference. 

In this chapter, we will first provide a brief background on DRAM
main memory organization and discuss previous proposals in
different related areas, namely memory interference mitigation,
DRAM optimizations to improve system performance, shared cache
capacity management, Quality of Service (QoS) and slowdown
estimation.

\section{DRAM Main Memory Organization}
\label{sec:dram-org}
\begin{figure}
    \centering
    \includegraphics[scale=0.9, angle=0]{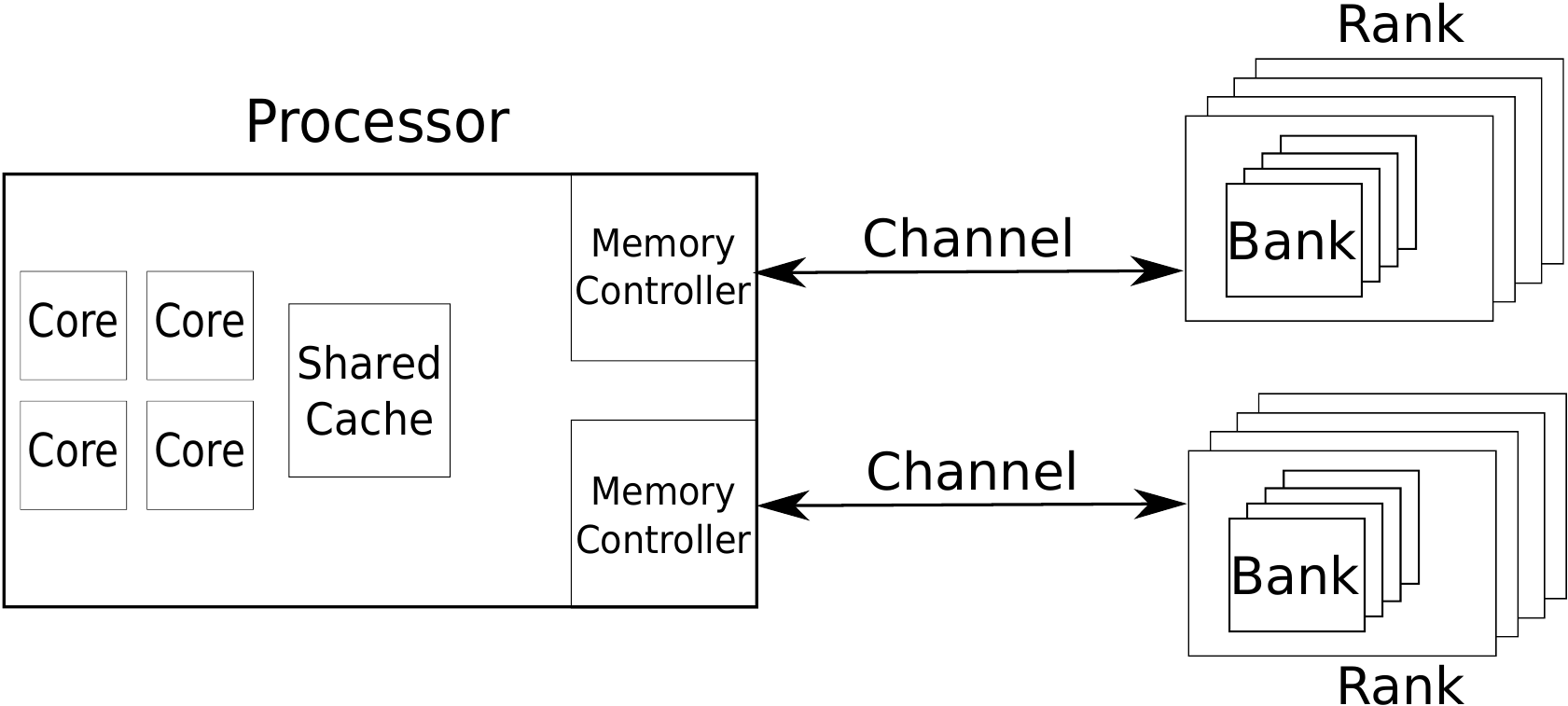}
    \caption{DRAM main memory organization}
    \label{fig:dram-main-memory}
%    \vspace{-6mm}
\end{figure}

The DRAM main memory system is organized as channels, ranks and
banks hierarchically as shown in
Figure~\ref{fig:dram-main-memory}. Channels are independent and
can operate completely in parallel. Each channel consists of ranks
(typically 1 - 4) that share the command and data bus of the
channel.

A rank consists of multiple banks. The banks can operate in
parallel. However, all banks within a channel share the command
and data bus of the channel. Each bank, in turn, is organized as
an array of rows and columns. On a data access, the entire row
containing the data is brought into an internal structure called
the row-buffer. Therefore, a subsequent access to the same row can
be served in the row-buffer itself and need not access the array.
This is called a row hit. On an access to a different row though,
the array needs to be accessed. Such an access is called a row
miss. A row hit is served
{\raise.17ex\hbox{$\scriptstyle\mathtt{\sim}$}}2x faster than a
row miss~\cite{jedec-ddr}. Please refer
to~\cite{salp,raidr,tldram} for more detail on DRAM operation. 

Commonly employed memory controllers employ a memory scheduling
policy called First Ready First Come First Served
(FRFCFS)~\cite{frfcfs-patent,frfcfs} that leverages the row buffer
by prioritizing row hits over row misses/conflicts. Older requests
are then prioritized over newer requests. FRFCFS aims to maximize
DRAM throughput by prioritizing row hits. However, it unfairly
prioritizes requests of applications that generate a large number
of requests to the same row (high-row-buffer-locality) and access
memory frequently (high-memory-intensity)~\cite{mph,stfm}.

\section{Related Work on Memory Scheduling}
\label{sec:memory-scheduling}
Much prior work has focused on mitigating this unfairness and
inter-application interference at the main memory, with the goals
of improving system performance and fairness, of which a
predominant solution direction is memory request scheduling.
Several previous
works~\cite{stfm,parbs,podc-08,atlas,tcm,rlmc,crit-scheduling-cornell}
have proposed application-aware memory scheduling techniques that
take into account the memory access characteristics of
applications and schedule requests appropriately in order to
mitigate inter-application interference and improve system
performance and fairness. 

Mutlu and Moscibroda propose PARBS~\cite{parbs}, an
application-aware memory scheduler that batches the oldest
requests from applications and prioritizes the batched requests,
with the goals of preventing starvation and improving fairness.
Within each batch, PARBS ranks individual applications based on
the number of outstanding requests from the application and, using
this total rank order, prioritizes requests of applications that
have low-memory-intensity to improve system throughput. Kim et
al.~\cite{atlas} observe that applications that receive low memory
service tend to experience interference from applications that
receive high memory service. Based on this observation, they
propose ATLAS, an application-aware memory scheduler that ranks
individual applications based on the amount of long-term memory
service each application receives and prioritizes applications
that receive low memory service, with the goal of improving
overall system throughput.

Another recently proposed memory scheduling technique, Thread
cluster memory scheduling (TCM)~\cite{tcm} ranks individual
applications by memory intensity such that low-memory-intensity
applications are prioritized over high-memory-intensity
applications (to improve system throughput). Kim et al.~\cite{tcm}
also observed that ranking all applications based on memory
intensity and prioritizing low-memory-intensity applications could
slow down the deprioritized high-memory-intensity applications
significantly and unfairly. This is because when all applications
are ranked by memory service, applications with high memory
intensities are ranked lower, as they inherently tend to have high
memory service, as compared to other applications. With the goal
of mitigating this unfairness, TCM clusters applications into low-
and high-memory-intensity clusters. In the low-memory-intensity
cluster, applications are ranked by memory- intensity, whereas, in
the high-memory-intensity cluster, applications' ranks are
shuffled randomly to provide fairness.  Both clusters employ a
total rank order among applications at any given time.

More recently, Ghose et al.~\cite{crit-scheduling-cornell} propose
a memory scheduler that aims to prioritize \emph{critical} memory
requests that stall the instruction window for long lengths of
time. The scheduler predicts the criticality of a load instruction
based on how long it has stalled the instruction window in the
past (using the instruction address (PC)) and prioritizes requests
from load instructions that have large total and maximum stall
times measured over a period of time. Although this scheduler is
not application-aware, we compare to it as it is the most recent
scheduler that aims to maximize performance by mitigating memory
interference. 

All these state-of-the-art schedulers incur significant hardware
complexity and cost to rank applications based on their memory
access characteristics and prioritize requests based on this
ranking. This results in significant increase in critical path
latency and area, as we discuss in Chapter~\ref{chap:blacklisting}.

\section{Related Complementary Memory Scheduling Proposals}

Parallel Application Memory Scheduling (PAMS)~\cite{pams} tackles
the problem of mitigating interference between different threads
of a multithreaded application, while Staged Memory Scheduling
(SMS)~\cite{sms} attempts to mitigate interference between the CPU
and GPU in CPU-GPU systems. Principles from our work can be
employed in both of these contexts to identify and deprioritize
interference-causing threads, thereby mitigating interference
experienced by vulnerable threads/applications. Complexity effective
memory access scheduling~\cite{complexity-effective} attempts to
achieve the performance of FRFCFS using a First Come First Served
scheduler in GPU systems, by preventing row-buffer locality from
being destroyed when data is transmitted over the on-chip network.
This proposal is complementary to our proposals and can be
combined with our techniques that prevent threads from hogging the
row-buffer and banks. Ipek et al.~\cite{rlmc}
propose a memory controller design that employs machine learning
techniques (reinforcement learning) to maximize DRAM throughput.
While such a policy could learn applications' memory access
characteristics over time and appropriately optimize its
scheduling policy to improve performance, implementing machine
learning techniques in the memory controller hardware could
increase complexity.

Several previous works have tackled the problem of scheduling
write back requests to memory. Stuecheli et al.~\cite{vwq} and Lee
et al.~\cite{llcwb} propose to schedule write backs such that
requests to the same row are scheduled together to exploit
row-buffer locality. Seshadri et al.~\cite{dbi} exploit their
proposed dirty-block index structure to identify dirty cache
blocks from the same row, enabling a simpler implementation of
row-locality-aware write back.  Zhao et al.~\cite{firm} propose
request scheduling mechanisms to tackle the problem of heavy write
traffic in persistent memory systems.  Our techniques can be
combined with these different write handling mechanisms to achieve
better fairness and performance.  

Previous works have also tackled the problem of memory management
and request scheduling in the presence of prefetch requests. Lee
et al.~\cite{changjoo-micro08} propose to dynamically
prioritize/deprioritize prefetch requests based on prefetcher
accuracy. Lee et al.~\cite{changjoo-micro09} also propose to
schedule requests accordingly to take advantage of the
memory-level parallelism in the system, in the presence of
prefetch requests. Ebrahimi et al.~\cite{eiman-isca11} propose to
incorporate prefetcher awareness, based on monitoring prefetcher
accuracy, into previously proposed fair memory schedulers such as
PARBS. These mechanisms can be combined with our proposals such as
BLISS and the memory bandwidth allocation policies we build on top
of MISE and ASM, to incorporate prefetch-awareness. 

\section{Other Related Work on Memory Interference Mitigation}
\label{sec:other-interf-mitigation}

While memory scheduling is a major solution direction towards
mitigating interference, previous works have also explored other
approaches such as address interleaving~\cite{mop}, memory
bank/channel
partitioning~\cite{mcp,bank-part,pact-bank-part,bank-part-hpca14,hyoseung-rtas14},
source
throttling~\cite{fst,selftuned,baydal05,hat,nychis,cc-hotnets10,kayiran-micro14}
and thread
scheduling~\cite{zhuravlev-thread-scheduling,tang-thread-scheduling,a2c,adrm}
to mitigate interference.

\noindent\textbf{Subrow Interleaving:} Kaseridis et al.~\cite{mop}
propose minimalist open page, a data mapping policy that
interleaves data at the granularity of a sub-row across channels
and banks such that applications with high row-buffer locality are
prevented from hogging the row buffer, while still preserving some
amount of row-buffer-locality. \\
%%% ONUR-8-24: Please keep the tense consistent. Present instead of present perfect. 
\noindent\textbf{Memory Channel/Bank Partitioning:} Previous
works~\cite{mcp,bank-part,pact-bank-part,bank-part-hpca14,hyoseung-rtas14}
propose techniques to mitigate inter-application interference by
partitioning channels/banks among applications such that the data
of interfering applications are mapped to different
channels/banks.\\ 
\noindent\textbf{Source Throttling:} Source throttling techniques
(e.g.,~\cite{fst,selftuned,baydal05,hat,nychis,cc-hotnets10,kayiran-micro14,rachata-sbacpad14})
propose to throttle the memory request injection rates of
interference-causing applications at the processor core itself
rather than regulating an application's access behavior at the
memory, unlike memory scheduling, partitioning or interleaving. Other previous work by Ebrahimi et
al.~\cite{eiman-isca11} proposes to tune shared resource
management policies such as FST~\cite{fst} to be aware of prefetch requests.\\
\noindent\textbf{OS Thread Scheduling:} Previous
works~\cite{zhuravlev-thread-scheduling,tang-thread-scheduling,adrm}
propose to mitigate shared resource contention by co-scheduling
threads that interact well and interfere less at the shared
resources. Such a solution relies on the presence of enough
threads with such symbiotic properties. Other
techniques~\cite{a2c} propose to map applications to cores to
mitigate memory interference.

Our proposals to mitigate memory interference, with the goals of
providing high performance and fairness, can be combined with
these solution approaches in a synergistic manner to achieve
better mitigation and consequently, higher performance and
fairness. 

%Furthermore, such thread scheduling policies and
%\bliss can be combined in a synergistic manner to further improve
%system performance and fairness. Other techniques to map
%applications to cores to mitigate memory interference, such
%as~\cite{a2c}, can be combined with BLISS.
%
%Prior work has also examined other solution
%approaches to mitigate memory interference, besides memory request
%scheduling. Ebrahimi et al. propose Fairness via Source
%Throttling (FST)~\cite{fst}. FST estimates the slowdown experienced by
%applications due to interference at the shared resources and if
%the slowdown exceeds a certain bound, throttles back the
%most-interference-causing applications. This approach also
%requires hardware support. Memory channel/bank partitioning
%proposals~\cite{bank-part,mcp} place the data of interfering
%applications on different memory partitions, in order to mitigate
%interference and improve performance. While this approach does not
%require hardware support, it causes high unfairness. Kaseridis et
%al. propose a memory interleaving scheme~\cite{mop} that
%interleaves data across different channels and banks such that
%inter-application interference is mitigated. As~\cite{mop}
%demonstrates, interleaving by itself does not achieve very high
%performance gains. However, it can be combined with memory request
%scheduling to achieve effective interference mitigation and
%high performance.

\section{Related Work on DRAM Optimizations to Improve
Performance} Several prior works have proposed optimizations to
DRAM (internals) to enable more parallelism within DRAM, thereby
improving performance. Kim et al.~\cite{salp} propose techniques
to enable access to multiple DRAM sub-arrays in parallel, thereby
overlapping the latencies of these parallel accesses. Lee et al.
in~\cite{tldram} observe that long bitlines contribute to high
access latencies and propose to split bitlines into two shorter
segments (using an isolation transistor), enabling faster access
to one of the shorter segments. More recently, Lee at
al.~\cite{aldram} propose to relax DRAM timing constraints in
order to optimize for performance in the common case. Multiple
previous works~\cite{mini-rank,rank-subsetting,mc-dimm} have
proposed to partition a DRAM rank, enabling parallel access to
these partitioned ranks. These techniques are complementary to
memory interference mitigation techniques and can be combined with
them to achieve high performance benefits.

\section{Related Work on Shared Cache Capacity Management}
The management of shared cache capacity among multiple contending
applications is a much explored area. A large body of previous
research has focused on improving the shared cache replacement
policy~\cite{iic,rrip,reuseprediction,zcache}. These proposals use
different techniques to predict which cache blocks would have high
reuse and try to retain such blocks in the cache. Furthermore,
some of these proposals also attempt to retain at least part of
the working set in the cache when an application's working set is
much larger than the cache size. A number of cache insertion
policies have also been studied by previous
proposals~\cite{rtb,eaf,suprediction,ship,tadip}. These policies use
information such as the memory region of an accessed address,
instruction pointer to predict the reuse behavior of a missed
cache block and insert blocks with higher reuse closer to the most
recently used position such that these blocks are not
evicted immediately. Other previous
works~\cite{ucp,pcasa,dynamicpartitioningofcache,cooperativecache,cqos,faircachesharing}
propose to partition the cache between applications such that
applications that have better utility for the cache are allocated
more cache space. While these previous proposals aim to improve
system performance, they are not designed with the objective of
providing controllable performance.

\section{Related Work on Coordinated Cache and Memory Management}

While several previous works have proposed techniques to manage
the shared cache capacity and main memory bandwidth independently,
there have been few previous works that have coordinated the
management of these resources. Bitirgen et
al.~\cite{bitirgen-micro08} propose a coordinated resource
management scheme that employs machine learning, specifically, an
artificial neural network, to predict each application's
performance for different possible resource allocations. Resources
are then allocated appropriately to different applications such
that a global system performance metric is optimized. More
recently, Wang et al.~\cite{xchange-hpca15} employ a
market-dynamics-inspired mechanism to coordinate allocation
decisions across resources. We take a different and more general
approach and propose a model that accurately estimates application
slowdowns. Our model can be used as an effective substrate to
build coordinated resource allocation policies that leverage our
slowdown estimates to achieve different goals such as high
performance, fairness and controllable performance.
 
\section{Related Work on Cache and Memory QoS}

Several prior works have attempted to provide QoS guarantees in
shared memory multicore systems. Previous works have proposed
techniques to estimate applications' sensitivity to
interference/propensity to cause interference by profiling
applications offline~(e.g.,
\cite{bubbleup,eklov-ispass2014,cache-pirate,bandwidth-bandit}).
However, in several scenarios, such offline profiling of
applications might not be feasible or accurate. For instance, in a
cloud service, where any user can run a job using the available
resources in a pay-as-you-go manner, profiling every application
offline to gain a priori application knowledge can be prohibitive.
In other cases, where the resource usage of an application is
heavily input set dependent, the profile may not be
representative. Mars et al.~\cite{bubbleflux} also attempt to
estimate applications' sensitivity to/propensity to cause
interference online. However, they assume that applications run by
themselves at different points in time, allowing for such
profiling, which might not necessarily be true for all
applications and systems. Our techniques, on the other hand,
strive to control and bound application slowdowns without relying on any
offline profiling and are therefore more generally applicable to
different systems and scenarios. 

Iyer et al.~\cite{ratebasedqos,cqos,qos-sigmetrics}, Guo et
al.~\cite{solihin-qos} propose mechanisms to provide guarantees
on shared cache space, memory bandwidth or IPC for different
applications. Kasture and Sanchez~\cite{ubik} propose to partition
shared caches with the goal of reducing the tail latency of
latency critical workloads. Nesbit et al.~\cite{fqm} propose a
mechanism to enforce a memory bandwidth allocation policy --
partition the available memory bandwidth across concurrently
running applications based on a given bandwidth allocation. Most
of these policies aim to provide guarantees on
resource allocation. Our goal, on the other hand, is to provide
soft guarantees on application slowdowns.

\section{Related Work on Storage QoS}
\label{sec:storage-qos}

A large body of previous work has tackled the challenge of
providing QoS in the presence of contention between different
applications for storage bandwidth. Several systems employ
bandwidth-based
throttling~(e.g.,~\cite{disk-qos-bandwidth-1,disk-qos-bandwidth-2,disk-qos-bandwidth-3,disk-qos-bandwidth-4})
to ensure that some applications do not hog storage bandwidth, at
the cost of degrading other applications' performance. One such
system, YFQ~\cite{disk-qos-bandwidth-1} controls the
proportions of bandwidth different applications receive by
assigning priority. Other systems such as
SLEDS~\cite{disk-qos-bandwidth-2} and
Zygaria~\cite{disk-qos-bandwidth-3} employ a leaky bucket type
model that controls the bandwidth of each workload, while
provisioning for some burstiness. 

Other systems employ deadline-based throttling
(e.g.,~\cite{disk-qos-deadline-1,disk-qos-deadline-2,disk-qos-deadline-3})
that attempts to provide latency guarantees for each request.
RT-FS~\cite{disk-qos-deadline-1} uses the notion of slack to
provide more resources to other applications.
Cello~\cite{disk-qos-deadline-2} deals with two kinds of
requests, ones that need to meet real-time latency requirements
and others that do not need to meet such requirements. Cello tries
to balance the needs of these two kinds of requests.
Facade~\cite{disk-qos-deadline-3} tailors its latency
guarantees depending on an application's demand in terms of number
of requests. More recent work such as Argon~\cite{argon} takes into account
that the system could be oversubscribed and determines feasibility
of meeting utilization requirements and then seeks to provide
guarantees in terms of utilization.

While all these previous works are effective in providing
different kinds of QoS at the storage, they do not take into
account main memory bandwidth and shared cache capacity
contention, which is the focus of our work.

\section{Related Work on Interconnect QoS}

Several previous works have tackled the problem of achieving QoS
in the context of both off-chip and on-chip networks. Fair
queueing~\cite{fair-queueing} emulates round-robin service order
among different flows. Virtual clock~\cite{virtual-clock} provides
a deadline-based scheme that effectively time-division multiplexes
slots among different flows. While these approaches are
rate-based, other previous works are frame-based. Time is divided
into epochs or frames and different flows reserve slots within a
frame. Some examples of frame-based policies are rotated combined
queueing~\cite{rcq} and globally synchronized frames~\cite{gsf}.
Other previous work~\cite{zhang-qos} proposes simple bandwidth allocation
schemes that reduce the complexity of allocation in the
intermediate router nodes.

Grot et al.~\cite{pvc} propose the preemptive virtual clock
mechanism that enables reclamation of idle resources, without
adding significant buffer overhead. This mechanism preempts
low-priority requests in order to provide better QoS to higher
priority requests. Grot et al. also propose
Kilo-NOC~\cite{kilonoc}, an NoC architecture designed to be
scalable to large systems. This proposal reduces the amount of
hardware changes required at every node, achieving low router
complexity. Das et al. in~\cite{stc} propose to employ stall time
criticality information to distinguish between and prioritize
different applications' packets at routers. Das et al. also
propose Aergia~\cite{aergia} to further distinguish between
packets of the same application, based on slack.

Our work on cache and memory QoS can be combined with these
previous works on interconnect QoS to achieve comprehensive and
effective QoS at the system level.

\section{Related Work on Online Slowdown Estimation}
\label{sec:related-work-slowdown-estimation}

Eyerman and Eeckhout~\cite{eeckhout-asplos2009} and Cazorla et
al.~\cite{pred-perf-smt} propose mechanisms to determine an
application's slowdown while it is running alongside other
applications on an SMT processor. Luque et al.~\cite{pred-perf-cache}
estimate application slowdowns in the presence of shared cache
interference. Both these studies assume a fixed latency for accessing
main memory, and hence do not take into account interference at the
main memory.

While a large body of previous work has focused on main memory and
shared cache interference reduction techniques, few previous works
have proposed techniques to estimate application slowdowns in the
presence of main memory and cache interference.

Li et al~\cite{yang-arxiv} propose a scheme to estimate the impact
of memory stall times on performance, for different applications,
in the context of hybrid memory system with DRAM and phase change
memory (PCM). The goal of this work is to leverage this
performance estimation scheme to map pages appropriately to DRAM
and PCM with the goal of improving performance. Hence, this scheme
does not focus much on very accurate performance estimation.

Stall Time Fair Memory Scheduling (STFM)~\cite{stfm} is one
previous work that attempts to estimate each application's
slowdown induced by memory interference, with the goal of
improving fairness by prioritizing the most slowed down
application. STFM estimates an application's slowdown as the ratio
of its memory stall time when it is run alone versus when it is
concurrently run alongside other applications. 

\begin{sloppypar}
Fairness via Source Throttling (FST)~\cite{fst} and Per-thread
cycle accounting (PTCA)~\cite{ptca} estimate application
slowdowns due to both shared cache capacity and main memory
bandwidth interference. They compute slowdown as the ratio of
alone and shared execution times and estimate alone execution time
by determining the number of cycles by which each request is
delayed.  Both FST and PTCA use a mechanism similar to STFM to
quantify interference at the main memory. To quantify interference
at the shared cache, both mechanisms determine which accesses of
an application miss in the shared cache but would have been hits
had the application been run alone on the system (contention misses),
and compute the number of additional cycles taken to serve
\emph{each contention miss}. The main difference between FST and
PTCA is in the mechanism they use to identify a contention miss.
FST uses a \emph{pollution filter} for each application that
tracks the blocks of the application that were evicted by other
applications. Any access that misses in the cache and hits in the
pollution filter is considered a contention miss. On the other
hand, PTCA uses an \emph{auxiliary tag store} for each application
that tracks the state of the cache had the application been
running alone on the system. PTCA classifies any access that
misses in the cache and hits in the auxiliary tag store as a
contention miss.
%Fairness via Source Throttling (FST)~\cite{fst} estimates
%application slowdowns due to inter-application interference at the
%shared caches and memory, as the ratio of uninterfered to
%interfered execution times. FST estimates uninterfered execution
%time by counting the number of excess cycles due to main memory
%interference, using a similar mechanism as STFM and excess cycles
%due to cache capacity interference using a pollution filter.
%Per-thread cycle accounting~\cite{cycle-accounting-taco} proposes
%a similar mechanism as FST, in order to determine an application's
%standalone execution time when it shares caches and memory with
%other applications in a multicore system.
\end{sloppypar}

The challenge in all these approaches is in determining the alone
stall time or execution time of an application {\em while} the
application is actually running alongside other applications.
STFM, FST and PTCA attempt to address this challenge by counting
the number of cycles by which each individual request that stalls execution
impacts execution time. This is fundamentally difficult and
results in high inaccuracies in slowdown estimation, as we will
describe in more detail in Chapters~\ref{chap:mise}
and~\ref{chap:asm}.

%% file: chapters/blacklisting.tex
\newcommand{\bliss}{BLISS\xspace}
\chapter{Mitigating Memory Bandwidth Interference Towards Achieving High Performance}
\label{chap:blacklisting}
\begin{sloppypar}
The prevalent solution direction to tackle the problem of memory
bandwidth interference is application-aware memory request
scheduling, as we describe in Chapter~\ref{chap:background}.
State-of-the-art application-aware memory schedulers attempt to
achieve two main goals - high system performance and high
fairness. However, previous schedulers have two major
shortcomings. First, these schedulers increase hardware complexity
in order to achieve high system performance and fairness.
Specifically, most of these schedulers rank individual
applications with a total order, based on their memory access
characteristics (e.g.,~\cite{parbs,podc-08,atlas,tcm}). Scheduling
requests based on a total rank order incurs high hardware
complexity, slowing down the memory scheduler significantly. For
instance, the critical path latency for TCM increases by 8x (area
increases by 1.8x) compared to an application-unaware FRFCFS
scheduler, as we demonstrate in
Section~\ref{sec:blacklisting-complexity}. Such high critical path
delays in the scheduler directly increase the time it takes to
schedule a request, potentially making the memory controller
latency a bottleneck. Second, a total-order ranking is unfair to
applications at the bottom of the ranking stack.  Even shuffling
the ranks periodically (like TCM does) does not fully mitigate the
unfairness and slowdowns experienced by an application when it is
at the bottom of the ranking stack, as we describe in more detail in
Section~\ref{sec:blacklisting-observations}.
\end{sloppypar}

\begin{figure}[h]
    \centering
    \includegraphics[scale=0.5, angle=0]{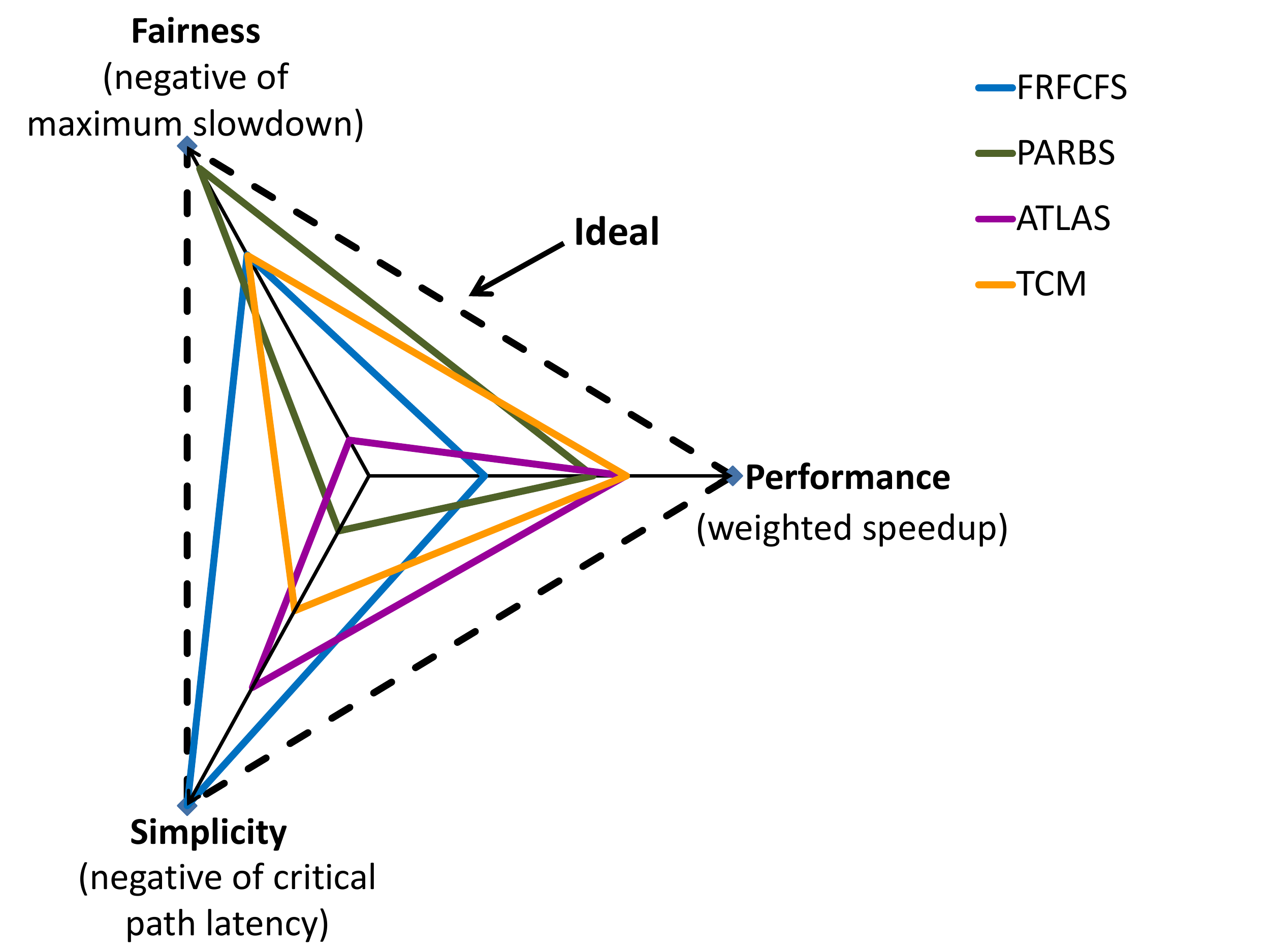}
    \caption{Performance vs. fairness vs. simplicity}
    \label{fig:perf-fairness-simplicity-initial}
\end{figure}

Figure~\ref{fig:perf-fairness-simplicity-initial} compares four
major previous schedulers using a three-dimensional plot with
performance, fairness and simplicity on three different
axes.\footnote{Results across 80 simulated workloads on a 24-core,
4-channel system. Section~\ref{sec:blacklisting-methodology}
describes our methodology and metrics.} On the fairness axis, we
plot the negative of maximum slowdown, and on the simplicity axis,
we plot the negative of critical path latency. Hence, the ideal
scheduler would have high performance, fairness and simplicity, as
indicated by the black triangle. As can be seen, previous
ranking-based schedulers, PARBS, ATLAS and TCM, increase
complexity significantly, compared to the currently employed
FRFCFS scheduler, in order to achieve high performance and/or
fairness.

\textbf{Our goal}, in this work, is to design a new memory
scheduler that does not suffer from these shortcomings: one that
achieves high system performance and fairness {\em while}
incurring low hardware cost and complexity. To this end, we seek
to overcome these shortcomings by exploring an alternative means
to protecting vulnerable applications from interference and
propose the {\em Blacklisting memory scheduler} ({\em \bliss}).

%To this end, we %propose the {\em Blacklisting memory scheduler} ({\em \bliss})
%based on two new observations described in the next section.

\input{blacklisting/observations}

\input{blacklisting/mechanism}
\input{blacklisting/implementation}
\input{blacklisting/methodology}

\input{blacklisting/evaluation}

\input{blacklisting/summary}

%% file: blacklisting/observations.tex
\section{Key Observations}
\label{sec:blacklisting-observations}

We build our Blacklisting memory scheduler (BLISS) based on two
key observations. 

%\begin{sloppypar}
%As we described in the previous section, several major
%state-of-the-art memory schedulers rank individual applications
%with a total order, to mitigate inter-application interference.
%While such ranking is one way to mitigate interference, it has
%shortcomings, as described in
%Section~\ref{sec:background-shortcomings}. We seek to overcome
%these shortcomings by exploring an alternative means to protecting
%vulnerable applications from interference. We make two key
%observations on which we build our new memory scheduling
%mechanism.
%\end{sloppypar}
\begin{sloppypar}
\textbf{Observation 1. }{\em Separating applications into only
two groups (interference-causing and vulnerable-to-interference),
without ranking individual applications using a total order, is sufficient to mitigate
inter-application interference. This leads to higher performance,
fairness and lower complexity, all at the same time.}
\end{sloppypar}

\begin{figure*} [ht!]
  \centering 
  \begin{subfigure}{0.31\textwidth} 
    \centering
    \includegraphics[scale=0.2, angle=270]{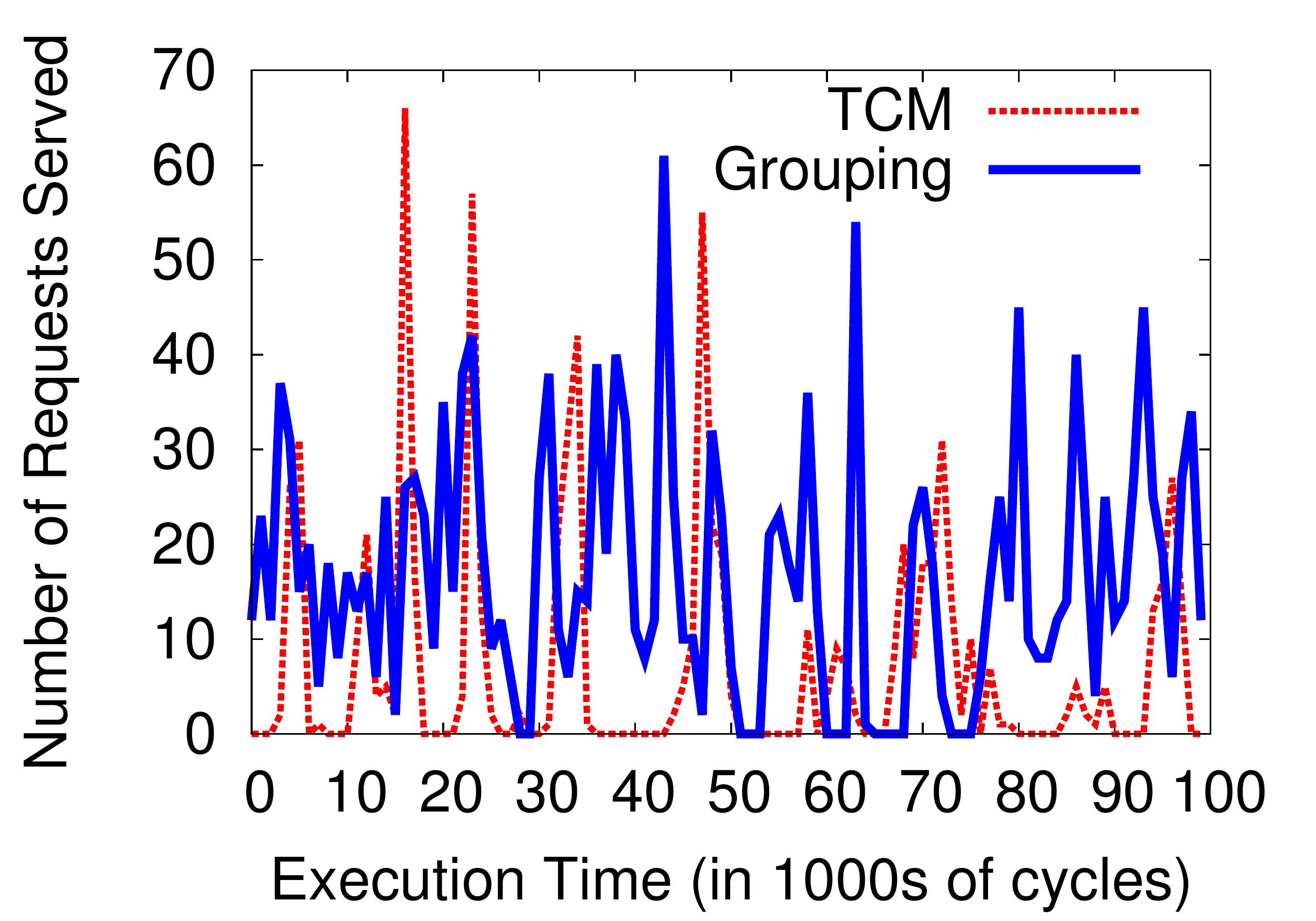}
    \caption{astar}
  \end{subfigure} 
  \begin{subfigure}{0.31\textwidth}
    \centering 
    \includegraphics[scale=0.2, angle=270]{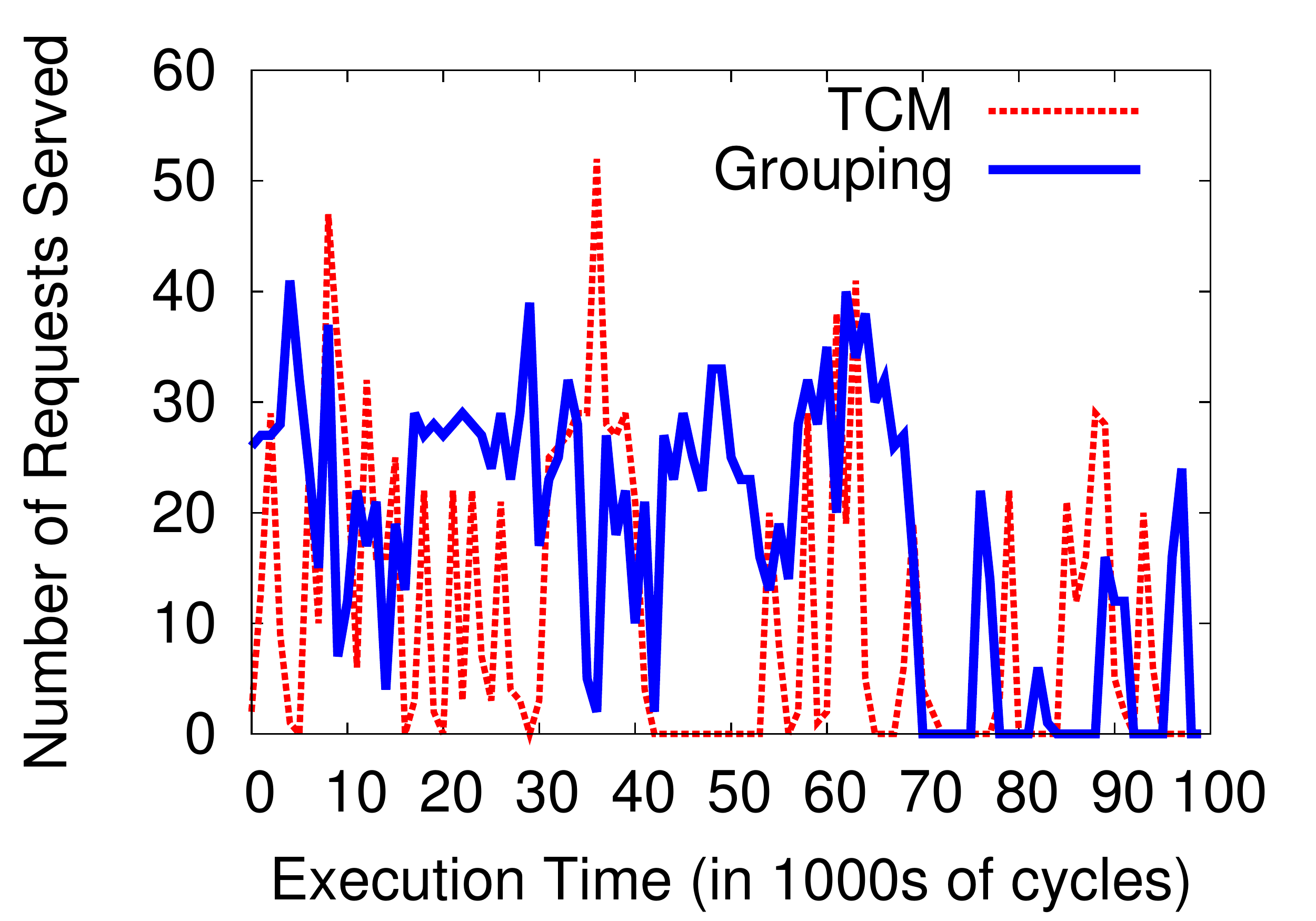}
    \caption{hmmer}
  \end{subfigure} 
  \begin{subfigure}{0.31\textwidth}
    \centering 
    \includegraphics[scale=0.2, angle=270]{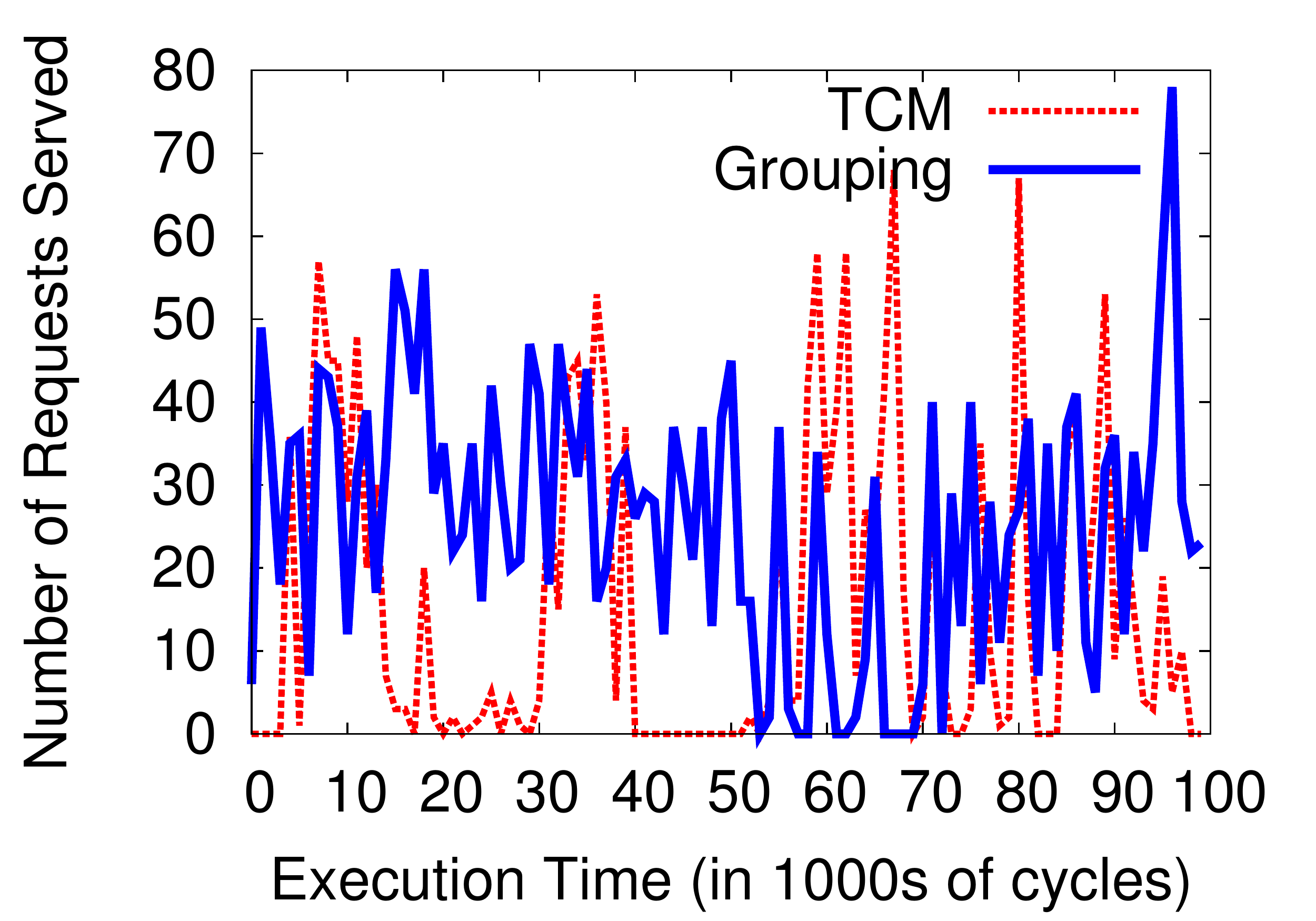}
    \caption{lbm}
  \end{subfigure} 
  \caption{Request service distribution over time with TCM and Grouping schedulers}
  \label{fig:request-dist}
\end{figure*}

\begin{sloppypar}
We observe that applications that are vulnerable to interference
can be protected from interference-causing applications by simply
separating them into two groups, one containing
interference-causing applications and another containing
vulnerable-to-interference applications, rather than ranking
individual applications with a total order as many
state-of-the-art schedulers do. To motivate this, we contrast
TCM~\cite{tcm}, which clusters applications into two groups and
employs a total rank order within each cluster, with a simple
scheduling mechanism ({\em Grouping}) that simply groups
applications only into two groups, based on memory intensity (as
TCM does), and prioritizes the low-intensity group {\em without}
employing ranking in each group. {\em Grouping} uses the FRFCFS
policy within each group. Figure~\ref{fig:request-dist} shows the
number of requests served during a 100,000 cycle period at
intervals of 1,000 cycles, for three representative applications,
astar, hmmer and lbm from the SPEC CPU2006 benchmark
suite~\cite{spec2006}, using these two schedulers.\footnote{All
these three applications are in the high-memory-intensity group.
We found very similar behavior in all other such applications we
examined.} These three applications are executed with other
applications in a simulated 24-core 4-channel system.\footnote{See
Section~\ref{sec:blacklisting-methodology} for our methodology.}
\end{sloppypar}

Figure~\ref{fig:request-dist} shows that TCM has high variance in
the number of requests served across time, with very few requests
being served during several intervals and many requests being
served during a few intervals. This behavior is seen in most
applications in the high-memory-intensity cluster since TCM ranks
individual applications with a total order. This ranking causes
some high-memory-intensity applications' requests to be
prioritized over \emph{other} high-memory-intensity applications'
requests, at any point in time, resulting in high interference.
Although TCM periodically shuffles this total-order ranking, we observe that
an application benefits from ranking \emph{only} during those periods
when it is ranked very high. These very highly ranked periods
correspond to the spikes in the number of requests served (for
TCM) in Figure~\ref{fig:request-dist} for that application. During
the other periods of time when an application is ranked lower
(i.e., most of the {\em shuffling intervals}), only a small number
of its requests are served, resulting in very slow progress.
Therefore, most high-memory-intensity applications experience high
slowdowns due to the total-order ranking employed by TCM.

\begin{figure*} [ht!]
  \centering 
  \begin{minipage}{0.47\textwidth}
    \centering 
    \includegraphics[scale=0.28, angle=270]{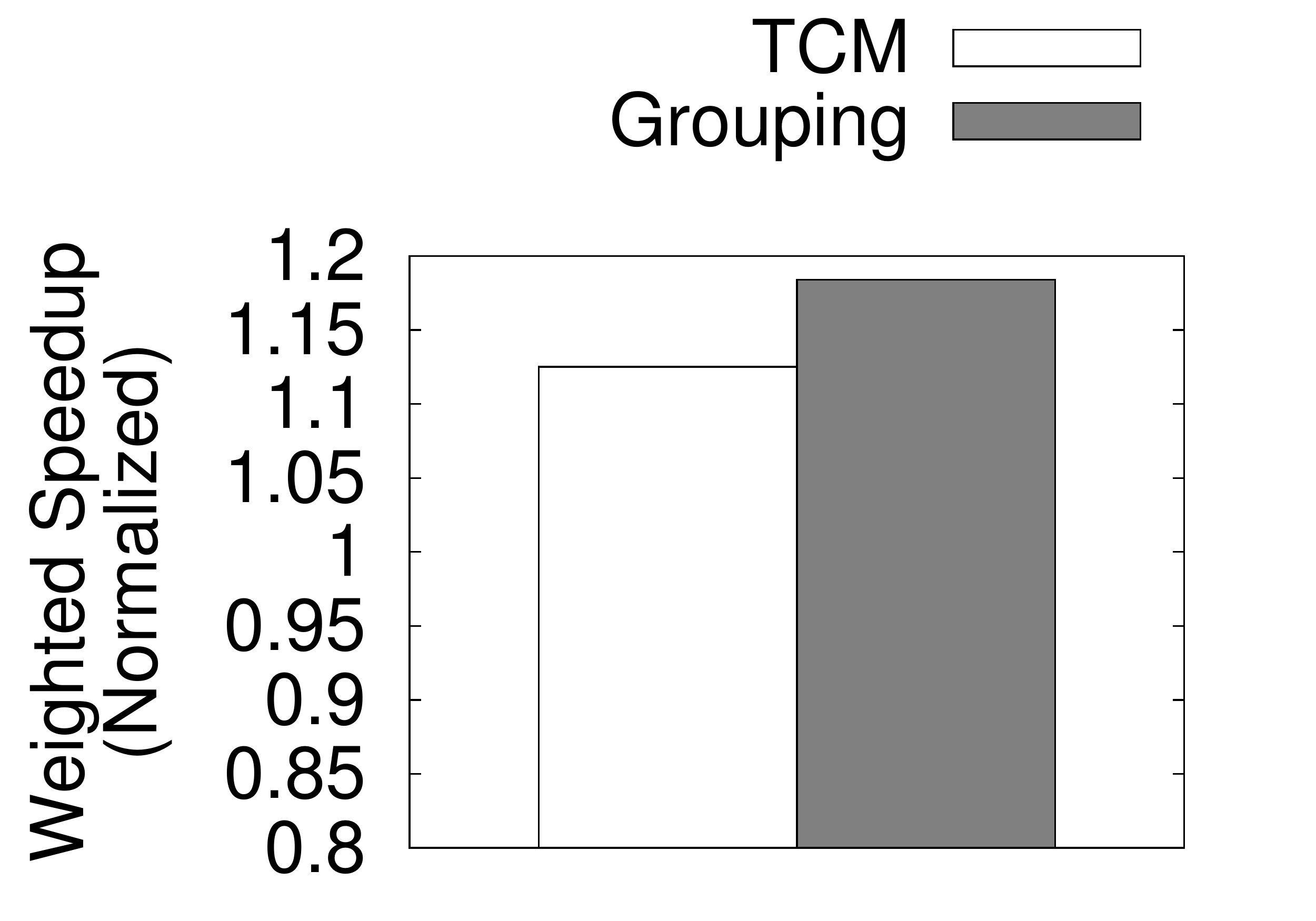}
  \end{minipage} 
%  \hspace{10pt}
  \begin{minipage}{0.47\textwidth} 
    \centering
    \includegraphics[scale=0.28, angle=270]{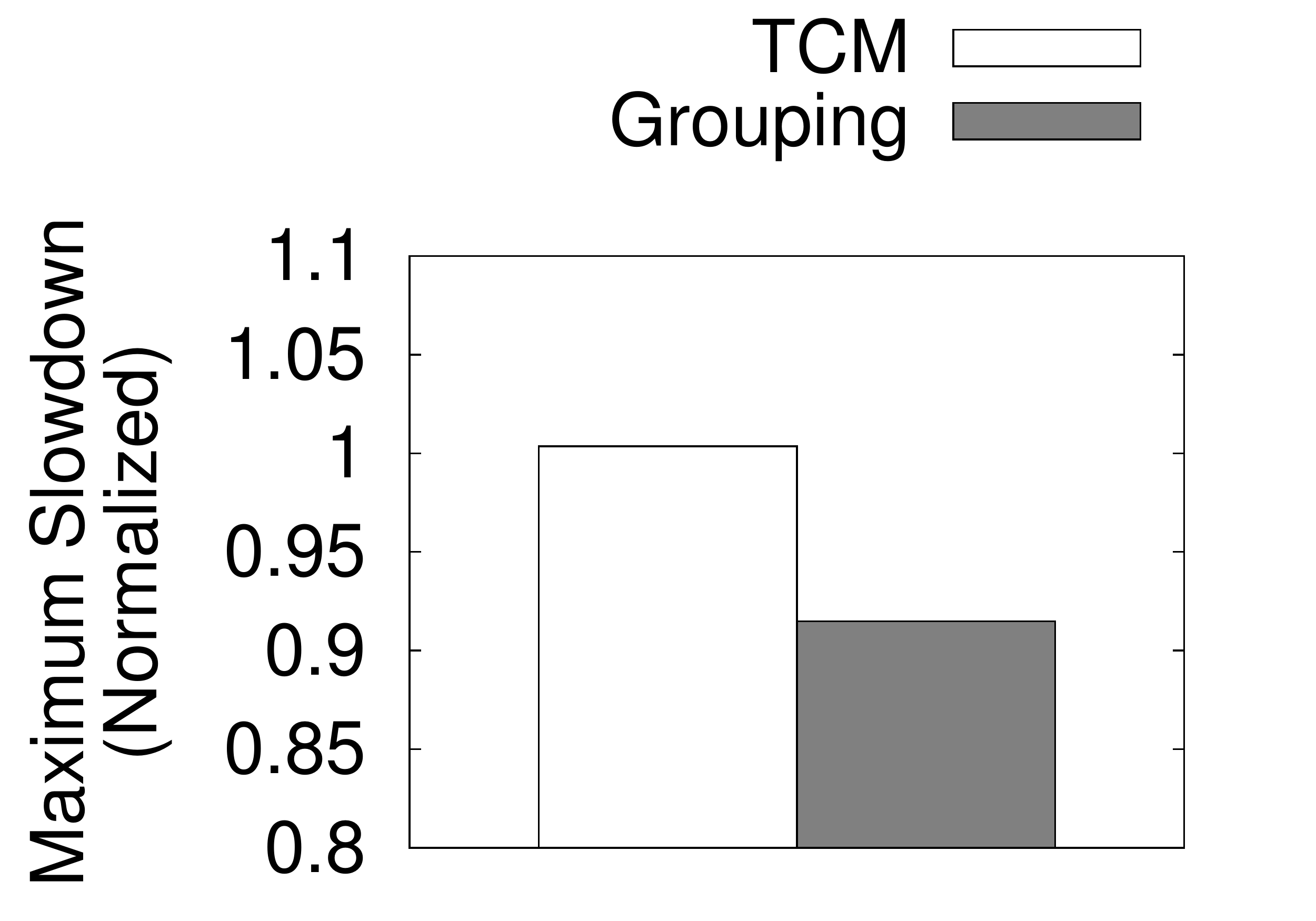}
  \end{minipage} 
  \caption{Performance and fairness of Grouping vs. TCM}
  \label{fig:tcm-clustering}
\end{figure*}

On the other hand, when applications are separated into only two
groups based on memory intensity and no per-application ranking is
employed within a group, some interference exists among
applications within each group (due to the application-unaware
FRFCFS scheduling in each group). In the high-memory-intensity
group, this interference contributes to the few
low-request-service periods seen for {\em Grouping} in
Figure~\ref{fig:request-dist}. However, the request service
behavior of {\em Grouping} is less spiky than that of TCM, resulting
in lower memory stall times and a more steady and overall higher
progress rate for high-memory-intensity applications, as compared
to when applications are ranked in a total order. In the
low-memory-intensity group, there is not much of a difference
between TCM and \emph{Grouping}, since applications anyway have low
memory intensities and hence, do not cause significant
interference to each other. Therefore, {\em Grouping} results in
higher system performance and significantly higher fairness than
TCM, as shown in Figure~\ref{fig:tcm-clustering} (across 80
24-core workloads on a simulated 4-channel system).

Grouping applications into two groups also requires much lower
hardware overhead than ranking-based schedulers that incur high
overhead for computing and enforcing a total rank order for all
applications. Therefore, grouping can not only achieve better
system performance and fairness than ranking, but it also can do
so while incurring lower hardware cost. However, classifying
applications into two groups at coarse time granularities, on the
order of a few million cycles, like TCM's clustering mechanism
does (and like what we have evaluated in
Figure~\ref{fig:tcm-clustering}), can still cause unfair
application slowdowns. This is because applications in one group
would be deprioritized for a long time interval, which is
especially dangerous if application behavior changes during the
interval. Our second observation, which we describe next,
minimizes such unfairness and at the same time reduces the
complexity of grouping even further.

\textbf{Observation 2. }{\em Applications can be classified into
\emph{interference-causing} and \emph{vulnerable-to-interference}
groups by monitoring the number of consecutive requests served
from each application at the memory controller. This leads to
higher fairness and lower complexity, at the same time, than
grouping schemes that rely on coarse-grained memory intensity
measurement.}

Previous work actually attempted to perform grouping, along with ranking, to
mitigate interference. Specifically, TCM~\cite{tcm} ranks applications
by memory intensity and classifies applications that make up a certain
fraction of the total memory bandwidth usage into a \emph{group} called the
\emph{low-memory-intensity cluster} and the remaining applications
into a second group called the \emph{high-memory-intensity cluster}. While employing such a grouping
scheme, without ranking individual applications, reduces hardware
complexity and unfairness compared to a total order based ranking
scheme (as we show in Figure~\ref{fig:tcm-clustering}), it i) \emph
{can still cause unfair slowdowns due to classifying applications into
  groups at coarse time granularities, which is especially dangerous
  if application behavior changes during an interval}, and ii)
\emph{incurs additional hardware overhead and scheduling latency to
  compute and rank applications by long-term memory intensity and total memory
  bandwidth usage}.

We propose to perform application grouping using a significantly
simpler, novel scheme: simply by counting the number of requests served from each
application in a short time interval. Applications that have a
large number (i.e., above a threshold value) of consecutive
requests served are classified as interference-causing (this
classification is periodically reset). The rationale behind this
scheme is that when an application has a large number of
consecutive requests served within a short time period, which is
typical of applications with high memory intensity or
high row-buffer locality, it delays other applications' requests,
thereby stalling their progress. Hence, identifying and
essentially \emph{blacklisting} such interference-causing
applications by placing them in a separate group and
deprioritizing requests of this blacklisted group can prevent such
applications from hogging the memory bandwidth. As a result, the
interference experienced by vulnerable applications is mitigated.
The blacklisting classification is cleared periodically, at short
time intervals (on the order of 1000s of cycles) in order not to
deprioritize an application for too long of a time period to cause
unfairness or starvation. Such clearing and re-evaluation of application
classification at short time intervals significantly reduces unfair
application slowdowns (as we quantitatively show in
Section~\ref{sec:tcm-clustering}), while reducing complexity
compared to tracking per-application metrics such as memory
intensity.

\textbf{Summary of Key Observations.} In summary, we make two key
novel observations that lead to our design in
Section~\ref{sec:blacklisting-mechanism}. First, separating applications into
only two groups can lead to a less complex and more fair and
higher performance scheduler. Second, the two application groups
can be formed seamlessly by monitoring the number of consecutive
requests served from an application and deprioritizing the ones
that have too many requests served in a short time interval.

%% file: blacklisting/mechanism.tex
\section{Mechanism}
\label{sec:blacklisting-mechanism}

The design of our Blacklisting scheduler (\bliss) is based on the
two key observations described in the previous section. The basic
idea behind \bliss is to observe the number of consecutive requests
served from an application over a short time interval and blacklist
applications that have a relatively large number of consecutive
requests served. The blacklisted (interference-causing) and
non-blacklisted (vulnerable-to-interference) applications are thus
separated into two different groups. The memory scheduler then
prioritizes the non-blacklisted group over the blacklisted group. The
two main components of \bliss are i) the blacklisting mechanism and
ii) the memory scheduling mechanism that schedules requests based on
the blacklisting mechanism. We describe each in turn.

\subsection{The Blacklisting Mechanism}
\label{sec:blacklist-mechanism}

The blacklisting mechanism needs to keep track of three quantities:
1) the application (i.e., hardware context) ID of the last scheduled
request (\textit{Application ID})\footnote{An application here denotes
  a hardware context. There can be as many applications executing
  actively as there are hardware contexts. Multiple hardware contexts
  belonging to the same application are considered separate
  applications by our mechanism, but our mechanism can be extended to
  deal with such multithreaded applications.}, 2) the number of
requests served from an application (\textit{\#Requests Served}), and 3)
the blacklist status of each application.

When the memory controller is about to issue a request, it
compares the application ID of the request with the
\textit{Application ID} of the \emph{last scheduled request}.
\begin{itemize}
\item 
If the application IDs of the two requests are the same, the
\textit{\#Requests Served} counter is incremented.
\item
If the application IDs of the two requests are not the same, the
\textit{\#Requests Served} counter is reset to zero and the
\textit{Application ID} register is updated with the application ID
of the request that is being issued.  
\end{itemize}

If the \textit{\#Requests Served} exceeds a \textit{Blacklisting
  Threshold} (4 in most of our evaluations): 
\begin{itemize}
\item
The application with ID \textit{Application ID} is blacklisted
(classified as interference-causing).
\item
The \textit{\#Requests Served} counter is reset to zero.
\end{itemize}

The blacklist information is cleared periodically after every
\textit{Clearing Interval} (set to 10000 cycles in our major
evaluations).

\subsection{Blacklist-Based Memory Scheduling}

Once the blacklist information is computed, it is used to determine
the scheduling priority of a request. Memory requests are prioritized
in the following order:
\begin{enumerate}
\item Non-blacklisted applications' requests
\item Row-buffer hit requests
\item Older requests
\end{enumerate}
Prioritizing requests of non-blacklisted applications over
requests of blacklisted applications mitigates interference.
Row-buffer hits are then prioritized to optimize DRAM bandwidth
utilization. Finally, older requests are prioritized over younger
requests for forward progress.

%% file: blacklisting/implementation.tex
\section{Implementation}
\label{sec:blacklisting-implementation}

The Blacklisting memory scheduler requires additional storage (flip
flops) and logic over an FRFCFS scheduler to 1) perform blacklisting
and 2) prioritize non-blacklisted applications' requests. We
analyze the storage and logic cost of it. 

\subsection{Storage Cost}
In order to perform blacklisting, the memory scheduler needs the
following storage components:
\begin{itemize}
\item one register to store \textit{Application ID} (5 bits for 24 applications)
\item one counter for \textit{\#Requests Served} (8 bits is more
than sufficient for the values of request count threshold
\textrm{N} that we observe achieves high performance and fairness.)
\item one register to store the \textit{Blacklisting Threshold} that
determines when an application should be blacklisted
\item a blacklist bit vector to indicate the blacklist status of each
application (one bit for each hardware context) (24 bits for 24 applications)
\end{itemize}

%\noindent
In order to prioritize non-blacklisted applications' requests, the
memory controller needs to store the application ID (hardware context
ID) of each request so it can determine the blacklist status of the
application and appropriately schedule the request.

\subsection{Logic Cost}
The memory scheduler requires comparison logic to
\begin{itemize}
\item
determine when an application's \textit{\#Requests Served} exceeds
the \textit{Blacklisting Threshold} and set the bit corresponding to the
application in the \textit{Blacklist} bit vector.
\item
prioritize non-blacklisted applications' requests.
\end{itemize}

%\noindent
We provide a detailed quantitative evaluation of the hardware area
cost and logic latency of implementing \bliss and previously
proposed memory schedulers, in Section~\ref{sec:blacklisting-complexity}.

%% file: blacklisting/methodology.tex
\newcommand{\aloneipc}{\textrm{IPC}_{i}^{\scriptstyle{alone}}\xspace}
\newcommand{\sharedipc}{\textrm{IPC}_{i}^{\scriptstyle{shared}}\xspace}

\section{Methodology}
\label{sec:blacklisting-methodology}

\subsection{System Configuration}
\begin{sloppypar}
We model the DRAM memory system using a cycle-level in-house
DDR3-SDRAM simulator. The simulator was validated against Micron's
behavioral Verilog model~\cite{ddr3verilog} and
DRAMSim2~\cite{dramsim2}. This DDR3 simulator is integrated with a
cycle-level in-house simulator that models out-of-order execution
cores, driven by a Pin~\cite{pin} tool at the frontend, Each core
has a private cache of 512 KB size. We present most of our results
on a system with the DRAM main memory as the only shared resource
in order to isolate the effects of memory bandwidth interference
on application performance. We also present results with shared
caches in Section~\ref{sec:sensitivity-system}. Table~\ref{tab:meth}
provides more details of our simulated system. We perform most of
our studies on a system with 24 cores and 4 channels. We provide a
sensitivity analysis for a wide range of core and channel counts,
in Section~\ref{sec:sensitivity-system}. Each channel has one
rank and each rank has eight banks. We stripe data across channels
and banks at the granularity of a row.
\end{sloppypar}

\input{blacklisting/tables/methodology}

%\vspace{10mm}
\subsection{Workloads}
We perform our main studies using 24-core multiprogrammed
workloads made of applications from the SPEC CPU2006
suite~\cite{spec2006}, TPC-C, Matlab and the NAS parallel
benchmark suite~\cite{nas}.\footnote{Each benchmark is single
threaded.} We classify a benchmark as memory-intensive if it has a
Misses Per Kilo Instruction (MPKI) greater than 5 and
memory-non-intensive otherwise. We construct four categories of
workloads (with 20 workloads in each category), with 25, 50, 75
and 100 percent of memory-intensive applications. This makes up a
total of 80 workloads with a range of memory intensities,
constructed using random combinations of benchmarks, modeling a
cloud computing like scenario where workloads of various types are
consolidated on the same node to improve efficiency. We also
evaluate 16-, 32- and 64- core workloads, with different memory
intensities, created using a similar methodology as described
above for the 24-core workloads. We simulate each workload for 100
million representative cycles, as done by previous studies in
memory scheduling~\cite{parbs,atlas,tcm}.

\subsection{Metrics}
We quantitatively compare \bliss with previous memory schedulers in terms of
system performance, fairness and complexity. We use the weighted
speedup~\cite{stc,weighted-speedup,symbjobscheduling} metric to measure system
performance.  We use the maximum slowdown
metric~\cite{stc,atlas,tcm,max-slowdown} to measure unfairness. We report the
harmonic speedup metric~\cite{harmonic-speedup} as another measure of system
performance. The harmonic speedup metric also serves as a measure of balance
between system performance and fairness~\cite{harmonic-speedup}. We report area
in $micro meter^2$ ($um^2$) and scheduler critical path latency in nanoseconds
(ns) as measures of complexity.

%\vspace{-6mm}
%
%\begin{small}
%  \begin{eqnarray*}
%    \textrm{Weighted Speedup} & = & \Sigma_i \frac{\sharedipc}{\aloneipc}\\
%    \textrm{Maximum Slowdown} & = & \textrm{max}\left(\frac{\aloneipc}{\sharedipc}\right)\\
%    \textrm{Harmonic Speedup} & = & N/\left(\Sigma_i\frac{\aloneipc}{\sharedipc}\right)\\
%  \end{eqnarray*}
%\end{small}
%
%\vspace{-8mm}

%%% ONUR-8-24: You can omit the above equations, especially if you
%%% need space to explain stuff related to BLISS.

\subsection{RTL Synthesis Methodology}

In order to obtain timing/area results for \bliss and previous
schedulers, we implement them in Register Transfer Level (RTL), using
Verilog. We synthesize the RTL implementations with a commercial 32 nm
standard cell library, using the Design Compiler tool from Synopsys.

\subsection{Mechanism Parameters}

%%% ONUR-8-24: If you name the thresholds and intervals well, you do
%%% not need to redefine what they are here. Can you?
For \bliss, we use a value of four for \textit{Blacklisting Threshold}, and a
value of 10000 cycles for \textit{Clearing Interval}. These values provide a
good balance between performance and fairness, as we observe from our
sensitivity studies in Section~\ref{sec:sensitivity-algorithm}. For the other
schedulers, we tuned their parameters to achieve high performance and fairness
on our system configurations and workloads. We use a \textit{Marking-Cap} of 5 for
PARBS, cap of 4 for FRFCFS-Cap, \textit{HistoryWeight} of 0.875 for ATLAS,
\textit{ClusterThresh} of 0.2 and \textit{ShuffleInterval} of 1000 cycles for
TCM.

%%% ONUR-8-24: Please use the names provided by the respective papers,
%%% above.

%%% ONUR-8-24: I was mistaken. Ignore the below.
%%% ONUR-8-24: Figure placement. Fig 5 is not references on Page 4,
%%% but it is on top of Page 4. Makes life harder for the reader. I am
%%% assuming you have not fixed the figure placement yet, so you
%%% probably will address this anyway.

%% file: blacklisting/tables/methodology.tex
\begin{table}[h!]
\begin{scriptsize}
  \vspace{-2mm}
  \centering
    \begin{tabular}{ll}
        \toprule
Processor           &  16-64 cores, 5.3GHz, 3-wide issue,\\ & 8 MSHRs, 128-entry instruction window     \\
        \cmidrule(rl){1-2}
Last-level cache    &  64B cache-line, 16-way associative,\\ & 512KB private cache-slice per core     \\
        \cmidrule(rl){1-2} Memory controller   &  128-entry read/write request queue per controller   \\
        \cmidrule(rl){1-2}
\multirow{2}[2]{*}{\centering Memory}              &  Timing: DDR3-1066 (8-8-8)~\cite{micron} \\
 & Organization: 1-8 channels, 1
 rank-per-channel,\\ & 8 banks-per-rank, 8 KB row-buffer \\ 
        \bottomrule
    \end{tabular}%
  \vspace{-1mm}
  \caption{Configuration of the simulated system}
  \label{tab:meth}%
  \vspace{-2mm}
\end{scriptsize}%
\end{table}%

%% file: blacklisting/evaluation.tex
\section{Evaluation}
\label{sec:blacklisting-evaluation}

\begin{figure*}[ht!]
  \centering
  \begin{minipage}{0.32\textwidth}
    \centering
    \includegraphics[scale=0.21, angle=270]{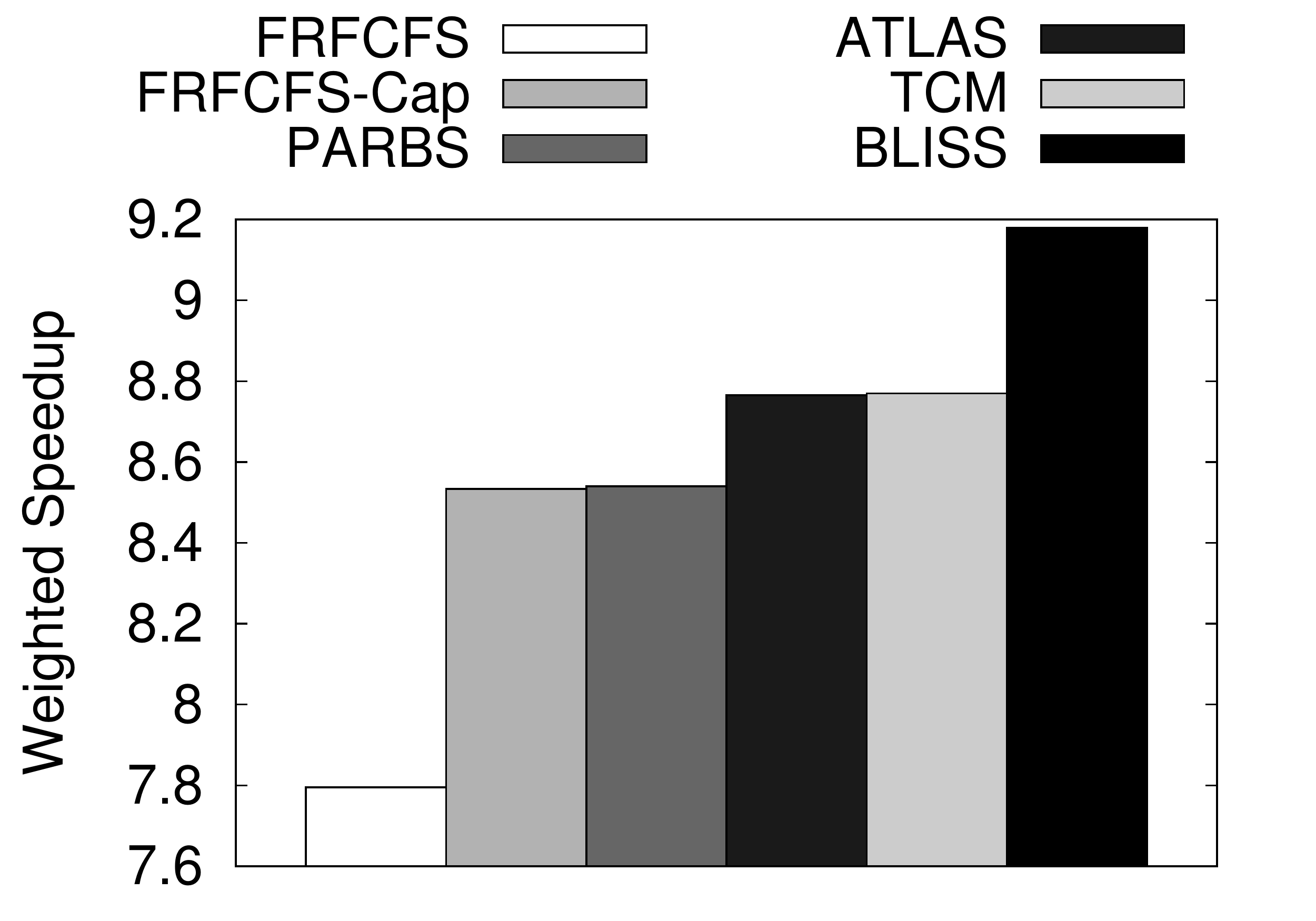}
  \end{minipage}
  \begin{minipage}{0.32\textwidth}
    \centering
    \includegraphics[scale=0.21, angle=270]{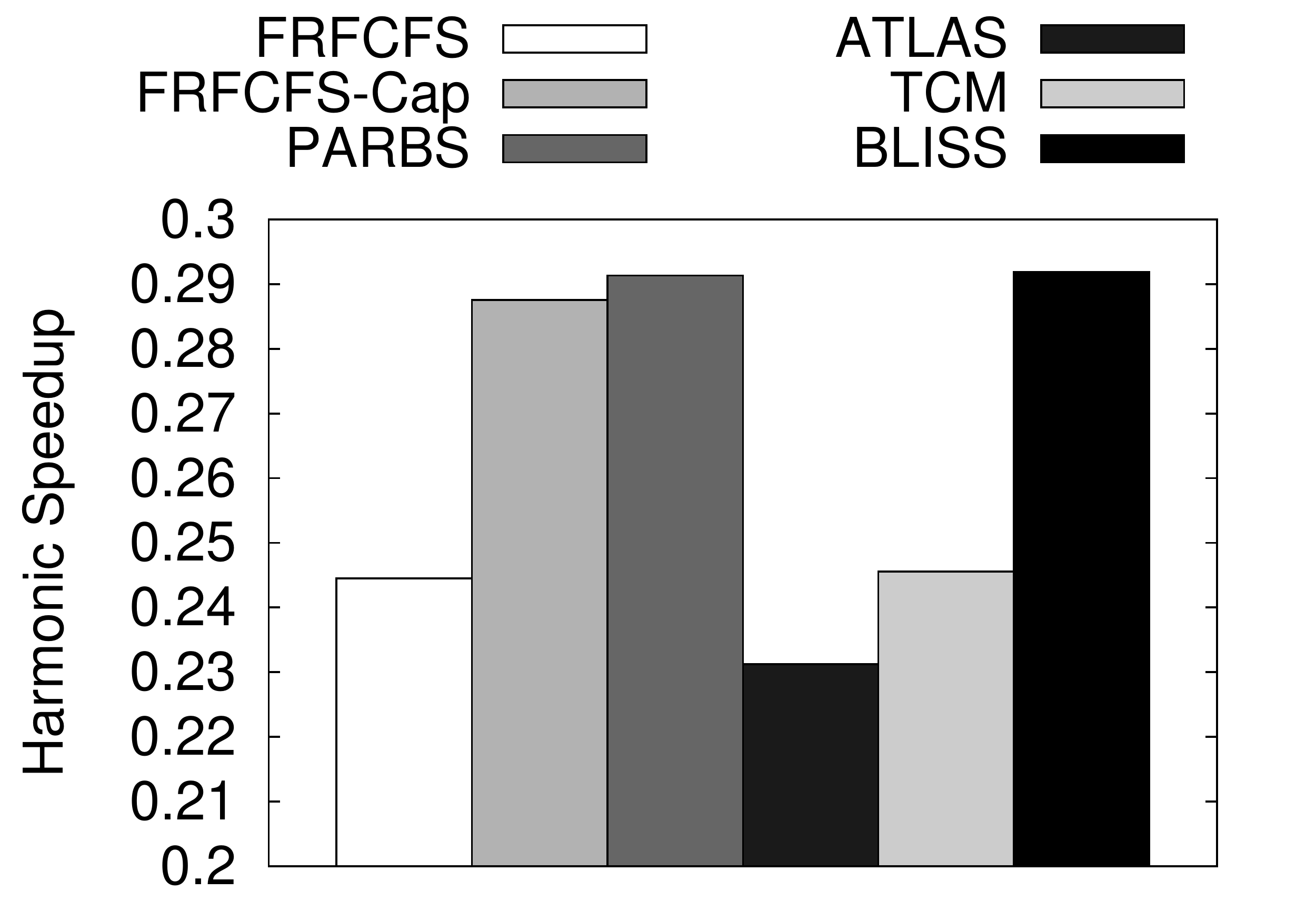}
  \end{minipage}
  \begin{minipage}{0.32\textwidth}
    \centering
    \includegraphics[scale=0.21, angle=270]{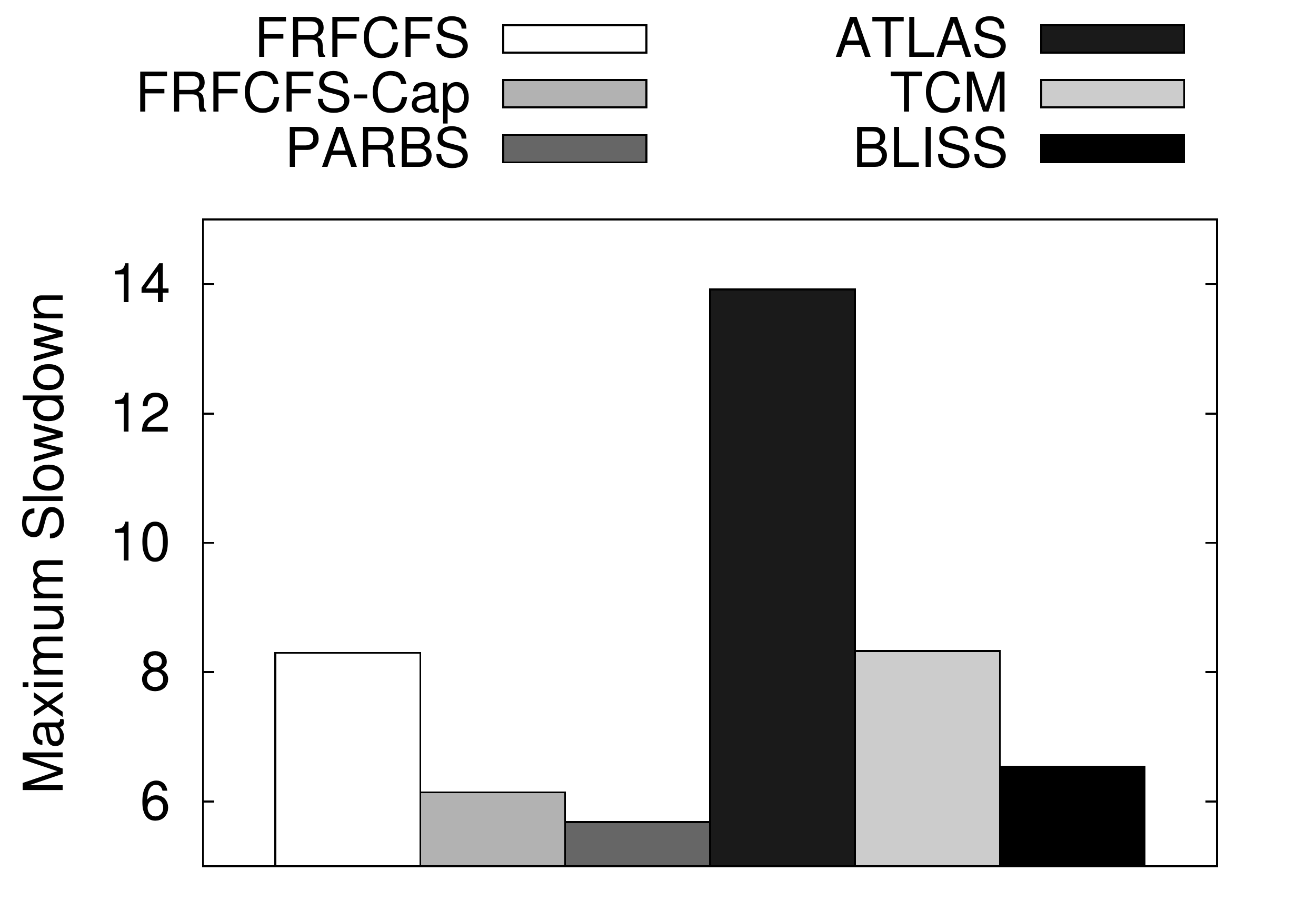}
  \end{minipage}
  \caption{System performance and fairness of \bliss compared to previous schedulers}
  \label{fig:main-results}
\end{figure*}

\begin{figure}[h]
    \centering
    \includegraphics[scale=0.3, angle=0]{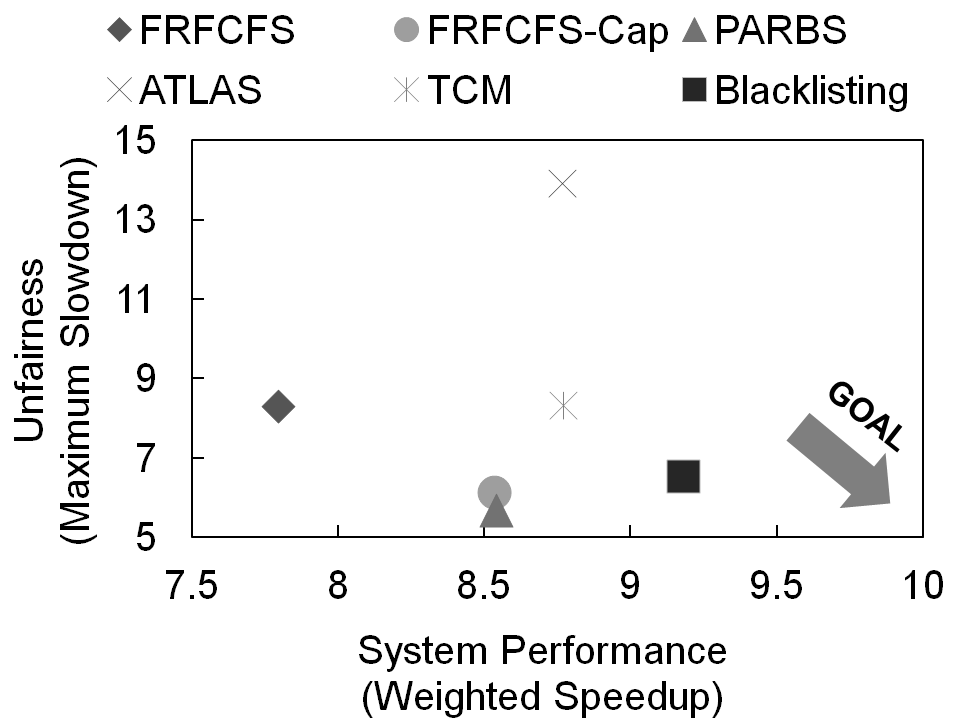}
    \caption{Pareto plot of system performance and fairness}
    \label{fig:pareto}
\end{figure}

We compare \bliss with five previously proposed memory schedulers,
FRFCFS, FRFCFS with a cap (FRFCFS-Cap)~\cite{stfm}, PARBS, ATLAS
and TCM.  FRFCFS-Cap is a modified version of FRFCFS that caps the
number of consecutive row-buffer hitting requests that can be
served from an application~\cite{stfm}.
Figure~\ref{fig:main-results} shows the average system performance
(weighted speedup and harmonic speedup) and unfairness (maximum
slowdown) across all our workloads. Figure~\ref{fig:pareto} shows
a pareto plot of weighted speedup and maximum slowdown. We draw
three major observations. First, \bliss achieves 5\% better
weighted speedup, 25\% lower maximum slowdown and 19\% better
harmonic speedup than the best performing previous scheduler
(in terms of weighted speedup), TCM, while reducing the critical path
and area by 79\% and 43\% respectively (as we will show in
Section~\ref{sec:blacklisting-complexity}). Therefore, we conclude that \bliss
achieves both high system performance and fairness, at low
hardware cost and complexity.

Second, \bliss significantly outperforms all these five previous
schedulers in terms of system performance, however, it has 10\%
higher unfairness than PARBS, the previous scheduler with the
least unfairness. PARBS creates request batches containing the
oldest requests from each application. Older batches are
prioritized over newer batches. However, within each batch,
individual applications' requests are ranked and prioritized based
on memory intensity. PARBS aims to preserve fairness by batching
older requests, while still employing ranking within a batch to
prioritize low-memory-intensity applications. We observe that the
batching aspect of PARBS is quite effective in mitigating
unfairness, although it increases complexity. This unfairness
reduction also contributes to the high harmonic speedup of PARBS.
However, batching restricts the amount of request reordering that
can be achieved through ranking. Hence, low-memory-intensity
applications that would benefit from prioritization via aggressive
request reordering have lower performance. As a result, PARBS has
8\% lower weighted speedup than \bliss. Furthermore, PARBS has a
6.5x longer critical path and \httilde2x greater area than \bliss,
as we will show in Section~\ref{sec:blacklisting-complexity}. Therefore, we
conclude that \bliss achieves better system performance than
PARBS, at much lower hardware cost, while slightly trading off
fairness.

Third, \bliss has 4\% higher unfairness than FRFCFS-Cap, but it
also 8\% higher performance than FRFCFS-Cap. FRFCFS-Cap has
higher fairness than \bliss since it restricts the length of
{\em only} the {\em ongoing} row hit streak, whereas blacklisting
an application can deprioritize the application {\em for a longer
time}, until the next clearing interval. As a result, FRFCFS-Cap
slows down high-row-buffer-locality applications to a lower degree
than \bliss. However, restricting \emph{only} the on-going streak
rather than blacklisting an interfering application for a longer
time causes more interference to other applications, degrading
system performance compared to \bliss. Furthermore, FRFCFS-Cap is
unable to mitigate interference due to applications with high
memory intensity yet low-row-buffer-locality, whereas \bliss is
effective in mitigating interference due to such applications as
well. Hence, we conclude that \bliss achieves higher performance
(weighted speedup) than FRFCFS-Cap, while slightly trading off
fairness.

\subsection{Analysis of Individual Workloads}
\label{sec:scurve-analysis}

\begin{figure*}[t!]
  \centering
  \begin{minipage}{0.48\textwidth}
    \centering
    \includegraphics[scale=0.3, angle=270]{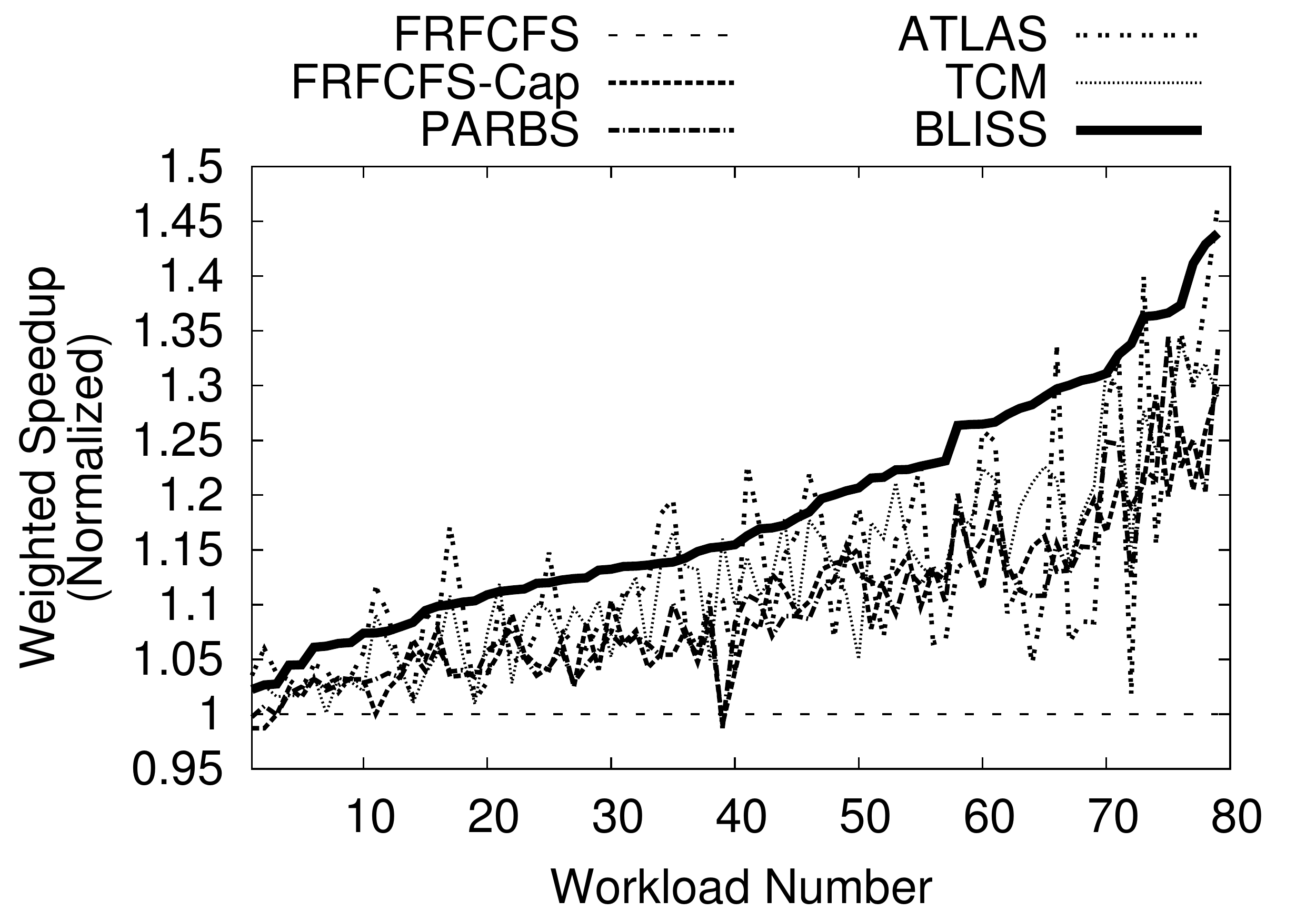}
  \end{minipage}
  \begin{minipage}{0.48\textwidth}
    \centering
    \includegraphics[scale=0.3, angle=270]{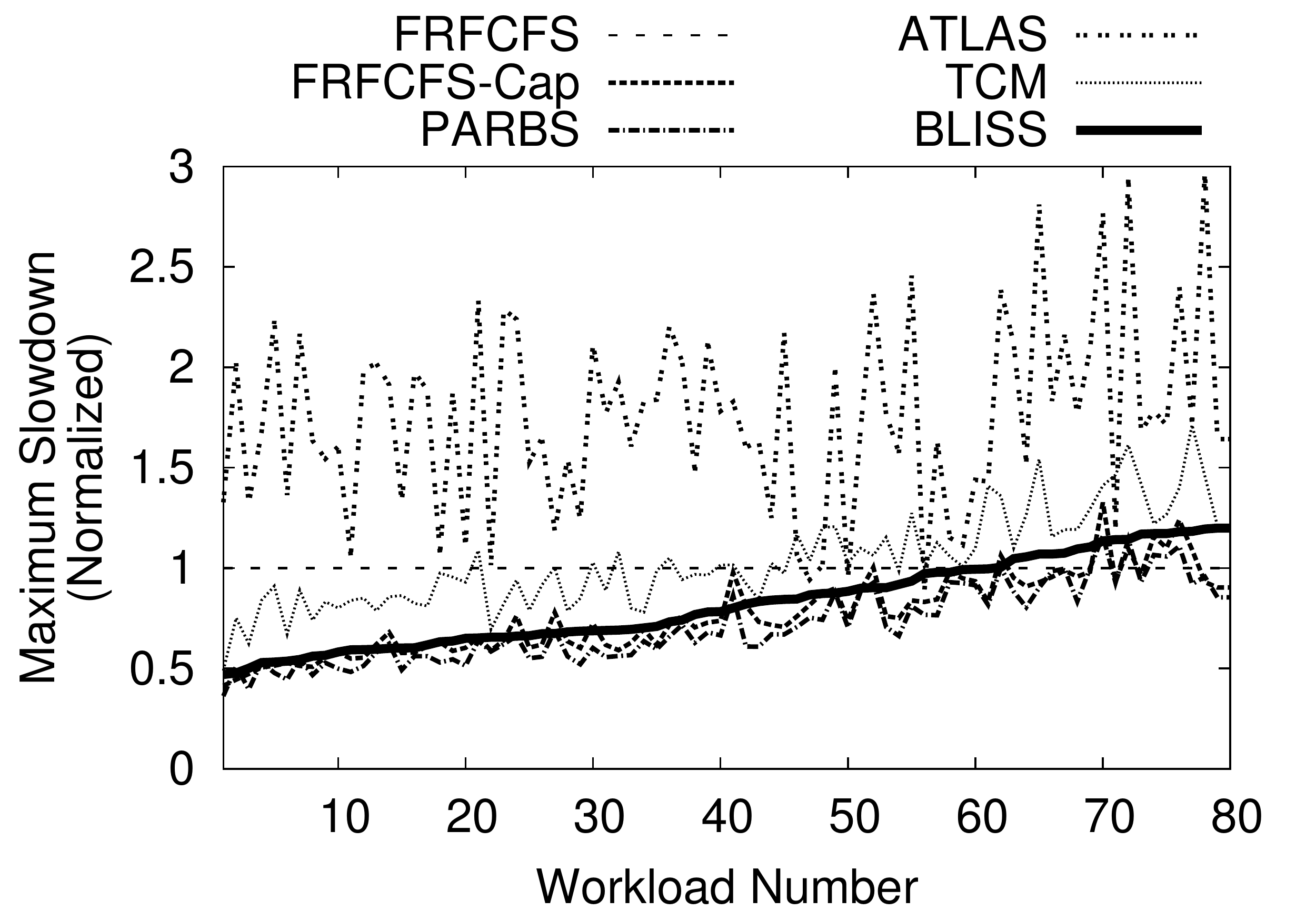}
  \end{minipage}
  \caption{System performance and fairness for all workloads}
  \label{fig:scurve}
\end{figure*}

In this section, we analyze the performance and fairness for
individual workloads, when employing different schedulers.
Figure~\ref{fig:scurve} shows the performance and fairness
normalized to the baseline FRFCFS scheduler for all our 80
workloads, for BLISS and previous schedulers, in the form of
S-curves~\cite{eaf}. The workloads are sorted based on the
performance improvement of BLISS. We draw three major
observations. First, BLISS achieves the best performance among all
previous schedulers for most of our workloads. For a few
workloads, ATLAS achieves higher performance, by virtue of always
prioritizing applications that receive low memory service.
However, always prioritizing applications that receive low memory
service can unfairly slow down applications with high memory
intensities, thereby degrading fairness significantly (as shown in
the maximum slowdown plot, Figure~\ref{fig:scurve} bottom).
Second, BLISS achieves significantly higher fairness than ATLAS
and TCM, the best-performing previous schedulers, while also
achieving higher performance than them and approaches the fairness
of the fairest previous schedulers, PARBS and FRFCFS-Cap. As
described in the analysis of average performance and fairness
results above, PARBS, by virtue of request batching and
FRFCFS-Cap, by virtue of restricting only the current row hit
streak achieve higher fairness (lower maximum slowdown) than BLISS
for a number of workloads. However, these schedulers achieve
higher fairness at the cost of lower system performance, as shown
in Figure~\ref{fig:scurve}. Third, for some workloads with very
high memory intensities, the default FRFCFS scheduler achieves the
best fairness. This is because memory bandwidth becomes a very
scarce resource when the memory intensity of a workload is very
high.  Hence, prioritizing row hits utilizes memory bandwidth
efficiently for such workloads, thereby resulting in higher
fairness. Based on these observations, we conclude that BLISS
achieves the best performance and a good trade-off between
fairness and performance for most of the workloads we examine.

\subsection{Hardware Complexity}
\label{sec:blacklisting-complexity}

\begin{figure}[ht]
  \centering
  \begin{minipage}{0.48\textwidth}
     \centering
     \includegraphics[scale=0.3, angle=270]{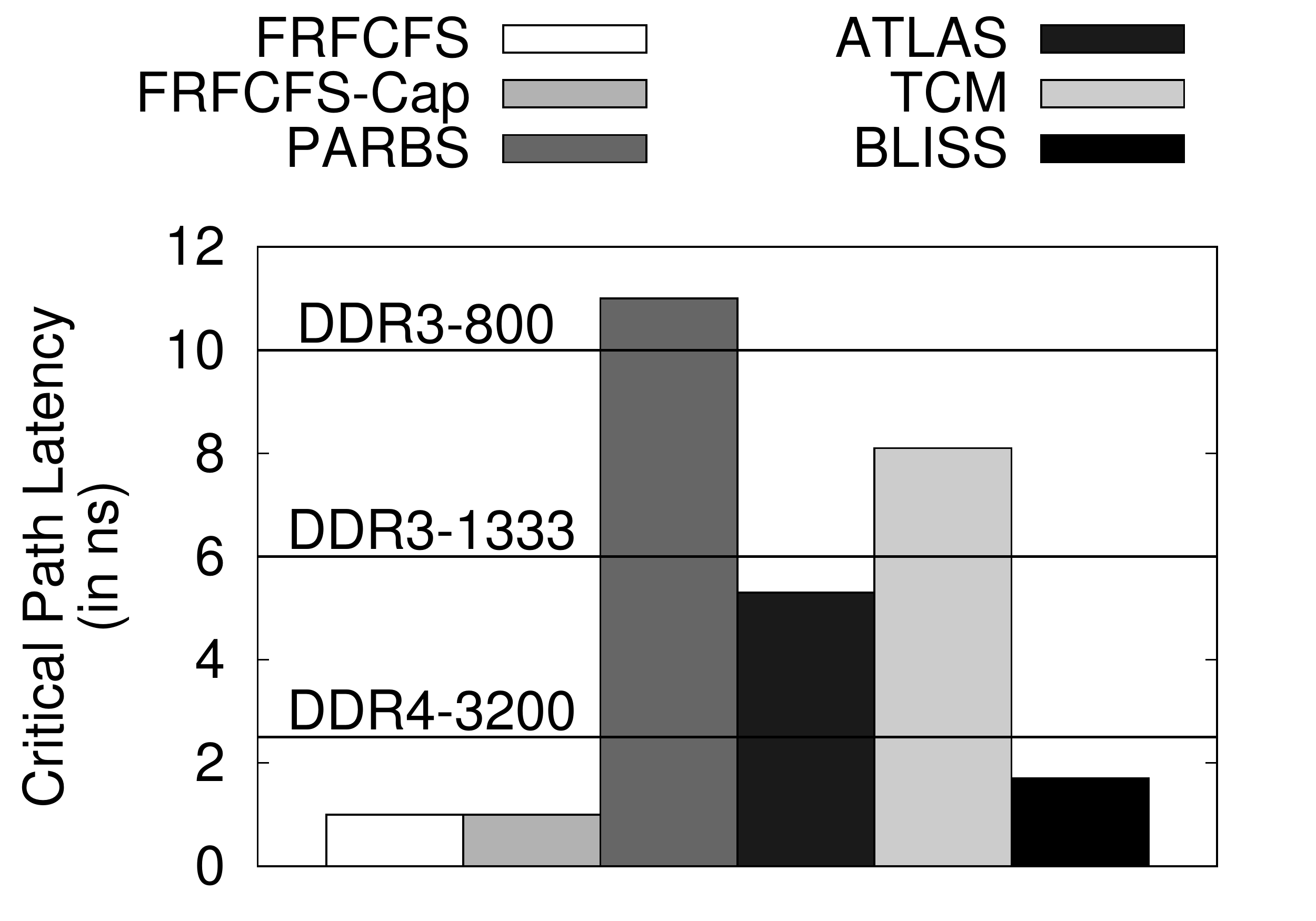}
     \caption{Critical path: \bliss vs. previous schedulers}
     \label{fig:crit-path-latency}
  \end{minipage}
  \begin{minipage}{0.48\textwidth}
%      \vspace{-1mm}
      \centering
      \includegraphics[scale=0.3, angle=270]{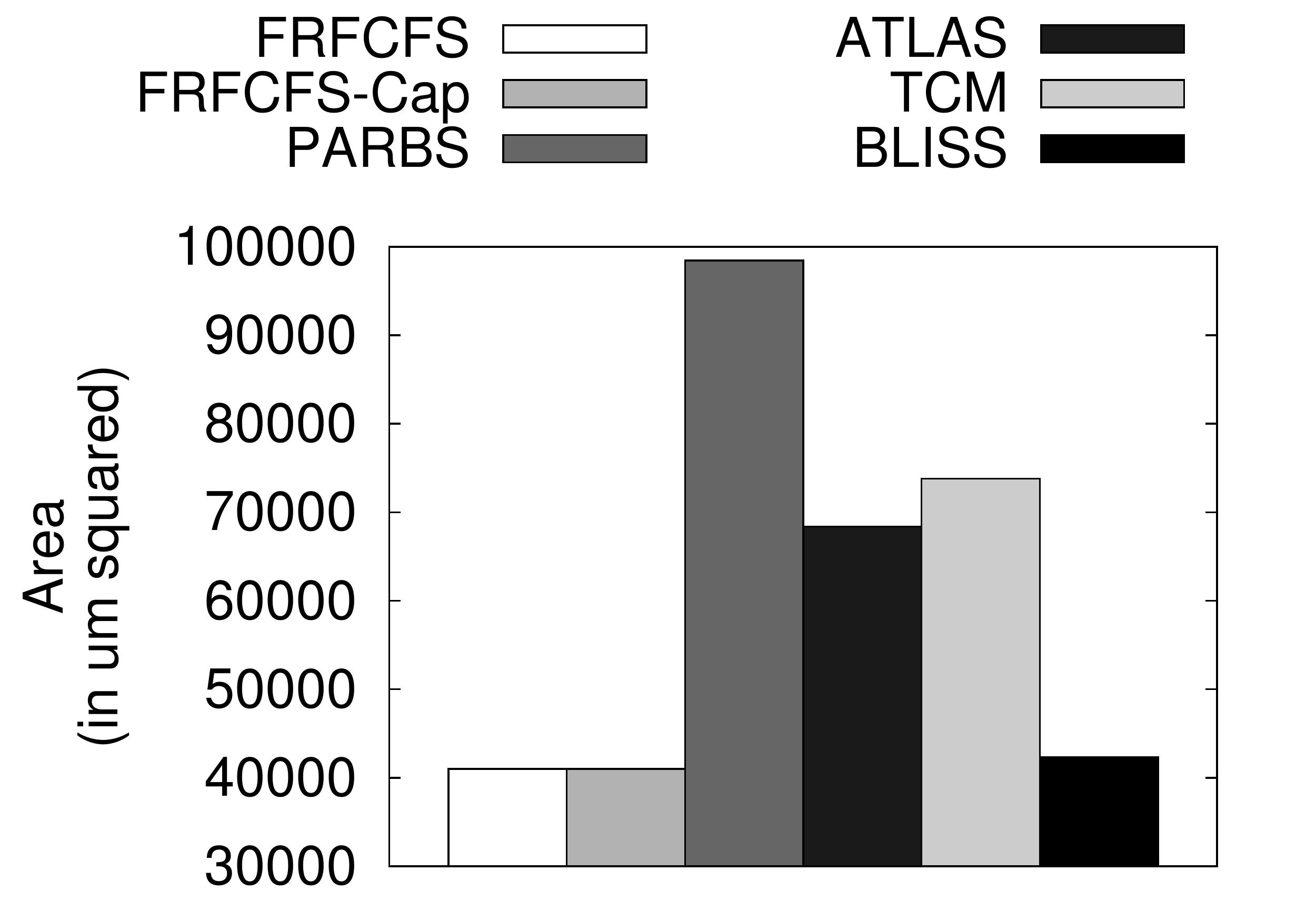}
      \caption{Area: \bliss vs. previous schedulers}
      \label{fig:area}
  \end{minipage}
\end{figure}

Figures~\ref{fig:crit-path-latency} and~\ref{fig:area} show the
critical path latency and area of five previous schedulers and
\bliss for a 24-core system for every memory channel. We draw two
major conclusions. First, previously proposed ranking-based
schedulers, PARBS/ATLAS/TCM, greatly increase the critical path
latency and area of the memory scheduler: by 11x/5.3x/8.1x and
2.4x/1.7x/1.8x respectively, compared to FRFCFS and FRFCFS-Cap,
whereas \bliss increases latency and area by only 1.7x and 3.2\%
over FRFCFS/FRFCFS-Cap.\footnote{The area numbers are for the
lowest value of critical path latency that the scheduler is able
to meet.} Second, PARBS, ATLAS and TCM cannot meet the stringent
worst-case timing requirements posed by the DDR3 and DDR4
standards~\cite{jedec-ddr3,jedec-ddr4}. In the case where every
request is a row-buffer hit, the memory controller would have to
schedule a request every read-to-read cycle time ($t_{CCD}$), the
minimum value of which is 4 cycles for both DDR3 and DDR4. TCM and
ATLAS can meet this worst-case timing only until DDR3-800
(read-to-read cycle time of 10 ns) and DDR3-1333 (read-to-read
cycle time of 6 ns) respectively, whereas \bliss can meet the
worst-case timing all the way down to the highest released
frequency for DDR4, DDR4-3200 (read-to-read time of 2.5 ns).
Hence, the high critical path latency of PARBS, ATLAS and TCM is a
serious impediment to their adoption in today's and future memory
interfaces. Techniques like pipelining could potentially be
employed to reduce the critical path latency of these previous
schedulers. However, the additional flops required for pipelining
would increase area, power and design effort significantly.
Therefore, we conclude that \bliss, with its greatly lower
complexity and cost as well as higher system performance and
competitive or better fairness, is a more effective alternative to
state-of-the-art application-aware memory schedulers.

\subsection{Trade-offs Between Performance, Fairness and Complexity}
\label{sec:tradeoff-performance-fairness-complexity}

\begin{figure}[h]
    \centering
    \includegraphics[scale=0.5, angle=0]{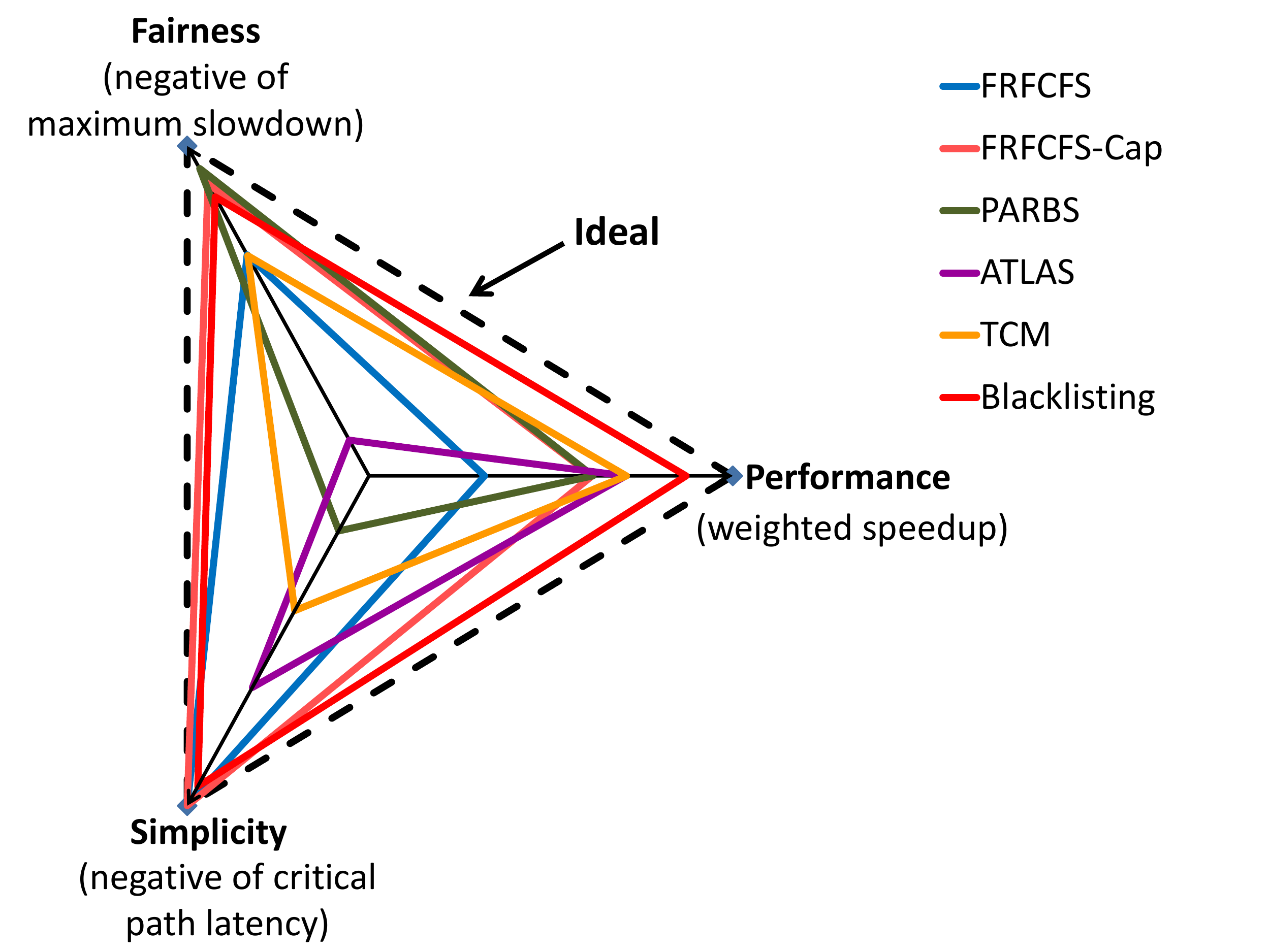}
    \caption{Performance, fairness and simplicity trade-offs}
    \label{fig:perf-fairness-simplicity}
\end{figure}

In the previous sections, we studied the performance, fairness and
complexity of different schedulers individually. In this section,
we will analyze the trade-offs between these metrics for
different schedulers. Figure~\ref{fig:perf-fairness-simplicity}
shows a three-dimensional radar plot with performance, fairness and
simplicity on three different axes. On the fairness axis, we plot
the negative of the maximum slowdown numbers, and on the simplicity axis, we
plot the negative of the critical path latency numbers. Hence, the ideal
scheduler would have high performance, fairness and simplicity, as
indicated by the encompassing, dashed black triangle. 

We draw three major conclusions about the different schedulers we
study. First, application-unaware schedulers, such as FRFCFS and
FRFCFS-Cap, are simple. However, they have low performance and/or
fairness. This is because, as described in our performance
analysis above, FRFCFS allows long streaks of row hits from one
application to cause interference to other applications.
FRFCFS-Cap attempts to tackle this problem by restricting the
length of the current row hit streak.  While such a scheme
improves fairness, it still does not improve performance
significantly. Second, application-aware schedulers, such as
PARBS, ATLAS and TCM, improve performance or fairness by ranking
based on applications' memory access characteristics.  However,
they do so at the cost of increasing complexity (reducing
simplicity) significantly, since they employ a full ordered
ranking across all applications. Third, BLISS, achieves high
performance and fairness, while keeping the design simple, thereby
approaching the ideal scheduler design (i.e., leading to a
triangle that is closer to the ideal triangle). This is because
BLISS requires only simple hardware changes to the memory
controller to blacklist applications that have long streaks of
requests served, which effectively mitigates interference.
Therefore, we conclude that BLISS achieves the best trade-off
between performance, fairness and simplicity.

\subsection{Understanding the Benefits of \bliss}
\label{sec:bliss-benefits}

\begin{figure*}[ht]
  \centering
  \captionsetup{justification=centering}
  \begin{subfigure}[t]{0.32\textwidth}
    \centering
    \includegraphics[scale=0.21, angle=270]{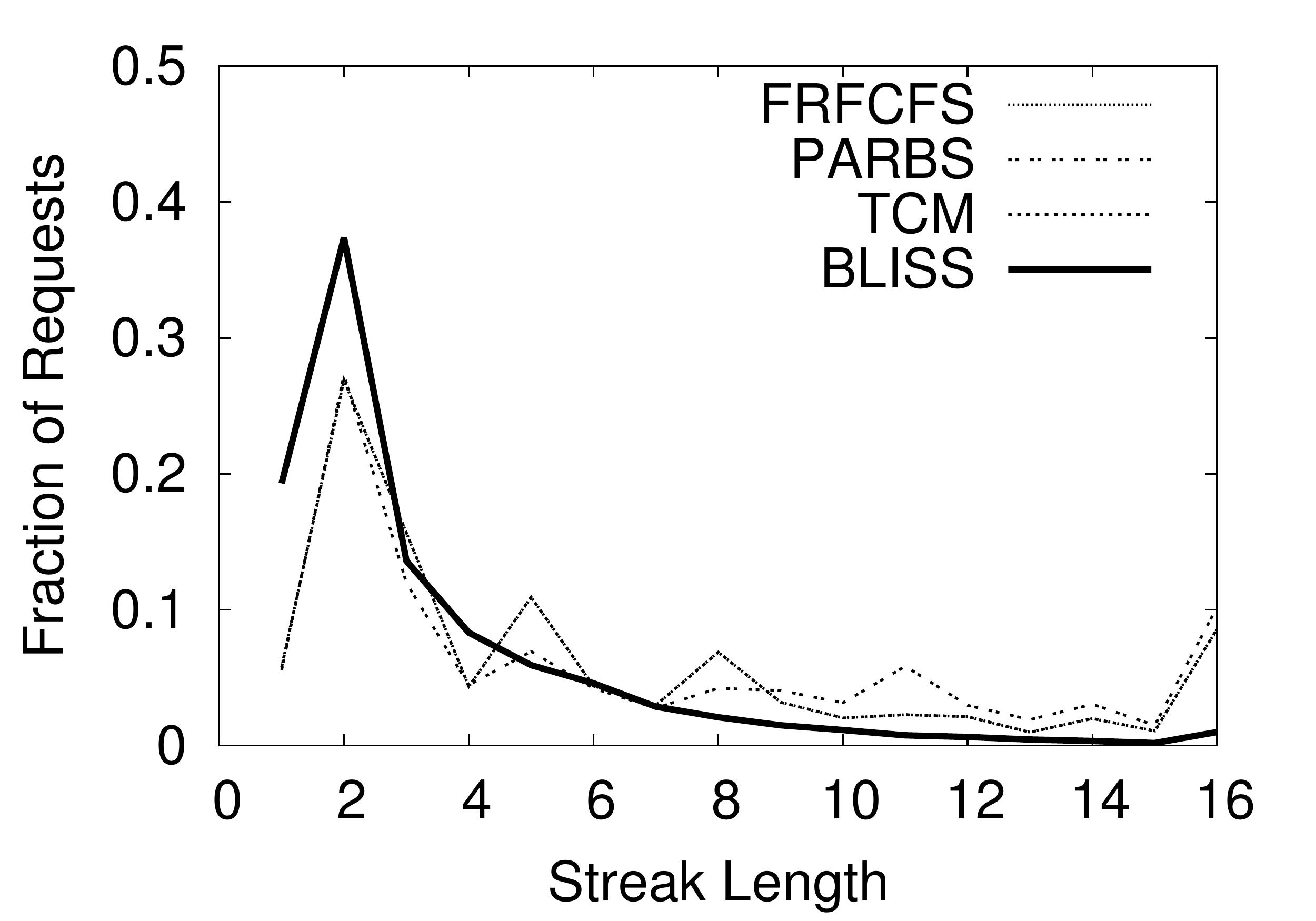}
    \caption{libquantum\\(MPKI: 52; RBH: 99\%)}
    \label{fig:libq}
  \end{subfigure}
  \begin{subfigure}[t]{0.32\textwidth}
    \centering
    \includegraphics[scale=0.21, angle=270]{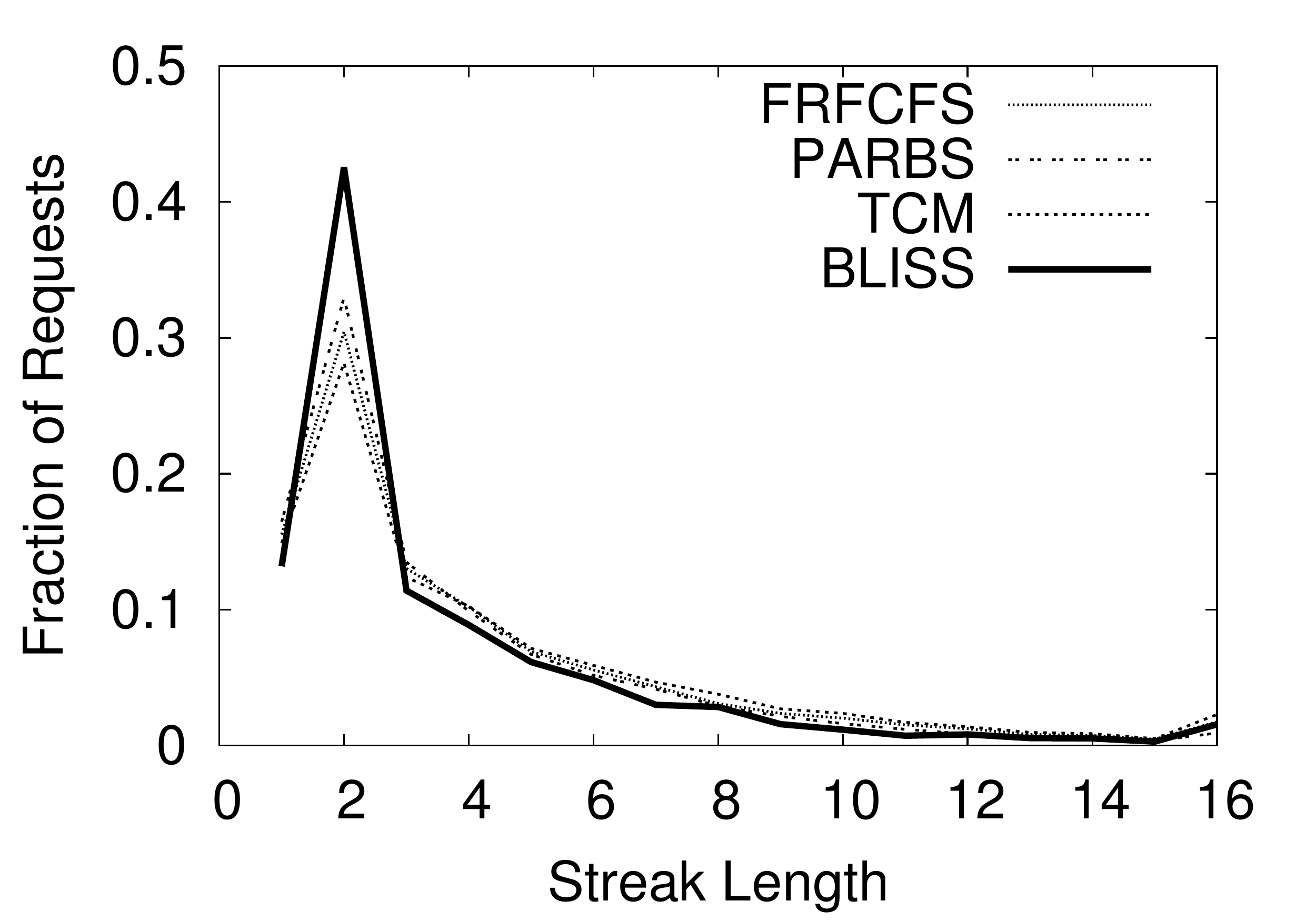}
    \caption{mcf\\(MPKI: 146; RBH: 40\%)}
    \label{fig:mcf}
  \end{subfigure}
  \begin{subfigure}[t]{0.32\textwidth}
    \centering
    \includegraphics[scale=0.21, angle=270]{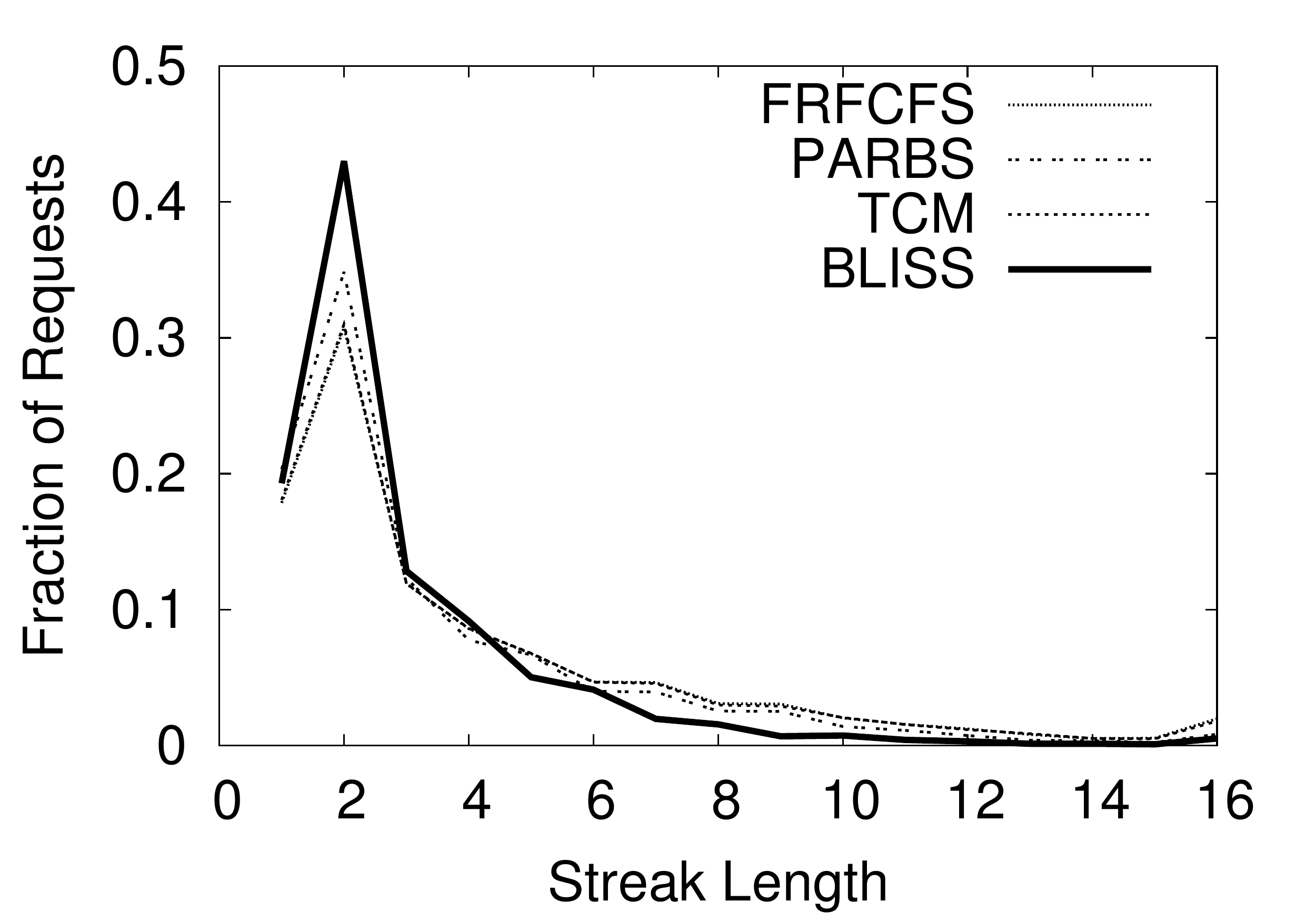}
    \caption{lbm\\(MPKI: 41; RBH: 89\%)}
    \label{fig:lbm}
    \vspace{5mm}
  \end{subfigure}
  \begin{subfigure}[t]{0.32\textwidth}
    \centering
    \includegraphics[scale=0.21, angle=270]{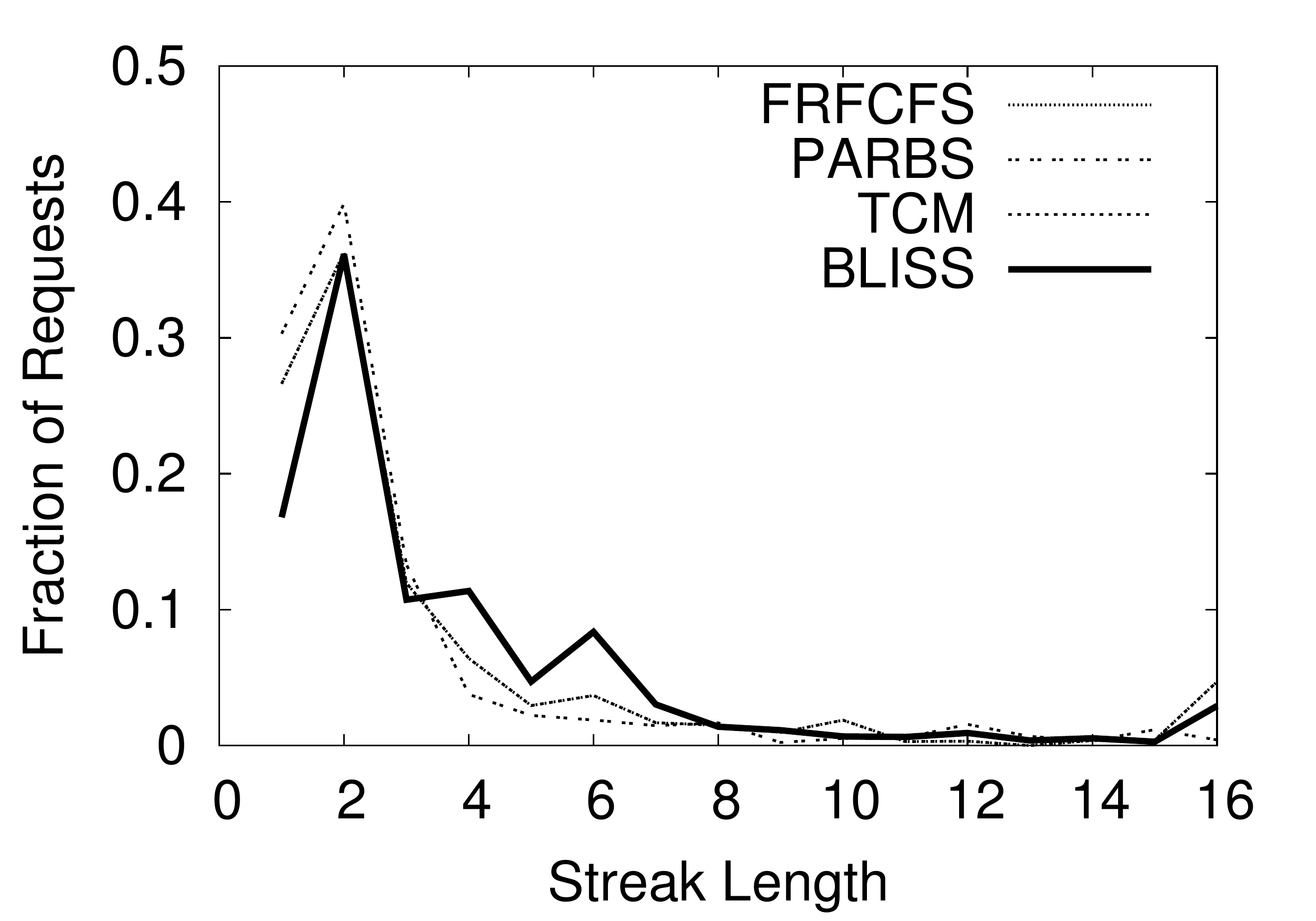}
    \caption{calculix\\(MPKI: 0.1; RBH: 85\%)}
    \label{fig:calculix}
  \end{subfigure}
  \begin{subfigure}[t]{0.32\textwidth}
    \centering
    \includegraphics[scale=0.21, angle=270]{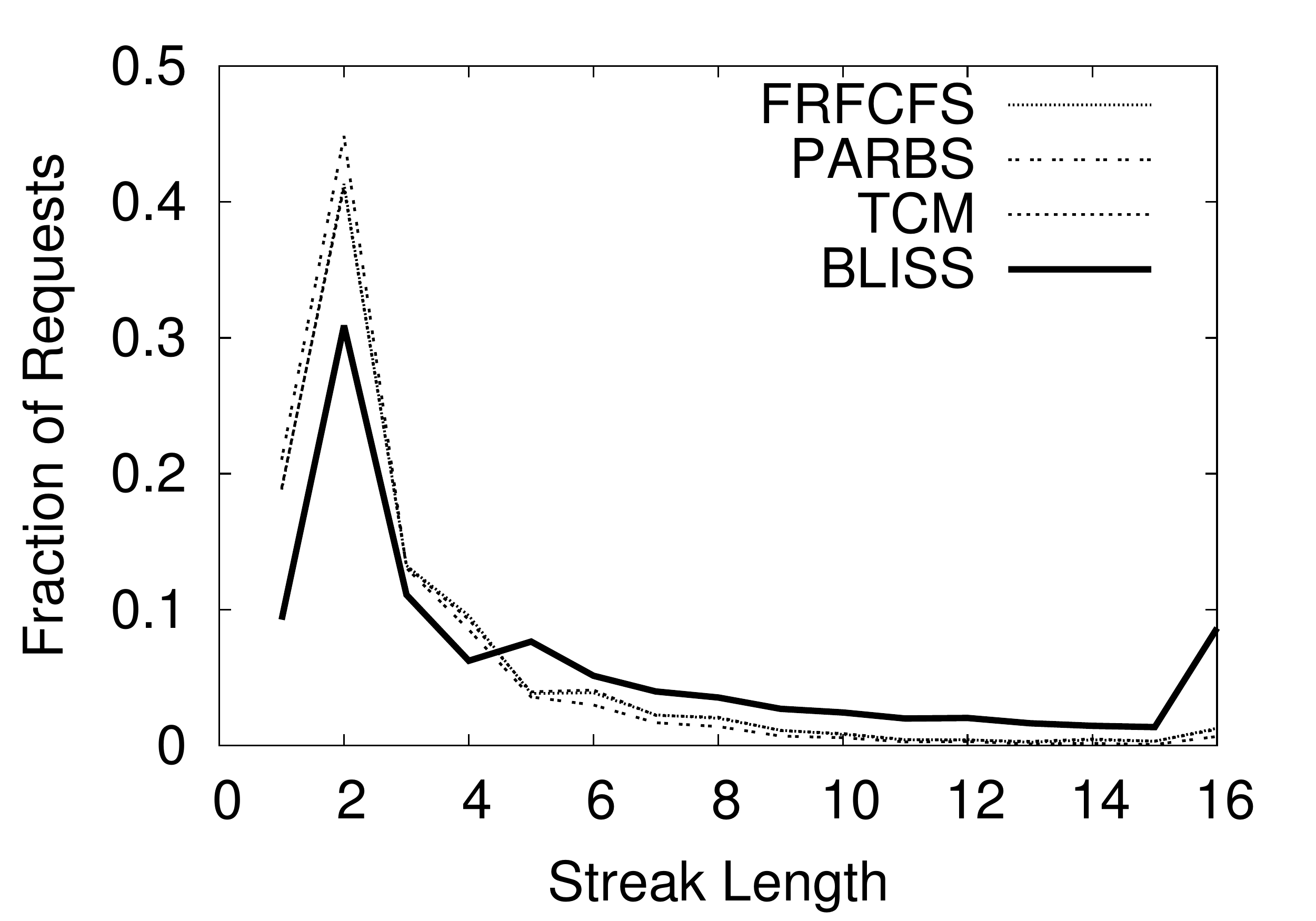}
    \caption{sphinx3\\(MPKI: 24; RBH: 91\%)}
    \label{fig:sphinx3}
  \end{subfigure}
  \begin{subfigure}[t]{0.32\textwidth}
    \centering
    \includegraphics[scale=0.21, angle=270]{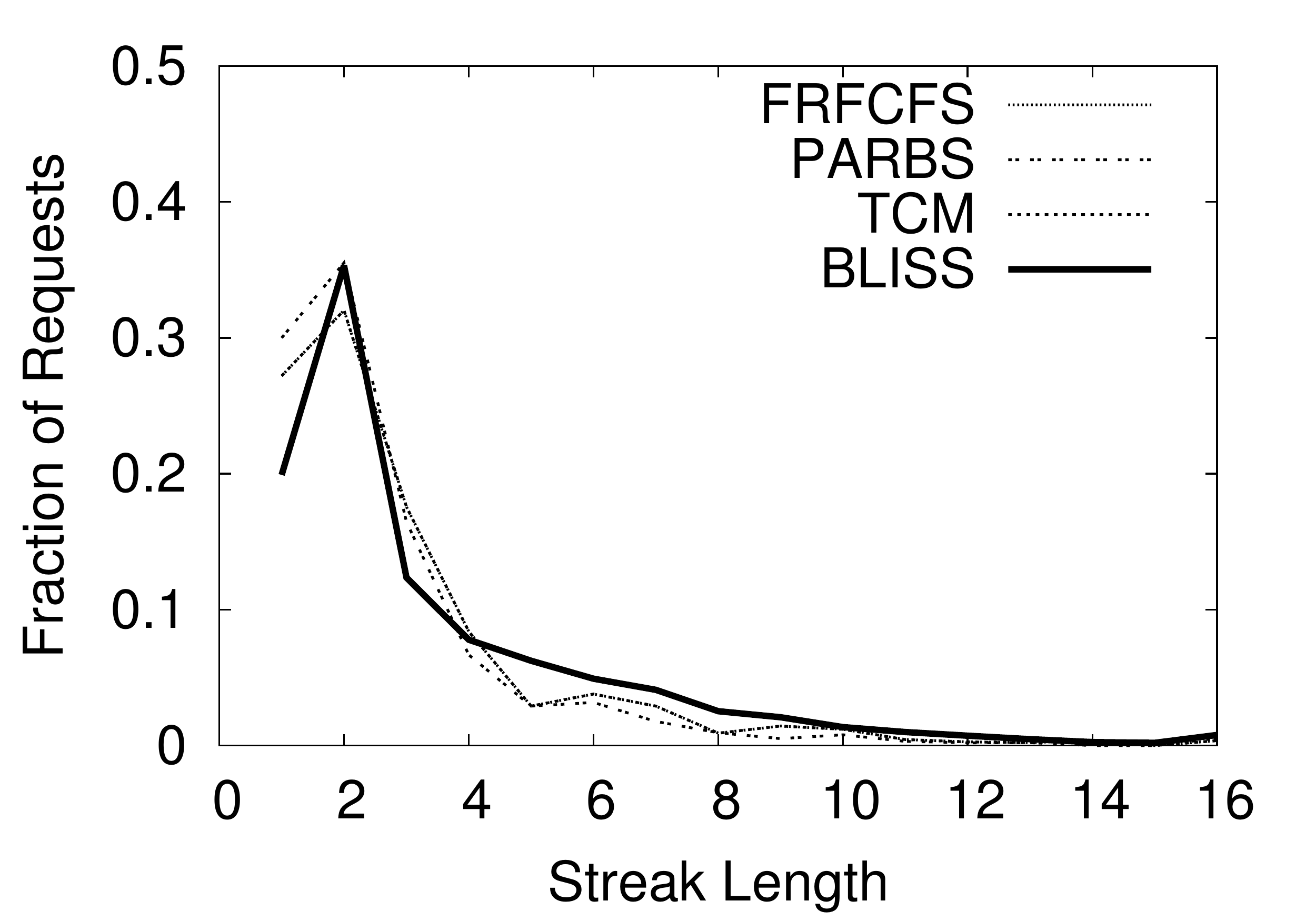}
    \caption{cactusADM\\(MPKI: 7; RBH: 49\%)}
    \label{fig:cactus}
  \end{subfigure}
  \caption{Distribution of streak lengths}
  \label{fig:streak_length}
\end{figure*}

We present the distribution of the number of consecutive requests
served (streaks) from individual applications to better understand
why \bliss effectively mitigates interference.
Figure~\ref{fig:streak_length} shows the distribution of requests
served across different streak lengths ranging from 1 to
16 for FRFCFS,
PARBS, TCM and \bliss for \emph{six representative applications
from the same 24-core workload}.\footnote{A value of 16 captures streak
lengths 16 and above.} The figure captions indicate the
memory intensity, in misses per kilo instruction (MPKI) and
row-buffer hit rate (RBH) of each application when it is run
alone. 

Figures~\ref{fig:libq},~\ref{fig:mcf} and~\ref{fig:lbm}
show the streak length distributions of applications that have a
tendency to cause interference ({\em libquantum}, {\em mcf} and
{\em lbm}). All these applications have high memory intensity
and/or high row-buffer locality.
Figures~\ref{fig:calculix},~\ref{fig:sphinx3} and~\ref{fig:cactus}
show applications that are vulnerable to interference ({\em
calculix}, {\em cactusADM} and {\em sphinx3}). These applications
have lower memory intensities and row-buffer localities, compared
to the interference-causing applications. We observe that \bliss
shifts the distribution of streak lengths towards the left for the
interference-causing applications, while it shifts the streak
length distribution to the right for the interference-prone
applications. Hence, \bliss breaks long streaks of consecutive
requests for interference-causing applications, while enabling
longer streaks for vulnerable applications. This enables such
vulnerable applications to make faster progress, thereby resulting
in better system performance and fairness. We have observed
similar results for most of our workloads.

\subsection{Average Request Latency}
\label{sec:average-request-latency}

\begin{figure}[h!]
  \centering
  \includegraphics[scale=0.25, angle=270]{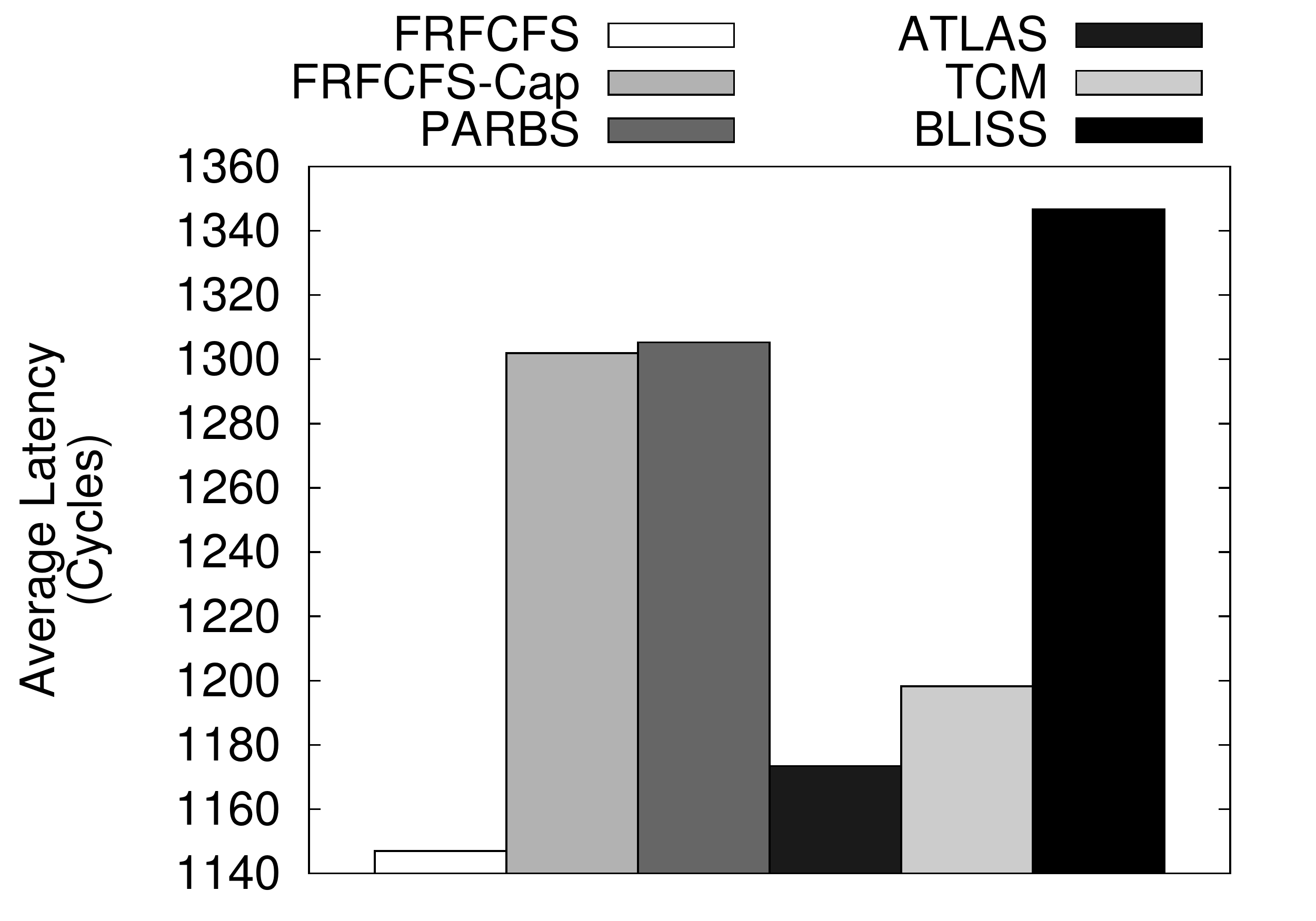}
  \caption{The Average Request Latency Metric}
  \label{fig:latency}
\end{figure}

In this section, we evaluate the average memory request latency
(from when a request is generated until when it is served)
metric and seek to understand its correlation with performance and
fairness. Figure~\ref{fig:latency} presents the average memory request
latency (from when the request is generated until when it is served)
for the five previously proposed memory schedulers and \bliss. Two
major observations are in order. First, FRFCFS has the lowest
average request latency among all the schedulers. This is
expected since FRFCFS maximizes DRAM throughput by prioritizing
row-buffer hits. Hence, the number of requests served is maximized
overall (across all applications). However, maximizing throughput
(i.e., minimizing overall average
request latency) degrades the performance of low-memory-intensity
applications, since these applications' requests are often delayed
behind row-buffer hits and older requests. This results in
degradation in system performance and fairness, as shown in
Figure~\ref{fig:main-results}. 

Second, ATLAS and TCM, memory schedulers that prioritize requests
of low-memory-intensity applications by employing a full ordered
ranking achieve relatively low average latency. This is because
these schedulers reduce the latency of serving requests from
latency-critical, low-memory-intensity applications significantly.
Furthermore, prioritizing low-memory-intensity applications'
requests does not increase the latency of high-memory-intensity
applications significantly. This is because high-memory-intensity
applications already have high memory access latencies (even when
run alone) due to queueing delays. Hence, average request latency does
not increase much from deprioritizing requests of such
applications. However, always prioritizing such latency-critical
applications results in lower memory throughput for
high-memory-intensity applications, resulting in unfair slowdowns
(as we show in Figure~\ref{fig:main-results}). Third, memory
schedulers that provide the best fairness, PARBS, FRFCFS-Cap and
\bliss have high average memory latencies. This is because these
schedulers, while employing techniques to prevent requests of
vulnerable applications with low memory intensity and low
row-buffer locality from being delayed, also avoid unfairly
delaying requests of high-memory-intensity applications. As a
result, they do not reduce the request service latency of
low-memory-intensity applications significantly, at the cost of
denying memory throughput to high-memory-intensity applications,
unlike ATLAS or TCM. Based on these observations, we conclude that while some
applications benefit from low memory access latencies, other
applications benefit more from higher memory throughput than lower
latency. Hence, average memory latency is {\em not} a suitable
metric to estimate system performance or fairness.

\subsection{Impact of Clearing the Blacklist Asynchronously}
\label{sec:blacklist-asynchronous-clearing}

The Blacklisting scheduler design we have presented and evaluated so far
clears the blacklisting information periodically (every 10000
cycles in our evaluations so far), such that \emph{all}
applications are removed from the blacklist at the end of a
\textit{Clearing Interval}. In this section, we evaluate an
alternative design where an individual application is removed from
the blacklist \textit{Clearing Interval} cycles after it has been
blacklisted (independent of the other applications). In order to
implement this alternative design, each application would need an
additional counter to keep track of the number of remaining cycles
until the application would be removed from the blacklist. This
counter is set (to the \textit{Clearing Interval}) when an
application is blacklisted and is decremented every cycle until it
becomes zero. When it becomes zero, the corresponding application
is removed from the blacklist. We use a \textit{Clearing Interval}
of 10000 cycles for this alternative design as well.     

Table~\ref{tab:individual-clearing} shows the system performance
and fairness of the original BLISS design (BLISS) and the
alternative design in which individual applications are removed
from the blacklist asynchronously (BLISS-Individual-Clearing). As
can be seen, the performance and fairness of the two designs are
similar. Furthermore, the first design (BLISS) is simpler since it
does not need to maintain an additional counter for each
application. We conclude that the original BLISS design is
more efficient, in terms of performance, fairness and complexity.

\begin{table}[h!]
  \centering
  \input{blacklisting/tables/individual-clearing}
  \caption{Clearing the blacklist asynchronously}
  \label{tab:individual-clearing}
\end{table}

\subsection{Comparison with TCM's Clustering Mechanism}
\label{sec:tcm-clustering}

\begin{figure*}[hbt!]
  \centering
  \begin{minipage}{0.32\textwidth}
    \centering
    \includegraphics[scale=0.21, angle=270]{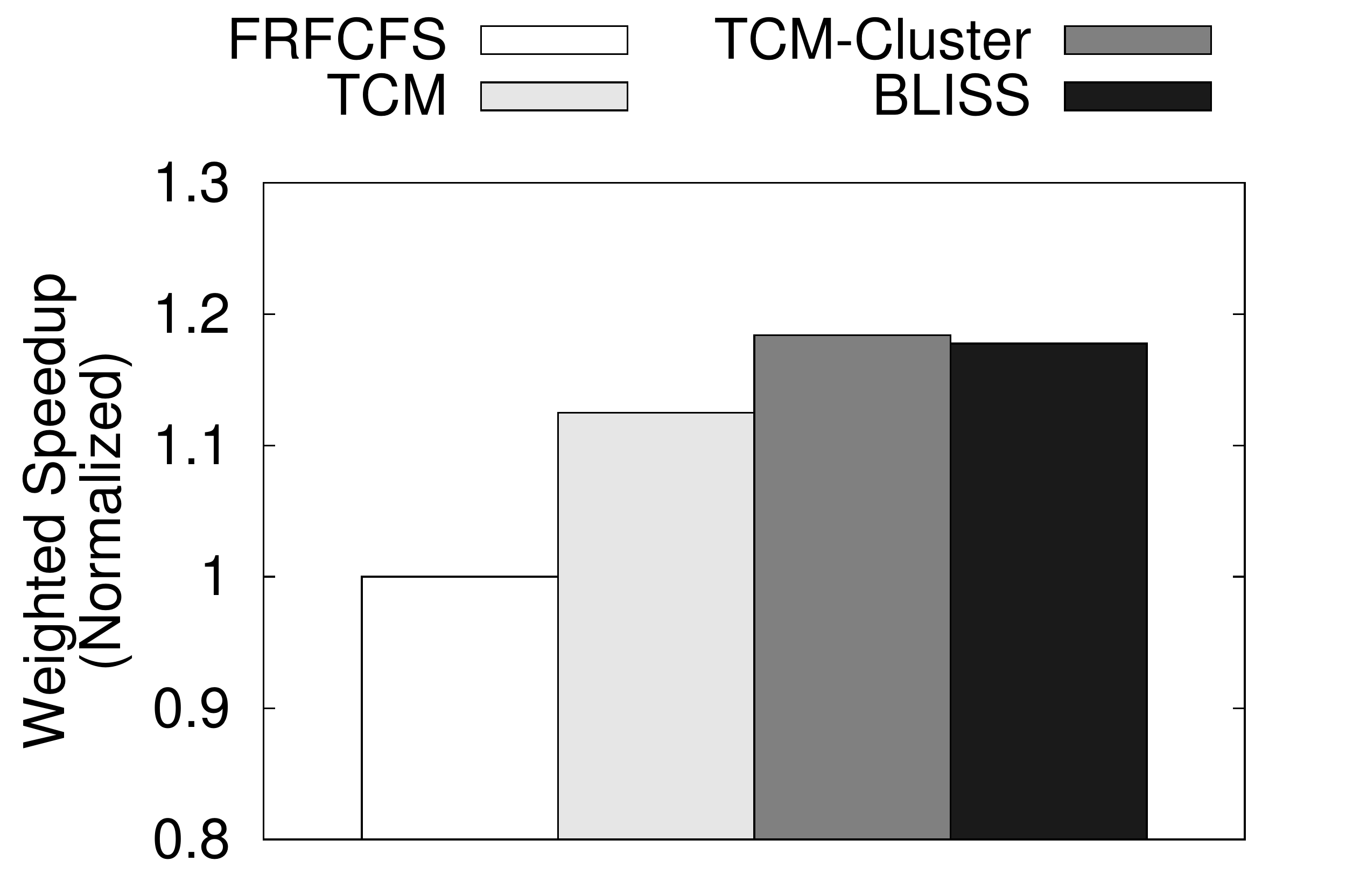}
  \end{minipage}
  \begin{minipage}{0.32\textwidth}
    \centering
    \includegraphics[scale=0.21, angle=270]{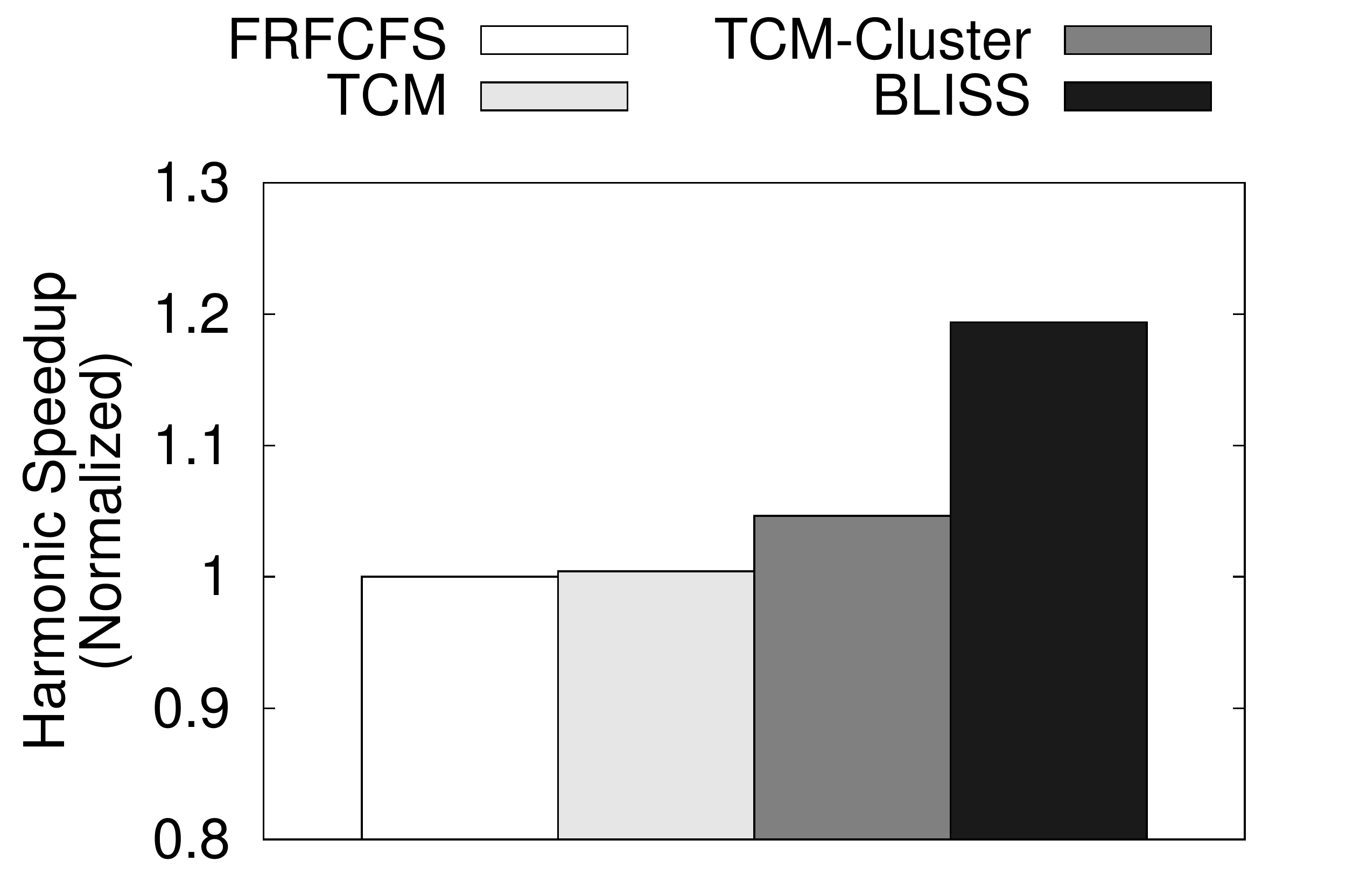}
  \end{minipage} 
  \begin{minipage}{0.32\textwidth}
    \centering
    \includegraphics[scale=0.21, angle=270]{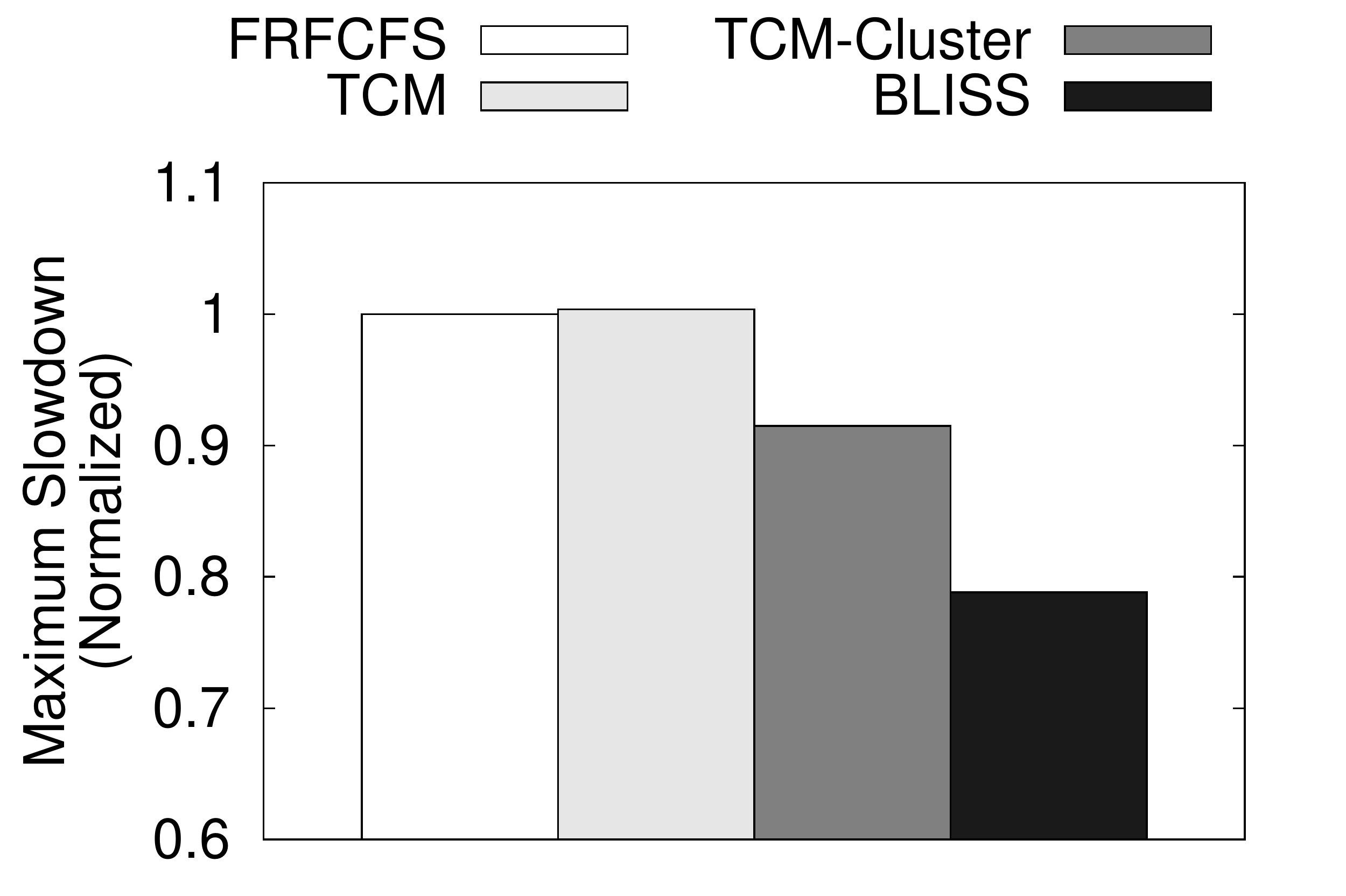}
  \end{minipage} 
  \caption{Comparison with TCM's clustering mechanism}
  \label{fig:tcm-no-ranking}
\end{figure*}

Figure~\ref{fig:tcm-no-ranking} shows the system performance and
fairness of \bliss, TCM and TCM's clustering mechanism
(TCM-Cluster). TCM-Cluster is a modified version of TCM that
performs clustering, but does not rank applications within each
cluster. We draw two major conclusions. First, TCM-Cluster has
similar system performance as \bliss, since both \bliss and
TCM-Cluster prioritize vulnerable applications by separating them
into a group and prioritizing that group rather than ranking
individual applications. Second, TCM-Cluster has significantly
higher unfairness compared to \bliss. This is because TCM-Cluster
always deprioritizes high-memory-intensity applications,
regardless of whether or not they are causing interference (as
described in Section~\ref{sec:blacklisting-observations}).
\bliss, on the other hand, observes an application at fine time
granularities, independently at every memory channel and
blacklists an application at a channel {\em only when} it is
generating a number of consecutive requests (i.e., potentially
causing interference to other applications).

\subsection{Evaluation of Row Hit Based Blacklisting}
\label{sec:blacklisting-frfcfs-cap}

\bliss, by virtue of restricting the number of consecutive
requests that are served from an application, attempts to mitigate
the interference caused by both high-memory-intensity and high-
row-buffer-locality applications. In this section, we attempt to
isolate the benefits from restricting consecutive row-buffer
hitting requests vs. non-row-buffer hitting requests. To this end,
we evaluate the performance and fairness benefits of a mechanism
that places an application in the blacklist when a certain number
of row-buffer hitting requests (\textrm{N}) to the same row have
been served for an application (we call this
FRFCFS-Cap-Blacklisting as the scheduler essentially is FRFCFS-Cap
with blacklisting). We use an \textrm{N} value of 4 in our
evaluations.

\begin{figure*}[ht!]
  \centering
  \begin{minipage}{0.32\textwidth}
    \centering
    \includegraphics[scale=0.21, angle=270]{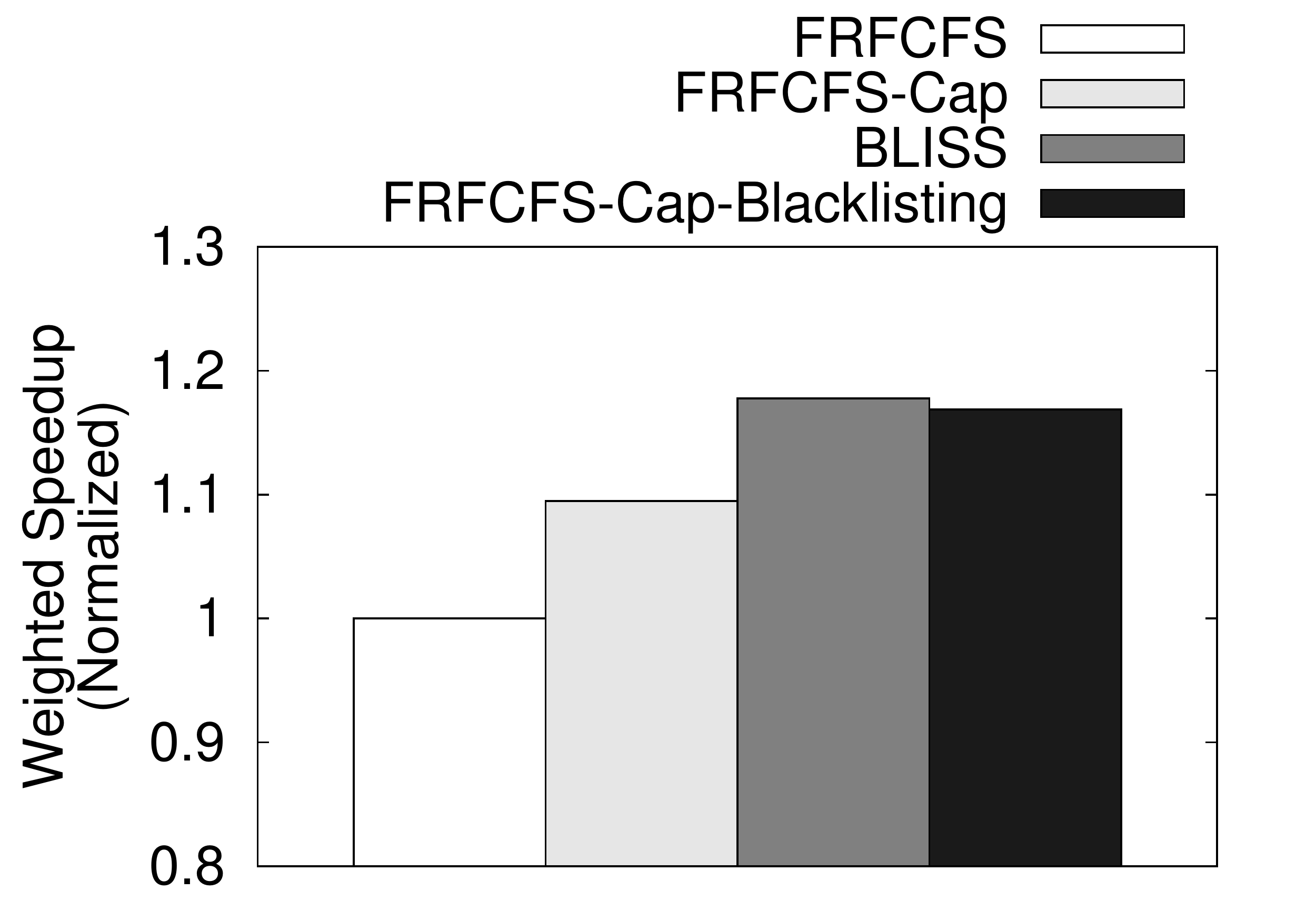}
  \end{minipage}
  \begin{minipage}{0.32\textwidth}
    \centering
    \includegraphics[scale=0.21, angle=270]{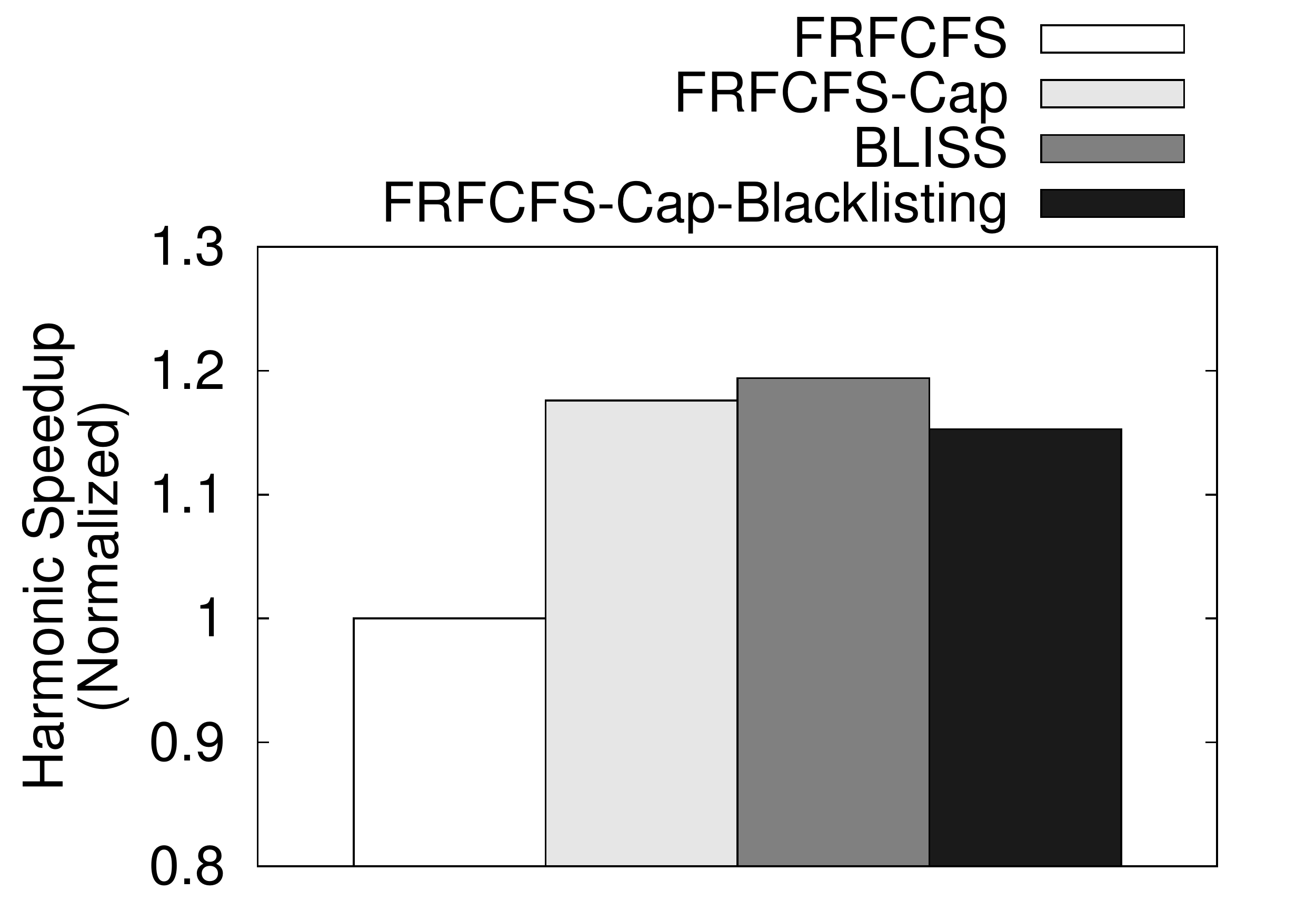}
  \end{minipage}
  \begin{minipage}{0.32\textwidth}
    \centering
    \includegraphics[scale=0.21, angle=270]{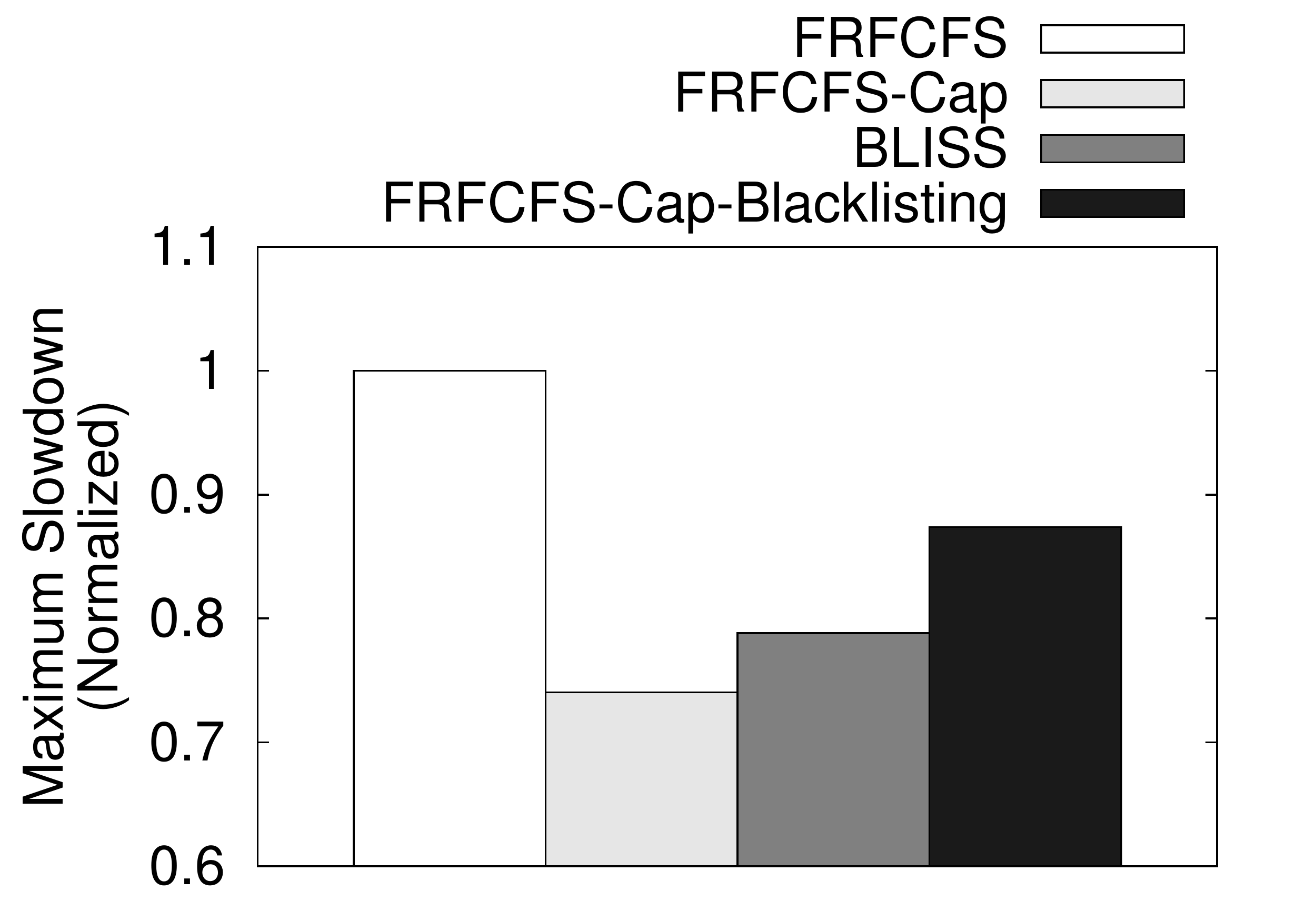}
  \end{minipage}
  \caption{Comparison with FRFCFS-Cap combined with blacklisting}
  \label{fig:frfcfs-cap-blacklisting}
\end{figure*}

\begin{sloppypar}
Figure~\ref{fig:frfcfs-cap-blacklisting} compares the system
performance and fairness of \bliss with FRFCFS-Cap-Blacklisting.
We make three major observations. First, FRFCFS-Cap-Blacklisting
has similar system performance as \bliss.  On further analysis of
individual workloads, we find that FRFCFS-Cap-Blacklisting
blacklists only applications with high row-buffer locality,
causing requests of non-blacklisted high-memory-intensity
applications to interfere with requests of low-memory-intensity
applications. However, the performance impact of this interference
is offset by the performance improvement of high-memory-intensity
applications that are not blacklisted. Second,
FRFCFS-Cap-Blacklisting has higher unfairness (higher maximum
slowdown and lower harmonic speedup) than \bliss. This is because
the high-memory-intensity applications that are not blacklisted
are prioritized over the blacklisted high-row-buffer-locality
applications, thereby interfering with and slowing down the
high-row-buffer-locality applications significantly. Third,
FRFCFS-Cap-Blacklisting requires a per-bank counter to count and
cap the number of row-buffer hits, whereas \bliss needs only one
counter per-channel to count the number of consecutive requests
from the same application. Therefore, we conclude that \bliss is
more effective in mitigating unfairness while incurring lower
hardware cost, than the FRFCFS-Cap-Blacklisting scheduler that we
build combining principles from FRFCFS-Cap and BLISS.
\end{sloppypar}

\subsection{Comparison with Criticality-Aware Scheduling}

\begin{figure}[ht!]
  \centering
  \begin{minipage}{0.48\textwidth}
    \centering
    \includegraphics[scale=0.3, angle=270]{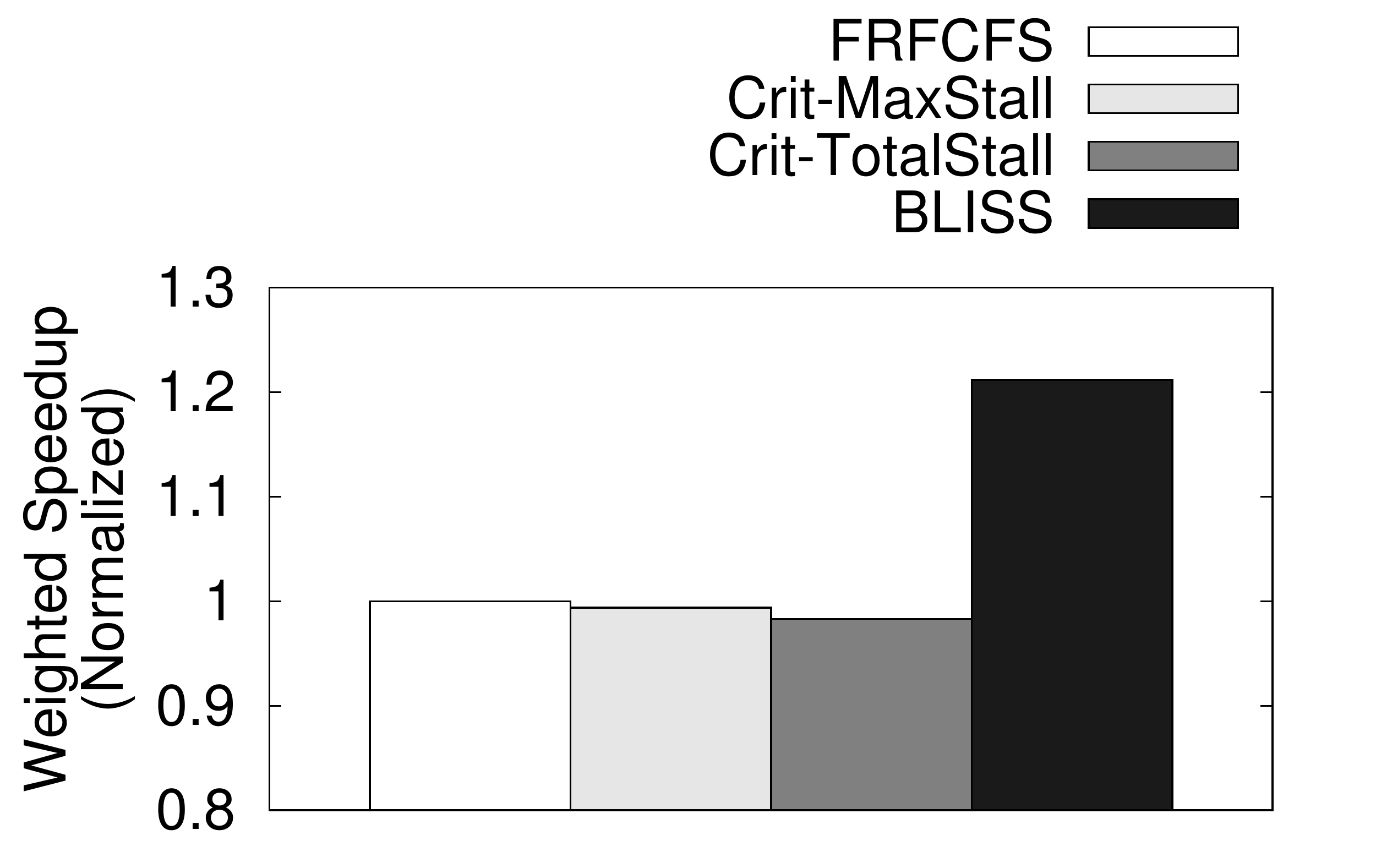}
  \end{minipage}
  \begin{minipage}{0.48\textwidth}
   \centering
   \includegraphics[scale=0.3, angle=270]{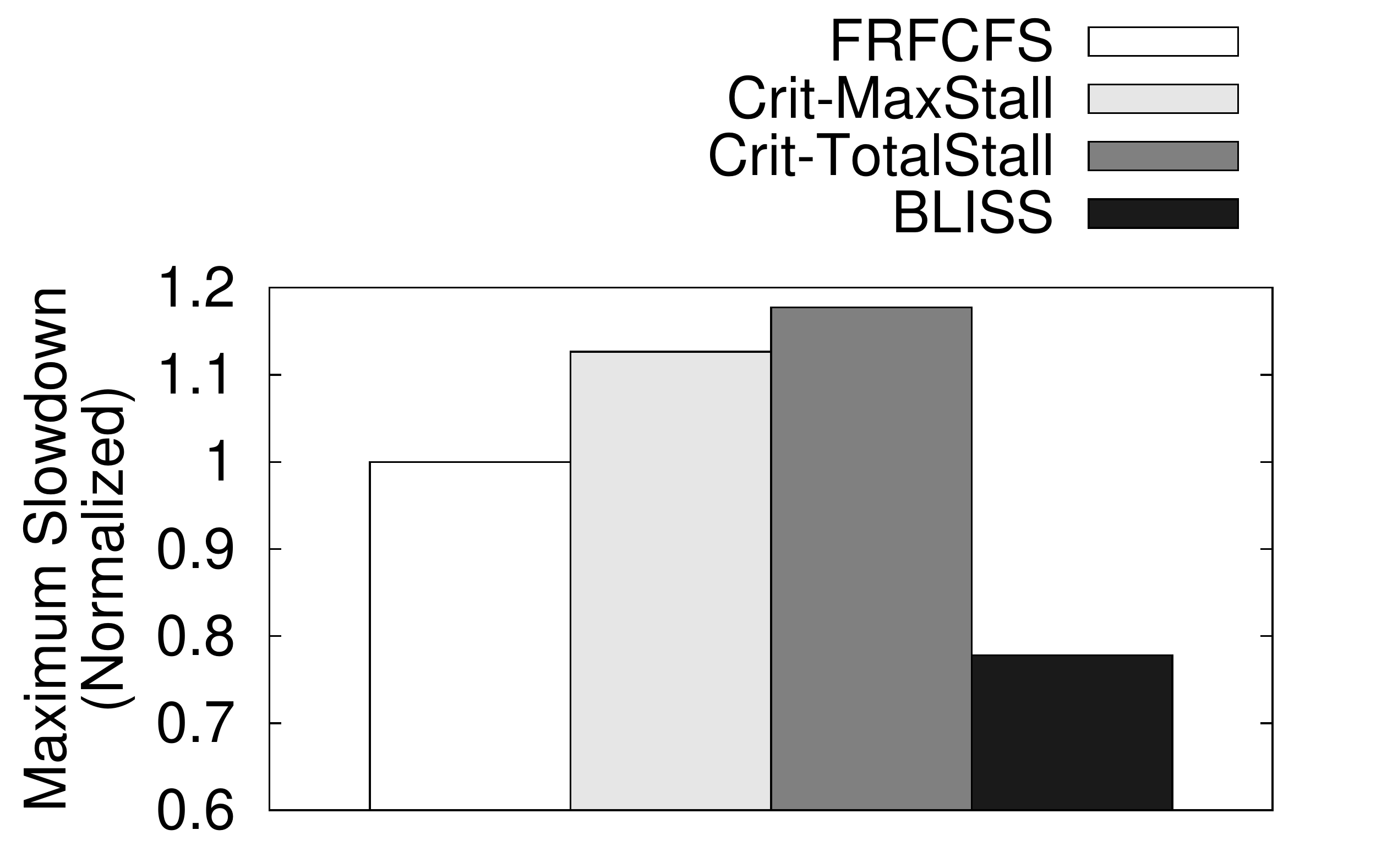}
  \end{minipage}
  \caption{Comparison with criticality-aware scheduling}
  \label{fig:crit-comparison}
\end{figure}

We compare the system performance and fairness of \bliss with
those of criticality-aware memory
schedulers~\cite{crit-scheduling-cornell}. The basic idea behind
criticality-aware memory scheduling is to prioritize memory
requests from load instructions that have stalled the instruction
window for long periods of time in the past. Ghose et
al.~\cite{crit-scheduling-cornell} evaluate prioritizing load
requests based on both maximum stall time (Crit-MaxStall) and
total stall time (Crit-TotalStall) caused by load instructions in
the past. Figure~\ref{fig:crit-comparison} shows the system
performance and fairness of \bliss and the criticality-aware
scheduling mechanisms, normalized to FRFCFS, across 40 workloads.
Two observations are in order. First, \bliss significantly
outperforms criticality-aware scheduling mechanisms in terms of
both system performance and fairness. This is because the
criticality-aware scheduling mechanisms unfairly deprioritize and
slow down low-memory-intensity applications that inherently
generate fewer requests, since stall times tend to be low for such
applications. Second, criticality-aware scheduling incurs hardware
cost to prioritize requests with higher stall times. Specifically,
the number of bits to represent stall times is on the order of
12-14, as described in~\cite{crit-scheduling-cornell}. Hence, the
logic for comparing stall times and prioritizing requests with
higher stall times would incur even higher cost than
per-application ranking mechanisms where the number of bits to
represent a core's rank grows only as as $log_2 Number Of Cores$
(e.g. 5 bits for a 32-core system). Therefore, we conclude that
\bliss achieves significantly better system performance and
fairness, while incurring lower hardware cost.

\subsection{Effect of Workload Memory Intensity and Row-buffer Locality}

In this section, we study the impact of workload memory intensity
and row-buffer locality on performance and fairness of \bliss and
five previous schedulers.\\

\begin{figure}[h!]
  \centering
  \begin{minipage}{0.48\textwidth}
    \centering
    \includegraphics[scale=0.3, angle=270]{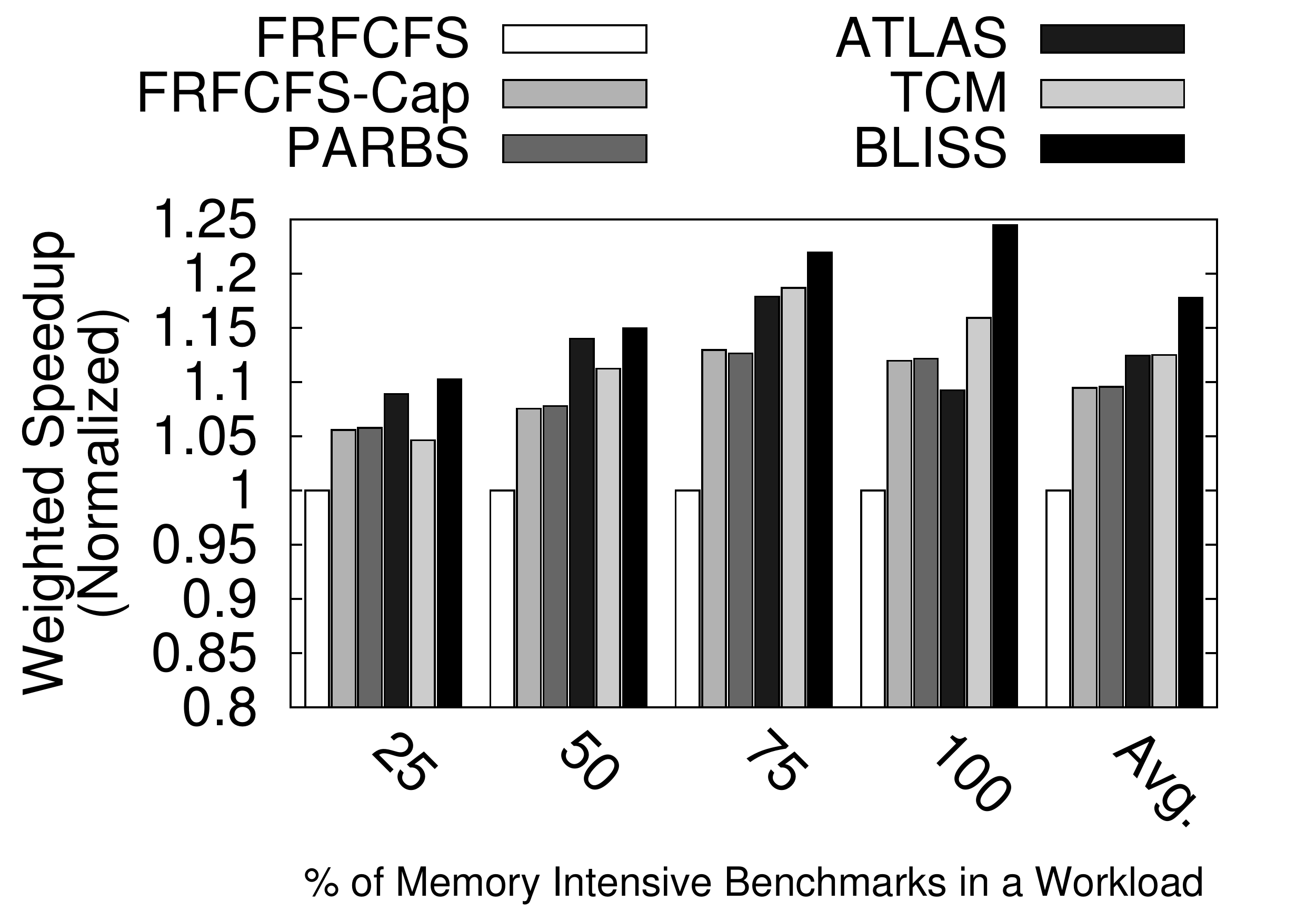}
  \end{minipage}
  \begin{minipage}{0.48\textwidth}  
    \centering
    \includegraphics[scale=0.3, angle=270]{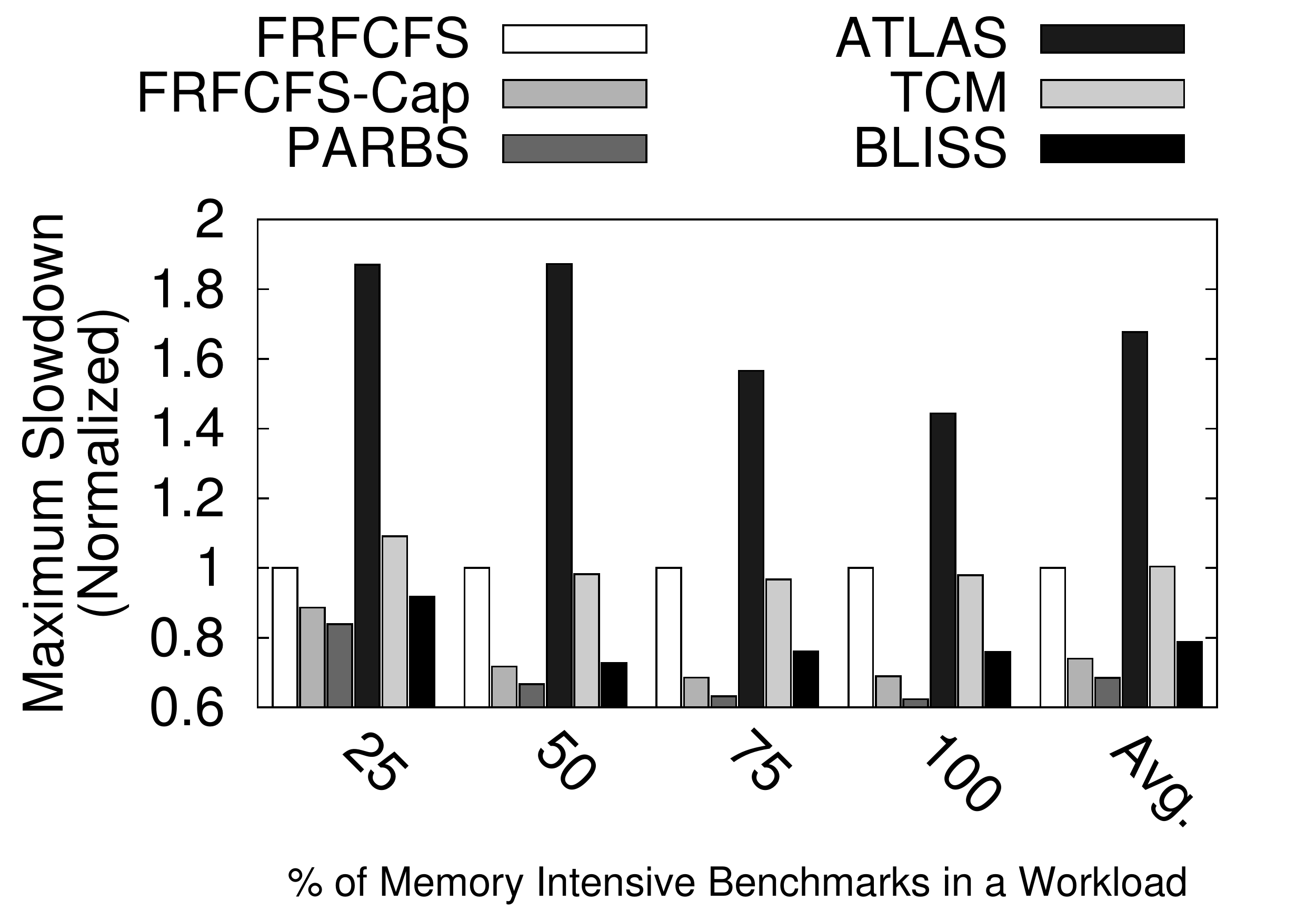}
  \end{minipage}
  \caption{Sensitivity to workload memory intensity}
  \label{fig:intensity-results}
\end{figure}

\noindent\textbf{Workload Memory Intensity.}
Figure~\ref{fig:intensity-results} shows system performance and
fairness for workloads with different memory intensities,
classified into different categories based on the fraction of
high-memory-intensity applications in a workload.\footnote{We
classify applications with MPKI less than 5 as
low-memory-intensity and the rest as high-memory-intensity.}  We
draw three major conclusions. First, \bliss outperforms previous
memory schedulers in terms of system performance across all
intensity categories. Second, the system performance benefits of
\bliss increase with workload memory intensity. This is because as
the number of high-memory-intensity applications in a workload
increases, ranking individual applications, as done by previous
schedulers, causes more unfairness and degrades system
performance. Third, \bliss achieves significantly lower unfairness
than previous memory schedulers, except FRFCFS-Cap and PARBS,
across all intensity categories. Therefore, we conclude that
\bliss is effective in mitigating interference and improving
system performance and fairness across workloads with different
compositions of high- and low-memory-intensity applications.

\begin{figure}[h!]
  \centering
  \begin{minipage}{0.48\textwidth}
    \centering
    \includegraphics[scale=0.3, angle=270]{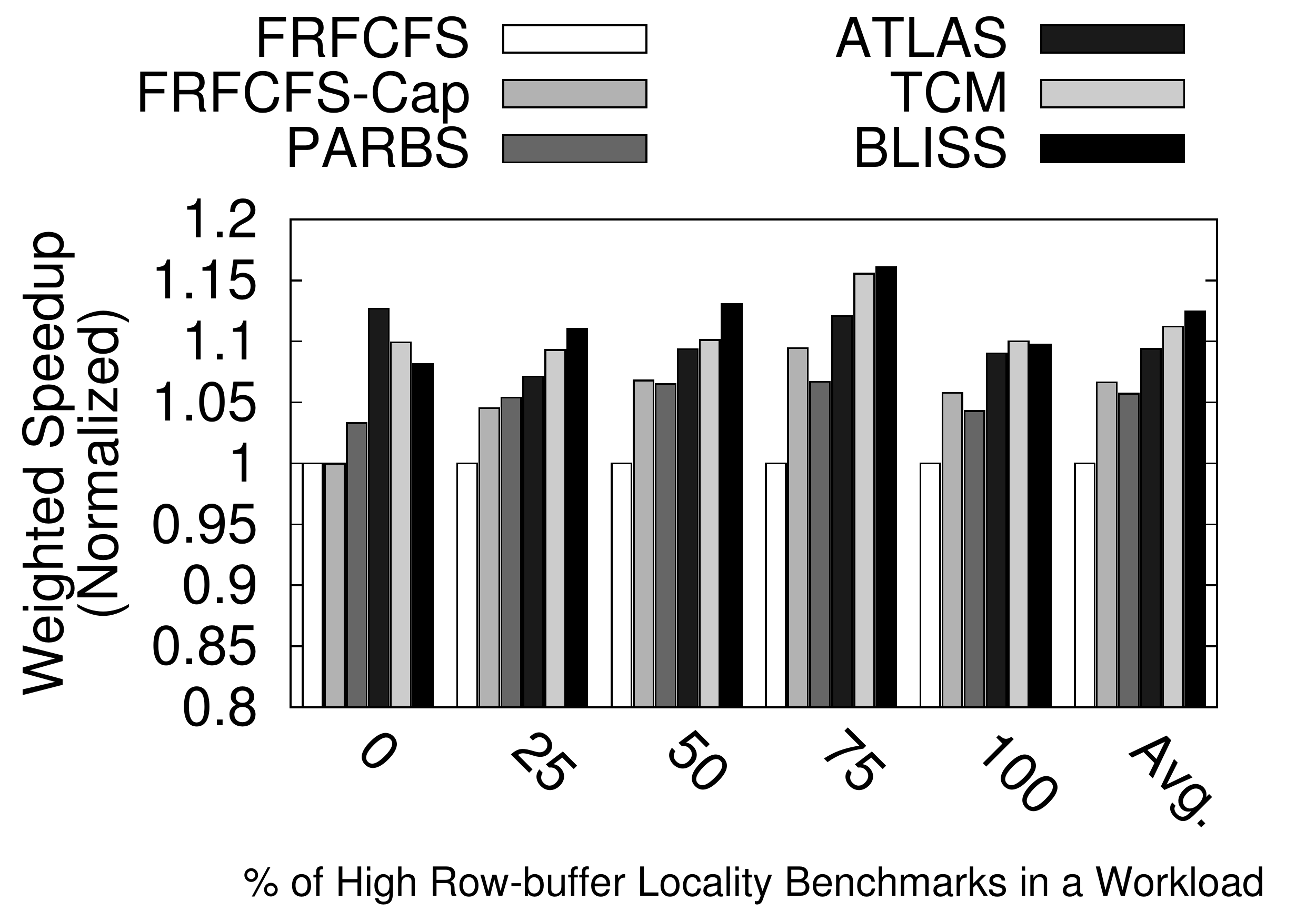}
  \end{minipage}
  \begin{minipage}{0.48\textwidth}  
    \centering
    \includegraphics[scale=0.3, angle=270]{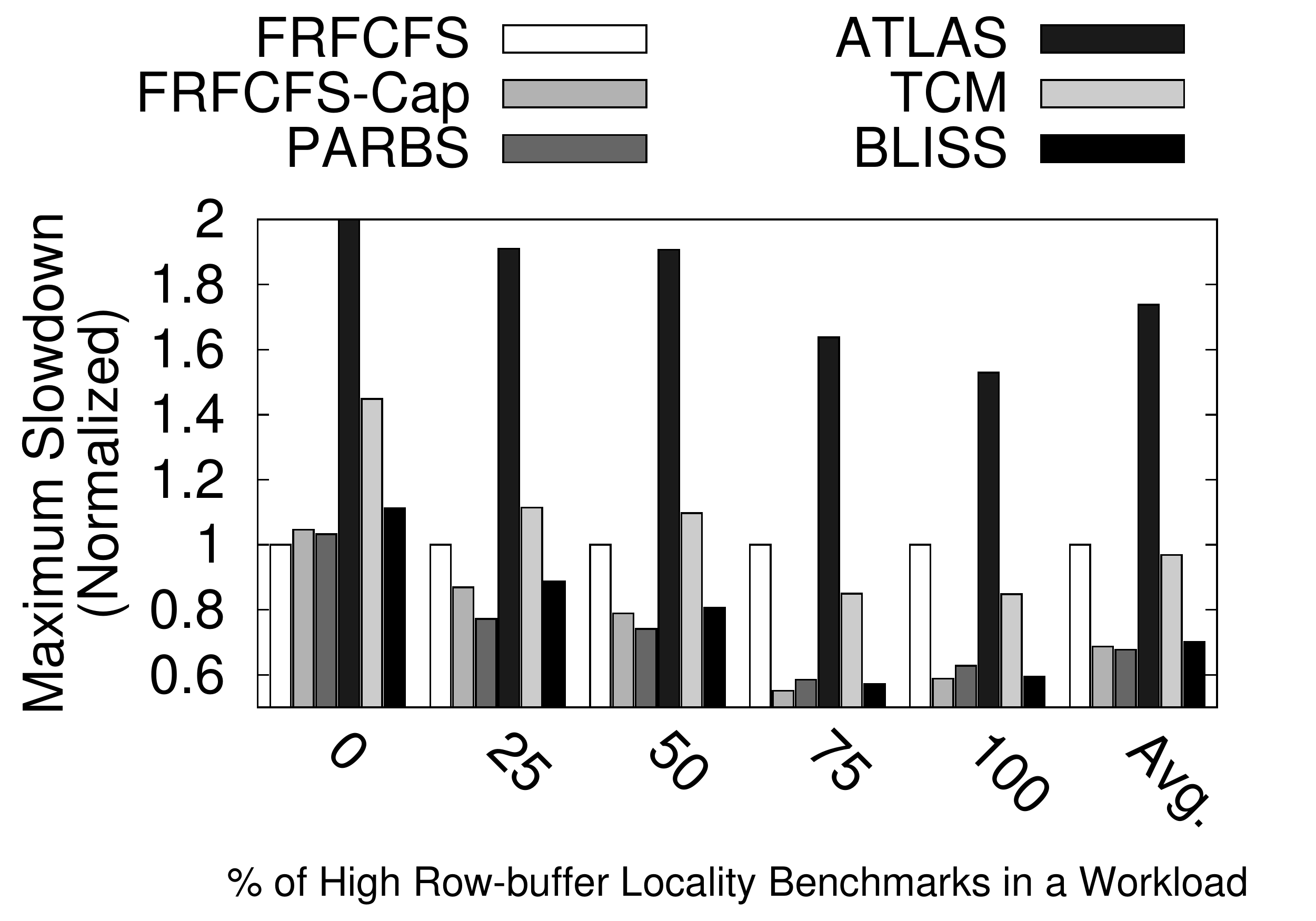}
  \end{minipage}
  \caption{Sensitivity to row-buffer locality}
  \label{fig:row-locality-results}
\end{figure}

\noindent\textbf{Workload Row-buffer Locality.}
Figure~\ref{fig:row-locality-results} shows the system performance
and fairness of five previous schedulers and BLISS when the number
of high row-buffer locality applications in a workload is
varied.\footnote{We classify an application as having high
row-buffer locality if its row-buffer hit rate is greater than
90\%.} We draw three observations. First, BLISS achieves the best
performance and close to the best fairness in most row-buffer
locality categories. Second, BLISS' performance and fairness
benefits over baseline FRFCFS increase as the number of
high-row-buffer-locality applications in a workload increases.
As the number of high-row-buffer-locality
applications in a workload increases, there is more interference to
the low-row-buffer-locality applications that are vulnerable.
Hence, there is more opportunity for BLISS to mitigate this
interference and improve performance and fairness. Third, when all
applications in a workload have high row-buffer locality (100\%),
the performance and fairness improvements of BLISS over baseline
FRFCFS are a bit lower than the other categories. This is because,
when all applications have high row-buffer locality, they each hog
the row-buffer in turn and are not as susceptible to interference
as the other categories in which there are vulnerable
low-row-buffer-locality applications. However, the
performance/fairness benefits of BLISS are still significant since BLISS is
effective in regulating how the row-buffer is shared among 
different applications. Overall, we conclude that BLISS is
effective in achieving high performance and fairness across
workloads with different compositions of high- and low-row-buffer-locality
applications.

\subsection{Sensitivity to System Parameters}
\label{sec:sensitivity-system}

\begin{figure}[h!]
    \centering
    \begin{minipage}{0.48\textwidth}
       \centering
       \includegraphics[scale=0.3, angle=270]{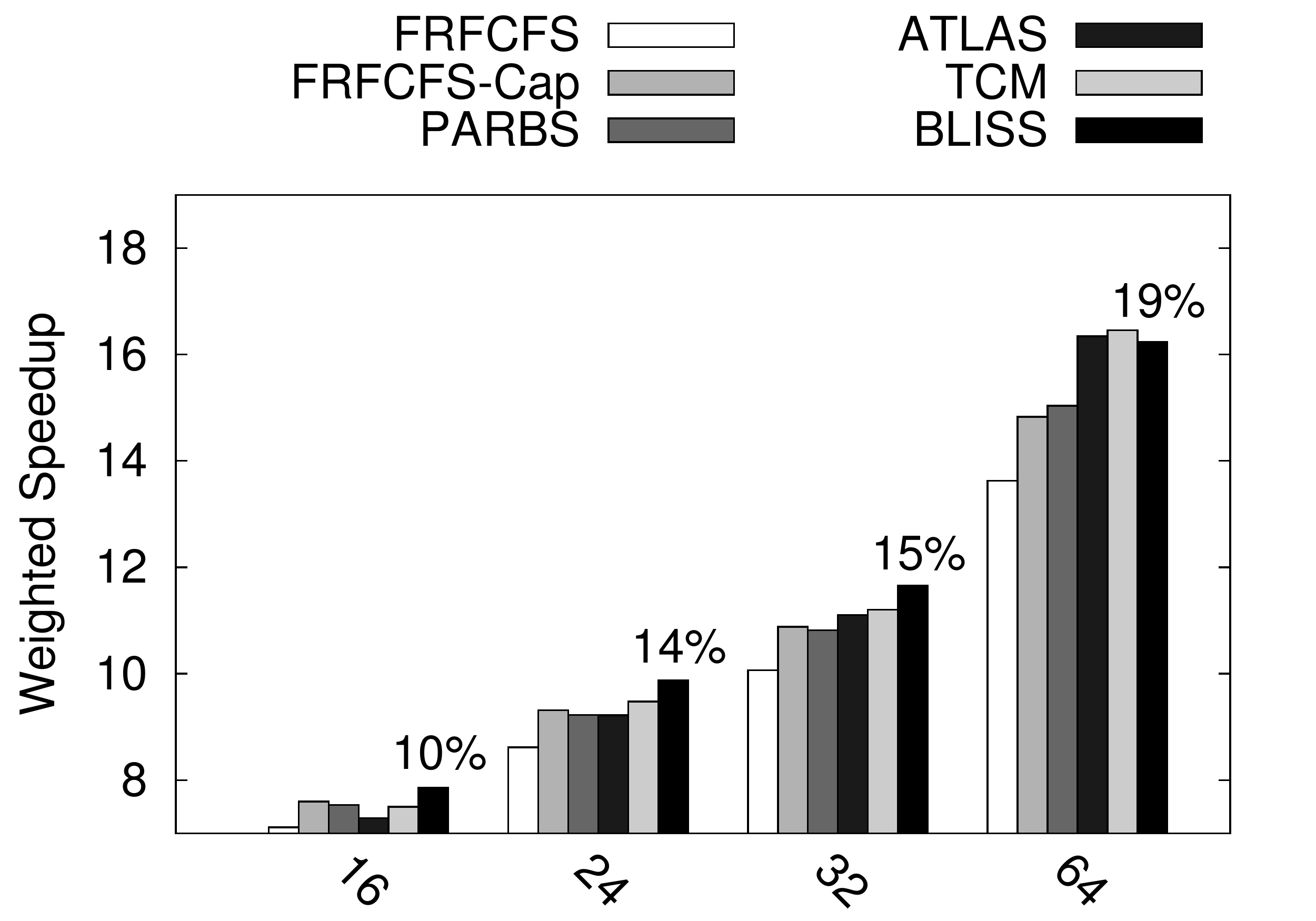}
    \end{minipage}
    \begin{minipage}{0.48\textwidth}
       \centering
        \includegraphics[scale=0.3, angle=270]{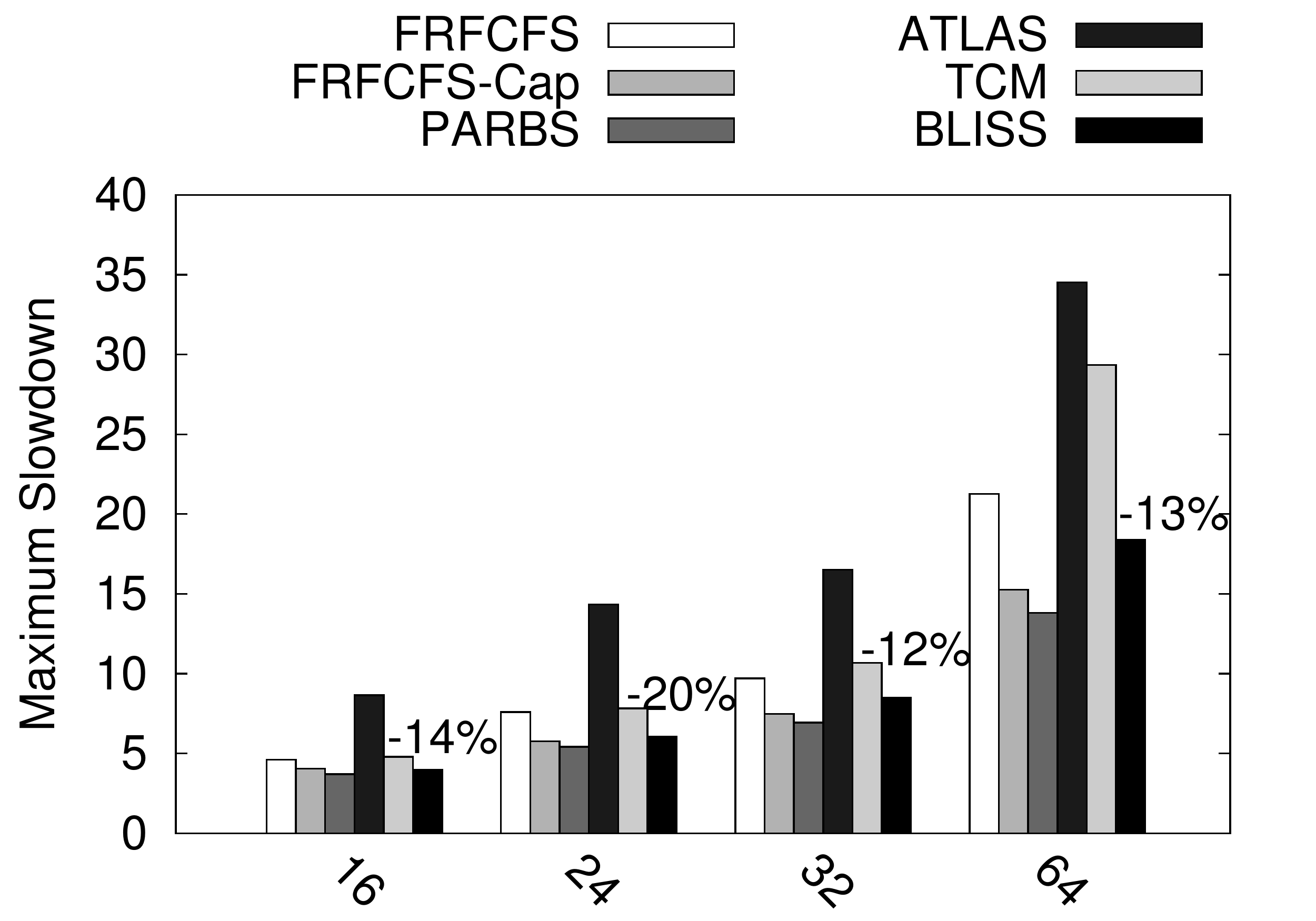}
    \end{minipage}
    \caption{Sensitivity to number of cores}
    \label{fig:sensitivity-num-cores}
\end{figure}

\begin{figure}[h!]
    \vspace{-7mm}
    \centering
    \begin{minipage}{0.48\textwidth}
       \centering
       \includegraphics[scale=0.3, angle=270]{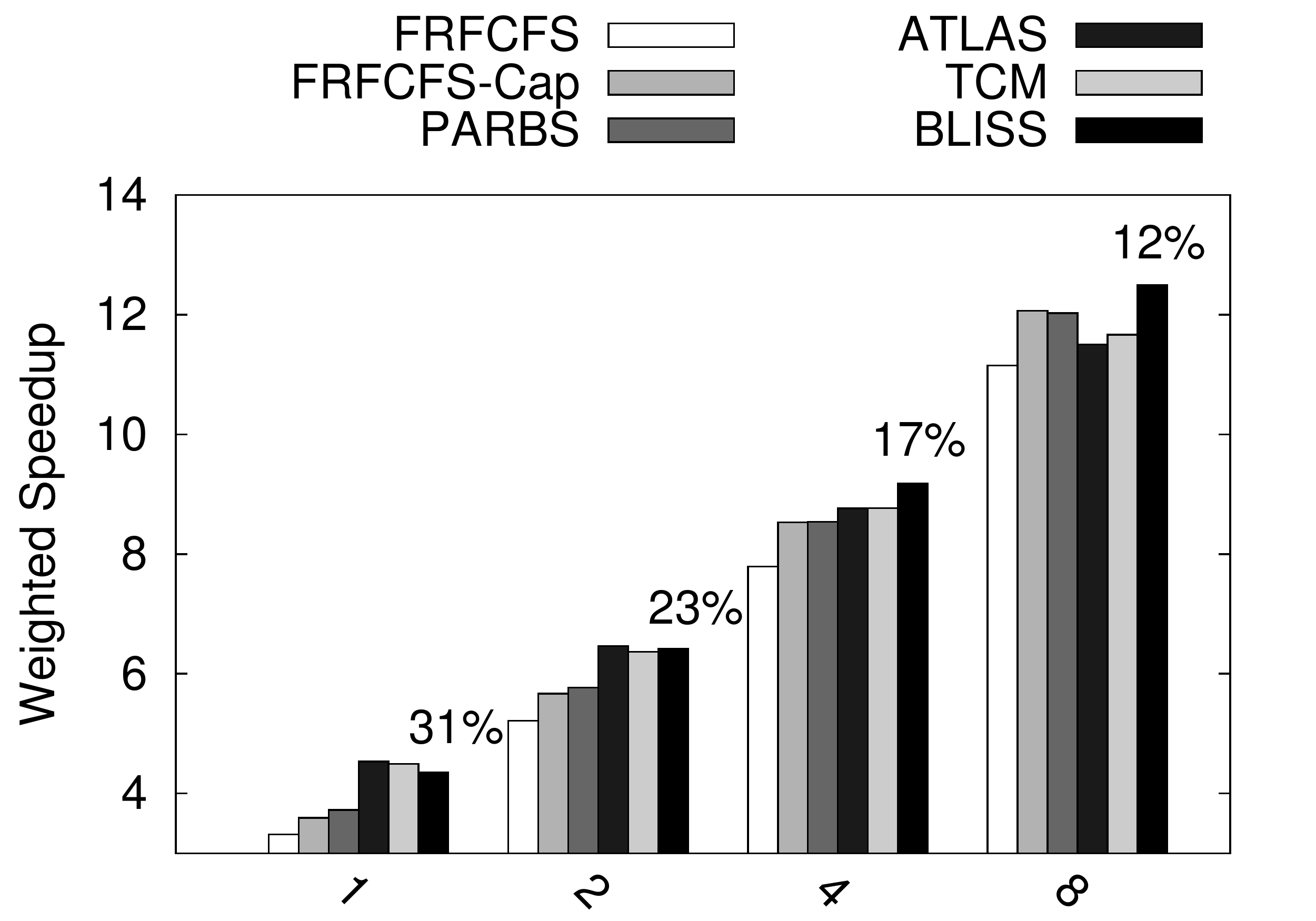}
    \end{minipage}
    \begin{minipage}{0.48\textwidth}
       \centering
        \includegraphics[scale=0.3, angle=270]{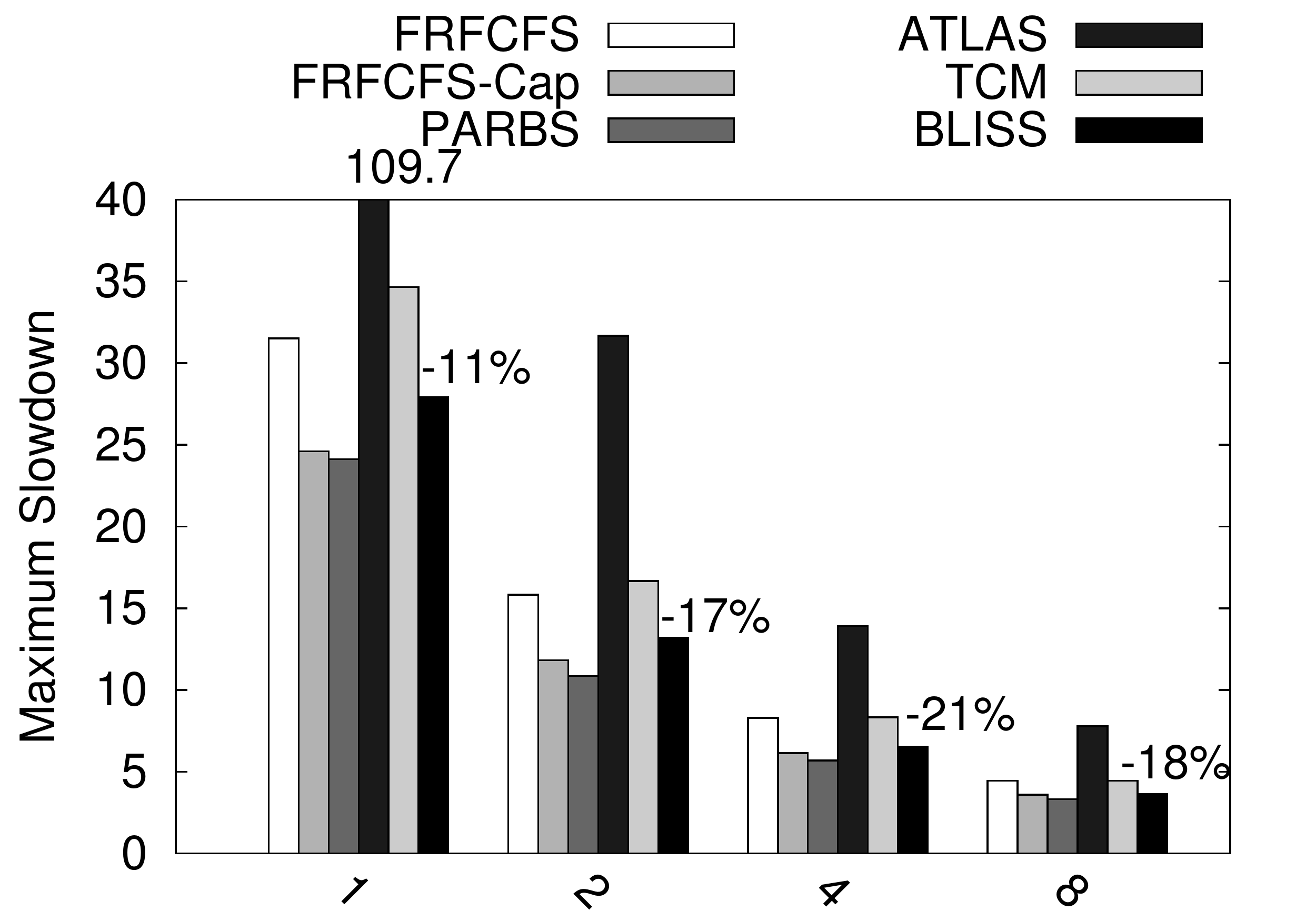}
    \end{minipage}
    \caption{Sensitivity to number of channels}
    \label{fig:sensitivity-num-channels}
    \vspace{-4mm}
\end{figure}

\noindent\textbf{Core and channel count.}
Figures~\ref{fig:sensitivity-num-cores}
and~\ref{fig:sensitivity-num-channels} show the system performance
and fairness of FRFCFS, PARBS, TCM and \bliss for different core
counts (when the channel count is 4) and different channel counts
(when the core count is 24), across 40 workloads for each
core/channel count. The numbers over the bars indicate percentage
increase or decrease compared to FRFCFS. We did not optimize the
parameters of different schedulers for each configuration as this
requires months of simulation time. We draw three major
conclusions. First, the absolute values of weighted speedup
increase with increasing core/channel count, whereas the absolute
values of maximum slowdown increase/decrease with increasing
core/channel count respectively, as expected. 
Second, \bliss achieves higher system performance and lower
unfairness than all the other scheduling policies (except PARBS,
in terms of fairness) similar to our results on the 24-core,
4-channel system, by virtue of its effective interference
mitigation. The only anomaly is that TCM has marginally higher
weighted speedup than \bliss for the 64-core system. However, this
increase comes at the cost of significant increase in unfairness.
Third, \bliss' system performance benefit (as indicated by the
percentages on top of bars, over FRFCFS) increases when the system
becomes more bandwidth constrained, i.e., high core counts and low
channel counts. As contention increases in the system, \bliss has
greater opportunity to mitigate it.\footnote{Fairness benefits
reduce at very high core counts and very low channel counts, since
memory bandwidth becomes highly saturated.}

\begin{figure}[ht!]
    \centering
    \begin{minipage}{0.48\textwidth}
       \centering
       \includegraphics[scale=0.3, angle=270]{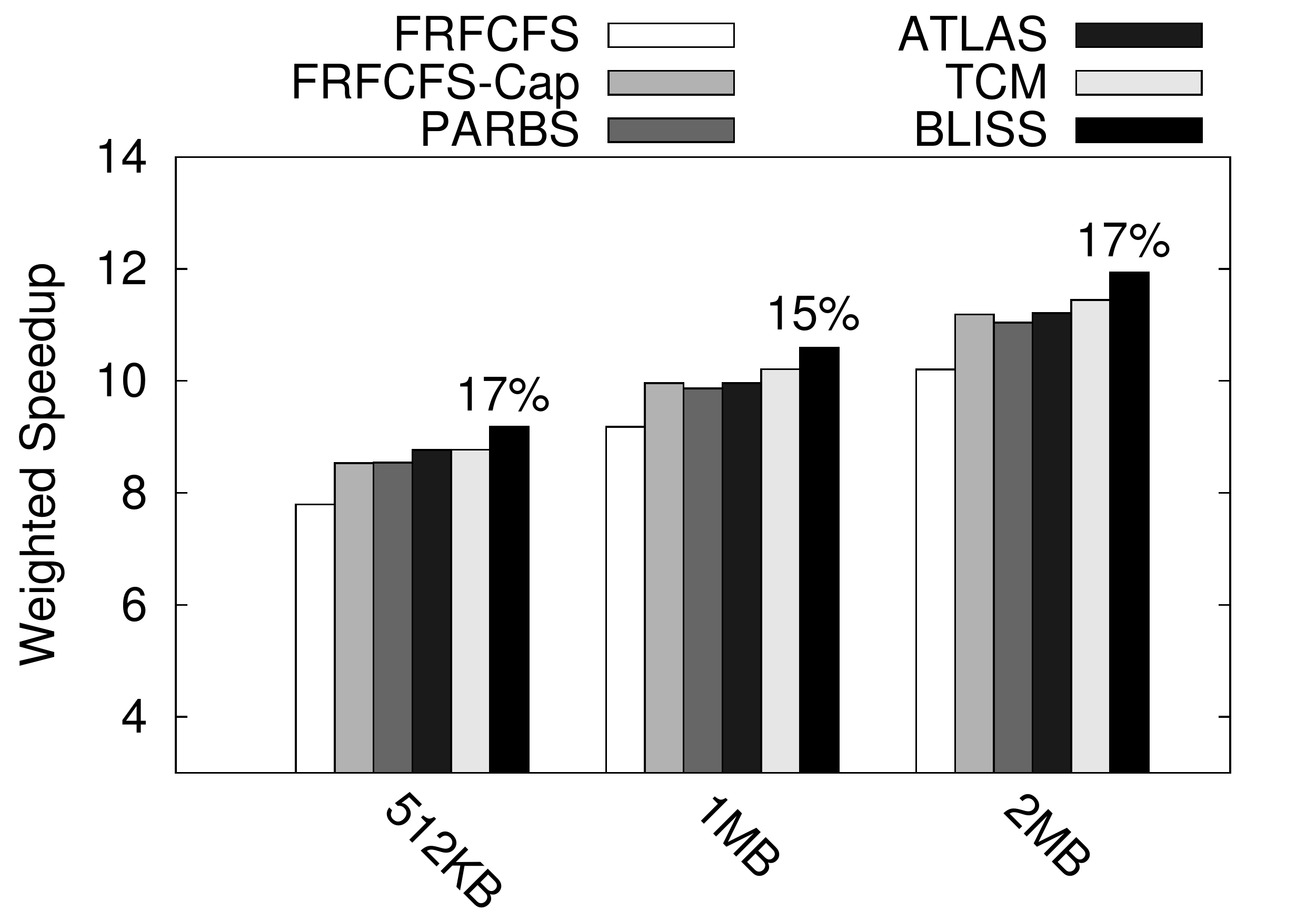}
    \end{minipage}
    \begin{minipage}{0.48\textwidth}
       \centering
        \includegraphics[scale=0.3, angle=270]{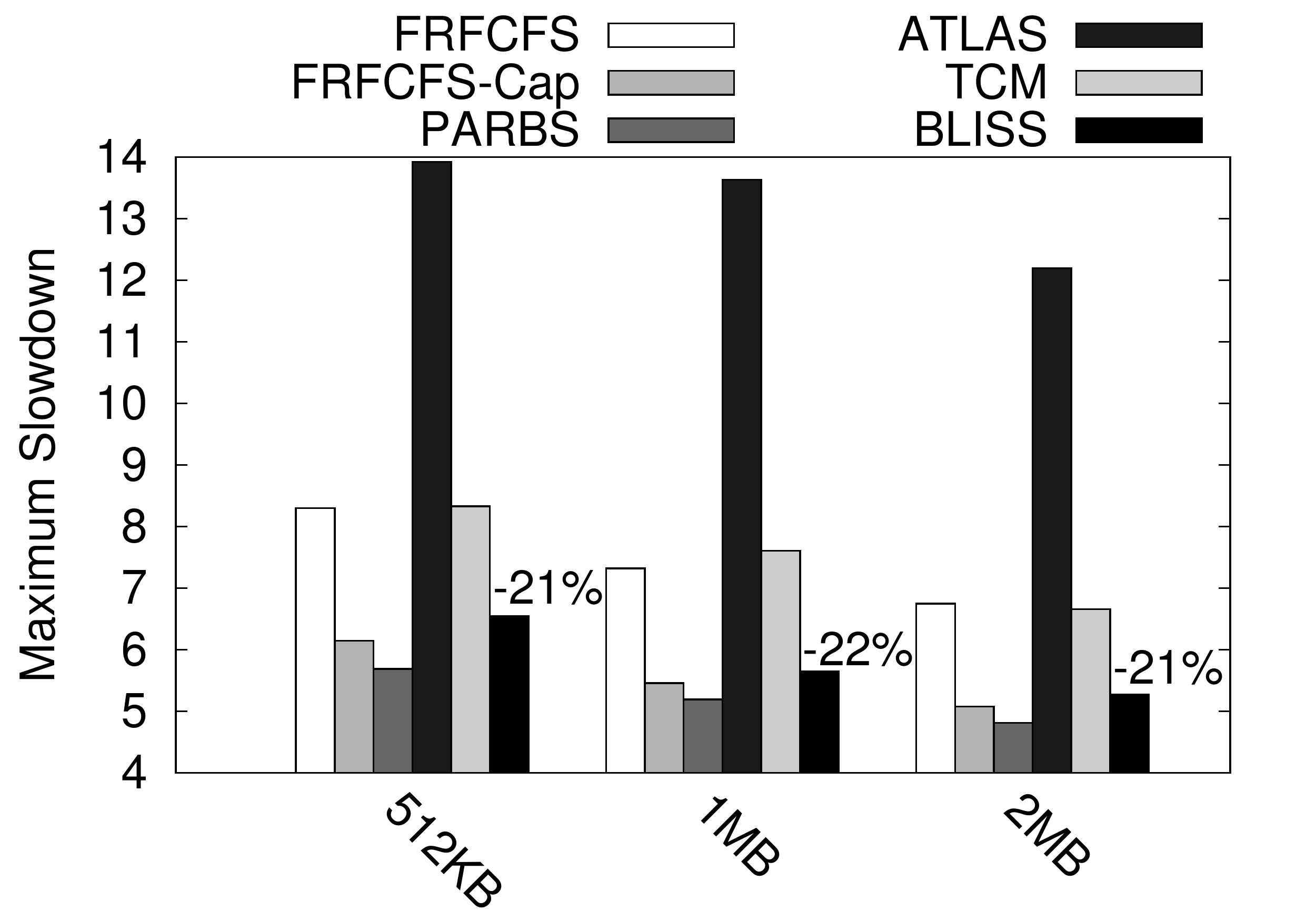}
    \end{minipage}
    \caption{Sensitivity to cache size}
    \label{fig:sensitivity-cache-size}
\end{figure}

\noindent\textbf{Cache size.}
Figure~\ref{fig:sensitivity-cache-size} shows the system
performance and fairness for five previous schedulers and BLISS
with different last level cache sizes (private to each core).  

We make two observations. First, the absolute values of weighted
speedup increase and maximum slowdown decrease, as the cache size
becomes larger for all schedulers, as expected. This is because
contention for memory bandwidth reduces with increasing cache
capacity, improving performance and fairness. Second, across all
the cache capacity points we evaluate, BLISS achieves significant
performance and fairness benefits over the best-performing
previous schedulers, while approaching close to the fairness of
the fairest previous schedulers.

\begin{figure}[ht!]
    \centering
    \begin{minipage}{0.48\textwidth}
       \centering
       \includegraphics[scale=0.3, angle=270]{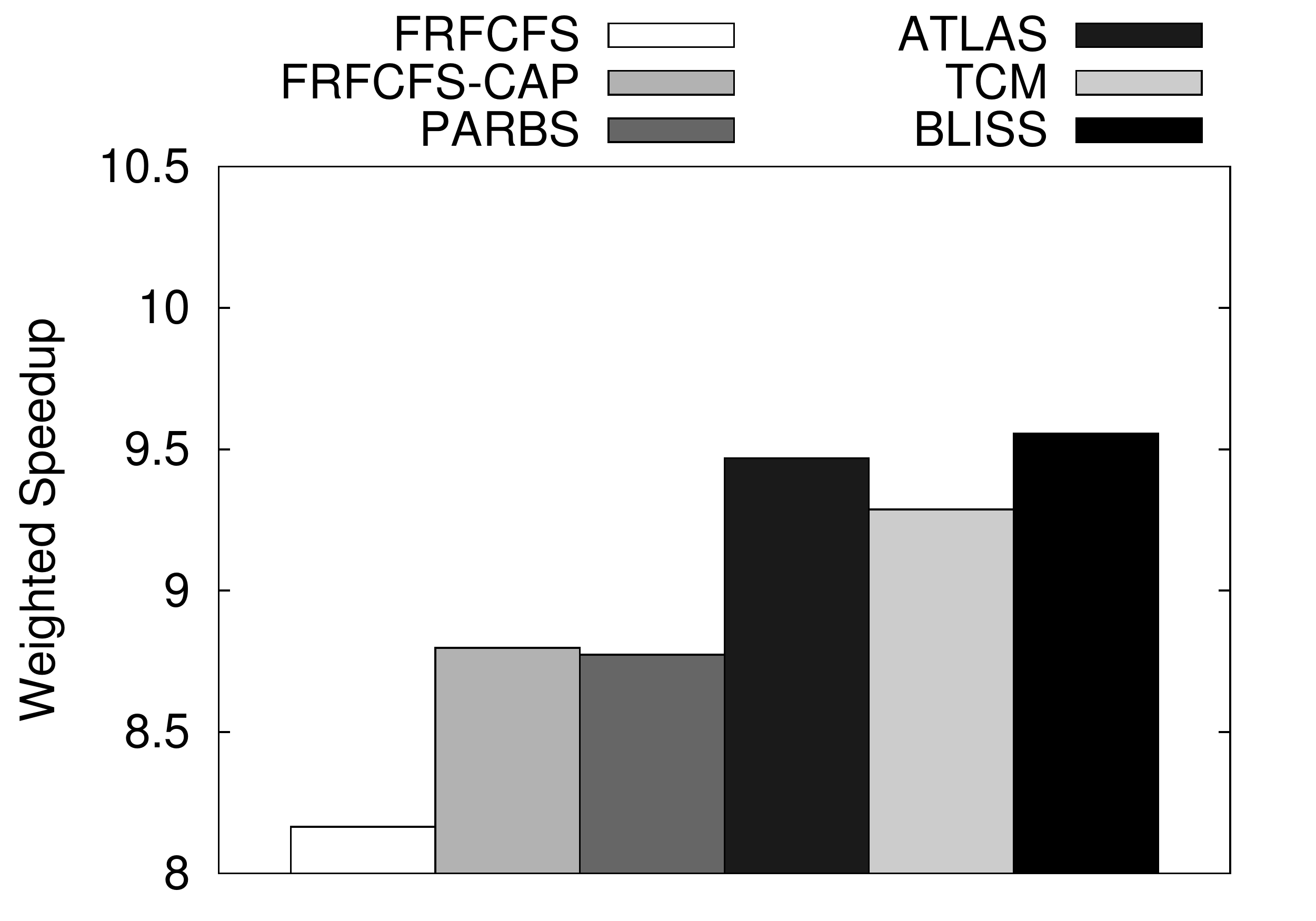}
    \end{minipage}
    \begin{minipage}{0.48\textwidth}
       \centering
        \includegraphics[scale=0.3, angle=270]{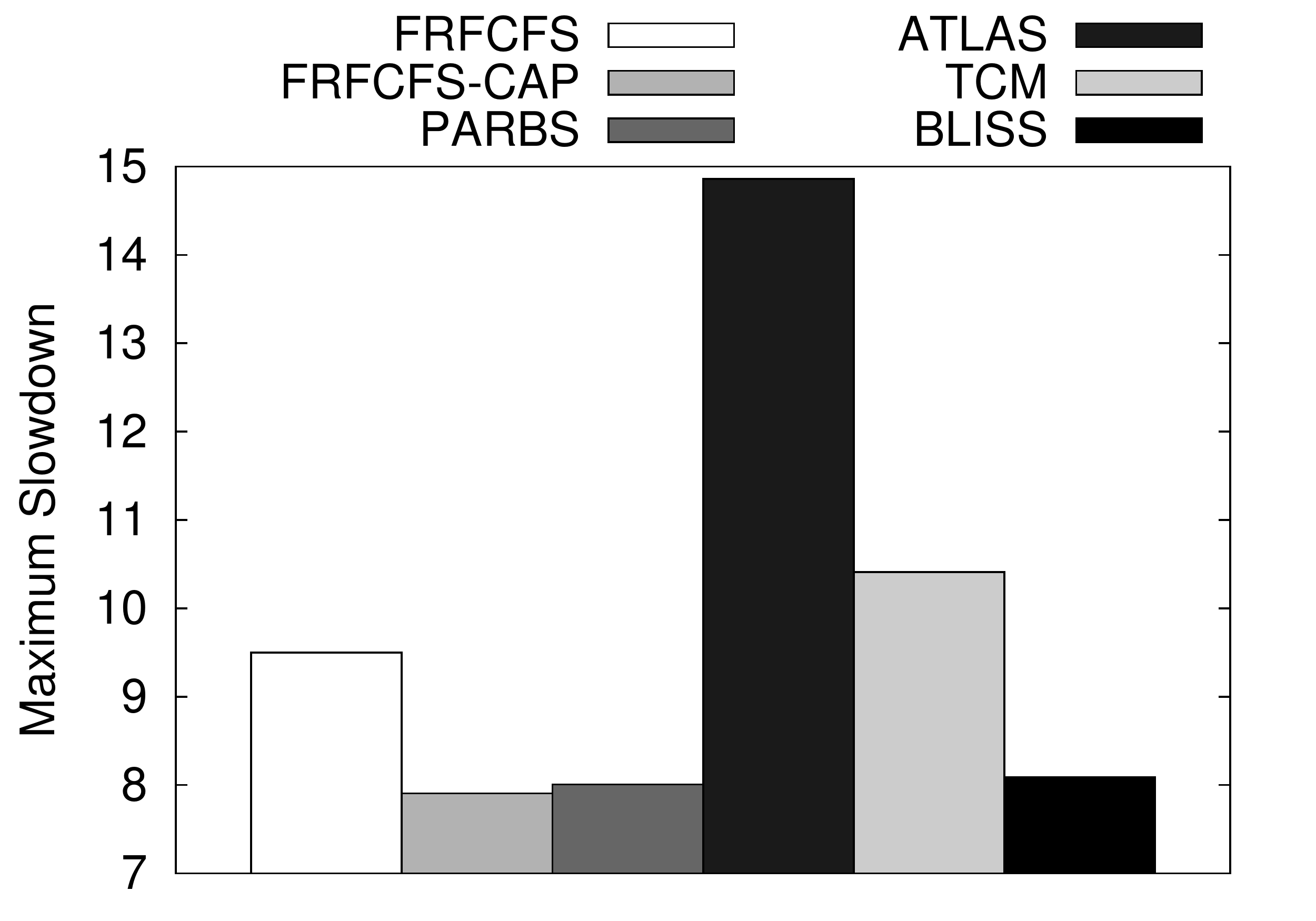}
    \end{minipage}
    \caption{Performance and fairness with a shared cache}
    \label{fig:shared-cache}
\end{figure}

\noindent\textbf{Shared Caches.} Figure~\ref{fig:shared-cache}
shows system performance and fairness with a 32 MB shared cache
(instead of the 512 KB per core private caches used in our other
experiments). \bliss achieves 5\%/24\% better performance/fairness
compared to TCM, demonstrating that \bliss is effective in
mitigating memory interference in the presence of large shared caches as
well.

\subsection{Sensitivity to Algorithm Parameters}
\label{sec:sensitivity-algorithm}

Tables~\ref{tab:sensitivity-ws} and \ref{tab:sensitivity-ms} show
the system performance and fairness respectively of \bliss for
different values of the \textit{Blacklisting Threshold} and
\textit{Clearing Interval}. Three major conclusions are in order.
First, a \textit{Clearing Interval} of 10000 cycles provides a
good balance between performance and fairness. If the blacklist is
cleared too frequently (1000 cycles), interference-causing
applications are not deprioritized for long enough, resulting in
low system performance. In contrast, if the blacklist is cleared
too infrequently, interference-causing applications are
deprioritized for too long, resulting in high unfairness. Second,
a \textit{Blacklisting Threshold} of 4 provides a good balance
between performance and fairness. When \textit{Blacklisting
Threshold} is very small, applications are blacklisted as soon as
they have very few requests served, resulting in poor interference
mitigation as too many applications are blacklisted. On the other
hand, when \textit{Blacklisting Threshold} is large, low- and
high-memory-intensity applications are not segregated effectively,
leading to high unfairness.

\begin{table}[ht]
  \centering
  \input{blacklisting/tables/sensitivity-ws}

  \caption{Performance sensitivity to threshold and interval}
  \label{tab:sensitivity-ws}
\end{table}

\begin{table}[ht]
  \centering
  \input{blacklisting/tables/sensitivity-ms}

  \caption{Unfairness sensitivity to threshold and interval}
  \label{tab:sensitivity-ms}
\end{table}

\subsection{Interleaving and Scheduling Interaction}
\label{sec:interleaving-interaction}

In this section, we study the impact of the address interleaving
policy on the performance and fairness of different schedulers.
Our analysis so far has assumed a row-interleaved policy, where
data is distributed across channels, banks and rows at the
granularity of a row. This policy optimizes for row-buffer
locality by mapping a consecutive row of data to the same channel,
bank, rank. In this section, we will consider two other
interleaving policies, cache block interleaving and sub-row
interleaving.

\noindent\textbf{Interaction with cache block interleaving.} In a
cache-block-interleaved system, data is striped across channels,
banks and ranks at the granularity of a cache block.  Such a
policy optimizes for bank level parallelism, by distributing data
at a small (cache block) granularity across channels, banks and
ranks. 

\begin{figure}[ht]
  \centering
  \begin{minipage}{0.48\textwidth}
    \centering
    \includegraphics[scale=0.3, angle=270]{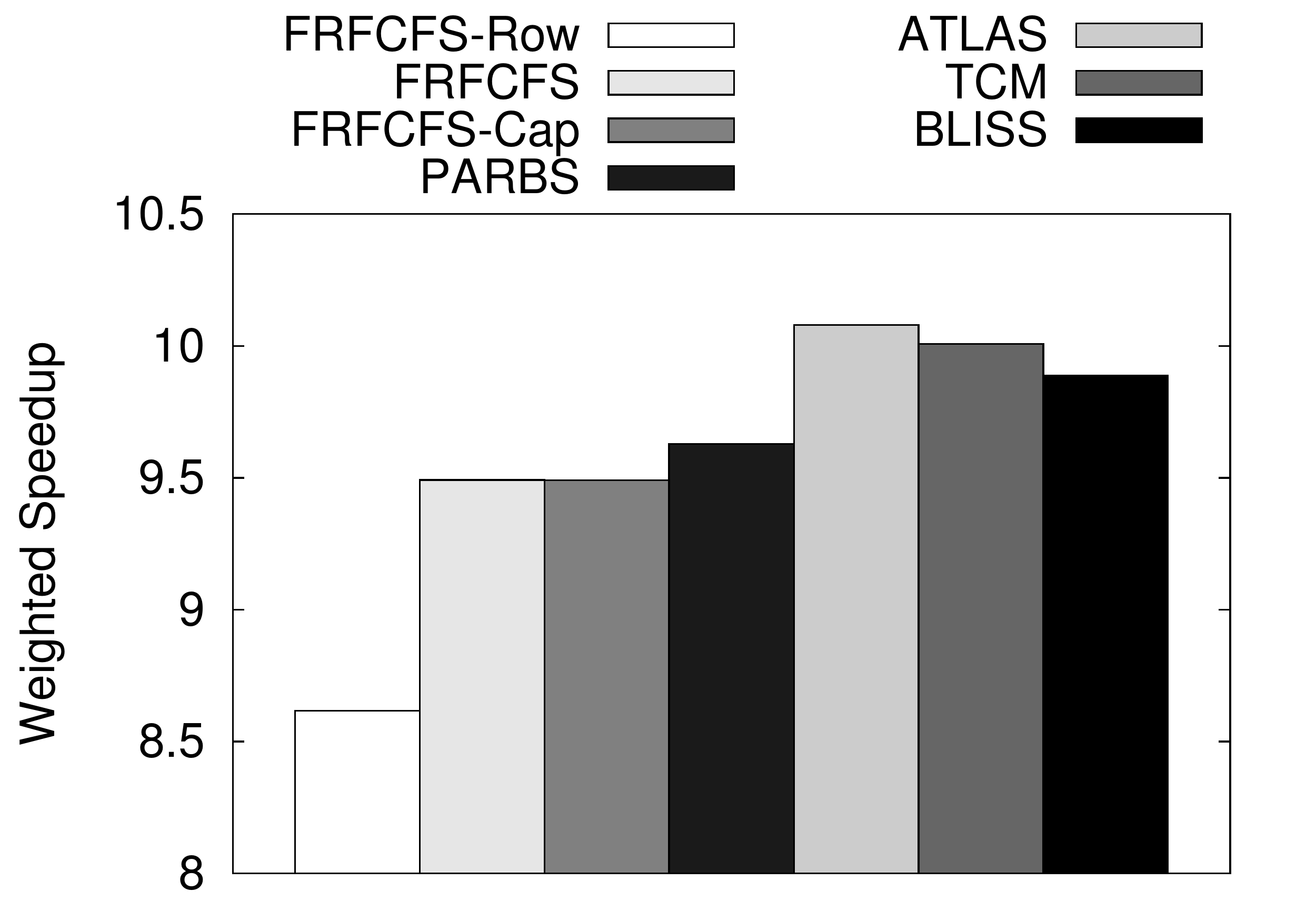}
  \end{minipage}
  \begin{minipage}{0.48\textwidth}
    \centering
    \includegraphics[scale=0.3, angle=270]{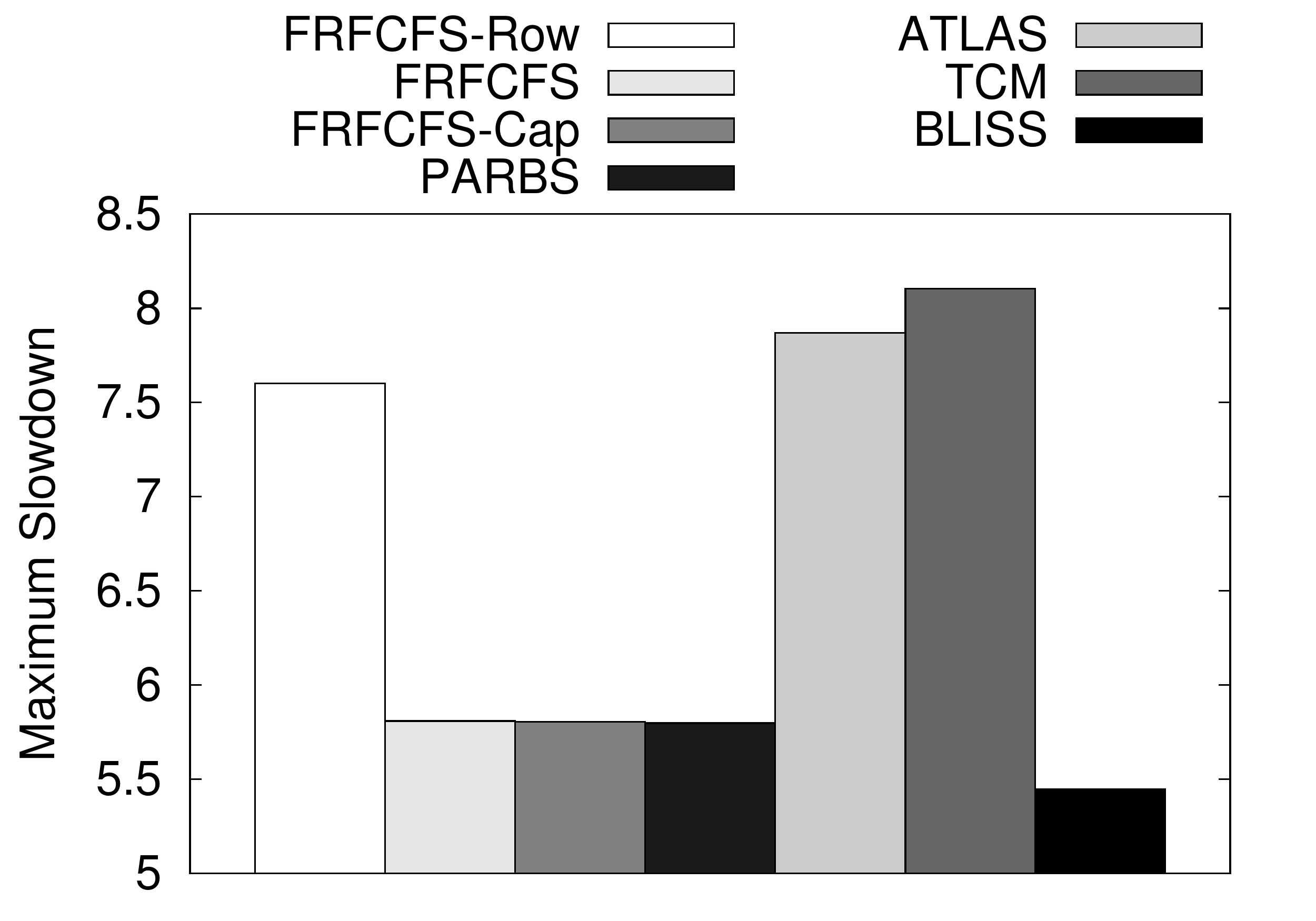}
  \end{minipage}
  \caption{Scheduling and cache block interleaving}
  \label{fig:cacheblockint-interaction}
\end{figure}

Figure~\ref{fig:cacheblockint-interaction} shows the system
performance and fairness of FRFCFS with row interleaving
(FRFCFS-Row), as a comparison point, five previous schedulers, and
BLISS with cache block interleaving. We draw three observations.
First, system performance and fairness of the baseline FRFCFS
scheduler improve significantly with cache block interleaving,
compared to with row interleaving. This is because cache block
interleaving enables more requests to be served in parallel at the
different channels and banks, by distributing data across channels
and banks at the small granularity of a cache block. Hence, most
applications, and particularly, applications that do not have very
high row-buffer locality benefit from cache block interleaving.

Second, as expected, application-aware schedulers such as ATLAS
and TCM achieve the best performance among previous schedulers, by
means of prioritizing requests of applications with low memory
intensities. However, PARBS and FRFCFS-Cap do not improve fairness
over the baseline, in contrast to our results with row
interleaving. This is because cache block interleaving already
attempts to provide fairness by increasing the parallelism in the
system and enabling more requests from across different
applications to be served in parallel, thereby reducing unfair
applications slowdowns. More specifically, requests that would
be row-buffer hits to the same bank, with row interleaving,
are now distributed across multiple channels and banks, with cache
block interleaving. Hence, applications' propensity to cause
interference reduces, providing lower scope for request capping
based schedulers such as FRFCFS-Cap and PARBS to mitigate
interference.  Third, BLISS achieves within 1.3\% of the
performance of the best performing previous scheduler (ATLAS),
while achieving 6.2\% better fairness than the fairest previous
scheduler (PARBS). BLISS effectively mitigates interference by
regulating the number of consecutive requests served from
high-memory-intensity applications that generate a large number of
requests, thereby achieving high performance and fairness.\\
\noindent\textbf{Interaction with sub-row interleaving.} While
memory scheduling has been a prevalent approach to mitigate memory
interference, previous work has also proposed other solutions, as
we describe in Chapter~\ref{chap:background}. One such previous
work by Kaseridis et al.~\cite{mop} proposes {\em minimalist open
page}, an interleaving policy that distributes data across
channels, ranks and banks at the granularity of a sub-row (partial
row), rather than an entire row, exploiting both row-buffer
locality and bank-level parallelism. We examine BLISS' interaction
with such a sub-row interleaving policy.

\begin{figure}[h]
  \centering
  \begin{minipage}{0.48\textwidth}
    \centering
    \includegraphics[scale=0.3, angle=270]{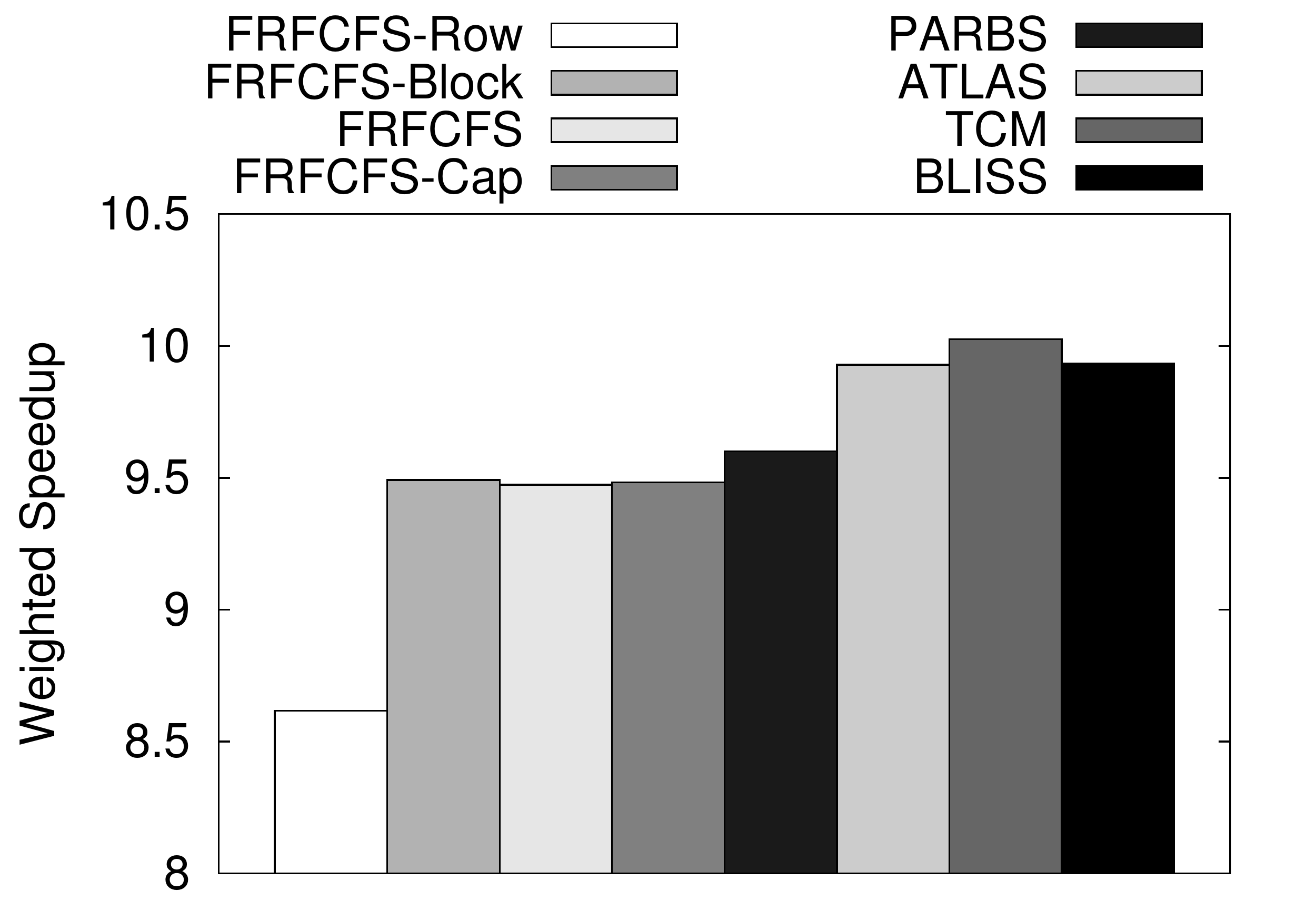}
  \end{minipage}
%  \vspace{-2mm}
  \begin{minipage}{0.48\textwidth}
    \centering
    \includegraphics[scale=0.3, angle=270]{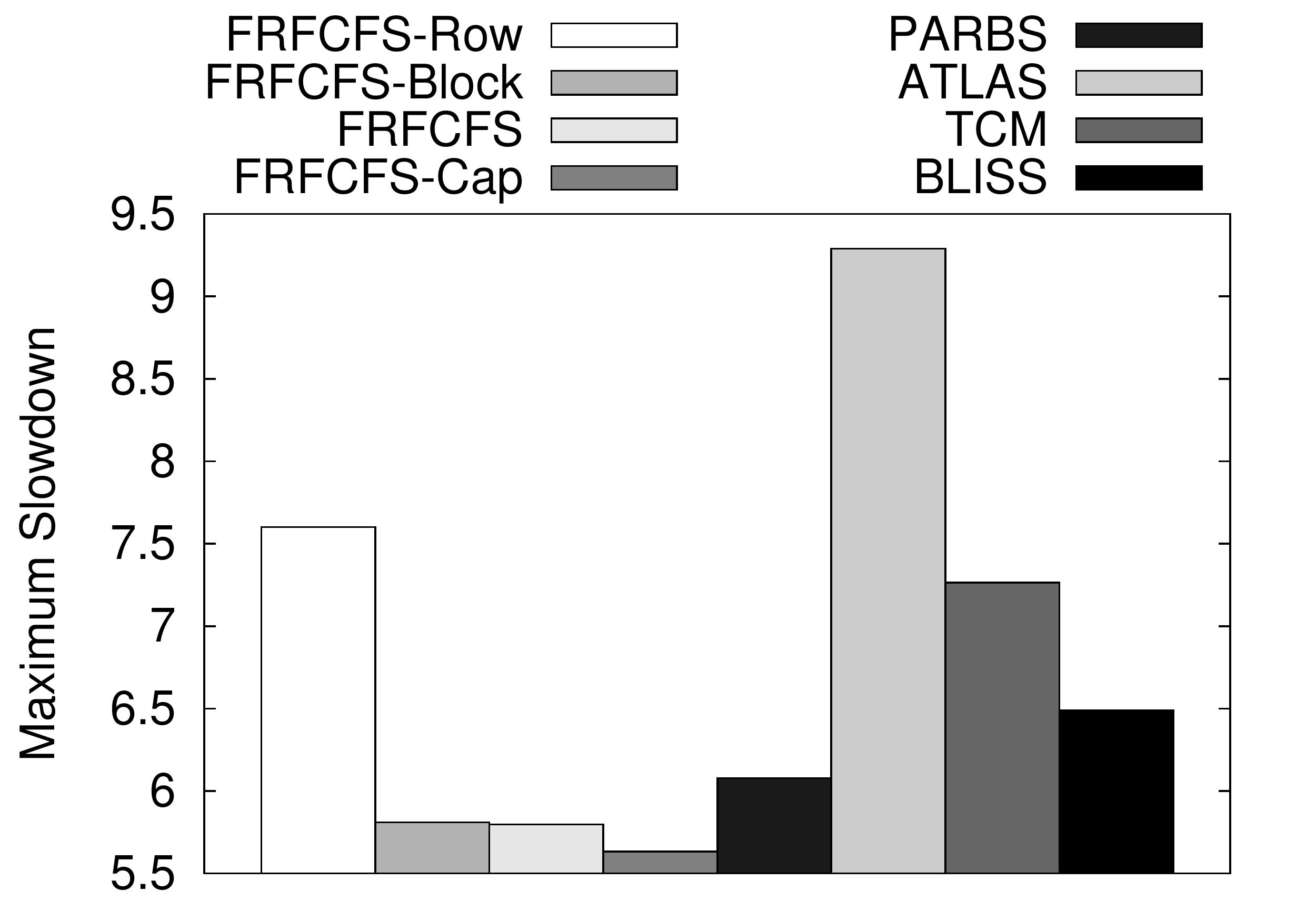}
  \end{minipage}
  \caption{Scheduling and sub-row interleaving}
  \label{fig:subrowint-interaction}
\end{figure}

Figure~\ref{fig:subrowint-interaction} shows the system
performance and fairness of FRFCFS with row interleaving
(FRFCFS-Row), FRFCFS with cache block interleaving (FRFCFS-Block)
and five previously proposed schedulers and \bliss, with sub-row
interleaving (when data is striped across channels, ranks and
banks at the granularity of four cache blocks). 

Three observations are in order. First, sub-row interleaving provides significant
benefits over row interleaving, as can be observed for FRFCFS (and
other scheduling policies by comparing with
Figure~\ref{fig:main-results}). This is because sub-row
interleaving enables applications to exploit both row-buffer
locality and bank-level parallelism, unlike row interleaving that
is mainly focused on exploiting row-buffer locality. Second,
sub-row interleaving achieves similar performance and fairness as
cache block interleaving. We observe that this is because cache
block interleaving enables applications to exploit parallelism
effectively, which makes up for the lost row-buffer locality from
distributing data at the granularity of a cache block across all
channels and banks. Third, BLISS achieves close to the performance
(within 1.5\%) of the best performing previous scheduler (TCM),
while reducing unfairness significantly and approaching the
fairness of the fairest previous schedulers. One thing to note is
that BLISS has higher unfairness than FRFCFS, when a
sub-row-interleaved policy is employed. This is because the
capping decisions from sub-row interleaving and BLISS could
collectively restrict high-row-buffer locality applications to a
large degree, thereby slowing them down and causing higher
unfairness. Co-design of the scheduling and interleaving policies
to achieve different goals such as performance/fairness is an
important area of future research. We conclude that a \bliss-like
scheduler, with its high performance and low complexity is a
significantly better alternative to schedulers such as ATLAS/TCM
in the pursuit of such scheduling-interleaving policy co-design.

%% file: blacklisting/tables/individual-clearing.tex
\begin{tabular}{|c|c|c|c|}
    \hline
    Metric & BLISS & BLISS-Individual-Clearing\\
    \hline
    Weighted Speedup & 9.18 & 9.12\\
    \hline
    Maximum Slowdown & 6.54 & 6.60\\
    \hline
\end{tabular}

%% file: blacklisting/tables/sensitivity-ws.tex
\begin{tabular}{|c|c|c|c|c|c|}
  \hline
  \backslashbox{Threshold}{Interval} & 1000 & 10000 & 100000\\
  \hline
  2 & 8.76 & 8.66 & 7.95\\
  \hline
  4 & 8.61 & 9.18 & 8.60\\
  \hline
  8 & 8.42 & 9.05 & 9.24\\
  \hline
\end{tabular}

%% file: blacklisting/tables/sensitivity-ms.tex
\begin{tabular}{|c|c|c|c|c|c|}
  \hline
  \backslashbox{Threshold}{Interval} & 1000 & 10000 & 100000\\
  \hline
  2 & 6.07 & 6.24 & 7.78\\
  \hline
  4 & 6.03 & 6.54 & 7.01\\
  \hline
  8 & 6.02 & 7.39 & 7.29\\
  \hline
\end{tabular}

%% file: blacklisting/summary.tex
\section{Summary}

In summary, the Blacklisting memory scheduler (\bliss) is a new and
simple approach to memory scheduling in systems with multiple threads. We
observe that the per-application ranking mechanisms employed by
previously proposed application-aware memory schedulers incur high
hardware cost, cause high unfairness, and lead to high scheduling
latency to the point that the scheduler cannot meet the fast
command scheduling requirements of state-of-the-art DDR protocols.
\bliss overcomes these problems based on the key observation that it is
sufficient to group applications into only two groups, rather than
employing a total rank order among all applications. Our
evaluations across a variety of workloads and systems demonstrate
that \bliss has better system performance and fairness than
previously proposed ranking-based schedulers, while incurring
significantly lower hardware cost and latency in making scheduling
decisions.

%% file: chapters/mise-model.tex
\newcommand{\arsr}{\emph{alone-request-service-rate}\xspace}
\newcommand{\srsr}{\emph{shared-request-service-rate}\xspace}
\newcommand{\arsrabb}{\texttt{ARSR}\xspace}
\newcommand{\srsrabb}{\texttt{SRSR}\xspace}
\newcommand{\miiseabb}{MISE\xspace}
\newcommand{\miise}{Memory Interference induced Slowdown Estimation\xspace}
\newcommand{\miisefair}{MISE-Fair\xspace}
\newcommand{\miiseqos}{MISE-QoS\xspace}

\chapter{Quantifying Application Slowdowns Due to Main Memory Interference}
\label{chap:mise}

In a multicore system, an application's performance and slowdowns
depend heavily on its corunning applications and the amount of
shared resource interference they cause, as we demonstrated and
discussed in Chapter~\ref{chap:introduction}. While the Blacklisting
Scheduler (BLISS) is able to achieve high system
performance and fairness at low hardware complexity in the
presence of main memory interference, it does not have
the ability to estimate and control application slowdowns.
%Hence, it cannot achieve predictable performance. 

The ability to accurately estimate application slowdowns can
enable several use cases. For instance, estimating the slowdown of each
application may enable a cloud service
provider~\cite{azure-billing,ec2-billing} to estimate the
performance provided to each application in the presence of
consolidation on shared hardware resources, thereby billing the
users appropriately. Perhaps more importantly, accurate slowdown
estimates may enable allocation of shared resources to different
applications in a slowdown-aware manner, thereby satisfying
different applications' performance requirements.

Mechanisms and models to accurately estimate application slowdowns
due to shared resource interference have not been explored as much
as shared resource interference mitigation techniques have.
Furthermore, the few previous works on slowdown estimation, STFM~\cite{stfm},
FST~\cite{fst} and PTCA~\cite{ptca} are inaccurate, as we briefly discuss
in Section~\ref{sec:related-work-slowdown-estimation}. These
works estimate slowdown as the ratio of uninterfered to interfered
stall/execution times. The uninterfered stall/execution times are computed by
estimating the number of cycles by which the interference
experienced by each individual request impacts execution time. Given the
abundant parallelism available in the memory subsystem, service of
different requests overlap significantly. As a result, accurately
estimating the number of cycles by which each request is delayed
due to interference is inherently difficult, thereby resulting in
high inaccuracies in the slowdown estimates.

We seek to accurately estimate application slowdowns due to memory
bandwidth interference, as a key step towards controlling
application slowdowns. Towards this end, we first build the Memory
Interference induced Slowdown Estimation (MISE) model to
accurately estimate application slowdowns in the presence of
memory bandwidth interference.

\input{mise/model}
\input{mise/implementation}
\input{mise/methodology}

\input{mise/comparison}

\input{mise/sensitivity}
\input{mise/summary}

%% file: mise/model.tex
\section{The \miiseabb Model}
\label{sec:mise-model}

In this section, we provide a detailed description of our \miise
(\miiseabb) model that estimates application slowdowns due to
memory bandwidth interference. For ease of understanding, we first
describe the observations that lead to a simple model for
estimating the slowdown of a memory-bound application when it is
run concurrently with other applications
(Section~\ref{sec:slowdown-memory-bound-app}). In
Section~\ref{sec:slowdown-non-memory-bound-app}, we describe how
we extend the model to accommodate non-memory-bound applications.
Section~\ref{sec:model-implementation} describes the detailed
implementation of our model in a memory controller.

\subsection{Memory-bound Application}
\label{sec:slowdown-memory-bound-app}

A memory-bound application is one that spends an overwhelmingly
large fraction of its execution time stalling on memory
accesses. Therefore, the rate at which such an application's
requests are served has significant impact on its
performance. More specifically, we make the following observation
about a memory-bound application.
\begin{quote}\parskip0pt
  \textbf{Observation 1:} \emph{The performance of a memory-bound
    application is roughly proportional to the rate at which its memory
    requests are served.}
\end{quote}
For instance, for an application that is bottlenecked at memory,
if the rate at which its requests are served is reduced by half,
then the application will take twice as much time to finish the
same amount of work. To validate this observation, we conducted a
real-system experiment where we ran memory-bound applications from
SPEC CPU2006~\cite{spec2006} on a 4-core Intel Core
i7~\cite{nehalem}. Each SPEC application was run along with three
copies of a microbenchmark whose memory intensity can be
varied.\footnote{The microbenchmark streams through a large region
of memory (one block at a time). The memory intensity of the
microbenchmark (LLC MPKI) is varied by changing the amount of
computation performed between memory operations.} By varying the
memory intensity of the microbenchmark, we can change the rate at which
the requests of the SPEC application are served.

\begin{figure}[h]
  \centering
  \includegraphics[scale=0.3, angle=270]{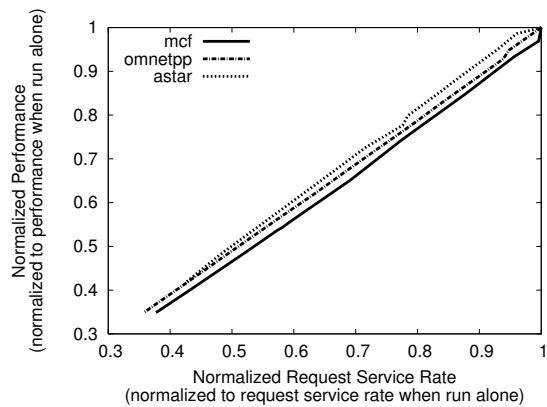}
  \caption{Request service rate vs. performance}
  \label{fig:bwfraction-speedup}
\end{figure}

Figure~\ref{fig:bwfraction-speedup} plots the results of this
experiment for three memory-intensive SPEC benchmarks, namely,
\emph{mcf}, \emph{omnetpp}, and \emph{astar}. The figure shows the
performance of each application vs. the rate at which its requests
are served. The request service rate and performance are
normalized to the request service rate and performance
respectively of each application when it is run alone on the same
system.

The results of our experiments validate our observation. The
performance of a memory-bound application is directly proportional
to the rate at which its requests are served. This suggests that
we can use the request-service-rate of an application as a proxy
for its performance. More specifically, we can compute the
slowdown of an application, i.e., the ratio of its performance
when it is run alone on a system vs. its performance when it is
run alongside other applications on the same system, as follows:
\begin{equation}
  \textrm{Slowdown of an App.} =
  \frac{\textrm{\arsr}}{\textrm{\srsr}}
  \label{eqn:slowdown-memory-bound}
\end{equation}
Estimating the \srsr (\srsrabb) of an application is
straightforward. It just requires the memory controller to keep
track of how many requests of the application are served in a
given number of cycles. However, the challenge is to
estimate the \arsr (\arsrabb) of an application \emph{while} it is
run alongside other applications. A naive way of estimating
\arsrabb of an application would be to prevent all other
applications from accessing memory for a length of time and
measure the application's \arsrabb.  While this would provide an
accurate estimate of the application's \arsrabb, this approach
would significantly slow down other applications in the system.
Our second observation helps us to address this problem.
\begin{quote}
  \textbf{Observation 2:} \emph{The \arsrabb of an application can
    be estimated by giving the requests of the application the
    highest priority in accessing memory.}
\end{quote}
Giving an application's requests the highest priority in accessing
memory results in very little interference from the requests of
other applications. Therefore, many requests of the application
are served as if the application were the only one running on the
system. Based on the above observation, the \arsrabb of an
application can be computed as follows:
\begin{equation}
  \textrm{\arsrabb of an App.} = \frac{\textrm{\# Requests with
      Highest Priority}}{\textrm{\# Cycles with Highest Priority}}
  \label{eqn:arsr}
\end{equation}
where \emph{\# Requests with Highest Priority} is the number of
requests served when the application is given highest priority,
and \emph{\# Cycles with Highest Priority} is the number of cycles
an application is given highest priority by the memory controller.

The memory controller can use Equation~\ref{eqn:arsr} to
periodically estimate the \arsrabb of an application and
Equation~\ref{eqn:slowdown-memory-bound} to measure the slowdown
of the application using the estimated
\arsrabb. Section~\ref{sec:model-implementation} provides a
detailed description of the implementation of our model inside a
memory controller.

\subsection{Non-memory-bound Application}
\label{sec:slowdown-non-memory-bound-app}

So far, we have described our \miiseabb model for a memory-bound
application. We find that the model presented above has low
accuracy for non-memory-bound applications. This is because a
non-memory-bound application spends a significant fraction of its execution
time in the {\em compute phase} (when the core is not stalled
waiting for memory). Hence, varying the request service rate for such an
application will not affect the length of the large compute phase.
Therefore, we take into account the duration of the compute phase
to make the model accurate for non-memory-bound applications.

Let $\alpha$ be the fraction of time spent by an application at
memory. Therefore, the fraction of time spent by the
application in the compute phase is $1-\alpha$. Since changing the
request service rate affects only the memory phase, we augment
Equation~\ref{eqn:slowdown-memory-bound} to take into account
$\alpha$ as follows:
\begin{equation}
  \emph{Slowdown of an App.} = (1 - \alpha) + \alpha
  \frac{\arsrabb}{\srsrabb}
  \label{eqn:slowdown-non-memory-bound}
\end{equation}
In addition to estimating \arsrabb and \srsrabb required by
Equation~\ref{eqn:slowdown-memory-bound}, the above equation
requires estimating the parameter $\alpha$, the fraction of time
spent in memory phase. However, precisely computing $\alpha$
for a modern out-of-order processor is a challenge since such a
processor overlaps computation with memory accesses. The processor
stalls waiting for memory only when the oldest instruction in the
reorder buffer is waiting on a memory request. For this reason,
we estimate $\alpha$ as the fraction of time the processor spends
stalling for memory.
\begin{equation}
  \alpha = \frac{\textrm{\# Cycles spent stalling on memory
      requests}}{\textrm{Total number of cycles}}
  \label{eqn:alpha}
\end{equation}
Setting $\alpha$ to 1 reduces
Equation~\ref{eqn:slowdown-non-memory-bound} to
Equation~\ref{eqn:slowdown-memory-bound}.  We find that even
when an application is moderately memory-intensive, setting
$\alpha$ to 1 provides a better estimate of slowdown. Therefore,
our final model for estimating slowdown takes into account the
stall fraction ($\alpha$) only when it is
low. Algorithm~\ref{alg:final-model} shows our final slowdown
estimation model.

%In fact, we find that
%when the fraction of time an application spends stalling for
%memory is high (close to 1), setting $\alpha$ to 1 provides a
%better estimate of slowdown. Therefore, our final model for
%estimating slowdown takes into account the stall fraction
%($\alpha$) only when it is low. Algorithm~\ref{alg:final-model}
%shows our final slowdown estimation model.

\begin{algorithm}
  \hrule\vspace{2mm}
  Compute $\alpha$\;
  \eIf{$\alpha < $ Threshold}{
    Slowdown $ = (1 - \alpha) + \alpha\frac{\arsrabb}{\srsrabb}$
  }{
    Slowdown $ = \frac{\arsrabb}{\srsrabb}$
  }
  \hrule\vspace{2mm}
  \caption{\textbf{The \miiseabb model}}
  \label{alg:final-model}
\end{algorithm}
\vspace{-10pt}

%% file: mise/implementation.tex
\vspace{-3mm}
\section{Implementation}
\label{sec:model-implementation}

In this section, we describe a detailed implementation of our
\miiseabb model in a memory controller. For each application in
the system, our model requires the memory controller to compute
three parameters: 1)~\srsr (\srsrabb), 2)~\arsr (\arsrabb), and
3)~$\alpha$ (stall fraction).\footnote{These three parameters need
to be computed only for the active applications in the system.
Hence, these need to be tracked only per hardware thread context.}
First, we describe the scheduling algorithm employed by the memory
controller. Then, we describe how the memory controller computes
each of the three parameters.

\newcommand{\interval}{\emph{interval}\xspace}
\newcommand{\epoch}{\emph{epoch}\xspace}
\newcommand{\intervals}{\emph{intervals}\xspace}
\newcommand{\epochs}{\emph{epochs}\xspace}

\subsection{Memory Scheduling Algorithm}
\label{sec:implementation-scheduling}

In order to implement our model, each application needs to be
given the highest priority periodically, such that its \arsr can be
measured. This can be achieved by simply assigning each
application's requests highest priority in a round-robin manner.
However, the mechanisms we build on top of our model allocate
bandwidth to different applications to achieve QoS/fairness.
Therefore, in order to facilitate the implementation of our
mechanisms, we employ a lottery-scheduling-like
approach~\cite{lottery-scheduling,lottery-scheduling-waldspurger}
to schedule requests in the memory controller. The basic idea of
lottery scheduling is to probabilistically enforce a given
bandwidth allocation, where each application is allocated a
certain share of the bandwidth. The exact bandwidth
allocation policy depends on the goal of the system -- e.g., QoS,
high performance, high fairness, etc. In this section, we describe
how a lottery-scheduling-like algorithm works to enforce a
bandwidth allocation.

The memory controller divides execution time into \intervals (of
$\mathcal{M}$ processor cycles each). Each interval is further divided
into small \epochs (of $\mathcal{N}$ processor cycles each). At the
beginning of each \interval, the memory controller estimates the
slowdown of each application in the system. Based on the slowdown
estimates and the final goal, the controller may change the
bandwidth allocation policy -- i.e., redistribute bandwidth
amongst the concurrently running applications. At the beginning of
each \epoch, the memory controller probabilistically picks a
single application and prioritizes all the requests of that
particular application during that epoch. The probability
distribution used to choose the prioritized application is such that an
application with higher bandwidth allocation has a higher
probability of getting the highest priority. For example, consider
a system with two applications, $A$ and $B$. If the memory
controller allocates $A$ 75\% of the memory bandwidth and $B$ the
remaining 25\%, then $A$ and $B$ get the highest priority with 
probability 0.75 and 0.25, respectively.

\subsection{Computing \srsr (\srsrabb)}
The \srsr of an application is the rate at which the application's
requests are served while it is running with other
applications. This can be directly computed by the memory
controller using a per-application counter that keeps track of the
number of requests served for that application. At the beginning
of each \interval, the controller resets the counter for each
application. Whenever a request of an application is served, the
controller increments the counter corresponding to that
application. At the end of each \interval, the \srsrabb of an
application is computed as
\begin{eqnarray*}
  \textrm{\srsrabb of an App} = \frac{\textrm{\# Requests served}}{\mathcal{M}\textrm{ (Interval Length)}}
\end{eqnarray*}
%where $\mathcal{M}$ is the length of each interval.

\subsection{Computing \arsr (\arsrabb)}
\label{sec:implementation-arsr}
The \arsr (\arsrabb) of an application is an estimate of the rate
at which the application's requests would have been served had it
been running alone on the same system. Based on our observation
(described in Section~\ref{sec:slowdown-memory-bound-app}), the
\arsrabb can be estimated by using the request-service-rate of the
application when its requests have the highest priority in
accessing memory. Therefore, the memory controller estimates the
\arsrabb of an application only during the \epochs in which the
application has the highest priority.

Ideally, the memory controller should be able to achieve this
using two counters: one to keep track of the number of \epochs
during which the application received highest priority and another
to keep track of the number of requests of the application served
during its highest-priority \epochs. However, it is possible that
even when an application's requests are given highest priority,
they may receive interference from other applications'
requests. This is because, our memory scheduling is {\em work
conserving} -- if there are no requests from the highest priority
application, it schedules a ready request from some other
application. Once a request is scheduled, it cannot be preempted
because of the way DRAM operates.

% However, even when an
%application's requests are given the highest priority, they may
%receive some interference from other applications' requests. This
%is because, the memory controller schedules other applications'
%requests when the highest priority application does not have any
%outstanding requests, in order to avoid bandwidth wastage.
%Furthermore, request scheduling is non-preemptive because of the
%way DRAM is accessed. Hence, if a new request from the highest
%priority application arrives to a bank that's servicing another
%application's request, it would have to wait until the current
%request is completed. 

In order to account for this interference, the memory controller
uses a third counter for each application to track the
number of cycles during which an application's request was blocked
due to some other application's request, in spite of the former
having highest priority. For an application with highest
priority, a cycle is deemed to be an \emph{interference cycle} if
during that cycle, a command corresponding to a request of that
application is waiting in the request buffer and the previous
command issued to any bank,
%to the same bank
was for a request from a different application.

Based on the above discussion, the memory controller keeps track
of three counters to compute the \arsrabb of an application:
1)~number of highest-priority \epochs of the application (\#~HPEs),
2)~number of requests of that application served during its
highest-priority \epochs (\#~HPE Requests), and 3)~number of
\emph{interference cycles} of the application during its
highest-priority \epochs (\#~Interference cycles). All these
counters are reset at the start of an \interval and the \arsrabb
is computed at the end of each interval as follows:
\begin{eqnarray*}
  \textrm{\arsrabb of an App.} = \frac{\textrm{\# HPE
      Requests}}{\mathcal{N}.(\textrm{\# HPEs}) - (\textrm{\#
      Interference cycles})}
\end{eqnarray*}
%where $\mathcal{N}$ is the length of an \epoch.

Our model does not take into account bank level parallelism (BLP)
or row-buffer interference when estimating \textrm{\# Interference
cycles}. We observe that this does not affect the accuracy of our
model significantly.  because we eliminate most of the
interference by measuring \arsrabb only when an application has
highest priority. We leave a study of the effects of bank-level
parallelism and row-buffer interference on the accuracy of our
model as part of future work.

\subsection{Computing \emph{stall-fraction} $\alpha$}
The \emph{stall-fraction} ($\alpha$) is the fraction of the cycles
spent by the application stalling for memory requests. The number
of stall cycles can be easily computed by the core and
communicated to the memory controller at the end of each interval.

\subsection{Hardware Cost}
\label{sec:cost-complexity}

Our implementation incurs additional storage cost due to 1)~the
counters that keep track of parameters required to compute
slowdown (five per hardware thread context), and 2)~a register
that keeps track of the current bandwidth allocation policy (one
per hardware thread context). We find that using four byte
registers for each counter is more than sufficient for the values
they keep track of. Therefore, our model incurs a storage cost of
at most 24 bytes per hardware thread context.

%The lottery-scheduling-like policy used by the memory controller adds
%an additional layer of arbitration to the memory scheduling logic.
%While picking a request to schedule, the memory controller first
%checks if there is a request from the highest priority
%application. If there is one, it schedules it, otherwise it
%schedules another application's request. 

% Depending on the result, the memory controller uses
%the base scheduling policy to either pick one of the
%highest-priority application's requests or one of the other
%applications' requests (if no request of the highest-priority
%application is present). In our experiments, we use the
%conventional FR-FCFS scheduling policy as the base policy. Note
%that the additional layer of arbitration required to check for a
%highest-priority request is similar to the logic required to check
%for a ``First-Ready'' request in the FR-FCFS policy. Therefore, we
%expect the additional complexity introduced by the new layer to be
%negligible.

%% file: mise/methodology.tex
\section{Methodology}
\label{sec:mise-methodology}

\textbf{Simulation Setup.} We model the memory system using an in-house cycle-accurate
DDR3-SDRAM simulator.
% that was validated against Micron's Verilog
%behavioral model~\cite{ddr3verilog} and
%DRAMSim2~\cite{dramsim2}. 
We have integrated this DDR3 simulator
into an in-house cycle-level x86 simulator with a Pin~\cite{pin}
frontend, which models out-of-order cores with a limited-size
instruction window. Each core has a 512 KB private cache. We
model main memory as the only shared resource, in order to isolate
and analyze the effect of memory interference on application
slowdowns. Table~\ref{tab:mise-meth} provides more details of
the simulated systems.

Unless otherwise specified, the evaluated systems consist of 4
cores and a memory subsystem with 1 channel, 1 rank/channel and 8
banks/rank. We use row-interleaving to map the physical address
space onto DRAM channels, ranks and banks. Data is striped across
different channels, ranks and banks, at the granularity of a
row. Our workloads are made up of 26 benchmarks from the SPEC
CPU2006~\cite{spec2006} suite.
%, which are
%listed along with their miss rate (LLC Misses Per Kilo Instruction
%(MPKI)) and IPC in Table~\ref{tab:workload-stats}. 

%\begin{table*}
%  \centering
%  \input{mise/tables/workload-stats}
%  \caption{Benchmark Characteristics when run on the baseline
%4-core, 1-channel system}
%  \label{tab:workload-stats}
%\end{table*}

\noindent
\textbf{Workloads.} We form multiprogrammed workloads using combinations of these 26
benchmarks. We extract a representative phase of each benchmark
using PinPoints~\cite{pinpoints} and run that phase for 200
million cycles.  We will provide more details about our workloads
as and when required.

\input{mise/tables/methodology}

\noindent
\textbf{Metrics.} 
We use average error to compare the
accuracy of MISE and previously proposed models. We compute
slowdown estimation error for each application, at the end of
every quantum ($Q$), as the absolute value of\\
\begin{eqnarray*}
    \textrm{Error} = \frac{\textrm{Estimated Slowdown} - \textrm{Actual Slowdown}}{\textrm{Actual Slowdown}} \times 100\%\\
\end{eqnarray*}
\vspace{-25mm}
\begin{eqnarray*}
    \textrm{Actual Slowdown} = \frac{IPC_{alone}}{IPC_{shared}}
\end{eqnarray*}
We compute $IPC_{alone}$ for the same amount of work as the shared
run for each quantum. For each application, we compute the average
slowdown estimation error across all quanta in a workload run and
then compute the average across all occurrences of the application
in all of our workloads.

%We compute both weighted speedup~\cite{symbjobscheduling} and
%harmonic speedup~\cite{harmonic-speedup} to measure
%system performance. However, since the goal of our mechanisms is
%to provide QoS/high fairness, while ensuring good system
%performance, we mainly present harmonic speedup throughout our
%evaluations, as the harmonic speedup metric provides a good
%balance between system performance and
%fairness~\cite{harmonic-speedup}. We use the maximum
%slowdown~\cite{stc,atlas,tcm} metric to measure unfairness.

%\begin{small}
%  \begin{eqnarray*}
%    \textrm{Weighted Speedup} & = & \Sigma_i \frac{\sharedipc}{\aloneipc}\\
%    \textrm{Harmonic Speedup} & = & N/\left(\Sigma_i\frac{\aloneipc}{\sharedipc}\right)\\
%    \textrm{Maximum Slowdown} & = & \textrm{max}\left(\frac{\aloneipc}{\sharedipc}\right)
%  \end{eqnarray*}
%\end{small}

\noindent
\textbf{Parameters.} We use an interval length ($\mathcal{M}$) of 5 million
cycles and an epoch length ($\mathcal{N}$) of 10000 cycles for all
our evaluations. Section~\ref{sec:sensitivity} evaluates
sensitivity of our model to these parameters.

%% \noindent$Weighted Speedup = \Sigma_{i} \frac {IPC_{i}^{\scriptstyle{shared}}} {IPC_{i}^{alone}}$,
%% $Harmonic Speedup = \Sigma_i \frac {N}{\frac{IPC_{i}^{alone}} {IPC_{i}^{shared}}}$,
%% $Maximum Slowdown = max_i  \frac
%%          {IPC_{i}^{alone}}{IPC_{i}^{shared}} $.

%We use weighted speedup~\cite{symbjobscheduling} and harmonic
%speedup~\cite{harmonic-speedup} to
%measure system performance and maximum slowdown~\cite{atlas, tcm}
%to measure fairness.

%We compute both weighted speedup and harmonic speedup. However,
%the harmonic speedup metric provides a good balance between system
%performance and fairness~\cite{harmonic-speedup}. As our goal is to provide QoS/high
%fairness, while ensuring good system performance, we present alone
%harmonic speedup in our evaluations.

%% file: mise/tables/methodology.tex
\begin{table}[h!]
\begin{footnotesize}
  \centering
    \begin{tabular}{ll}
        \toprule
Processor           &  4-16 cores, 5.3GHz, 3-wide issue,\\ & 8 MSHRs, 128-entry instruction window     \\
        \cmidrule(rl){1-2}
Last-level cache    &  64B cache-line, 16-way associative,\\ & 512KB private cache-slice per core     \\
        \cmidrule(rl){1-2} Memory controller   &  64/64-entry read/write request queues per controller   \\
        \cmidrule(rl){1-2}
\multirow{2}[2]{*}{\centering Memory}              &  Timing: DDR3-1066 (8-8-8)~\cite{micron} \\
 & Organization: 1 channel, 1
 rank-per-channel,\\ & 8 banks-per-rank, 8 KB row-buffer \\ 
        \bottomrule
    \end{tabular}%
  \caption{Configuration of the simulated system}
  \label{tab:mise-meth}%
\end{footnotesize}%
\end{table}%

%% file: mise/comparison.tex
\section{Comparison to STFM}
\label{sec:comparison-stfm}

Stall-Time-Fair Memory scheduling (STFM)~\cite{stfm} is one of the
few previous works that attempt to estimate main-memory-induced
slowdowns of individual applications when they are run
concurrently on a multicore system. As we described in
Section~\ref{sec:related-work-slowdown-estimation} and earlier in
this chapter, STFM estimates the slowdown of an application by
estimating the number of cycles it stalls due to interference from
other applications' requests. Other previous works on slowdown
estimation~\cite{fst,ptca} estimate slowdown due to memory
bandwidth interference in a similar manner as STFM and in
addition, estimate slowdown due to shared cache interference as
well. Since MISE's focus is on slowdown estimation in the presence
of memory bandwidth interference, we qualitatively and
quantitatively compare \miiseabb with STFM.

There are two key differences between \miiseabb and
STFM for estimating slowdown. First, \miiseabb uses
request service rates rather than stall times to estimate
slowdown. As we mentioned in Section~\ref{sec:mise-model}, the \arsr of
an application can be fairly accurately estimated by giving the
application highest priority in accessing memory. Giving the
application highest priority in accessing memory results in very
little interference from other applications. In contrast, STFM
attempts to estimate the alone-stall-time of an application while
it is receiving significant interference from other
applications. Second, \miiseabb takes into account the effect of the
compute phase for non-memory-bound applications. STFM, on the
other hand, has no such provision to account for the compute
phase. As a result, \miiseabb's slowdown estimates for
non-memory-bound applications are significantly more accurate than
STFM's estimates.

\begin{figure*}[h!]
  \centering
  \begin{subfigure}{0.32\textwidth}
      \centering
      \includegraphics[scale=0.21, angle=0]{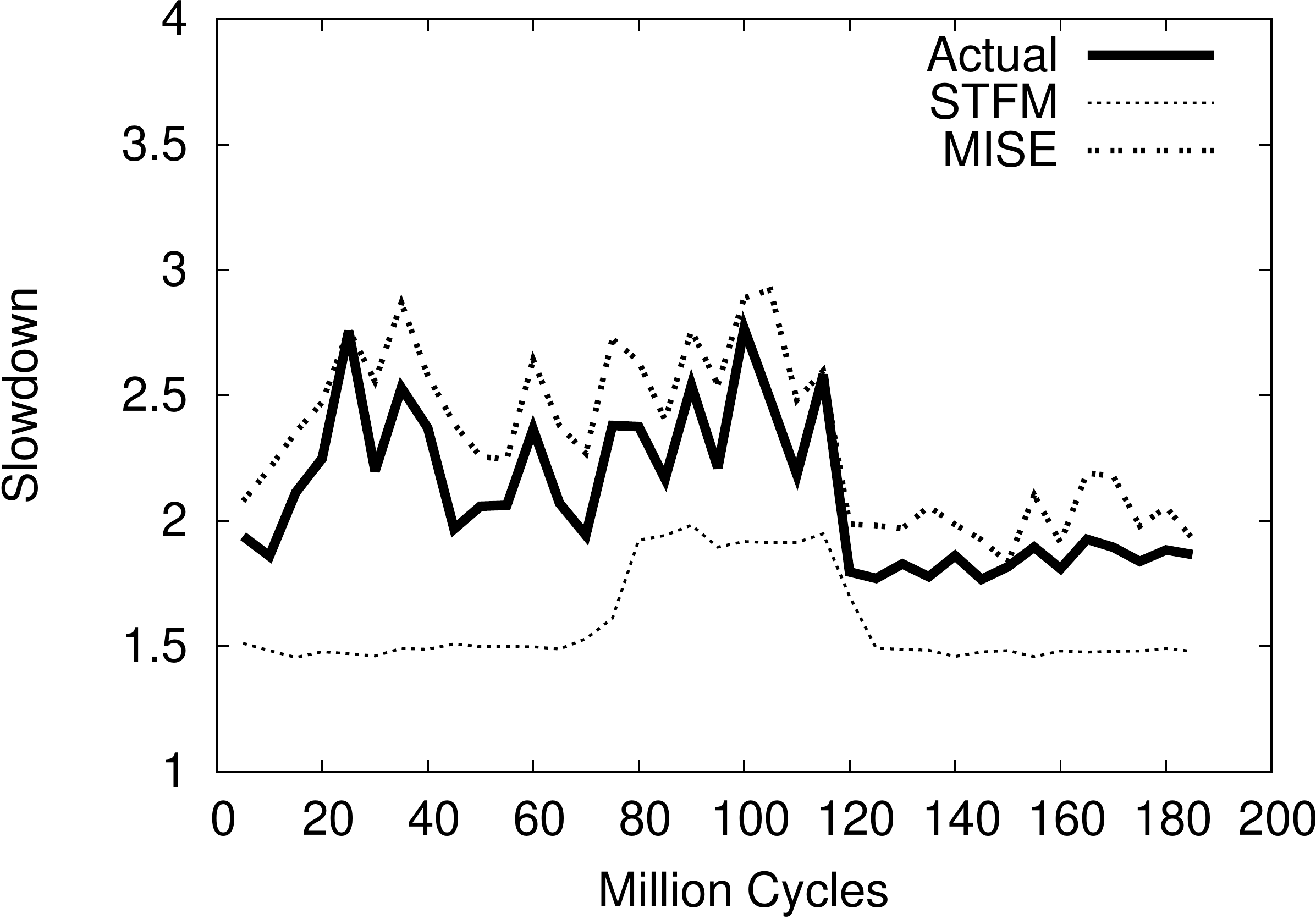}
      \subcaption{lbm}
      \label{fig:lbm-stfm-comp}
  \end{subfigure}
  \begin{subfigure}{0.32\textwidth}
      \centering
      \includegraphics[scale=0.21, angle=0]{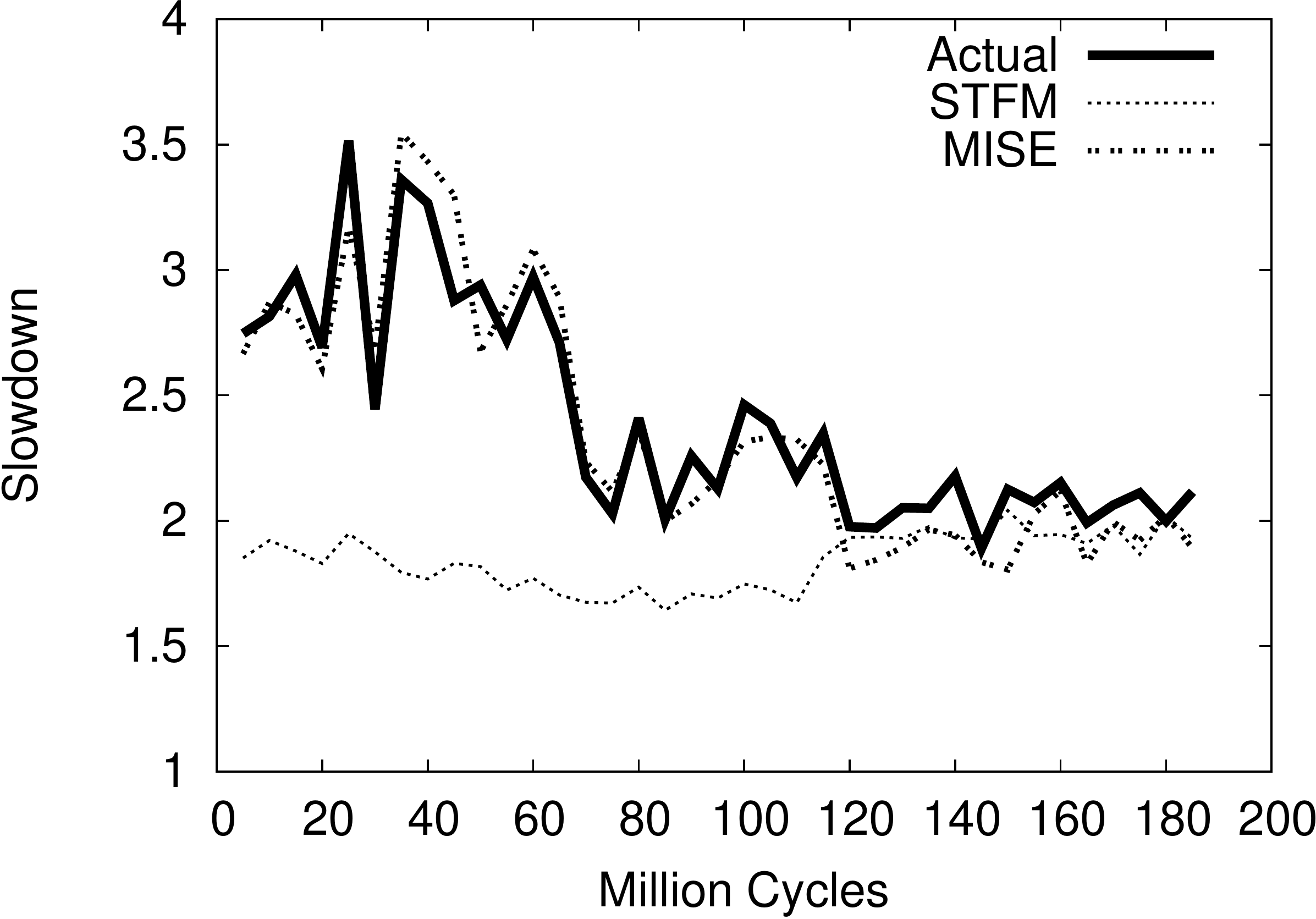}
      \subcaption{leslie3d}
      \label{fig:leslie-stfm-comp}
  \end{subfigure}
  \begin{subfigure}{0.32\textwidth}
      \centering
      \includegraphics[scale=0.21, angle=0]{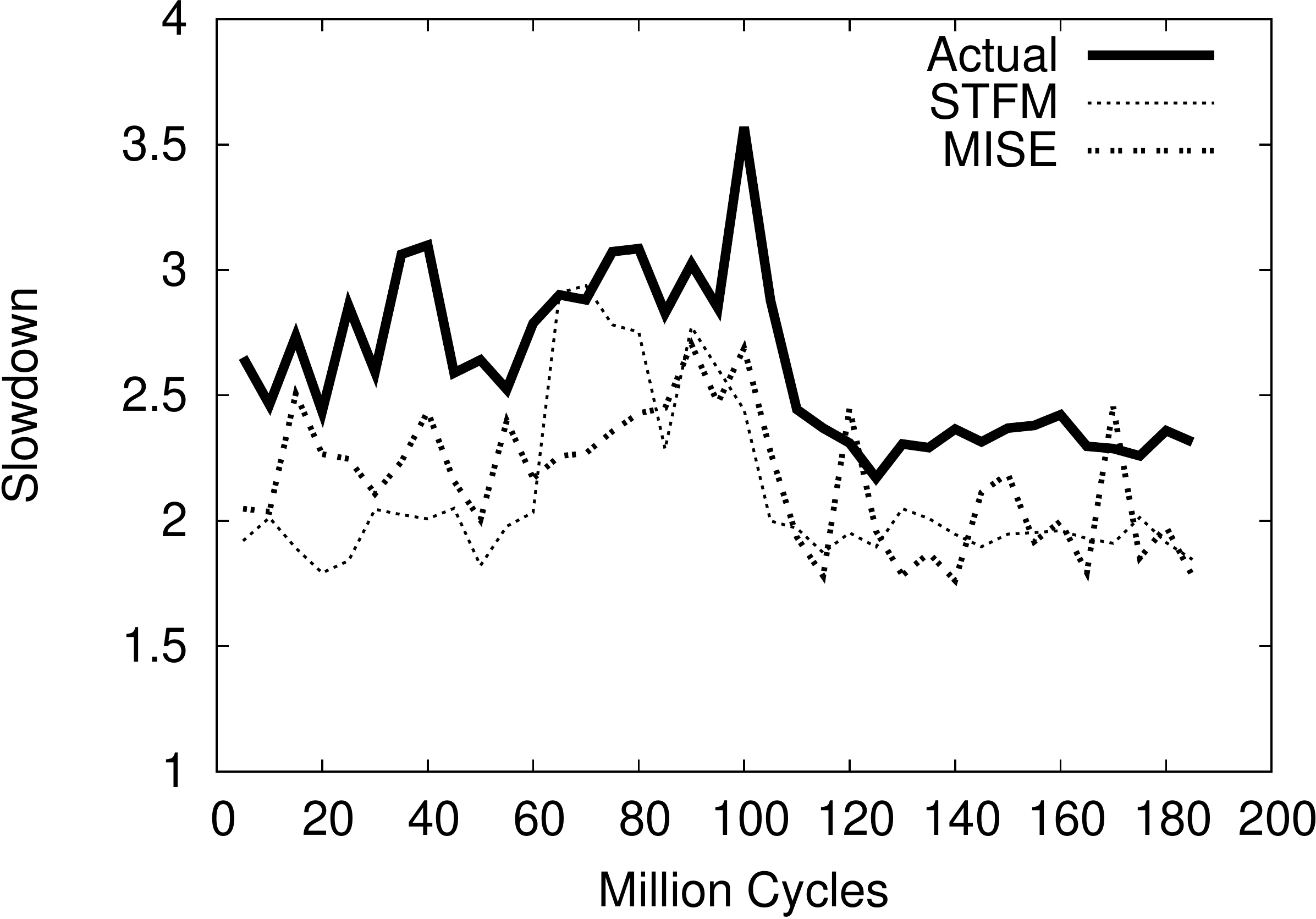}
      \subcaption{sphinx3}
      \label{fig:sphinx3-stfm-comp}
  \end{subfigure}
  \begin{subfigure}{0.32\textwidth}
      \centering
      \includegraphics[scale=0.21, angle=0]{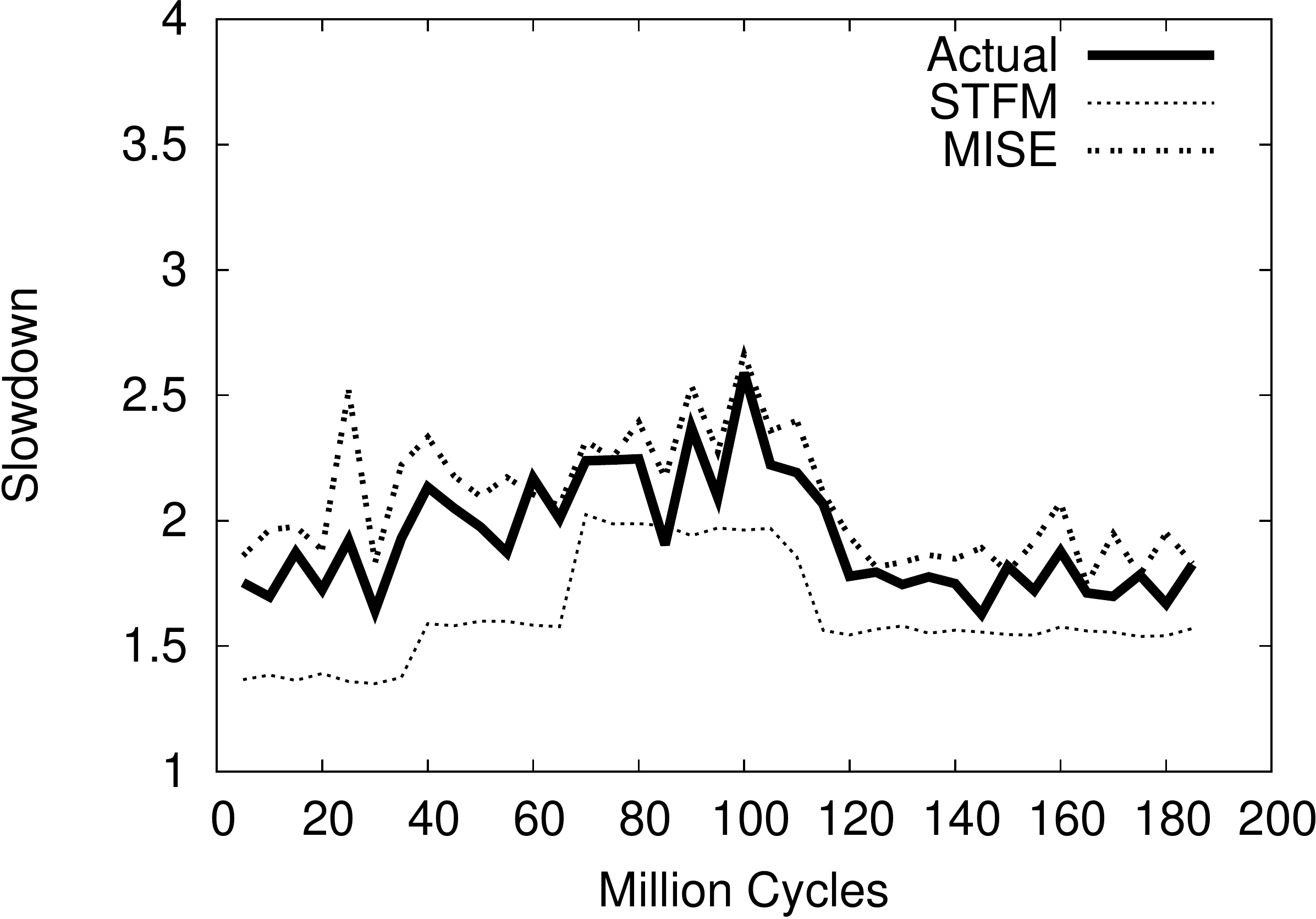}
      \subcaption{GemsFDTD}
      \label{fig:GemsFDTD-stfm-comp}
  \end{subfigure}
  \begin{subfigure}{0.32\textwidth}
      \centering
      \includegraphics[scale=0.21, angle=0]{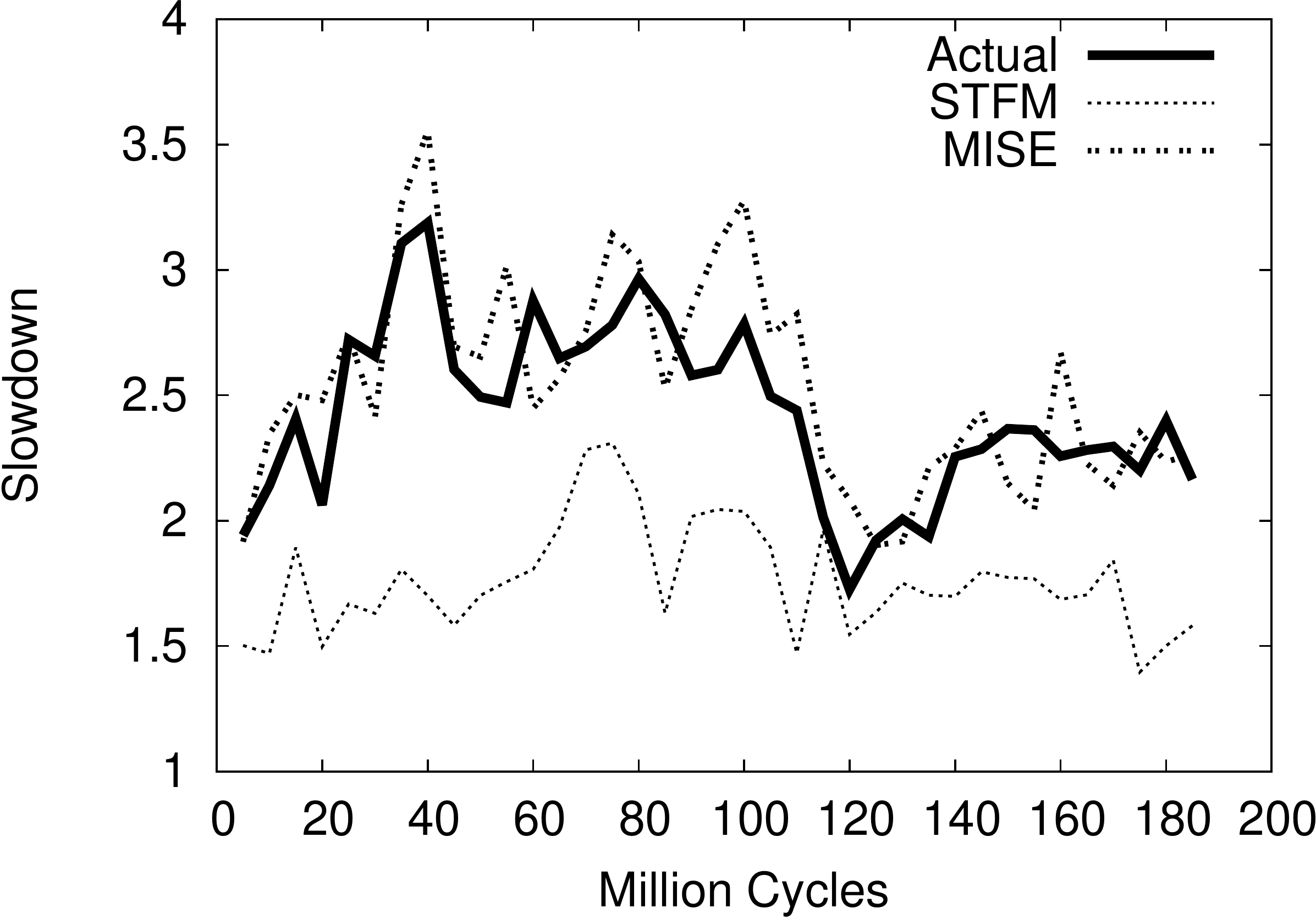}
      \subcaption{soplex}
      \label{fig:soplex-stfm-comp}
  \end{subfigure}
  \begin{subfigure}{0.32\textwidth}
      \centering
      \includegraphics[scale=0.21, angle=0]{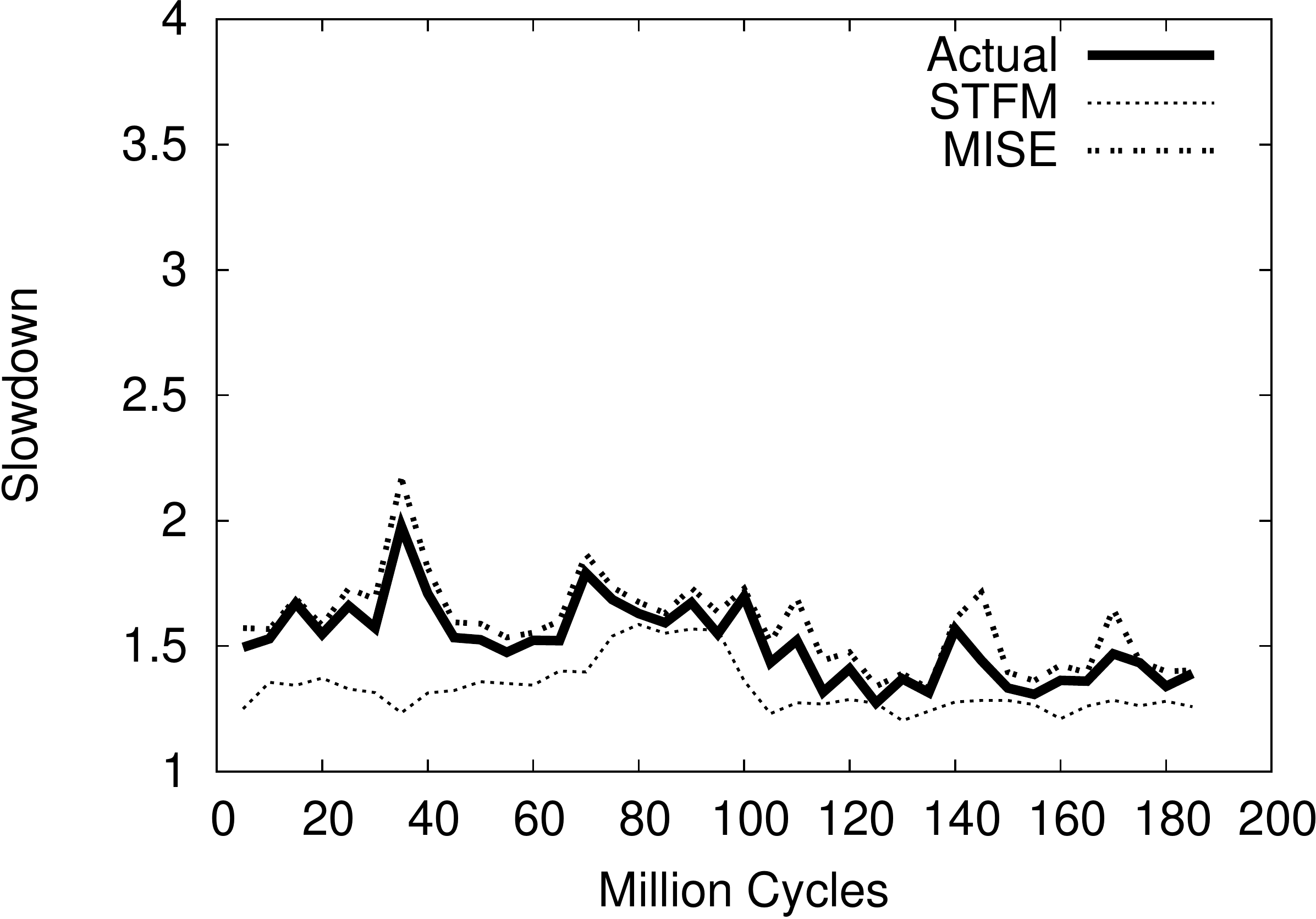}
      \subcaption{cactusADM}
      \label{fig:cactusADM-stfm-comp}
  \end{subfigure}

  \caption{Comparison of our \miiseabb model with STFM for representative memory-bound applications}
  \label{fig:comparison-stfm-membound}
  
\end{figure*}

Figure~\ref{fig:comparison-stfm-membound} compares the accuracy of the
\miiseabb model with STFM for six representative memory-bound
applications from the SPEC CPU2006 benchmark suite.  Each
application is run on a 4-core system along with three other
applications: \emph{sphinx3}, \emph{leslie3d}, and \emph{milc}.
The figure plots three curves: 1)~actual slowdown, 2)~slowdown
estimated by STFM, and 3)~slowdown estimated by \miiseabb. For
most applications in the SPEC CPU2006 suite, the slowdown
estimated by \miiseabb is significantly more accurate than STFM's
slowdown estimates. All applications whose slowdowns are shown in
Figure~\ref{fig:comparison-stfm-membound}, except sphinx3, are
representative of this behavior. For a few applications and
workload combinations, STFM's estimates are comparable to the
slowdown estimates from our model: \emph{sphinx3} is an example of
such an application.  However, as we will show below, across all
workloads, the \miiseabb model provides lower average slowdown
estimation error for all applications.

\begin{figure*}[h!]
  \centering
  \begin{subfigure}{0.32\textwidth}
      \centering
      \includegraphics[scale=0.21, angle=0]{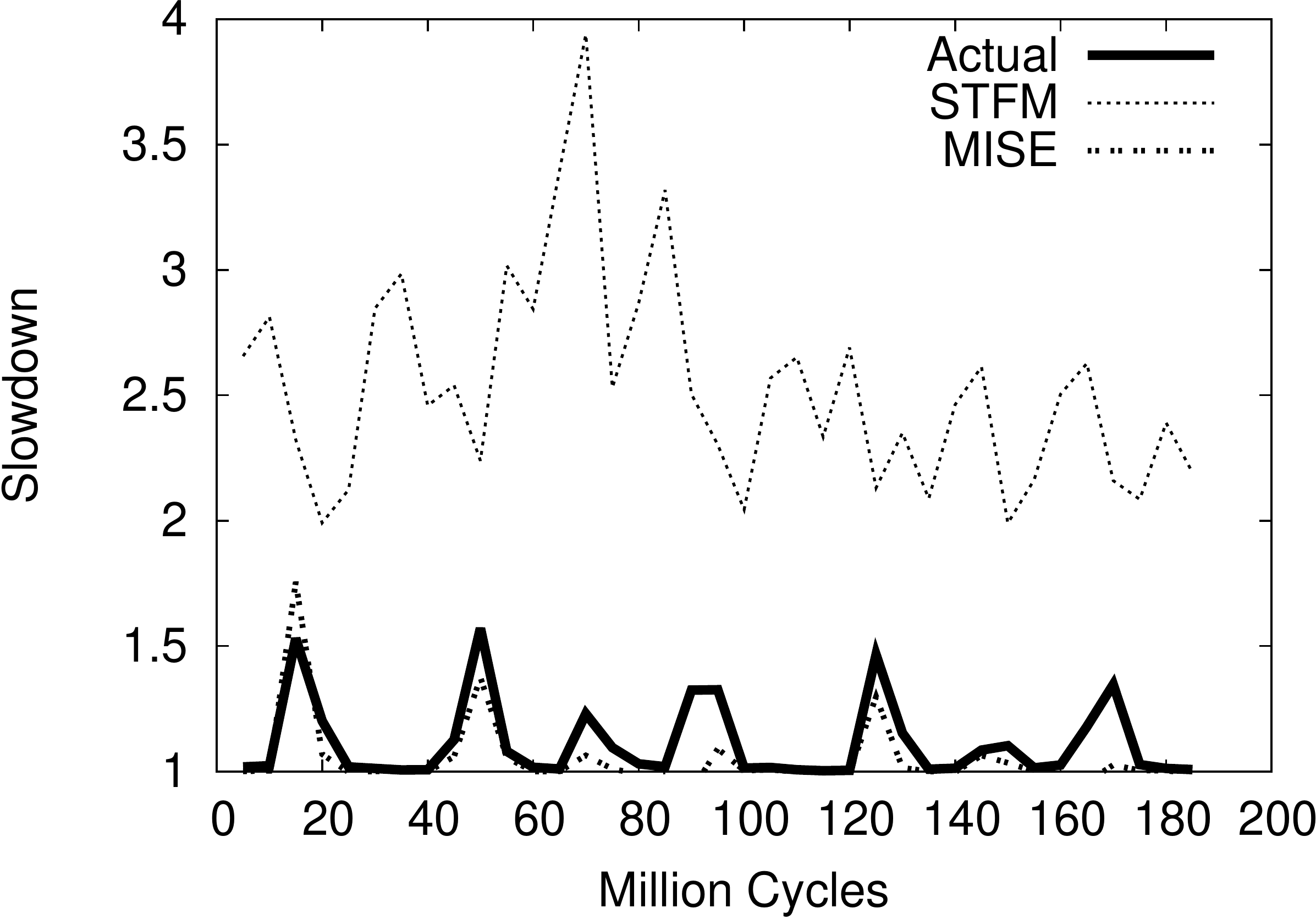}
      \subcaption{wrf}
      \label{fig:wrf-stfm-comp}
  \end{subfigure}
  \begin{subfigure}{0.32\textwidth}
      \centering
      \includegraphics[scale=0.21, angle=0]{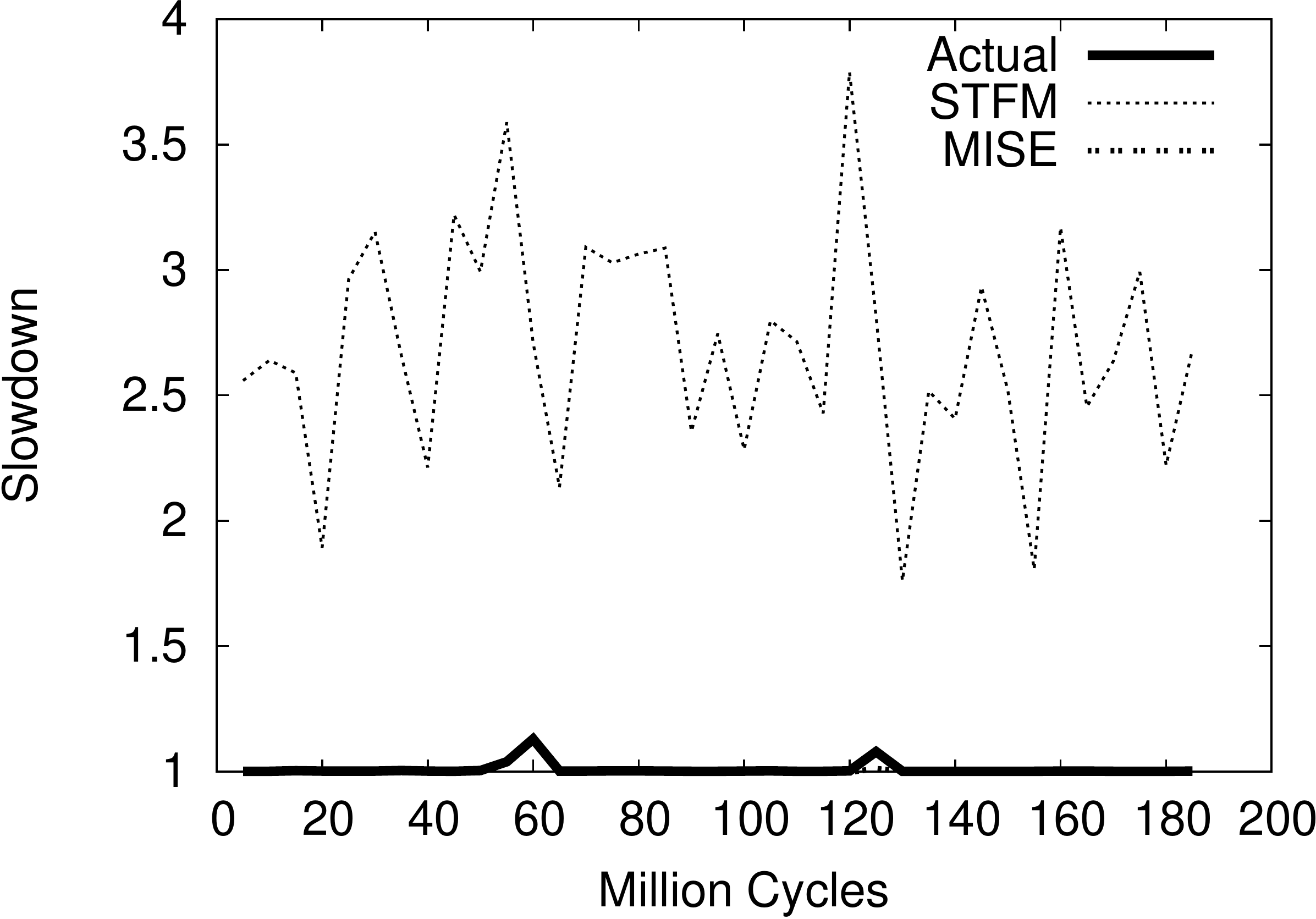}
      \subcaption{povray}
      \label{fig:povray-stfm-comp}
  \end{subfigure}
  \begin{subfigure}{0.32\textwidth}
      \centering
      \includegraphics[scale=0.21, angle=0]{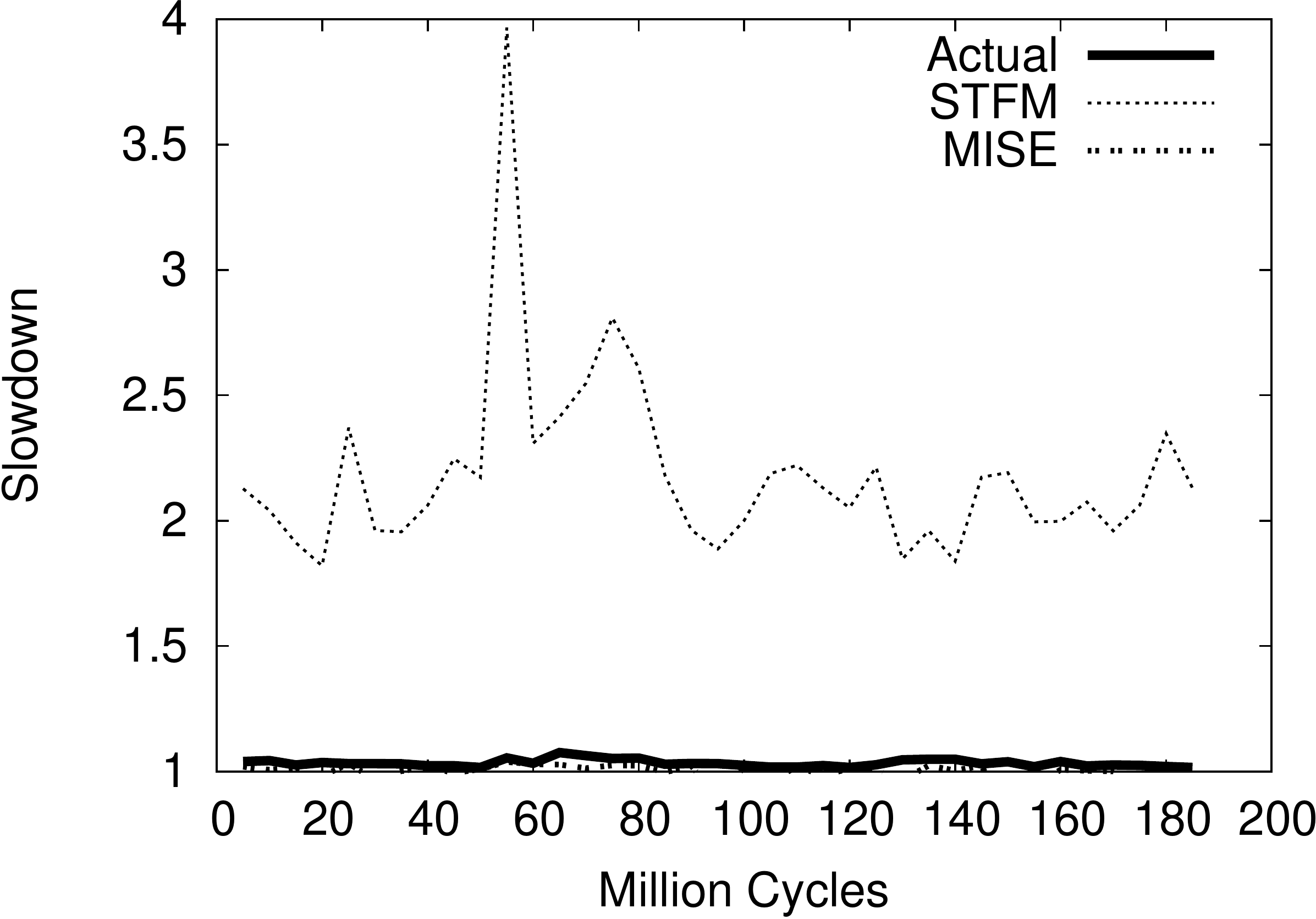}
      \subcaption{calculix}
      \label{fig:povray-stfm-comp}
  \end{subfigure}
  \caption{Comparison of our \miiseabb model with STFM for representative non-memory-bound applications}
  \label{fig:comparison-stfm-nonmembound}
  \vspace{-2mm}
\end{figure*}

Figure~\ref{fig:comparison-stfm-nonmembound} compares the accuracy
of \miiseabb with STFM for three representative
non-memory-bound applications, when each application is run on a
4-core system along with three other applications: \emph{sphinx3},
\emph{leslie3d}, and \emph{milc}. As shown in the figure,
\miiseabb's estimates are significantly more accurate compared to
STFM's estimates. As mentioned before, STFM does not account for
the compute phase of these applications. However, these
applications spend significant amount of their execution time in
the compute phase. This is the reason why our model, which takes
into account the effect of the compute phase of these
applications, is able to provide more accurate slowdown estimates
for non-memory-bound applications.

\begin{table}
  \centering
  \input{mise/tables/slowdown-errors}
  \caption{Average error for each benchmark (in \%)}
  \label{tab:slowdown-errors}
\end{table}
\newpage
Table~\ref{tab:slowdown-errors} shows the average slowdown
estimation error for each benchmark, with STFM and \miiseabb,
across all 300 4-core workloads of different memory
intensities.\footnote{See Table~\ref{tab:mise-qos-workloads} and
Section~\ref{sec:singleaoi} for more details about these 300
workloads.} As can be observed, \miiseabb's slowdown estimates have significantly
lower error than STFM's slowdown estimates across most benchmarks.
Across {\em 300} workloads, STFM's estimates deviate from the actual slowdown by 29.8\%,
whereas, our proposed \miiseabb model's estimates deviate from the
actual slowdown by only 8.1\%. Therefore, we conclude that our
slowdown estimation model provides better accuracy than STFM.

%% file: mise/tables/slowdown-errors.tex
\footnotesize{
\begin{tabular}{|c|c|c||c|c|c|}
  \hline
  \textbf{Benchmark} & \textbf{STFM} & \textbf{\miiseabb\xspace} & \textbf{Benchmark} & \textbf{STFM} & \textbf{\miiseabb\xspace}\\
%  \textbf{Benchmark} & \textbf{STFM - Avg. Error (in \%)} & \textbf{\miiseabb\xspace - Avg. Error (in \%)} & \textbf{Benchmark} & \textbf{STFM - Avg. Error (in \%)} & \textbf{\miiseabb\xspace  - Avg. Error (in \%)}\\
  \hline
  453.povray & 56.3 & 0.1 & 473.astar & 12.3 & 8.1\\
  \hline
  454.calculix & 43.5 & 1.3 & 456.hmmer & 17.9 & 8.1\\
  \hline
  400.perlbench	& 26.8 & 1.6 & 464.h264ref & 13.7 & 8.3\\
  \hline
  447.dealII & 37.5 & 2.4 & 401.bzip2 &	28.3 & 8.5\\
  \hline
  436.cactusADM	& 18.4 & 2.6 & 458.sjeng & 21.3 & 8.8\\
  \hline
  450.soplex & 29.8 & 3.5 & 433.milc & 26.4 & 9.5\\
  \hline
  444.namd & 43.6 & 3.7 & 481.wrf & 33.6 & 11.1\\
  \hline
  437.leslie3d & 26.4 & 4.3 & 429.mcf &	83.74 & 11.5\\
  \hline
  403.gcc & 25.4 & 4.5 & 445.gobmk & 23.1 & 12.5\\
  \hline
  462.libquantum & 48.9 & 5.3 & 483.xalancbmk &	18.0 & 13.6\\
  \hline
  459.GemsFDTD & 21.6 & 5.5 & 435.gromacs & 31.4 & 15.6\\
  \hline
  470.lbm & 6.9 & 6.3 & 482.sphinx3 & 21 & 16.8\\
  \hline
  473.astar & 12.3 & 8.1 & 471.omnetpp & 26.2 & 17.5\\
  \hline
  456.hmmer & 17.9 & 8.1 & 465.tonto & 32.7 & 19.5\\
  \hline
\end{tabular}
}

%% file: mise/sensitivity.tex
\section{Sensitivity to Algorithm Parameters}
\label{sec:sensitivity}
We evaluate the sensitivity of the \miiseabb
model to epoch and interval lengths.
Table~\ref{tab:mise-alg-sensitivity} presents the average error (in \%)
of the \miiseabb model for different values of epoch and interval
lengths. Two major conclusions are in order. First, when the interval
length is small (1 million cycles), the error is very high. This
is because the request service rate is not stable at such small
interval lengths and varies significantly across intervals.
Therefore, it cannot serve as an effective proxy for performance.
On the other hand, when the interval length is larger, request
service rate exhibits a more stable behavior and can serve as an
effective measure of application slowdowns. Therefore, we conclude
that except at very low interval lengths, the \miiseabb model is
robust. Second, the average error is
high for high epoch lengths (1 million cycles) because the number
of epochs in an interval reduces. As a result, some applications
might not be assigned highest priority for any epoch during an
interval, preventing estimation of their \arsr. Note that the effect of this is
mitigated as the interval length increases, as with a larger
interval length the number of epochs in an interval increases.
For smaller epoch length values, however, the average error of
\miiseabb does not exhibit much variation and is robust. The lowest average
error of 8.1\% is achieved at an interval length of 5 million
cycles and an epoch length of 10000 cycles. Furthermore, we
observe that estimating slowdowns at an interval
length of 5 million cycles also enables enforcing QoS at fine
time granularities, although, higher interval lengths exhibit
similar average error. Therefore, we use
these values of interval and epoch lengths for our evaluations.

\begin{table}[ht]
  \centering
  \input{mise/tables/interval-epoch-sensitivity}
  \caption{Sensitivity of average error to epoch and interval lengths}
  \label{tab:mise-alg-sensitivity}
\end{table}

%% file: mise/tables/interval-epoch-sensitivity.tex
\footnotesize{
%\begin{tabular}{|c|c|c|c|}
%  \hline
%   Epoch Length & 1000 & 10000 & 100000\\
%  \hline
%   Avg. Error (in \%) & 9.1 & 8.1 & 11.2\\
%  \hline
%  \hline
%  Interval Length & 1 mil & 5 mil & 10 mil\\
%  \hline
%  Avg. Error (in \%) & 64.1 & 8.1 & 9.6 \\
%  \hline
%\end{tabular}

\begin{tabular}{|c|c|c|c|c|c|}
  \hline
  \backslashbox{Epoch\\Length}{Interval\\Length} & 1 mil. & 5 mil. & 10 mil. & 25 mil. & 50 mil.\\
  \hline
  1000 & 65.1\% & 9.1\% & 11.5\% & 10.7\% & 8.2\%\\
  \hline
  10000 & 64.1\% & 8.1\% & 9.6\% & 8.6\% & 8.5\%\\
  \hline
  100000 & 64.3\% & 11.2\% & 9.1\% & 8.9\% & 9\%\\
  \hline
  1000000 & 64.5\% & 31.3\% & 14.8\% & 14.9\% & 11.7\%\\
  \hline
\end{tabular}
}

%% file: mise/summary.tex
\section{Summary}
In summary, we propose \miiseabb, a new and simple model to
estimate application slowdowns due to inter-application
interference in main memory. \miiseabb is based on two simple
observations: 1)~the rate at which an application's memory
requests are served can be used as a proxy for the application's
performance, and 2)~the uninterfered request-service-rate of an
application can be accurately estimated by giving the
application's requests the highest priority in accessing main
memory. Compared to state-of-the-art approaches for estimating
main memory slowdowns, \miiseabb is simpler and more accurate, as
our evaluations show.

%% file: chapters/applications-mise-model.tex
\chapter{Applications of the MISE Model}
\label{chap:mise-applications}

Accurate slowdown estimates from the MISE model can be
leveraged in multiple possible ways. On the one hand, they can be
leveraged in hardware, to perform allocation of memory bandwidth
to different applications, such that the overall system
performance/fairness is improved or different applications'
performance guarantees are met. On the other hand, MISE's slowdown
estimates can be communicated to the system software/hypervisor,
enabling virtual machine migration and admission control schemes.
%Furthermore, accurate slowdown estimates can be used to drive fair
%pricing schemes based on slowdowns, rather than just resource
%allocation or virtual machine migration, in a cloud computing
%setting~\cite{azure,ec2}.

We propose and evaluate two such use cases of MISE: 1)~a mechanism to
provide soft QoS guarantees (\miiseqos) and 2)~a mechanism that
attempts to minimize maximum slowdown to improve overall system
fairness (\miisefair).

\input{applications-mise-model/application1.tex}

\input{applications-mise-model/application2.tex}

\input{applications-mise-model/summary.tex}

%% file: applications-mise-model/application1.tex
\newcommand{\aoi}{AoI\xspace}
\newcommand{\aois}{AoIs\xspace}

\section{\miiseqos: Providing Soft QoS Guarantees}
\label{sec:app1}

\miiseqos is a mechanism to provide soft QoS guarantees to one or
more applications of interest in a workload with many
applications, while trying to maximize overall performance for the
remaining applications. By {\em soft QoS guarantee}, we mean that
the applications of interest (\aois) should not be slowed down by
more than an operating-system-specified bound. One way of
achieving such a soft QoS guarantee is to always prioritize the
\aois. However, such a mechanism has two shortcomings. First, it
would work when there is only one \aoi. With more than one \aoi,
prioritizing all AoIs will cause them to interfere with each other
making their slowdowns uncontrollable. Second, even with just one
\aoi, a mechanism that always prioritizes the \aoi may
unnecessarily slow down other applications in the system.
\miiseqos addresses these shortcomings by using slowdown estimates
of the \aois to allocate them just enough memory bandwidth to meet
their specified slowdown bound. We present the operation of
\miiseqos with one \aoi and then describe how it can be extended
to multiple \aois.

%\vspace{-3pt}

\subsection{Mechanism Description}
\label{sec:app1-description}
The operation of \miiseqos with one AoI is simple. As we describe in
Section~\ref{sec:implementation-scheduling}, the memory controller
divides execution time into intervals of length $\mathcal{M}$. The
controller maintains the current bandwidth allocation for the AoI.
At the end of each interval, it estimates the slowdown of the AoI
and compares it with the specified bound, say $B$. If the
estimated slowdown is less than $B$, then the controller reduces
the bandwidth allocation for the AoI by a small amount (2\% in our
experiments). On the other hand, if the estimated slowdown is more
than $B$, the controller increases the bandwidth allocation for
the AoI (by 2\%).\footnote{We found that 2\% increments in memory
bandwidth work well empirically, as our results indicate. Better
techniques that dynamically adapt the increment are
possible and are a part of our future work.} The remaining
bandwidth is used by all other applications in the system in a
free-for-all manner. The above mechanism attempts to ensure that
the AoI gets just enough bandwidth to meet its target slowdown
bound. As a result, the other applications in the system are not
\emph{unnecessarily} slowed down.

%In some cases, it is possible that the target bound \emph{cannot}
%be met even by allocating all the memory bandwidth to the AoI --
%i.e., prioritizing its request 100\% of the time. This is because,
%our memory scheduling is work conserving -- if there are no
%requests from the highest priority application, it schedules a
%ready request from some other application. Once a request is
%scheduled, it cannot be preempted because of the way DRAM
%operates. As a result, even the application with the highest
%priority (AoI) will face some interference from other
%applications, slowing it down by some factor. Therefore, in
%scenarios when it is not possible to meet the target bound for the
%AoI, the memory controller conveys this information to the
%operating system, which can then take appropriate action (e.g.,
%deschedule some other applications from the machine).

In some cases, it is possible that the target bound \emph{cannot}
be met even by allocating all the memory bandwidth to the AoI --
i.e., prioritizing its requests 100\% of the time. This is because,
even the application with the highest priority (AoI) could be
subject to interference, slowing it down by some factor, as we
describe in Section~\ref{sec:implementation-arsr}. Therefore, in
scenarios when it is not possible to meet the target bound for the
AoI, the memory controller can convey this information to the
operating system, which can then take appropriate action (e.g.,
deschedule some other applications from the machine).

\subsection{\miiseqos with Multiple AoIs}
\label{sec:multiple-aoi}
The above described \miiseqos mechanism can be easily extended to
a system with multiple AoIs. In such a system, the memory
controller maintains the bandwidth allocation for each AoI. At the
end of each interval, the controller checks if the slowdown
estimate for each AoI meets the corresponding target bound. Based
on the result, the controller either increases or decreases the
bandwidth allocation for each AoI (similar to the mechanism in
Section~\ref{sec:app1-description}).

With multiple AoIs, it may not be possible to meet the specified
slowdown bound for any of the AoIs.  Our mechanism concludes that
the specified slowdown bounds cannot be met if: 1)~all the
available bandwidth is partitioned only between the AoIs -- i.e.,
no bandwidth is allocated to the other applications, and 2)~any of
the AoIs does not meet its slowdown bound after R intervals (where
R is empirically determined at design time). Similar to the
scenario with one AoI, the memory controller can convey this
conclusion to the operating system (along with the estimated
slowdowns), which can then take an appropriate action. Note that
other potential mechanisms for determining whether slowdown bounds
can be met are possible.

%The above two conditions indicate that it is not possible to further
%increase the bandwidth allocation of all the AoIs.

\newcommand{\ap}{\emph{AlwaysPrioritize}\xspace}
\subsection{Evaluation with Single AoI}\label{sec:singleaoi}
To evaluate \miiseqos with a single AoI, we run each benchmark as
the AoI, alongside 12 different workload mixes shown in
Table~\ref{tab:mise-qos-workloads}. We run each workload with 10 different
slowdown bounds for the AoI: $\frac{10}{1}, \frac{10}{2}, ...,
\frac{10}{10}$. These slowdown bounds are chosen so as to have
more data points between the bounds of $1\times$ and
$5\times$.\footnote{Most applications are not slowed down by more
than $5\times$ for our system configuration.} In all, we present
results for 3000 data points with different workloads and slowdown
bounds. We compare \miiseqos with a mechanism that always
prioritizes the AoI~\cite{qos-sigmetrics} (\ap).

\begin{table}
  \centering
  \input{applications-mise-model/tables/workloads}
  \caption{Workload mixes}
  \label{tab:mise-qos-workloads}
\end{table}

%% The goals of \miiseqos are two-fold. Its primary goal is to ensure
%% that the AoI meets the specified slowdown bound. The secondary goal is
%% to maximize the performance for the remaining applications. First, we
%% evaluate \miiseqos with respect to its primary goal: if the
%% slowdown bound is met.

%present the results related to the primary goal of \miiseqos: if the
%slowdown bound is met.

Table~\ref{tab:perf-guarantees} shows the effectiveness of
\miiseqos in meeting the prescribed slowdown bounds for the 3000
data points. As shown, for approximately 79\% of the workloads,
\miiseqos meets the specified bound and correctly estimates that
the bound is met. However, for 2.1\% of the workloads,
\miiseqos\xspace {\em does} meet the specified bound but it
incorrectly estimates that the bound is not met. This is because,
in some cases, \miiseqos slightly overestimates the slowdown of
applications.  Overall, \miiseqos meets the specified slowdown
bound for close to 80.9\% of the workloads, as compared to \ap
that meets the bound for 83\% of the workloads. Therefore, we
conclude that \miiseqos meets the bound for 97.5\% of the
workloads where \ap meets the bound. Furthermore, \miiseqos
correctly estimates whether or not the bound was met for 95.7\% of
the workloads, whereas \ap has no provision to estimate whether or
not the bound was met.

\begin{table}[htb*]
\centering 
\input{applications-mise-model/tables/perf-guarantees}
\caption{Effectiveness of \miiseqos} 
\label{tab:perf-guarantees}
\end{table}

\begin{figure}[h]
  \centering
  \includegraphics[scale=0.6, angle=270]{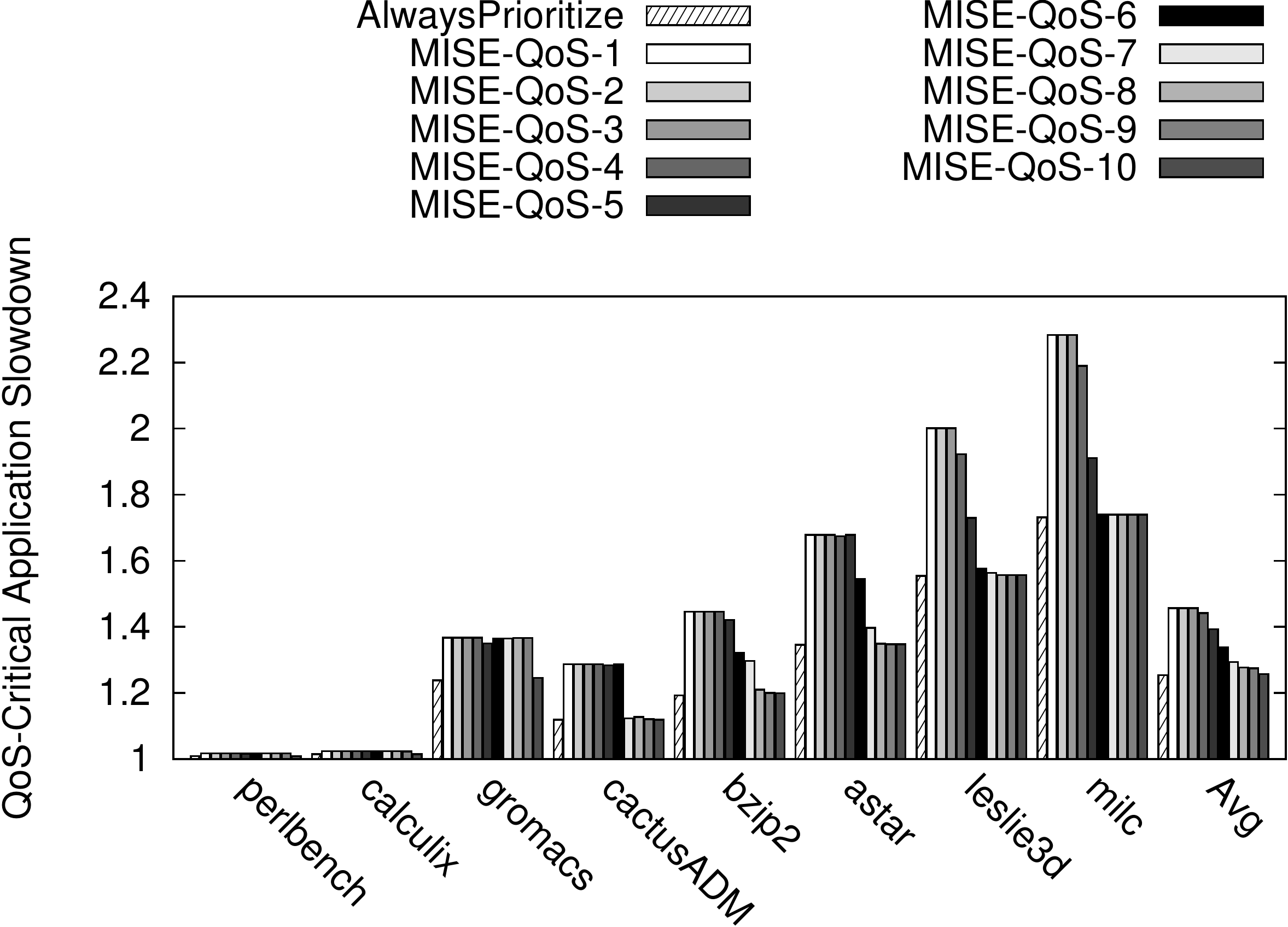}
  \caption{\aoi performance: \miiseqos vs. \emph{AlwaysPrioritize}}
  \label{fig:ind-app-slowdown}
\end{figure}

To show the effectiveness of \miiseqos, we compare the AoI's slowdown
due to \miiseqos and the mechanism that always prioritizes the AoI
(\ap)~\cite{qos-sigmetrics}. Figure~\ref{fig:ind-app-slowdown}
presents representative results for 8 different AoIs when they are run
alongside Mix 1 (Table~\ref{tab:mise-qos-workloads}). The label \miiseqos-n
corresponds to a slowdown bound of $\frac{10}{n}$. (Note that \ap does
not take into account the slowdown bound). Note that the slowdown
bound decreases (i.e., becomes tighter) from left to right for
each benchmark in Figure~\ref{fig:ind-app-slowdown} (as well as in
other figures). We draw three conclusions from the results.

First, for most applications, the slowdown of \ap is considerably more
than one. As described in Section~\ref{sec:app1-description}, always
prioritizing the AoI does not completely prevent other applications
from interfering with the AoI.

Second, as the slowdown bound for the AoI is decreased (left to
right), \miiseqos gradually increases the bandwidth allocation for
the AoI, eventually allocating all the available bandwidth to the
AoI. At this point, \miiseqos performs very similarly to the
\ap mechanism.

Third, in almost all cases (in this figure and across all our 3000
data points), \miiseqos meets the specified slowdown bound
\emph{if} \ap is able to meet the bound. One exception to this is
benchmark \emph{gromacs}. For this benchmark, \miiseqos meets the
slowdown bound for values ranging from $\frac{10}{1}$ to
$\frac{10}{6}$.\footnote{Note that the slowdown bound becomes
tighter from left to right.} For slowdown bound values of
$\frac{10}{7}$ and $\frac{10}{8}$, \miiseqos does not meet the
bound even though allocating all the bandwidth for \emph{gromacs}
would have achieved these slowdown bounds (since \ap can meet the
slowdown bound for these values). This is because our \miiseabb
model underestimates the slowdown for \emph{gromacs}. Therefore,
\miiseqos incorrectly assumes that the slowdown bound is met for
\emph{gromacs}.

Overall, \miiseqos accurately estimates the slowdown of the AoI and
allocates just enough bandwidth to the AoI to meet a slowdown
bound. As a result, \miiseqos is able to significantly improve the
performance of the other applications in the system (as we show next).

%\begin{figure*}[!ht] 
%  \centering 
%  \begin{minipage}{0.45\textwidth}
%    \centering 
%    \includegraphics[scale=0.25, angle=270]{applications-mise-model/plots/avg_hs_file}
%  \end{minipage} 
%  \begin{minipage}{0.45\textwidth}
%    \centering
%    \includegraphics[scale=0.25, angle=270]{applications-mise-model/plots/avg_ws_file}
%  \end{minipage}
%  \caption{Average system performance across
%    300 workloads of different memory intensities}
%  \label{fig:sef-avg-hs-ws}
%\end{figure*}
%
%\begin{figure}[!ht]
%  \centering
%  \includegraphics[scale=0.25, angle=270]{applications-mise-model/plots/avg_ms_file}
%  \caption{Average fairness across 300 workloads of different memory intensities}
%  \label{fig:sef-avg-ms}
%\end{figure}

\begin{figure*}[!ht] 
  \centering 
  \begin{minipage}{0.48\textwidth}
    \centering 
    \includegraphics[scale=0.3, angle=270]{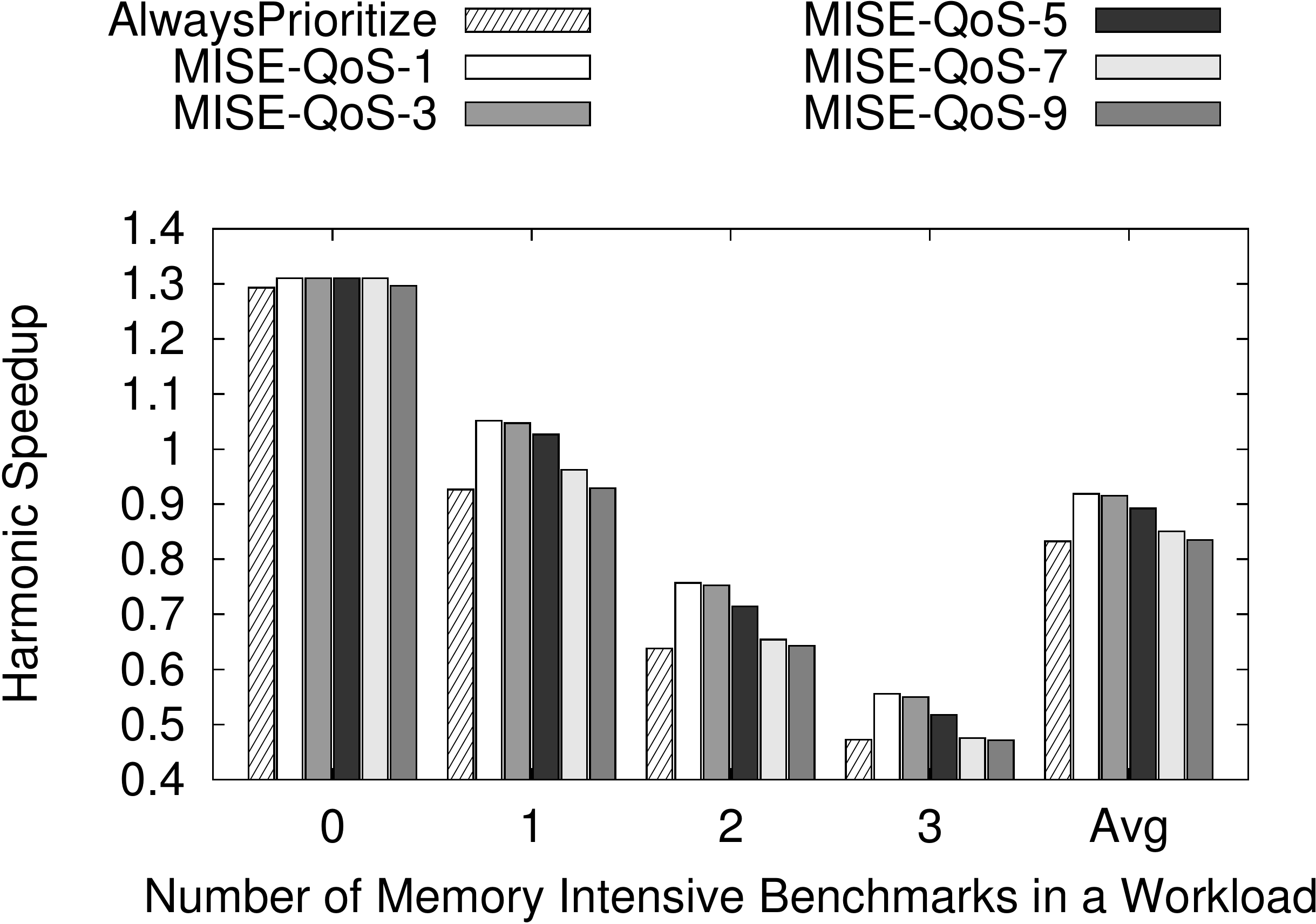}
  \end{minipage} 
  \begin{minipage}{0.48\textwidth}
    \centering
    \includegraphics[scale=0.3, angle=270]{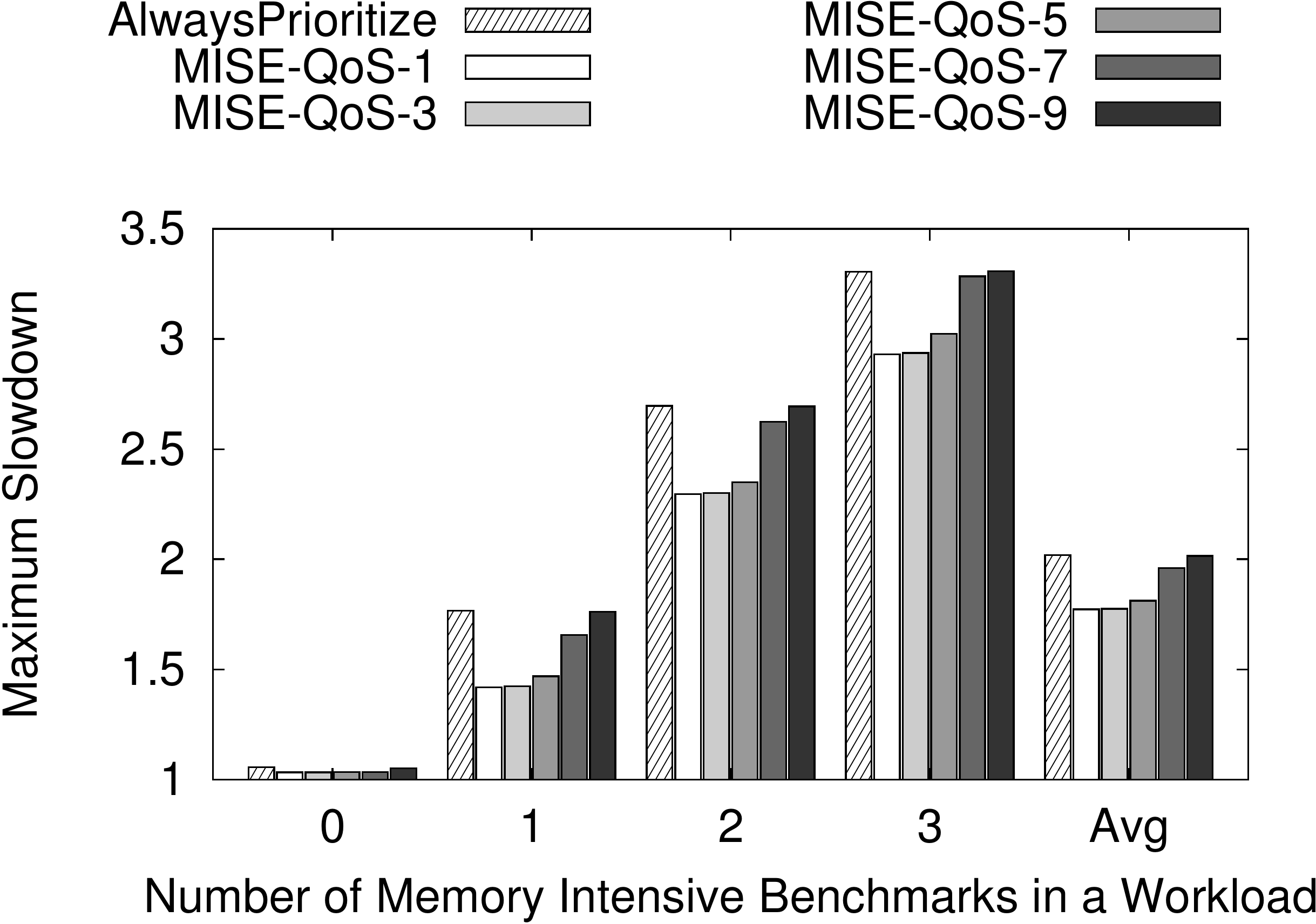}
  \end{minipage}
  \caption{Average system performance and fairness across
    300 workloads of different memory intensities}
  \label{fig:sef-avg-hs-ms}
\end{figure*}

\textbf{System Performance and Fairness.}
Figure~\ref{fig:sef-avg-hs-ms} compares the system performance
(harmonic speedup) and fairness (maximum slowdown) of \miiseqos
and \ap for different values of the bound. We omit the AoI from
the performance and fairness calculations. The results are
categorized into four workload categories (0, 1, 2, 3) indicating
the number of memory-intensive benchmarks in the workload. For
clarity, the figure shows results only for a few slowdown
bounds. Three conclusions are in order.

First, \miiseqos significantly improves performance compared to
\ap, especially when the slowdown bound for the AoI is large. On
average, when the bound is $\frac{10}{3}$, \miiseqos improves
harmonic speedup by 12\% and weighted speedup by 10\% (not shown
due to lack of space) over \ap, while reducing maximum slowdown by
13\%. Second, as expected, the performance and fairness of
\miiseqos approach that of \ap as the slowdown bound is decreased
(going from left to right for a set of bars). Finally, the
benefits of \miiseqos increase with increasing memory
intensity because always prioritizing a memory intensive
application will cause significant interference to other
applications.

Based on our results, we conclude that \miiseqos can effectively
ensure that the AoI meets the specified slowdown bound while
achieving high system performance and fairness across the other
applications. In Section~\ref{sec:two-aoi}, we discuss a case study of
a system with two AoIs.

\begin{figure}[h!] 
  \centering 
    \includegraphics[scale=0.35, angle=270]{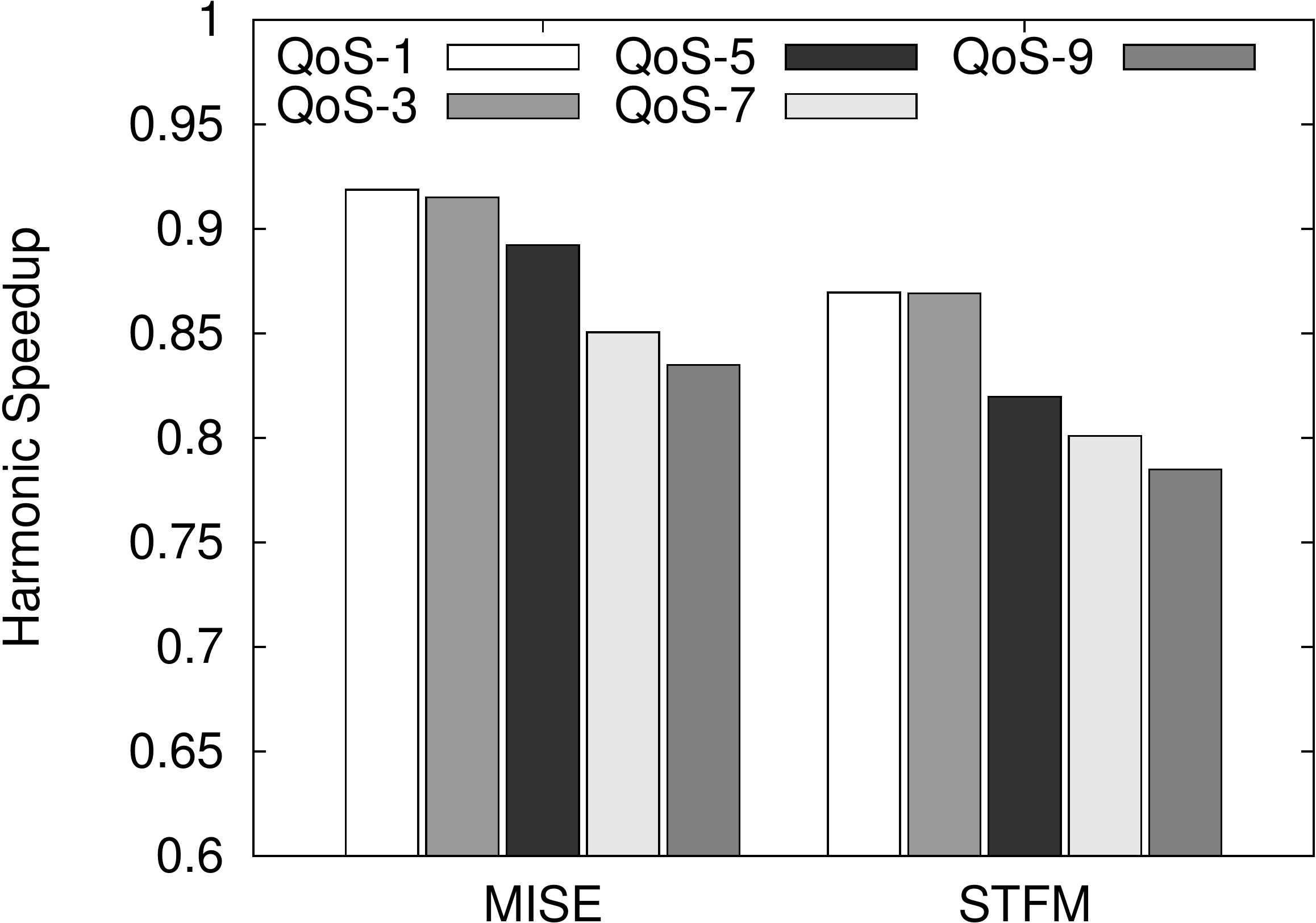}
    \caption{Average system performance using MISE and STFM's slowdown
estimation models (across 300 workloads)}
  \label{fig:stfm-mise-avg-hs}
\end{figure}

\textbf{Using STFM's Slowdown Estimates to Provide QoS.}
We evaluate the effectiveness of STFM in providing slowdown
guarantees, by using slowdown estimates from STFM's model to drive
our QoS-enforcement mechanism.
Table~\ref{tab:perf-guarantees-stfm} shows the effectiveness of
STFM's slowdown estimation model in meeting the prescribed
slowdown bounds for the 3000 data points. We draw two major
conclusions. First, the slowdown bound is met and estimated as met
for only 63.7\% of the workloads, whereas \miiseqos meets the
slowdown bound and estimates it right for 78.8\% of the workloads
(as shown in Table~\ref{tab:perf-guarantees}). The reason is
STFM's high slowdown estimation error. Second, the percentage of
workloads for which the slowdown bound is met/not-met and is
estimated wrong is 18.4\%, as compared to 4.3\% for \miiseqos.
This is because STFM's slowdown estimation model overestimates the
slowdown of the \aoi and allocates it more bandwidth than is
required to meet the prescribed slowdown bound.  Therefore,
performance of the other applications in a workload suffers, as
demonstrated in Figure~\ref{fig:stfm-mise-avg-hs} which shows the
system performance for different values of the prescribed slowdown
bound, for \miiseabb and STFM. For instance, when the slowdown
bound is $\frac{10}{3}$, STFM-QoS has 5\% lower average system
performance than \miiseqos.  Therefore, we conclude that the
proposed \miiseabb model enables more effective enforcement of QoS
guarantees for the \aoi, than the STFM model, while providing
better average system performance.

\begin{table}[htb*]
\centering 
\input{applications-mise-model/tables/perf-guarantees-stfm}
\caption{Effectiveness of STFM-QoS} 
\label{tab:perf-guarantees-stfm}
\end{table}

%\begin{figure}[!ht] 
%  \centering 
%    \includegraphics[scale=0.25, angle=270]{applications-mise-model/plots/avg_hs_file_stfm}
%    \caption{Average system performance using STFM's slowdown
%estimation model (across 300 workloads)}
%  \label{fig:stfm-avg-hs}
%\end{figure}

\input{applications-mise-model/app1-case-study}

%% file: applications-mise-model/tables/workloads.tex
\footnotesize{
\begin{tabular}{|c|c|c|c|}
  \hline
  \textbf{Mix No.} & \textbf{Benchmark 1} & \textbf{Benchmark 2} & \textbf{Benchmark 3}\\
  \hline
  1 & sphinx3 & leslie3d & milc \\
  \hline
  2 & sjeng & gcc & perlbench \\
  \hline
  3 & tonto & povray & wrf \\
  \hline
  4 & perlbench & gcc & povray \\
  \hline
  5 & gcc & povray & leslie3d \\
  \hline
  6 & perlbench & namd & lbm \\
  \hline
  7 & hef & bzip2 & libquantum\\
  \hline
  8 & hmmer & lbm & omnetpp \\
  \hline
  9 & sjeng & libquantum & cactusADM \\
  \hline
  10 & namd & libquantum & mcf \\
  \hline
  11 & xalancbmk & mcf & astar \\
  \hline
  12 & mcf & libquantum & leslie3d \\
  \hline
\end{tabular}
}

%% file: applications-mise-model/tables/perf-guarantees.tex
\footnotesize{
\begin{tabular}{|l|c|c|}
  \hline
  \textbf{Scenario} & \textbf{\# Workloads} & \textbf{\% Workloads} \\
  \hline
  Bound Met and Predicted Right & 2364 & 78.8\% \\
  \hline
  Bound Met and Predicted Wrong & 65 & 2.1\% \\
  \hline
  Bound Not Met and Predicted Right & 509 & 16.9\%  \\
  \hline
  Bound Not Met and Predicted Wrong & 62 & 2.2\%\\
  \hline
\end{tabular}
}  

%% file: applications-mise-model/tables/perf-guarantees-stfm.tex
\footnotesize{
\begin{tabular}{|l|c|c|}
  \hline
  \textbf{Scenario} & \textbf{\# Workloads} & \textbf{\% Workloads} \\
  \hline
  Bound Met and Predicted Right & 1911 & 63.7\% \\
  \hline
  Bound Met and Predicted Wrong & 480 & 16\% \\
  \hline
  Bound Not Met and Predicted Right & 537 & 17.9\%  \\
  \hline
  Bound Not Met and Predicted Wrong & 72 & 2.4\%\\
  \hline
\end{tabular}
}  

%% file: applications-mise-model/app1-case-study.tex
\newcommand{\bp}{\emph{EqualBandwidth}\xspace}

\subsection{Case Study: Two AoIs}
\label{sec:two-aoi}
So far, we have discussed and analyzed the benefits of \miiseqos
for a system with one AoI. However, there could be scenarios with
multiple AoIs each with its own target slowdown bound. One can
think of two naive approaches to possibly address this
problem. In the first approach, the memory controller can
prioritize the requests of all AoIs in the system. This is similar
to the \ap mechanism described in the previous section. In the
second approach, the memory controller can equally partition the
memory bandwidth across all AoIs. We call this
approach \bp. However, neither of these mechanisms can guarantee
that the AoIs meet their target bounds. On the other hand, using
the mechanism described in
Section~\ref{sec:multiple-aoi}, \miiseqos can be used to achieve
the slowdown bounds for multiple AoIs.

\begin{figure}[h]
  \centering
  \includegraphics[scale=0.4, angle=270]{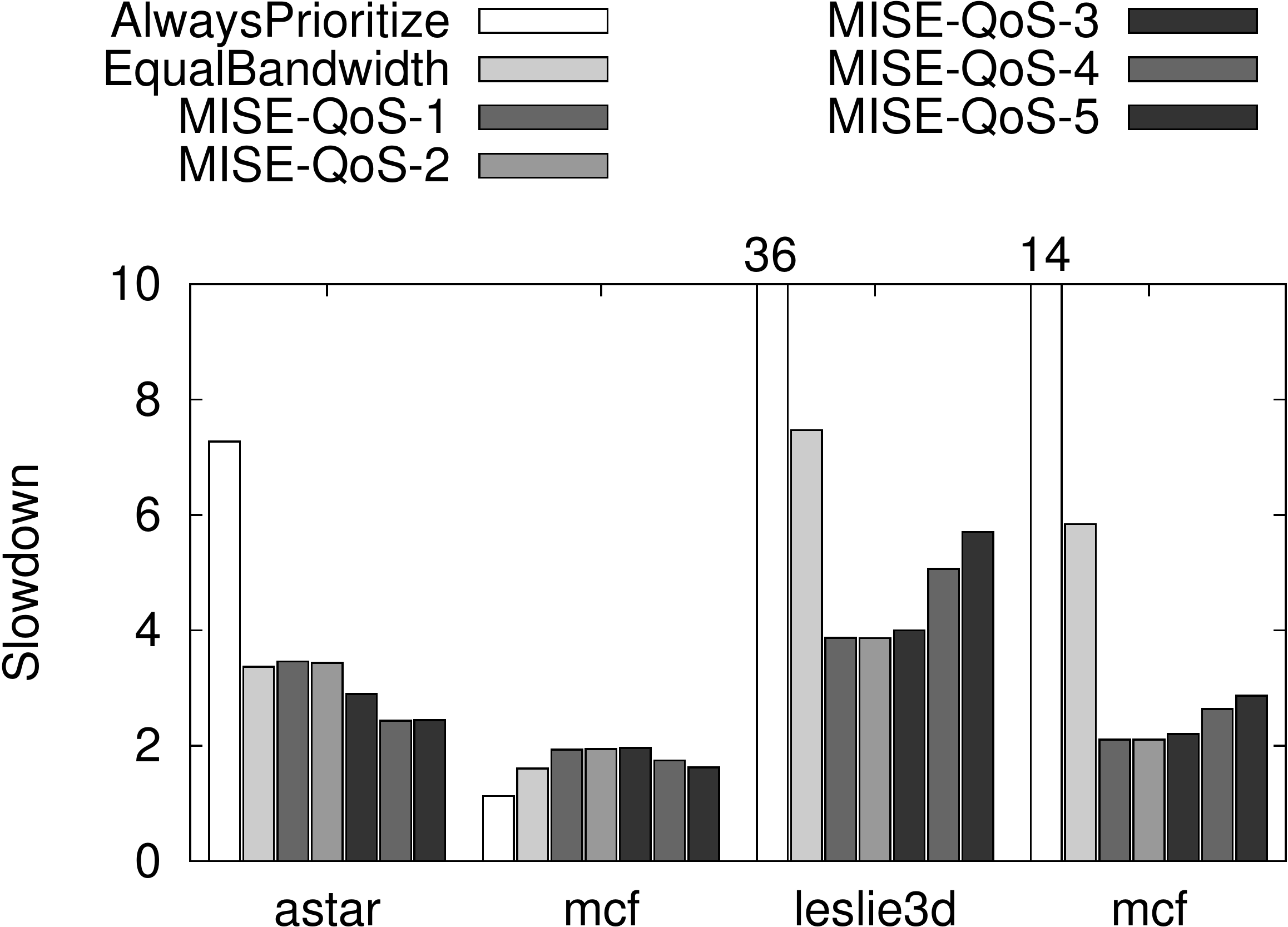}
  \caption{Meeting a target bound for two applications}
  \label{fig:casestudy-2app}
\end{figure}

To show the effectiveness of \miiseqos with multiple AoIs, we present
a case study with two AoIs. The two AoIs, \emph{astar}
and \emph{mcf} are run in a 4-core system with \emph{leslie} and
another copy of \emph{mcf}. Figure~\ref{fig:casestudy-2app} compares
the slowdowns of each of the four applications with the different
mechanisms. The same slowdown bound is used for both AoIs.

%\footnote{A more detailed evaluation of \miiseqos with multiple
%applications is left for future work.}

Although \ap prioritizes both AoIs, \emph{mcf} (the more
memory-intensive AoI) interferes significantly with \emph{astar}
(slowing it down by more than $7\times$). \bp mitigates this interference
problem by partitioning the bandwidth between the two
applications. However, \miiseqos intelligently partitions the
available memory bandwidth equally between the two applications to
ensure that both of them meet a more stringent target bound. For
example, for a slowdown bound of $\frac{10}{4}$, \miiseqos allocates
more than 50\% of the bandwidth to \emph{astar}, thereby reducing
\emph{astar}'s slowdown below the bound of $2.5$, while \bp can only
achieve a slowdown of $3.4$ for astar, by equally partitioning the
bandwidth between the two \aois. Furthermore, as a result of its
intelligent bandwidth allocation, \miiseqos significantly reduces the
slowdowns of the other applications in the system compared to \ap and
\bp (as seen in Figure~\ref{fig:casestudy-2app}).

We conclude, based on the evaluations presented above, that \miiseqos
manages memory bandwidth efficiently to achieve both high system
performance and fairness while meeting performance guarantees for one
or more applications of interest.

%% file: applications-mise-model/application2.tex
\section{\miiseabb-Fair: Minimizing Maximum Slowdown}
\label{sec:app2}

The second mechanism we build on top of our \miiseabb model is
one that seeks to improve overall system fairness. Specifically, this
mechanism attempts to minimize the maximum slowdown across all
applications in the system. Ensuring that no application is unfairly
slowed down while maintaining high system performance is an important
goal in multicore systems where co-executing applications are
similarly important.

%modern cloud computing
%systems where multiple users run their own applications in a
%shared system.

\subsection{Mechanism}
At a high level, our mechanism works as follows. The memory controller
maintains two pieces of information: 1)~a target slowdown bound ($B$)
for all applications, and 2)~a bandwidth allocation policy that
partitions the available memory bandwidth across all applications.
The memory controller enforces the bandwidth allocation policy using
the lottery-scheduling technique as described in
Section~\ref{sec:implementation-scheduling}. The controller attempts
to ensure that the slowdown of all applications is within the bound
$B$. To this end, it modifies the bandwidth allocation policy so that
applications that are slowed down more get more memory
bandwidth. Should the memory controller find that bound $B$ is not
possible to meet, it increases the bound. On the other hand, if the
bound is easily met, it decreases the bound. We describe the two
components of this mechanism: 1)~bandwidth redistribution policy, and
2)~modifying target bound ($B$).

\textbf{Bandwidth Redistribution Policy.} As described in
Section~\ref{sec:implementation-scheduling}, the memory controller
divides execution into multiple \intervals. At the end of each
interval, the controller estimates the slowdown of each application
and possibly redistributes the available memory bandwidth amongst the
applications, with the goal of minimizing the maximum
slowdown. Specifically, the controller divides the set of applications
into two clusters. The first cluster contains those applications whose
estimated slowdown is less than $B$. The second cluster contains those
applications whose estimated slowdown is more than $B$. The memory
controller steals a small fixed amount of bandwidth allocation (2\%)
from each application in the first cluster and distributes it equally
among the applications in the second cluster. This ensures that the
applications that do not meet the target bound $B$ get a larger share
of the memory bandwidth.

\textbf{Modifying Target Bound.}  The target bound $B$ may depend
on the workload and the different phases within each workload.
This is because different workloads, or phases within a workload,
have varying demands from the memory system. As a result, a target
bound that is easily met for one workload/phase may not be
achievable for another workload/phase. Therefore, our mechanism
dynamically varies the target bound $B$ by predicting whether or
not the current value of $B$ is achievable. For this purpose, the
memory controller keeps track of the number of applications that
met the slowdown bound during the past $N$ intervals (3 in our
evaluations). If all the applications met the slowdown bound in
all of the $N$ intervals, the memory controller predicts that the
bound is easily achievable. In this case, it sets the new bound to
a slightly lower value than the estimated slowdown of the
application that is the most slowed down (a more competitive
target). On the other hand, if more than half the applications did
not meet the slowdown bound in all of the $N$ intervals, the
controller predicts that the target bound is not achievable. It
then increases the target slowdown bound to a slightly higher
value than the estimated slowdown of the most slowed down
application (a more achievable target).

\subsection{Interaction with the OS} As we will show in
Section~\ref{sec:app2-eval}, our mechanism provides the best
fairness compared to three state-of-the-art approaches for memory
request scheduling~\cite{atlas,tcm,stfm}. In addition to this,
there is another benefit to using our approach. Our mechanism,
based on the \miiseabb model, can accurately estimate the slowdown
of each application. Therefore, the memory controller can
potentially communicate the estimated slowdown information to the
operating system (OS). The OS can use this information to make
more informed scheduling and mapping decisions so as to further
improve system performance or fairness. Since prior memory
scheduling approaches do not explicitly attempt to minimize
maximum slowdown by accurately estimating the slowdown of
individual applications, such a mechanism to interact with the OS
is not possible with them. Evaluating the benefits of the
interaction between our mechanism and the OS is beyond the scope
of this thesis.

\subsection{Evaluation}
\label{sec:app2-eval}

\begin{figure}[!ht]
  \centering
  \includegraphics[scale=0.4, angle=270]{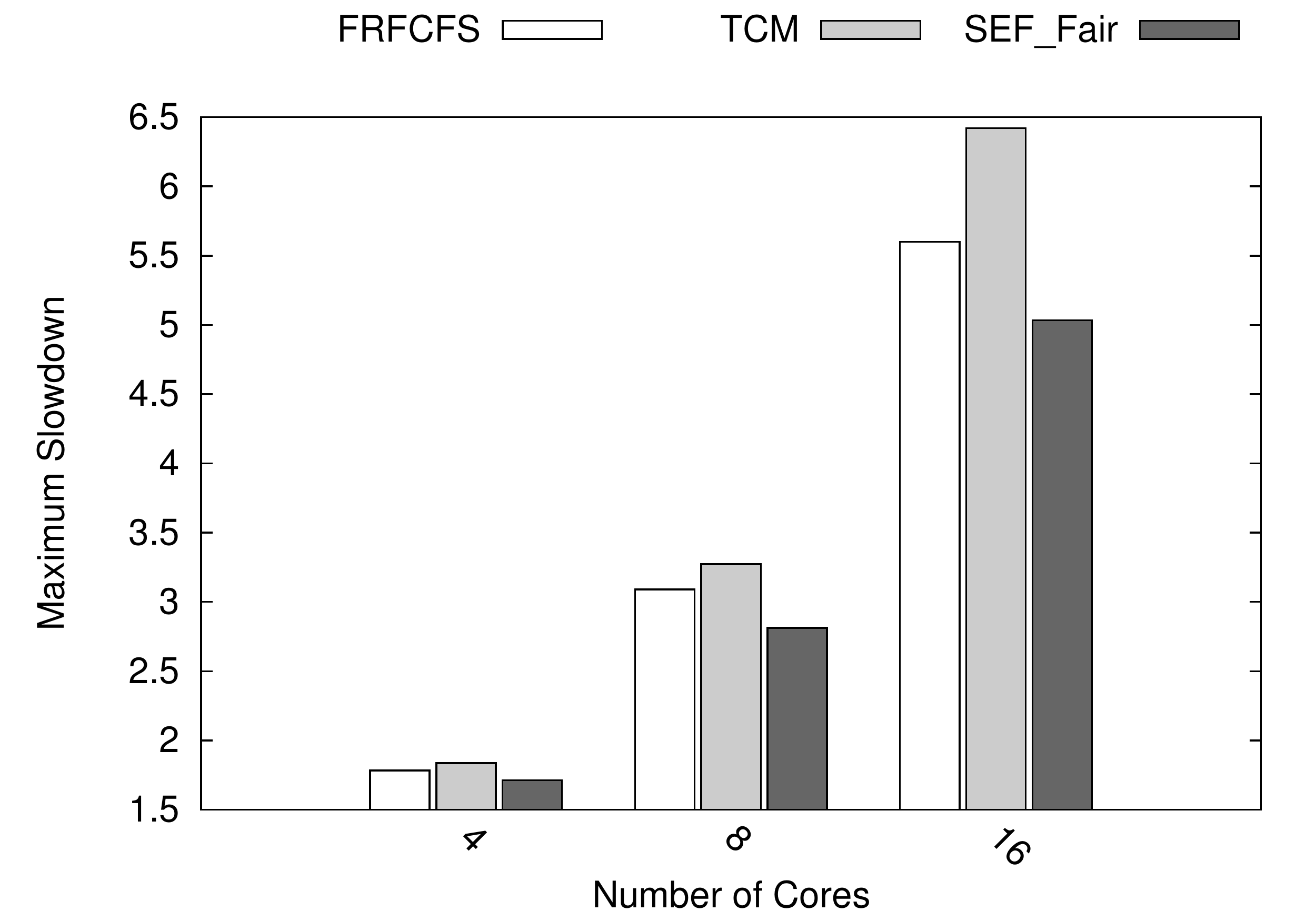}
  \caption{Fairness with different core counts}
  \label{fig:sef-app2-core-scalability-ms}
\end{figure}

Figure~\ref{fig:sef-app2-core-scalability-ms} compares the system
fairness (maximum slowdown) of different mechanisms with increasing
number of cores. The figure shows results with four previously
proposed memory scheduling policies (FRFCFS~\cite{frfcfs,
  frfcfs-patent}, ATLAS~\cite{atlas}, TCM~\cite{tcm}, and
STFM~\cite{stfm}), and our proposed mechanism using the \miiseabb
model (\miiseabb-Fair). We draw three conclusions from our results.

First, \miiseabb-Fair provides the best fairness compared to all
other previous approaches. The reduction in the maximum slowdown
due to \miiseabb-Fair when compared to STFM (the best previous
mechanism) increases with increasing number of cores. With 16
cores, \miiseabb-Fair provides 7.2\% better fairness compared to
STFM.

Second, STFM, as a result of prioritizing the most slowed down
application, provides better fairness than all other previous
approaches. While the slowdown estimates of STFM are not as
accurate as those of our mechanism, they are good enough to
identify the most slowed down application. However, as the number
of concurrently-running applications increases, simply
prioritizing the most slowed down application may not lead to
better fairness. \miiseabb-Fair, on the other hand, works towards
reducing maximum slowdown by stealing bandwidth from those
applications that are less slowed down compared to others. As a
result, the fairness benefits of \miiseabb-Fair compared to STFM
increase with increasing number of cores.

Third, ATLAS and TCM are more unfair compared to FRFCFS. As shown in
prior work~\cite{atlas,tcm}, ATLAS trades off fairness to obtain
better performance. TCM, on the other hand, is designed to provide
high system performance and fairness. Further analysis showed us that
the cause of TCM's unfairness is the strict ranking employed by
TCM. TCM ranks all applications based on its clustering and shuffling
techniques~\cite{tcm} and strictly enforces these rankings. We found
that such strict ranking destroys the row-buffer
locality of low-ranked applications. This increases the slowdown of
such applications, leading to high maximum slowdown.

\begin{figure}[h!]
      \centering
      \includegraphics[scale=0.4, angle=270]{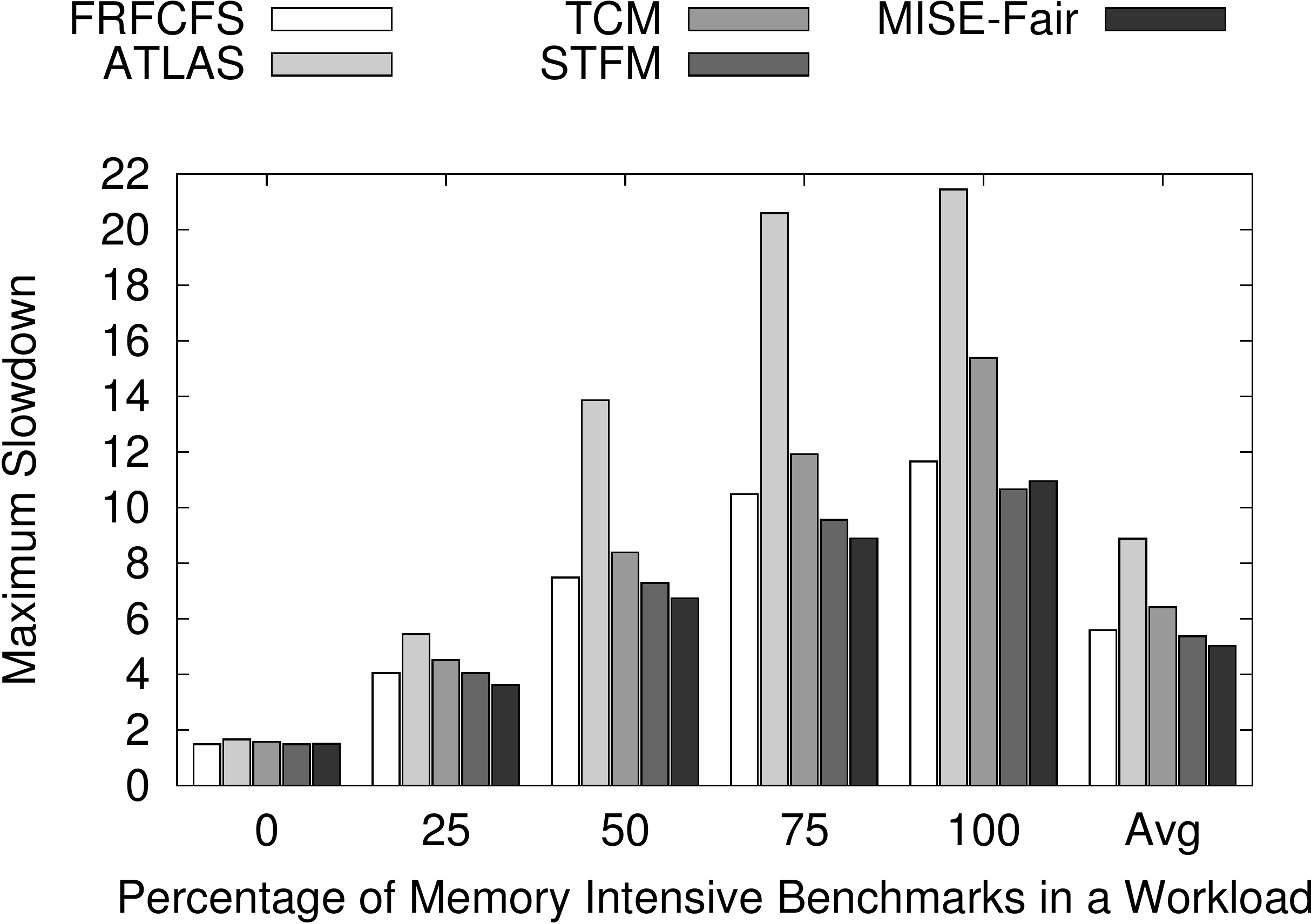}
      \caption{Fairness for 16-core workloads}
      \label{fig:sef-app2-16core-ms}
\end{figure}

\textbf{Effect of Workload Memory Intensity on Fairness.}
Figure~\ref{fig:sef-app2-16core-ms} shows the maximum slowdown of the
16-core workloads categorized by workload intensity. While most trends
are similar to those in Figure~\ref{fig:sef-app2-core-scalability-ms},
we draw the reader's attention to a specific point: for workloads with
non-memory-intensive applications (25\%, 50\% and 75\% in the figure),
STFM is more unfair than \miisefair. As shown in
Figure~\ref{fig:comparison-stfm-nonmembound}, STFM significantly
overestimates the slowdown of non-memory-bound
applications. Therefore, for these workloads, we find that STFM
prioritizes such non-memory-bound applications which are not the most
slowed down. On the other hand, \miisefair, with its more accurate
slowdown estimates, is able to provide better fairness for these
workload categories.

\begin{figure}[h!]
  \centering
  \includegraphics[scale=0.4, angle=270]{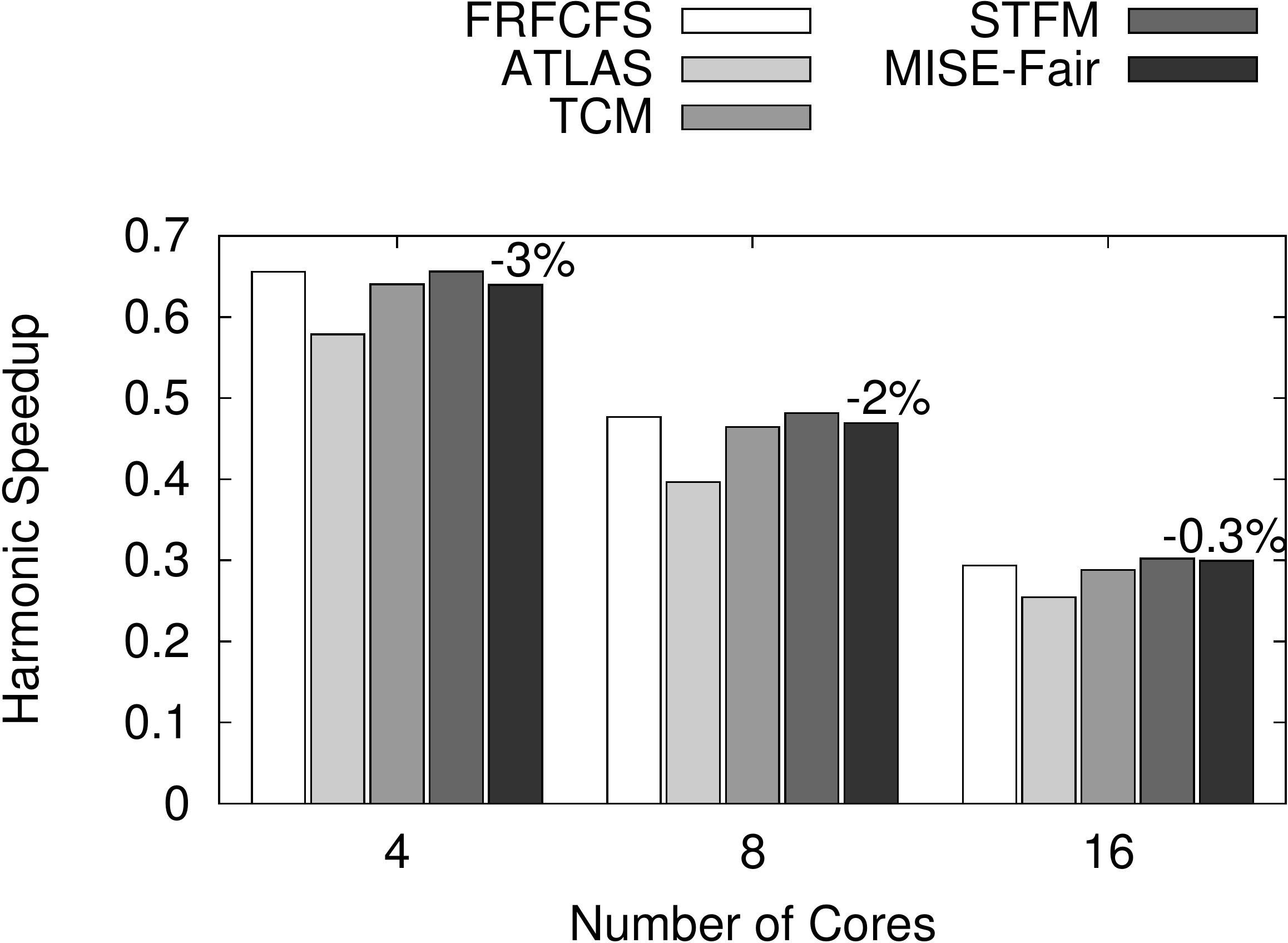}
  \caption{Harmonic speedup with different core counts}
  \label{fig:sef-app2-core-scalability-hs}
\end{figure}

\textbf{System Performance.}
Figure~\ref{fig:sef-app2-core-scalability-hs} presents the
harmonic speedup of the four previously proposed mechanisms
(FRFCFS, ATLAS, TCM, STFM) and \miisefair, as the number of cores
is varied.  The results indicate that STFM provides the best
harmonic speedup for 4-core and 8-core systems. STFM achieves this
by prioritizing the most slowed down application. However, as the
number of cores increases, the harmonic speedup of \miisefair
matches that of STFM.  This is because, with increasing number of
cores, simply prioritizing the most slowed down application can be
unfair to other applications. In contrast, \miisefair takes
into account slowdowns of all applications to manage memory
bandwidth in a manner that enables good progress for all
applications.  We conclude that \miisefair achieves the best
fairness compared to prior approaches, without significantly
degrading system performance.

%% file: applications-mise-model/summary.tex
\section{Summary}
We present two new main memory request scheduling mechanisms that
use \miiseabb to achieve two different goals: 1)~\miiseqos aims to provide soft
QoS guarantees to one or more applications of interest while ensuring high
system performance, 2)~\miisefair attempts to minimize maximum slowdown to
improve overall system fairness. Our evaluations show that our proposed
mechanisms are more effective than the state-of-the-art memory scheduling
approaches~\cite{qos-sigmetrics,atlas,tcm,stfm} in achieving their respective
goals, thereby demonstrating the MISE model's effectiveness in
estimating and controlling application slowdowns.

%% file: chapters/asm-model.tex
\newcommand{\asmfull}{Application Slowdown Model\xspace}
\newcommand{\asm}{ASM\xspace}
\newcommand{\alonecar}{alone-CAR\xspace}
\newcommand{\sharedcar}{shared-CAR\xspace}
\newcommand{\acar}{\emph{cache-access-rate}$_{\textrm{alone}}$\xspace}
\newcommand{\scar}{\emph{cache-access-rate}$_{\textrm{shared}}$\xspace}
\newcommand{\scarn}{\emph{cache-access-rate-N ways}\xspace}
\newcommand{\acarabb}{\emph{CAR}$_{\textrm{alone}}$\xspace}
\newcommand{\scarabb}{\emph{CAR}$_{\textrm{shared}}$\xspace}
\newcommand{\hitshp}{\emph{hits-high-priority}\xspace}
\newcommand{\misseshp}{\emph{misses-high-priority}\xspace}
\newcommand{\hitsats}{\emph{hits-ats}\xspace}
\newcommand{\missesats}{\emph{misses-ats}\xspace}
\newcommand{\hitsalone}{\emph{hits-alone}\xspace}
\newcommand{\missesalone}{\emph{misses-alone}\xspace}
\newcommand{\hits}{\emph{hits}\xspace}
\newcommand{\hitscurrent}{\emph{hits-current}\xspace}

\newcommand{\asmcache}{ASM-Cache\xspace}
\newcommand{\asmmem}{ASM-Mem\xspace}

\newcommand{\hpcycles}{\emph{high-priority-cycles}\xspace}
\newcommand{\alonecycles}{\emph{alone-cycles}\xspace}
\newcommand{\excesscycles}{\emph{excess-cycles}\xspace}
\newcommand{\hitfraction}{\emph{hit-fraction}\xspace}
\newcommand{\missfraction}{\emph{miss-fraction}\xspace}
\newcommand{\avghpmisstime}{\emph{avg-high-priority-miss-time}\xspace}
\newcommand{\avghphittime}{\emph{avg-high-priority-hit-time}\xspace}
\newcommand{\avgmisstime}{\emph{avg-miss-time}\xspace}
\newcommand{\avghittime}{\emph{avg-hit-time}\xspace}
\newcommand{\avgqueueingtime}{\emph{avg-queueing-time}\xspace}
\newcommand{\totalaccesses}{\emph{total-accesses-high-priority}\xspace}
\newcommand{\queueingcycles}{\emph{queueing-cycles}\xspace}

\chapter{Quantifying Application Slowdowns Due to Both Shared
Cache Interference and Shared Main Memory Interference}
\label{chap:asm}

In a multicore system, the shared cache is a key source of
contention among applications. Applications that share the cache
contend for its limited capacity. The shared cache capacity
allocated to an application directly determines its memory
intensity and hence, the degree of memory interference in a
system. 

\begin{figure*} [ht!]
  \centering
  \begin{subfigure}{0.45\textwidth}
    \centering
    \includegraphics[scale=0.25, angle=270]{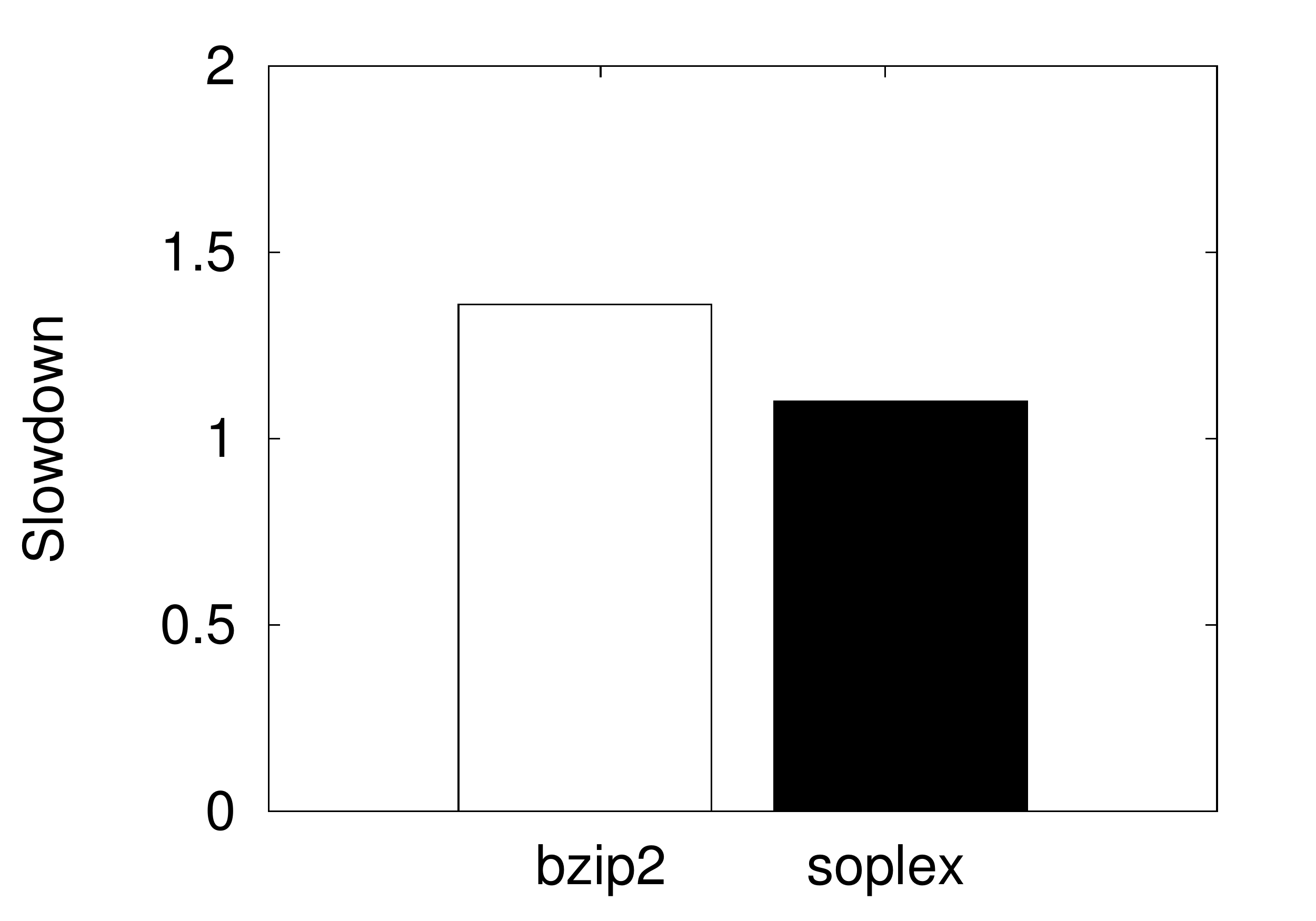}
    \caption{Shared main memory}
  \end{subfigure}
  \begin{subfigure}{0.45\textwidth}
    \centering
    \includegraphics[scale=0.25, angle=270]{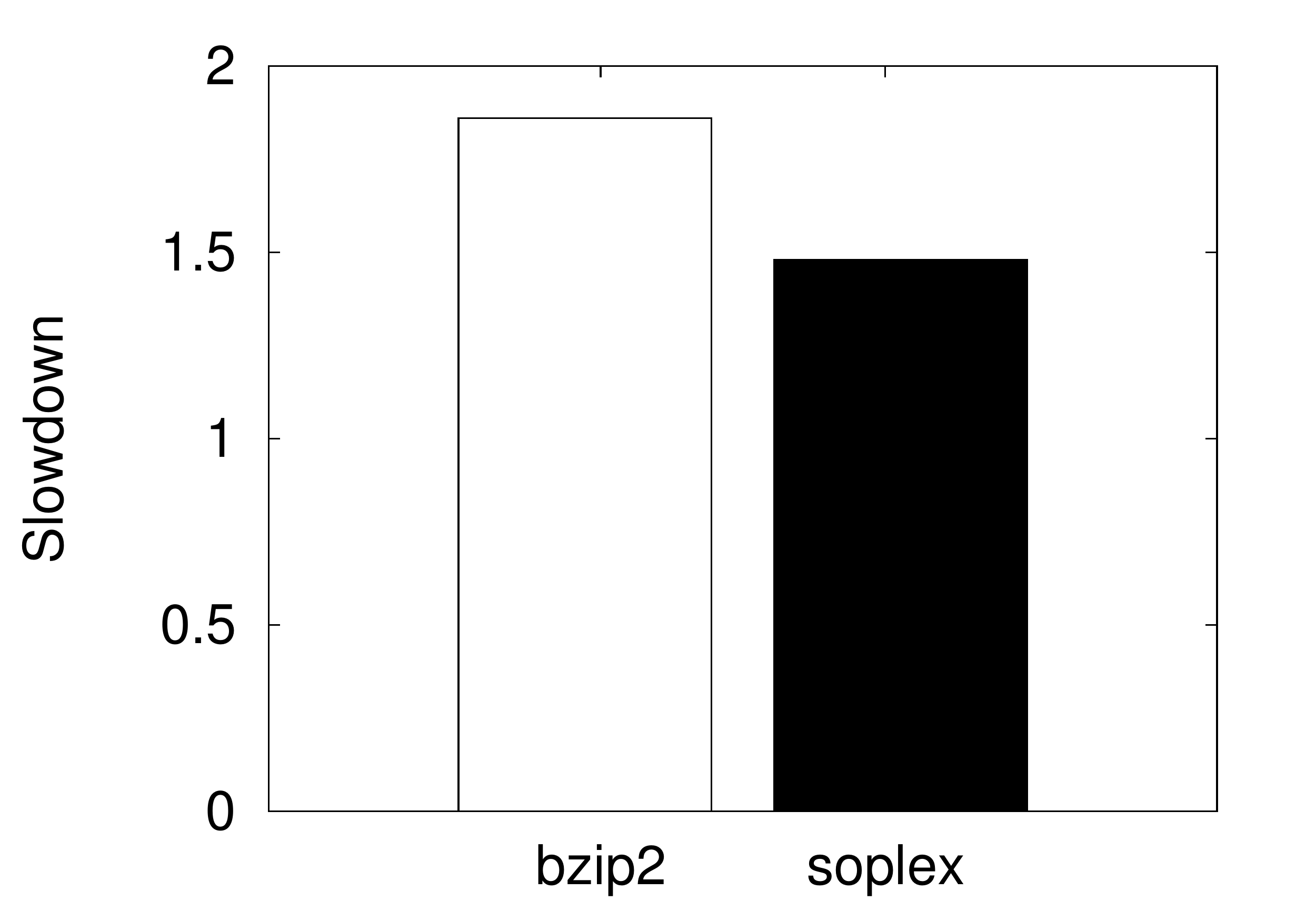}
    \caption{Shared main memory and cache}
  \end{subfigure}
  \caption{Impact of shared cache interference on application slowdowns}
  \label{fig:cache-slowdown-plot}
\end{figure*}

Figure~\ref{fig:cache-slowdown-plot} shows the slowdown of two
representative applications, bzip2 and soplex, when they share
main memory alone and when they share both shared caches and main
memory. As can be seen, when the two applications share the cache,
their slowdown increases significantly compared to when they share
main memory alone. We observe such shared cache interference
across several applications and workloads.

While the MISE model focuses on estimating slowdowns due to
contention for main memory bandwidth, it does not take into account
interference at the the shared caches. We propose to take into
account the effect of shared cache capacity interference, in
addition to main memory bandwidth interference, in estimating
application slowdowns.

Previous works, FST~\cite{fst} and PTCA~\cite{ptca} attempt to
estimate slowdown due to both shared cache and main memory
interference. However, they are inaccurate, since they quantify
the impact of interference at a per-request granularity, as we
described in Chapters~\ref{chap:introduction} and \ref{chap:mise}.
The presence of a shared cache only makes the problem worse as the
request stream of an application to main memory could be
\emph{completely different} depending on whether or not the
application shares the cache with other applications. We strive to
estimate an application' slowdown accurately in the presence of
interference at both the shared cache and the main memory. Towards
this end, we propose the Application Slowdown Model (ASM). 

\input{asm/observations}

\input{asm/model}

\input{asm/methodology}
\input{asm/model-evaluation}
\input{asm/summary}

%% file: asm/observations.tex
\section{Overview of the Application Slowdown Model (\asm)}
\label{sec:observations}

In contrast to prior works which quantify interference at a
per-request granularity, \asm uses \emph{aggregate request behavior} to
quantify interference, based on the following observation.

\subsection{Observation: Access rate as a proxy for performance}  
\label{sec:observation1}

\begin{quote}\parskip0pt\itemsep0pt
   \textit{The performance of each application is proportional to
     the rate at which it accesses the shared cache.}
   \vspace{-4mm}
\end{quote}

Intuitively, an application can make progress when its data
accesses are served. The faster its accesses are served, the
faster it makes progress. In the steady state, the rate at which
an application's accesses are served (service rate) is almost the
same as the rate at which it generates accesses (access rate).
Therefore, if an application can generate more accesses to the cache
in a given period of time (higher access rate), then it can make
more progress during that time (higher performance).

MISE observes that the performance of a \emph{memory-bound
application} is proportional to the rate at which its main memory
accesses are served. However, this observation is stronger than
MISE's observation because this observation relates performance to
the shared cache access rate and not just main memory access rate,
thereby accounting for the impact of both shared cache and main
memory interference. Hence, it holds for a broader class of
applications that are sensitive to cache capacity and/or main
memory bandwidth, and not just memory-bound applications.

To validate our observation, we conducted an experiment in which
we run each application of interest alongside a hog
program on an Intel Core-i5 processor with 6MB shared cache. The
cache and memory access behavior of the hog can be varied
to cause different amounts of interference to the main
program. Each application is run multiple times with the
hog with different characteristics. During each run, we
measure the performance and shared cache
access rate of the application.

\begin{figure}[h]
  \centering
  \includegraphics[scale=0.4, angle=270]{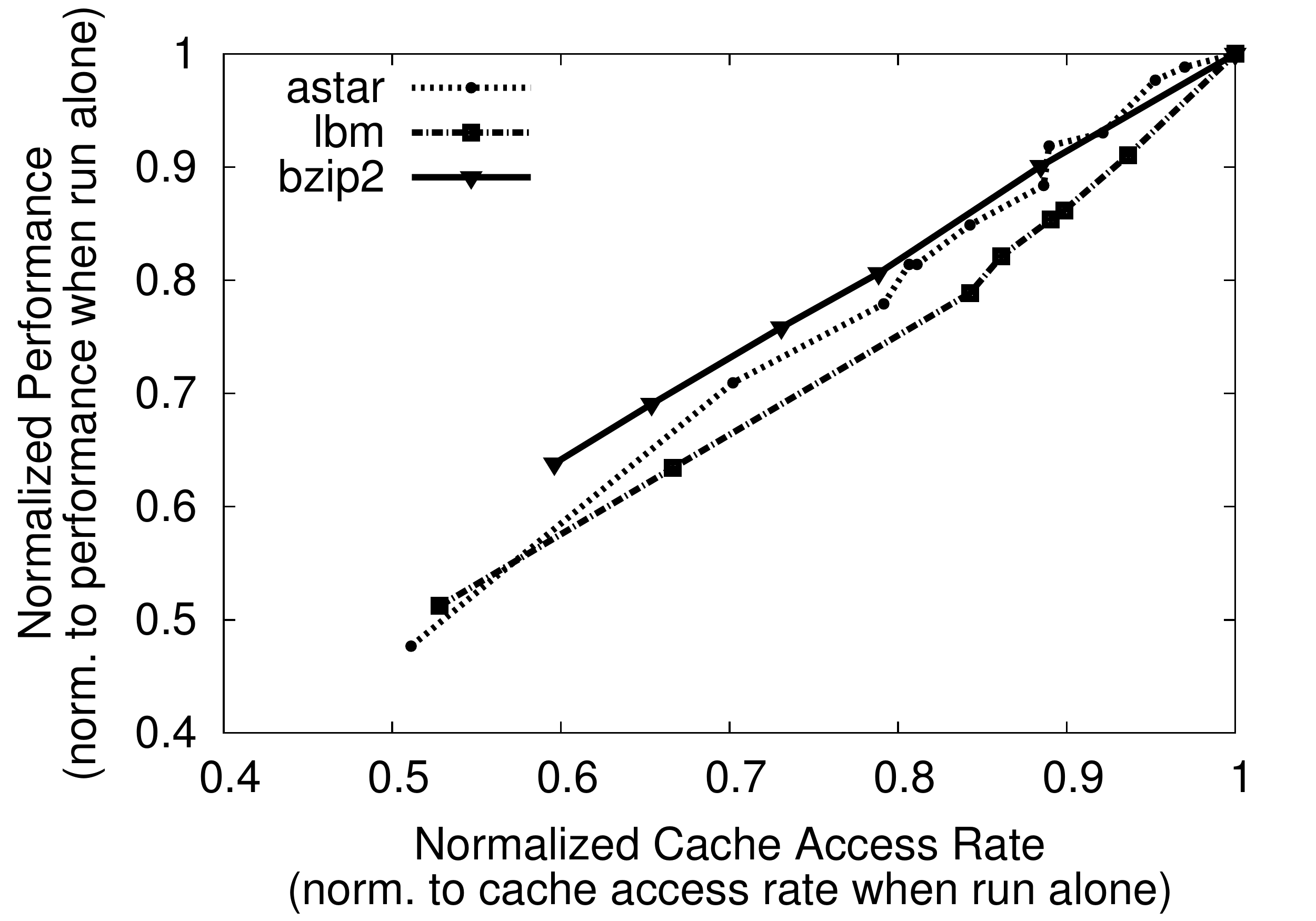}
  \caption{Cache access rate vs. performance}
  \label{fig:bwfraction-speedup}
\end{figure}

Figure~\ref{fig:bwfraction-speedup} plots the results of our
experiment for three applications from the SPEC CPU2006
suite~\cite{spec2006}. The plot shows cache access rate vs.
performance of the application normalized to when it is run alone.
As our results indicate, the performance of each application is
indeed proportional to the cache access rate of the application,
validating our observation. We observed the same behavior for a
wide range of applications.

%Figure~\ref{fig:bwfraction-speedup} plots the results of our
%experiment for three applications from the SPEC CPU2006 benchmark
%suite~\cite{spec2006}. The x-axis plots the cache access rate
%normalized to the cache access rate of the application when it is
%run alone on the system. The y-axis plots the performance of the
%application normalized to when it is run alone. As our results
%indicate, the performance of each application is indeed
%proportional to the cache access rate of the application,
%validating our observation. We observed the same behavior for a
%wide range of applications.

\newcommand{\perf}{\emph{performance}}
\newcommand{\carate}{\emph{cache-access-rate}\xspace}
\newcommand{\caratealone}{$\textrm{\emph{CAR}}_{\textrm{alone}}$\xspace}
    
\asm exploits our observation to estimate slowdown as a ratio of
cache access rates, instead of as a ratio of
performance.\vspace{0mm}
\begin{eqnarray}\belowdisplayskip-2pt
  \textrm{\perf} & \propto & \textrm{\carate~(\emph{CAR})}\nonumber\\
  \vspace{-2mm}
  \textrm{Slowdown} & = &
  \frac{\textrm{\perf}_{\textrm{alone}}}{\textrm{\perf}_{\textrm{shared}}}
  = \frac{\textrm{\emph{CAR}}_{\textrm{alone}}}{\textrm{\emph{CAR}}_{\textrm{shared}}}\nonumber
%  \vspace{-2mm}
\end{eqnarray} 
%  \vspace{-2mm}
\begin{sloppypar}
While $\textrm{\emph{CAR}}_{\textrm{shared}}$/
$\textrm{\perf}_{\textrm{shared}}$ are both easy to measure. the
challenge is in estimating $\textrm{\perf}_{\textrm{alone}}$ or
$\textrm{\emph{CAR}}_{\textrm{alone}}$.
\end{sloppypar}

%\textbf{Why \carate instead of performance?}
%$\textrm{\perf}_{\textrm{shared}}$ and
%$\textrm{\emph{CAR}}_{\textrm{shared}}$ can be estimated in a
%straightforward manner when an application is run with other
%applications. However, it is not straightforward to estimate
%$\textrm{\perf}_{\textrm{alone}}$ or
%$\textrm{\emph{CAR}}_{\textrm{alone}}$ without running an
%application alone. We observe that it is easier to estimate
%$\textrm{\emph{CAR}}_{\textrm{alone}}$ more accurately than
%$\textrm{\perf}_{\textrm{alone}}$, which is why we use \emph{CAR}
%instead of performance. This is enabled by the observation made by
%several previous works(e.g.,~\cite{Sherwood-ASPLOS02,Isci-WCC6}) that
%applications' phase behavior does not change significantly over
%short periods of time (on the order of a few million cycles).
%Hence, $\textrm{\emph{CAR}}_{\textrm{alone}}$ can be estimated periodically during short
%periods of time within a given interval, rather than throughout
%execution. In contrast, previously proposed models~\cite{ptca,fst}
%that estimate slowdown as the ratio of
%$\textrm{\perf}_{\textrm{alone}}$ and
%$\textrm{\perf}_{\textrm{shared}}$ need to estimate
%$\textrm{\perf}_{\textrm{alone}}$ by monitoring interference
%throughout execution. We will describe how we estimate
%$\textrm{\emph{CAR}}_{\textrm{alone}}$, in the next section.

\textbf{$\textrm{\emph{CAR}}_{\textrm{alone}}$ vs.
$\textrm{\perf}_{\textrm{alone}}$.} In order to estimate an
application's slowdown during a given interval, prior works such
as FST and PTCA estimate its alone execution time
($\textrm{\perf}_{\textrm{alone}}$) by tracking the interference
experienced by \emph{each of the application's requests served
during this interval} and subtracting these interference cycles
from the application's shared execution time
($\textrm{\perf}_{\textrm{shared}}$). This approach leads to
inaccuracy, since estimating per-request interference is difficult
due to the parallelism in the memory system.
$\textrm{\emph{CAR}}_{\textrm{alone}}$, on the other hand, can be
estimated more accurately by exploiting the observation made by
several prior works that applications' phase behavior does not
change significantly over time scales on the order of a few
million cycles (e.g.,~\cite{Sherwood-ASPLOS02,Isci-WCC6}). Hence,
$\textrm{\emph{CAR}}_{\textrm{alone}}$ can be estimated
periodically over short time periods during which main memory
interference is minimized (thereby implicitly accounting for
memory level parallelism) and shared cache interference is
quantified, rather than throughout execution. We describe this in
detail in the next section.

\subsection{Challenge: Accurately Estimating \caratealone}
\label{sec:overview-challenge}

A naive way of estimating \caratealone of an application
periodically is to run the application by itself for short periods
of time and measure \caratealone. While such a scheme would
eliminate main memory interference, it would not eliminate shared
cache interference, since the caches cannot be warmed up at will
in a short time duration. Hence, it is not possible to take this
approach to estimate \caratealone accurately. Therefore, \asm
takes a hybrid approach to estimate \caratealone for each
application by 1)~minimizing interference at the main memory, and
2)~quantifying interference at the shared cache.

\textbf{Minimizing main memory interference.} 
\asm minimizes interference for each application at the main
memory by simply giving each application's requests the highest
priority in the memory controller periodically for short lengths
of time, similar to MISE. This has two benefits. First, it
eliminates most of the impact of main memory interference when
\asm is estimating \caratealone for the application (remaining
minimal interference accounted for in
Section~\ref{sec:memory-queueing}). Second, it provides \asm an
accurate estimate of the cache miss service time for the
application in the absence of main memory interference. This
estimate will be used in the next step, in quantifying shared
cache interference for the application.  
%Third, it gives each
%application highest priority for only short time periods, thereby
%preventing any application from causing interference to other
%applications for a long time (negligible change in
%performance/fairness compared to baseline). 
%\footnote{The change in performance and fairness with this prioritization scheme
%compared to baseline FRFCFS~\cite{frfcfs} are negligible.}.

\textbf{Quantifying shared cache interference.} 
%Similar to the first step, we can potentially minimize interference for an
%application at the shared cache by giving all the shared cache
%capacity to the application. This may allow us to directly
%estimate \caratealone. However, we do not take this approach
%because, unlike memory scheduling, where an application can
%immediately benefit from getting high priority, giving all the
%cache space to an application will not immediately benefit the
%application as the cache will not be warm. Furthermore, this could
%degrade other applications' performance. Therefore,
%we propose to quantify the effect of shared cache interference on
%\caratealone of each application, rather than minimizing it.
To quantify the effect of cache interference, we need to identify
the excess cycles that are spent in serving shared cache misses
that are \emph{contention misses}---those that would have otherwise hit
in the cache had the application run alone on the system. We use
an auxiliary tag store for each application to first identify
contention misses. Once we determine the aggregate number of
contention misses, we use the average cache miss service time
(computed in the previous step) and average cache hit service time
to estimate the excess number of cycles spent serving the
contention misses---essentially quantifying the effect of shared
cache interference.

\subsection{\asm vs. Prior Work} 
\asm is better than prior work due to three reasons. First, as we
describe in
Section~\ref{sec:related-work-slowdown-estimation} and in the
beginning of this chapter, prior works aim to
estimate the effect of main memory interference on each contention
miss individually, which is difficult and inaccurate. In contrast,
our approach eliminates most of the main memory interference for
an application by giving the application's requests the highest
priority, which also allows \asm to gather a good estimate of
the \emph{average cache miss service time}. Second, to quantify
the effect of shared cache interference, \asm only needs to
identify the \emph{number} of contention misses, unlike prior
approaches that need to determine whether or not every individual
request is a contention miss. This makes \asm more amenable to
hardware-overhead-reduction techniques like set sampling (more
details in Sections~\ref{sec:aux-sampling}
and~\ref{sec:hardware-cost}). In other
words, the error introduced by set sampling in estimating the
number of contention misses is \emph{far lower} than the error it
introduces in estimating \emph{the actual number of cycles by which each
contention miss is delayed due to interference}. Third, as we
describe in Section~\ref{sec:asm-cache}, \asm
enables estimation of slowdowns for different
cache allocations in a straightforward manner, which is
non-trivial using prior models.

In summary, \asm estimates application slowdowns as a ratio of
cache access rates. \asm
overcomes the challenge of estimating $\textrm{\emph{CAR}}_{\textrm{alone}}$ by minimizing
interference at the main memory and quantifying interference at
the shared cache. In the next section, we describe the
implementation of \asm.

%% file: asm/model.tex
\section{Implementing \asm}
\label{sec:asm-model}

\setlength{\belowdisplayskip}{2pt} \setlength{\belowdisplayshortskip}{2pt}
\setlength{\abovedisplayskip}{2pt}
\setlength{\abovedisplayshortskip}{2pt}

\newcommand{\qcycles}{$Q$\xspace}
\newcommand{\ecycles}{$E$\xspace}
\newcommand{\ehits}{\textrm{\emph{epoch-hits}}\xspace}
\newcommand{\emisses}{\textrm{\emph{epoch-misses}}\xspace}
\newcommand{\eaccesses}{\textrm{\emph{epoch-accesses}}\xspace}
\newcommand{\ecount}{\textrm{\emph{epoch-count}}\xspace}
\newcommand{\ehittime}{\textrm{\emph{epoch-hit-time}}\xspace}
\newcommand{\emisstime}{\textrm{\emph{epoch-miss-time}}\xspace}
\newcommand{\eatshits}{\textrm{\emph{epoch-ATS-hits}}\xspace}
\newcommand{\eatsmisses}{\textrm{\emph{epoch-ATS-misses}}\xspace}
\newcommand{\samatshits}{\textrm{\emph{sampled-ATS-hits}}\xspace}
\newcommand{\samatsmisses}{\textrm{\emph{sampled-ATS-misses}}\xspace}
\newcommand{\eexcess}{\textrm{\emph{epoch-excess-cycles}}\xspace}
\newcommand{\avgqueue}{\textrm{\emph{avg-queueing-delay}}\xspace}
\newcommand{\amisstime}{\textrm{\emph{avg-miss-time}}\xspace}
\newcommand{\ahittime}{\textrm{\emph{avg-hit-time}}\xspace}
\newcommand{\atshitfrac}{\textrm{\emph{ats-hit-fraction}}\xspace}
\newcommand{\atsmissfrac}{\textrm{\emph{ats-miss-fraction}}\xspace}

%Many applications have multiple phases. As a result, the slowdown
%of each application due to shared cache and main memory
%interference could vary with time. To account for this, 
\asm divides execution into multiple quanta, each of length \qcycles
cycles (a few million cycles). At the end of
each quantum, \asm 1)~measures \scarabb, and 2)~estimates \acarabb
for each application, and reports the slowdown of each application
as the ratio of the application's \acarabb and \scarabb.

\subsection{Measuring \scarabb}

Measuring \scarabb for each application is fairly
straightforward. \asm keeps a per-application counter that
tracks the number of shared cache accesses for the
application. The counter is cleared at the beginning of each
quantum and is incremented whenever there is a new shared cache
access for the application. At the end of each quantum, the
\scarabb for each application can be computed as
\begin{eqnarray}
    \textrm{\scar} = \frac{\textrm{\# Shared Cache Accesses}}{Q}\nonumber
\end{eqnarray}
\subsection{Estimating \acarabb}

As we described in Section~\ref{sec:overview-challenge}, during
each quantum, \asm periodically estimates the \acarabb of each
application by minimizing interference at the main memory and
quantifying the interference at the shared cache. Towards this
end, \asm divides each quantum into epochs of length \ecycles
cycles (thousands of cycles), similar to MISE. Each epoch is
probabilistically assigned to one of the co-running applications.
During each epoch, \asm collects information for the corresponding
application that will later be used to estimate \acarabb for the
application. Each application has equal probability of being
assigned an epoch. Assigning epochs to applications in a
round-robin fashion could also achieve similar effects. However,
we build mechanisms on top of \asm that allocate bandwidth to
applications in a slowdown-aware manner
(Section~\ref{sec:asm-mem}), similar to MISE-QoS and MISE-Fair.
Therefore, in order to facilitate building such mechanisms on top
of \asm, we employ a policy that probabilistically assigns an
application to each epoch.

At the beginning of each epoch, \asm communicates the ID of the
application assigned to the epoch to the memory
controller. During that epoch, the memory controller gives the
corresponding application's requests the highest priority in
accessing main memory. 

To track contention misses, \asm maintains an auxiliary tag store
for each application that tracks the state of the cache had the
application been running alone. The auxiliary tag store of an
application holds the tag entries alone (not the data) of
cache blocks. When a request from another application evicts an
application's block from the shared cache, the tag entry
corresponding to the evicted block still remains in the
application's auxiliary tag store. Hence, the auxiliary tag store
effectively tracks the state of the cache had the application been
running alone on the system.

In this section, we will assume a full auxiliary tag store for
ease of description. However, as we will describe in
Section~\ref{sec:aux-sampling}, our final implementation uses set
sampling to significantly reduce the overhead of the auxiliary tag
store with negligible loss in accuracy.

Table~\ref{table:quantities} lists the quantities that are
measured by \asm for each application during the epochs that are
assigned to the application. At the end of each quantum, \asm uses
these quantities to estimate the \acarabb of the
application. These metrics can be measured using a
counter for each quantity while the application is running with
other applications.

\begin{table}
  \begin{small}
    \begin{tabular}{lp{4.6in}}
      \toprule
      \textbf{Name} & \textbf{Definition}\\
      \midrule
      \ecount & Number of epochs assigned to the application\\
      \midrule
      \multirow{2}{*}{\ehits} & Total number of shared cache hits for the application
      during its assigned epochs\\
      \midrule
      \multirow{2}{*}{\emisses} & Total number of shared cache misses for the application
      during its assigned epochs\\
      \midrule
      \multirow{3}{*}{\ehittime} & Number of cycles during which the application has at
      least one outstanding hit during its assigned epochs\\
      \midrule
      \multirow{3}{*}{\emisstime} & Number of cycles during which the application has
      at least one outstanding miss during its assigned epochs\\
      \midrule
      \multirow{2}{*}{\eatshits} & Number of auxiliary tag store hits for the
      application during its assigned epochs\\
      \midrule
      \multirow{2}{*}{\eatsmisses} & Number of auxiliary tag store misses for the
      application during its assigned epochs\\
      \bottomrule
    \end{tabular}
  \end{small}
  \caption{Quantities measured by \asm for each application to
    estimate \acarabb}
  \label{table:quantities}
\end{table}

The \acarabb of an application is given by,
\begin{eqnarray*}
  \textrm{\emph{CAR}}_{\textrm{alone}} & = & \frac{\textrm{\#
      Requests served during application's
      epochs}}{\textrm{Time to serve above requests when run alone}}\\
  & = & \frac{\ehits + \emisses}{(\ecount * E) - \eexcess}
\end{eqnarray*}
where, \ecount $*$ \ecycles represents the actual time the system
spent serving those requests from the application, and \eexcess is
the number of excess cycles spent serving the application's
contention misses---those that would have been hits had the
application run alone.

At a high level, for each contention miss, the system spends the
time of serving a miss as opposed to a hit had the application
been running alone. Therefore,\\
\noindent
\begin{tabular}{cccc}
  \multirow{1}{*}{\eexcess} & \multirow{1}{*}{$=$} & \textrm{(\# Contention Misses}) $\times$&(\amisstime $-$ \ahittime)
\end{tabular}

where, \amisstime is the average miss service time and \ahittime
is the average hit service time for the application for requests
served during the application's epochs. Each of these terms can be
computed using the quantities measured by \asm, as follows.
\begin{eqnarray*}
  \textrm{\# Contention Misses} &=& \eatshits - \ehits\\
  \textrm{Average Miss Service Time} &=& \frac{\emisstime}{\emisses}\\
  \textrm{Average Hit Service Time} &=& \frac{\ehittime}{\ehits}
\end{eqnarray*}
\subsection{Accounting for Memory Queueing}
\label{sec:memory-queueing}
During each epoch, when there are no requests from the
highest priority application, the memory controller may schedule
requests from other applications. If a high priority request
arrives after another application's request is scheduled, it may
be delayed. To address this problem, we apply a similar mechanism
as the interference cycle estimation mechanism in MISE
(Section~\ref{sec:implementation-arsr}), wherein \asm
measures the number of queueing cycles for each application using
a counter. A cycle is deemed a queueing cycle if a request from
the highest priority application is outstanding and the previous
command issued by the memory controller was from another
application. At the end of each quantum, the counter represents
the queueing delay for all \emisses. However, since \asm has
already accounted for the queueing delay of the contention misses
during its previous estimate by removing the \eexcess taken to
serve contention misses, it only needs to account for the queueing
delay for the remaining true misses, i.e., \eatsmisses. In order
to do this, \asm computes the average queueing cycle for each miss
from the application.
\begin{eqnarray*}
  \avgqueue = \frac{\textrm{\# queueing cycles}}{\emisses}
\end{eqnarray*}

\noindent and computes its final \acarabb estimate as

\noindent
\begin{tabular}{llc}
  \multirow{2}{*}{\acarabb} & \multirow{2}{*}{$=$} & {\centering \ehits + \emisses}\\
  \cmidrule{3-3}
  &&$(\ecount * E) - \eexcess - (\eatsmisses * \avgqueue)$
\end{tabular}

\subsection{Sampling the Auxiliary Tag Store}
\label{sec:aux-sampling}

As we mentioned before, in our final implementation, we use set
sampling to reduce the overhead of the auxiliary tag store
(ATS). Using this approach, the ATS is maintained only for a few
sampled sets. The only two quantities that are affected by
sampling are \eatshits and \eatsmisses. With sampling enabled, we
first measure the fraction of hits/misses in the sampled ATS. We
then compute \eatshits/\eatsmisses as a product of the hit/miss
fraction with the total number of cache accesses.
\begin{eqnarray*}
    \eatshits = \atshitfrac \times \eaccesses\\
    \eatsmisses = \atsmissfrac \times \eaccesses
\end{eqnarray*}
where \eaccesses = \ehits + \emisses

\subsection{Hardware Cost}
\label{sec:hardware-cost}

\asm tracks the seven quantities in Table~\ref{table:quantities}
and \textrm{\# queueing cycles} using registers. We find that
using a four byte register for each of these counters is more than
sufficient for the values they keep track of. Hence, the counter
overhead is 32 bytes for each application.  In addition to these
counters, an auxiliary tag store (ATS) is maintained for each
application. The ATS size depends on the number of sets that are
sampled. For 64 sampled sets and 16 ways per set, assuming four
bytes for each entry, the overhead is 4KB \emph{per-application},
which is 0.2\% the size of a 2MB cache (used in our main
evaluations).

%% file: asm/methodology.tex
\section{Methodology}

\noindent\textbf{System Configuration.} We model the main memory
system using a cycle-level in-house DDR3-SDRAM simulator. We
validated the simulator against DRAMSim2~\cite{dramsim2} and
Micron's behavioral Verilog model~\cite{ddr3verilog}. We integrate
our DRAM simulator with an in-house simulator that models
out-of-order cores with a Pin~\cite{pin} frontend. Each system
consists of a per-core private L1 cache and a shared L2
cache. Table~\ref{tab:meth} lists the main system parameters. Our
main evaluations use a 4-core system with a 2MB shared cache and
1-channel main memory.

\input{asm/tables/methodology}

\noindent\textbf{Workloads.} For our multiprogrammed workloads, we
use applications from the SPEC CPU2006~\cite{spec2006} and NAS
Parallel Benchmark~\cite{nas} suites (run
single-threaded). We construct workloads with varying
memory intensity---applications for each workload are chosen
randomly. We run each workload for 100 million cycles. In all, we
present results for 100 4-core, 100 8-core and 100 16-core workloads.

\noindent\textbf{Metrics.} We use average error to compare the
accuracy of \asm and previously proposed models (similar to MISE). We compute
slowdown estimation error for each application, at the end of
every quantum ($Q$), as the absolute value of
\begin{eqnarray*}
    \textrm{Error} = \frac{\textrm{Estimated Slowdown} - \textrm{Actual Slowdown}}{\textrm{Actual Slowdown}} \times 100\%\\
\end{eqnarray*}
\vspace{-7mm}
\begin{eqnarray*}
    \textrm{Actual Slowdown} = \frac{IPC_{alone}}{IPC_{shared}}
\end{eqnarray*}
We compute $IPC_{alone}$ for the same amount of work as the shared
run for each quantum. For each application, we compute the average
slowdown estimation error across all quanta in a workload run and
then compute the average across all occurrences of the application
in all of our workloads.

%In later sections, where we leverage \asm to
%build resource management schemes, we use the harmonic
%speedup~\cite{harmonic-speedup} metric to measure system
%performance and the maximum slowdown
%metric~\cite{stc,atlas,tcm,max-slowdown} to measure unfairness.

\noindent\textbf{Parameters.} We compare \asm with two previous
slowdown estimation models: Fairness via Source Throttling
(FST)~\cite{fst} and Per-Thread Cycle Accounting
(PTCA)~\cite{ptca}. For \asm, we set the quantum length ($Q$) to
5,000,000 cycles and the epoch length ($E$) to 10,000 cycles. For
\asm and PTCA, we present results both with sampled and unsampled
auxiliary tag stores (ATS). For FST, we present results with
various pollution filter sizes that match the size of the ATS.
Section~\ref{sec:sensitivity-algo-parameters} evaluates the
sensitivity of \asm to sampling, quantum and epoch lengths.

%% We use a quantum length,
%% $Q$, of 5000000 cycles and an epoch length, $E$, of 10000 cycles
%% for our primary evaluations. For \asm and PTCA, we evaluate both
%% sampled auxiliary tag stores (4KB overhead, 0.2\% of a 2MB cache
%% that we employ in our primary evaluations) and full auxiliary tag
%% stores that are maintained for all sets (128KB overhead, 6.25\% of
%% a 2MB cache). For FST, we evaluate pollution filters of size 32K
%% and 1024K bits (so as to incur the same overhead as the sampled
%% and unsampled auxiliary tag stores respectively). We present
%% sensitivity to auxiliary tag store size and epoch/quantum lengths
%% in Section~\ref{sec:sensitivity-algo-parameters}.

%% file: asm/tables/methodology.tex
\begin{table}[h!]
%\vspace{-3mm}
  \begin{small}
    \centering
    \begin{tabular}{p{0.3\linewidth}p{0.62\linewidth}}
      \toprule
      \multirow{1}{*}{\textbf{Processor}} &  4-16 cores, 5.3GHz, 3-wide issue, 128-entry instruction window\\
      \midrule
      \multirow{2}{*}{\textbf{L1 cache}} &  64KB, private, 4-way associative, LRU, line size = 64B,\\
       & latency = 1 cycle\\
      \midrule
      \multirow{2}{*}{\textbf{Last-level cache}} & 1MB-4MB, shared, 16-way associative, LRU, line size = 64B,\\
       & latency = 20 cycles\\
      \midrule
      \multirow{2}{*}{\textbf{Memory controller}} &  128-entry request buffer per
      controller, FR-FCFS~\cite{frfcfs} scheduling policy\\
      \midrule
      \multirow{2}{*}{\textbf{Main Memory}} & DDR3-1333 (10-10-10)~\cite{micron}, 1-4 channels,
      \mbox{1 rank/channel},\\
       & 8 banks/rank, 8KB rows\\
      \bottomrule
    \end{tabular}
    \vspace{-1mm}
    \caption{Configuration of the simulated system}
    \label{tab:meth}
  \end{small}
\vspace{-2mm}
\end{table}

%% file: asm/model-evaluation.tex
\section{Evaluation of the Model}
\label{sec:model-evaluation}

%In this section, we compare \asm against prior work on
%slowdown estimation and evaluate the
%sensitivity of \asm and prior models to different system and
%algorithm parameters.

%% In this section, we quantitatively compare \asm against previously proposed
%% slowdown estimation schemes, Fairness via Source Throttling (FST)
%% and Per-thread cycle accounting (PTCA).
%% As described in Section~\ref{sec:background}, FST and PTCA
%% estimate slowdown by tracking individual request behavior and
%% estimating the impact of delaying individual requests on
%% application performance. \asm, on the other hand, estimates
%% slowdown as a function of aggregate request behavior, based on the
%% observation that an application's progress is closely correlated
%% with the rate at which it accesses the shared cache.
%% (Section~\ref{sec:observation1}).

\subsection{Slowdown Estimation Accuracy}
\label{sec:main-model-results}

\begin{figure*}[h!]
  \centering
  \includegraphics[scale=0.25,angle=270]{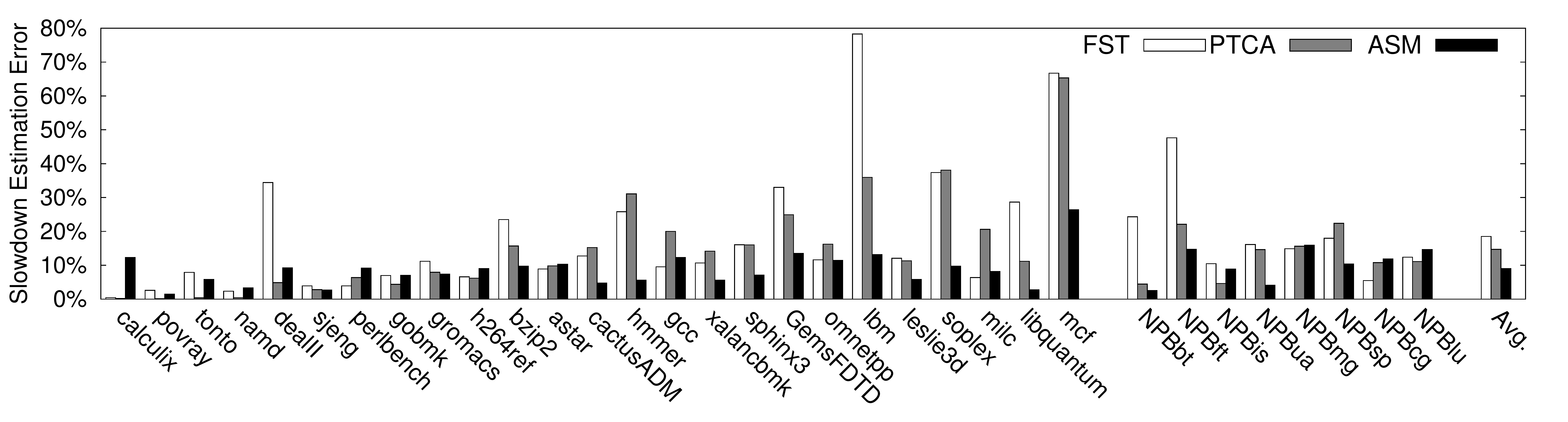}
  \caption{Slowdown estimation accuracy with no sampling}
  \label{fig:avg-error-nosampling}
\end{figure*}

\begin{figure*}[h!]
  \centering
  \includegraphics[scale=0.25,angle=270]{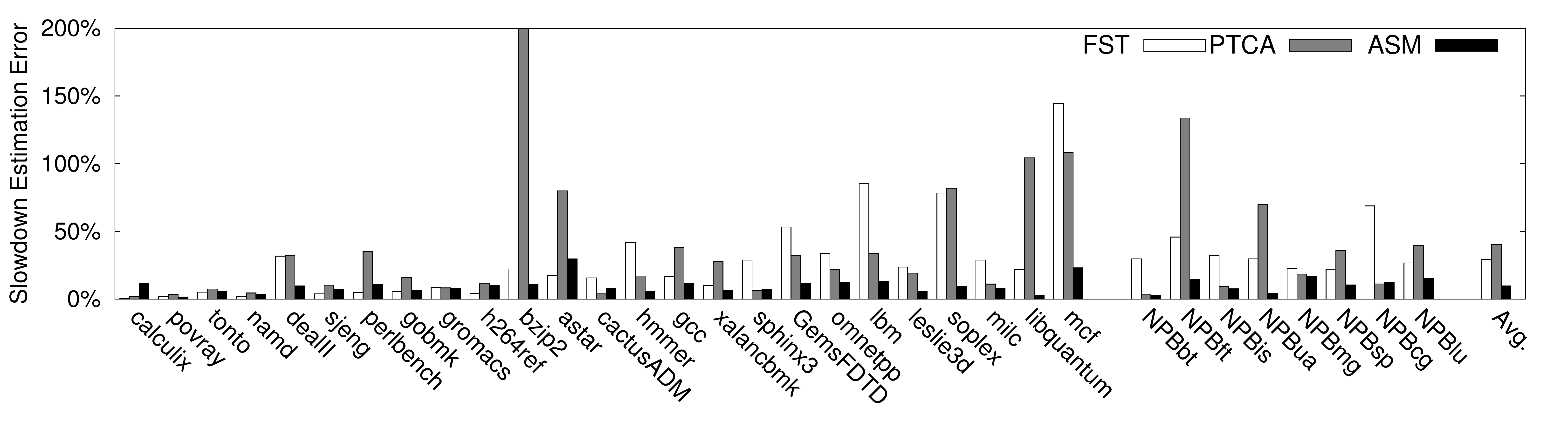}
  \caption{Slowdown estimation accuracy with sampling}
  \label{fig:avg-error-sampling}
\end{figure*}

Figure~\ref{fig:avg-error-nosampling} compares the average
slowdown estimation error from FST, PTCA, and \asm,
\emph{with no sampling} in the auxiliary tag store for PTCA and
\asm, and equal-overhead pollution filter for FST. The benchmarks
on the left are from SPEC CPU2006 suite and those on the right are
from NAS benchmark suite. Benchmarks within each suite are sorted
based on memory intensity. Figure~\ref{fig:avg-error-sampling}
presents the corresponding results with a sampled auxiliary tag
store (64 cache sets) for PTCA and \asm, and an equal-size
pollution filter for FST. 


We draw three major conclusions. First, even without 
sampling, \asm has significantly lower slowdown estimation
error (9\%) compared to FST (18.5\%) and PTCA (14.7\%) (error in
estimating $\textrm{perf}_{alone}$ is very similar). This is
because, as described in Section~\ref{sec:asm-model}, prior works
attempt to quantify the effect of interference on a per-request
basis, which is inherently difficult and inaccurate given the
abundant parallelism in the memory subsystem. \asm, in
contrast, uses aggregate request behavior to quantify
the effect of interference, and hence is more accurate. Our error estimates for PTCA are higher than what is
reported in their paper~\cite{ptca}. This is because they use 
first-come-first-served scheduling at
the memory controller. It is easier to estimate the effect of
interference using this policy than with the FR-FCFS~\cite{frfcfs}
policy which can dynamically reorder requests from different
applications.

Second, sampling the auxiliary tag store and reducing the size of
the pollution filter significantly increase the slowdown
estimation error of PTCA and FST respectively, while it has
negligible impact on the slowdown estimation error of \asm. PTCA's
error increases from 14.7\% to 40.4\% and FST's error increases
from 18.5\% to 29.4\%, whereas \asm's error increases from 9\% to
only 9.9\%. PTCA's error increases from sampling is because it
estimates the number of cycles by which each contention miss (from
the sampled sets) is delayed, and scales up this number to the
entire cache. However, since different requests may experience
different levels of interference, this \emph{scaling} introduces
more error in PTCA's estimates. FST's slowdown estimation error
also increases from sampling, but the increase is not as
significant as PTCA's increase from sampling, because it uses a
pollution filter that is implemented using a Bloom
filter~\cite{bloomfilter}, which is robust to size reductions.
ASM's slowdown estimation error does not increase much from
sampling, since the slowdown of an application is estimated only
when an application has highest priority and is experiencing
minimal interference at the main memory. Quantifying the impact of
shared cache interference using aggregate contention miss counts
and average high priority miss service time estimates, is easier
and more accurate when an application's memory interference is
minimized, rather than tracking per-request interference when an
application is experiencing interference at both the shared cache
and main memory. Section~\ref{sec:sensitivity-algo-parameters}
presents more detailed evaluations demonstrating the impact of
sampling.

Third, FST and PTCA's slowdown estimates are particularly
inaccurate for applications with high memory intensity (e.g.,
\textit{soplex}, \textit{libquantum}, \textit{mcf}) and high cache
sensitivity (e.g., \textit{NPBft}, \textit{dealII},
\textit{bzip2}). This is because applications with high memory
intensity generate a large number of requests to memory, and
accurately modeling the overlap in service of such large number of
requests is difficult, resulting in inaccurate slowdown estimates.
Similarly, an application with high cache sensitivity is severely
affected by shared cache interference. Hence, the request streams
to main memory of the application will be drastically different
when it is run alone vs. when it shares the cache with other
applications. This makes it hard to estimate per-request
interference. \asm simplifies the problem by minimizing main
memory interference and \emph{tracking aggregate rather than per-request
behavior when memory interference is minimized}, resulting in
significantly lower error than prior work for applications with
high memory intensity and/or cache sensitivity.

In summary, with reasonable hardware overhead, \asm estimates
slowdowns more accurately than prior work and is more robust to
applications with varying access behaviors.

%In summary, with reasonable hardware overhead, \asm estimates
%application slowdowns more accurately than prior work (3X better
%than FST). Unlike prior approaches, \asm is more robust to
%applications with varying access behaviors. 

%Since PTCA's slowdown
%estimation error degrades significantly with sampling, from now
%on, we present results for prior works (PTCA and FST) with no
%sampling. However, for \asm, we still present results with a
%sampled auxiliary tag store.

\subsection{Distribution of Slowdown Estimation Error}

\begin{figure}[h]
  \centering
  \includegraphics[scale=0.43,angle=270]{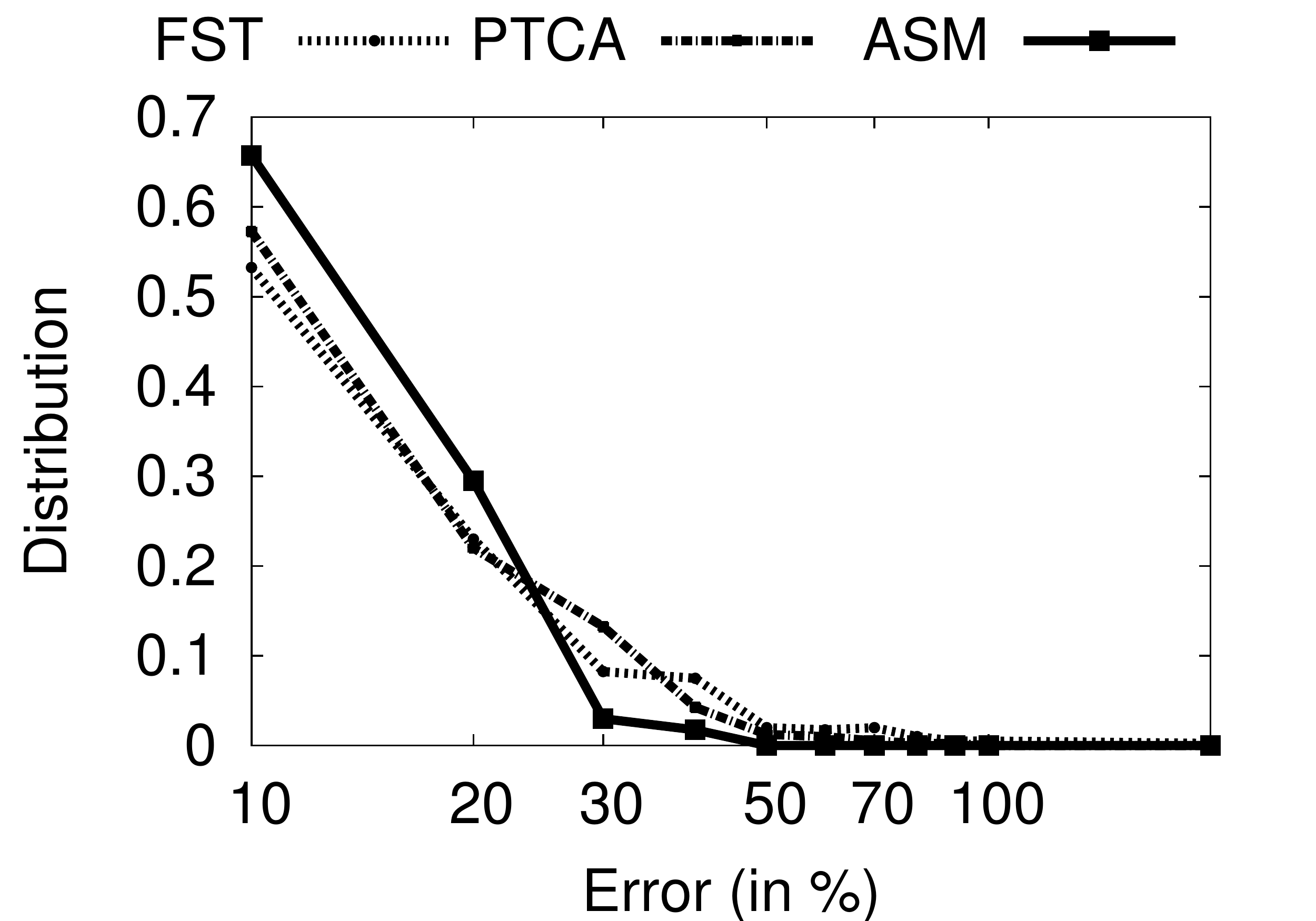}
  \caption{Error distribution}
  \label{fig:error-distribution}
\end{figure}

Figure~\ref{fig:error-distribution} shows a distribution of
slowdown estimation error for FST, PTCA (unsampled) and \asm
(sampled), across all the 400 instances of different applications
in our 100 4-core workloads. The x-axis shows error ranges and the
y-axis shows what fraction of points lie in each range.  Two
observations are in order. First, 95.25\% of \asm's slowdown
estimates have an error less than 20\%, whereas only 76.25\% and
79.25\% of FST and PTCA's estimates respectively lie within the
20\% mark. Second, \asm's maximum error is only 36\%, while
FST and PTCA have maximum errors of 133\% and 87\% respectively.
We observe that \asm's maximum error is for \textit{astar}, which has moderate memory
intensity. When \textit{astar} is run with other higher intensity
applications, it is difficult to account for memory queueing
accurately for \textit{astar}, despite employing the technique
described in Section~\ref{sec:memory-queueing}, leading to
inaccuracy. The highest errors for FST/PTCA are for \textit{lbm}/\textit{mcf} respectively, which are
among the most memory-intensive of the applications we evaluate.
As described in Section~\ref{sec:main-model-results}, the request
overlap when run alone vs. together is particularly dissimilar for
such applications, making it difficult to estimate alone behavior
accurately by tracking per-request interference. We conclude that
\asm's slowdown estimates have much lower variance than FST and
PTCA's estimates and are more robust.

%the request overlap behavior when run alone vs. run together
%is particularly dissimilar for high memory-intensity applications,
%making it difficult to estimate alone behavior accurately by
%tracking per-request interference. We conclude that \asm's slowdown
%estimates have much lower variance than FST and PTCA's estimates and are more robust.

%\begin{figure}[h]
%  \vspace{-4mm}
%  \centering
%  \includegraphics[scale=0.17,angle=270]{asm/plots/distribution_line_plot}
%  \vspace{-1.5mm}
%  \caption{Distribution of slowdown estimation error}
%  \label{fig:error-distribution}
%  \vspace{-7mm}
%\end{figure}

\subsection{Impact of Prefetching}

\begin{figure}[h]
  \centering
  \includegraphics[scale=0.35,angle=270]{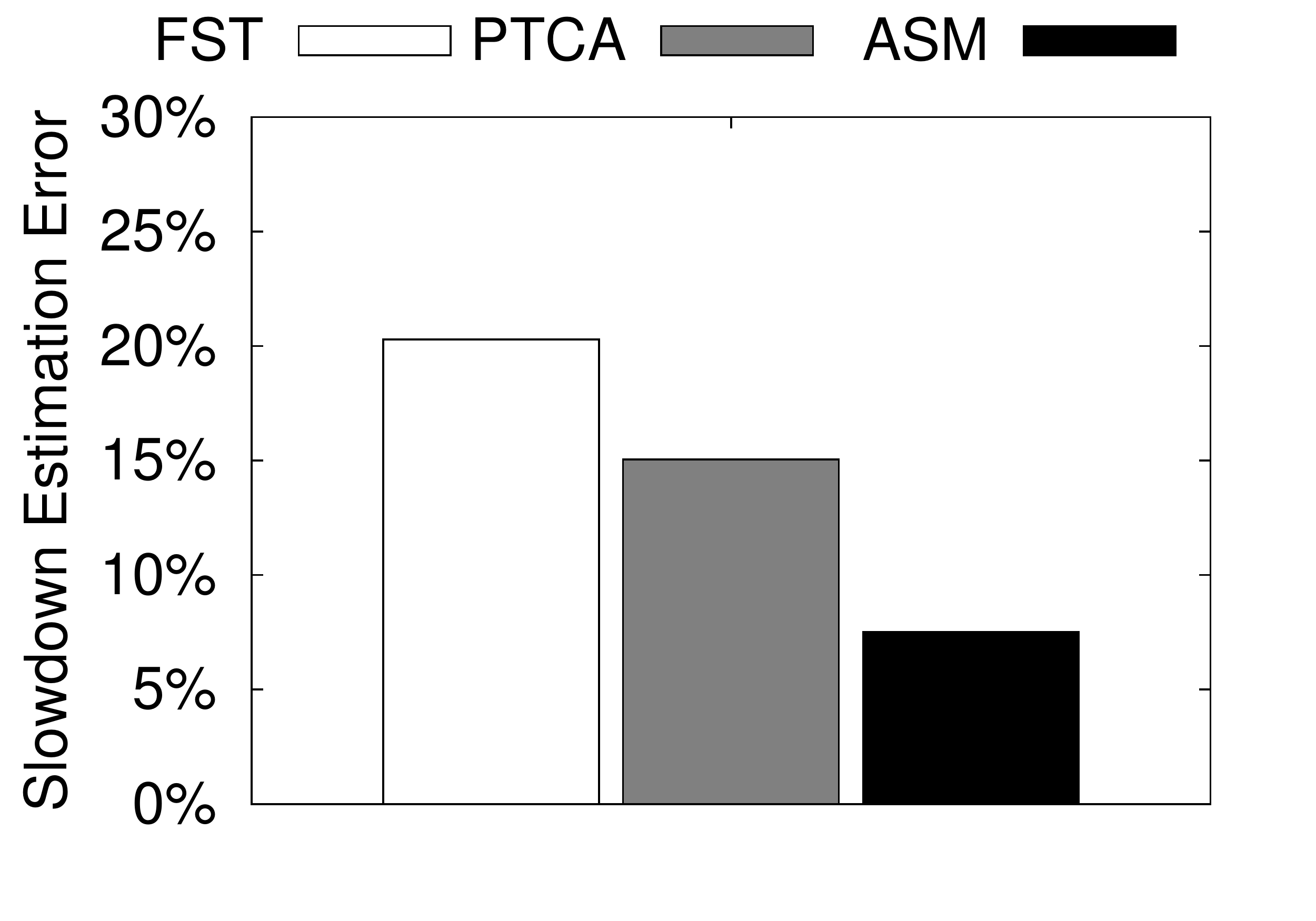}
  \caption{Prefetching impact}
  \label{fig:error-prefetching}
\end{figure}

Figure~\ref{fig:error-prefetching} shows the average slowdown
estimation error for FST, PTCA and \asm, across 100 4-core
workloads (unsampled), with a stride
prefetcher~\cite{stride-prefetching} of degree four. \asm achieves
a significantly low error of 7.5\%, compared to 20\% and 15\% for
FST and PTCA respectively. \asm's error reduces compared to not
employing a prefetcher, since memory interference induced stalls
reduce with prefetching, thereby reducing the amount of
interference whose impact on slowdowns needs to be estimated. This
reduction in interference is true for FST and PTCA as well.
However, their error increases slightly compared to not employing
a prefetcher, since they estimate interference at a per-request
granularity. The introduction of prefetch requests causes more
disruption and hard-to-estimate overlap behavior among requests
going to main memory, making it even more difficult to estimate
interference at a per-request granularity. In contrast, \asm uses
aggregate request behavior to estimate slowdowns, which is more
robust, resulting in more accurate slowdown estimates with
prefetching.

%\begin{figure}[h]
%  \vspace{-3mm}
%  \centering
%  \includegraphics[scale=0.155,angle=270]{asm/plots/error_prefetching}
%  \vspace{-2mm}
%  \caption{Impact of prefetching}
%  \label{fig:error-prefetching}
%  \vspace{-7mm}
%\end{figure}

%\vspace{-2mm}
\subsection{Sensitivity to System Parameters}
\label{sec:sens-core-count}

\begin{figure*}[ht]
  \begin{minipage}{0.48\textwidth}
  \centering
  \includegraphics[scale=0.3,angle=270]{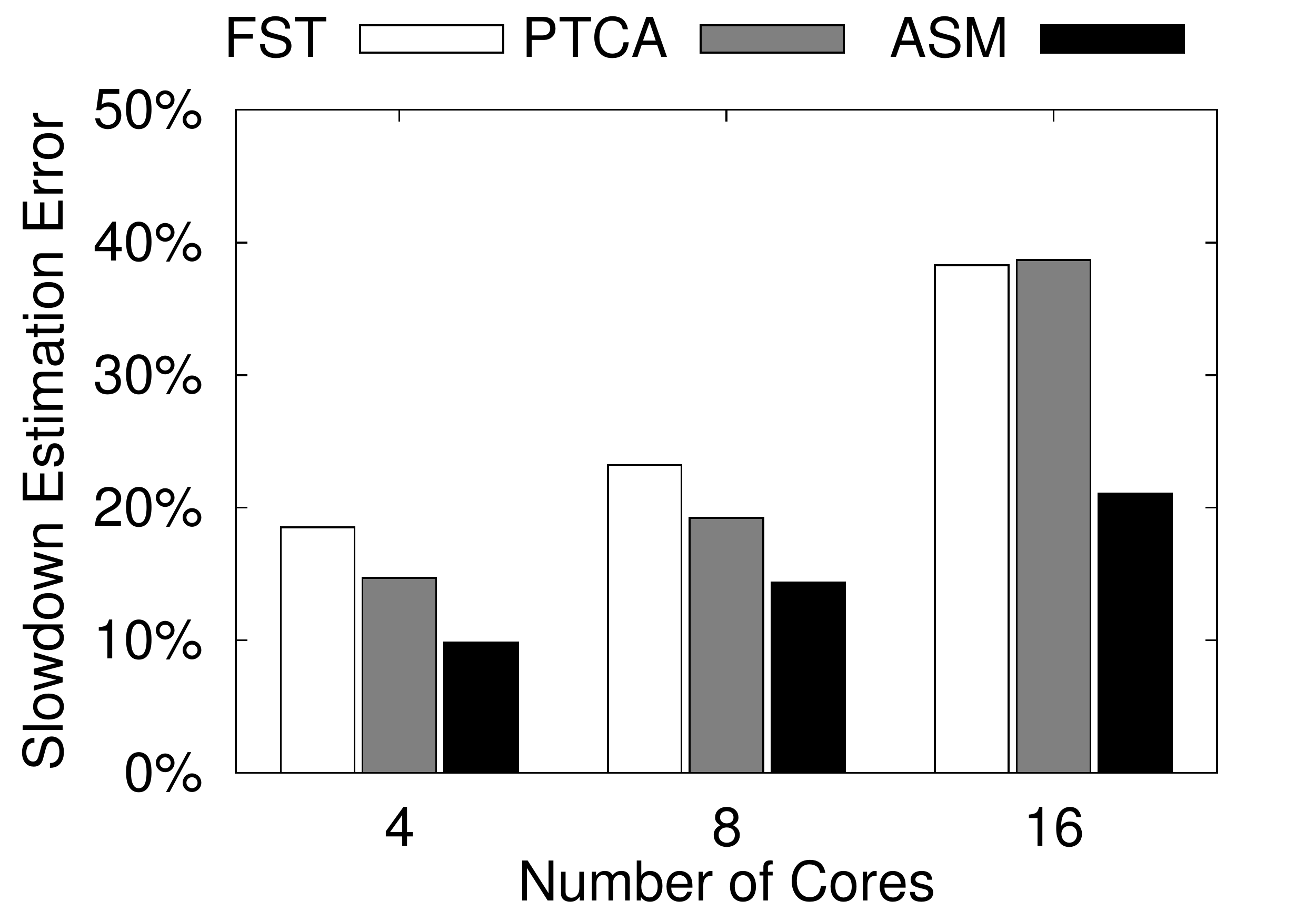}
  \caption{Sensitivity to core count}
  \label{fig:sensitivity-core-count}
  \end{minipage}
  \begin{minipage}{0.48\textwidth}
  \centering
  \includegraphics[scale=0.3,angle=270]{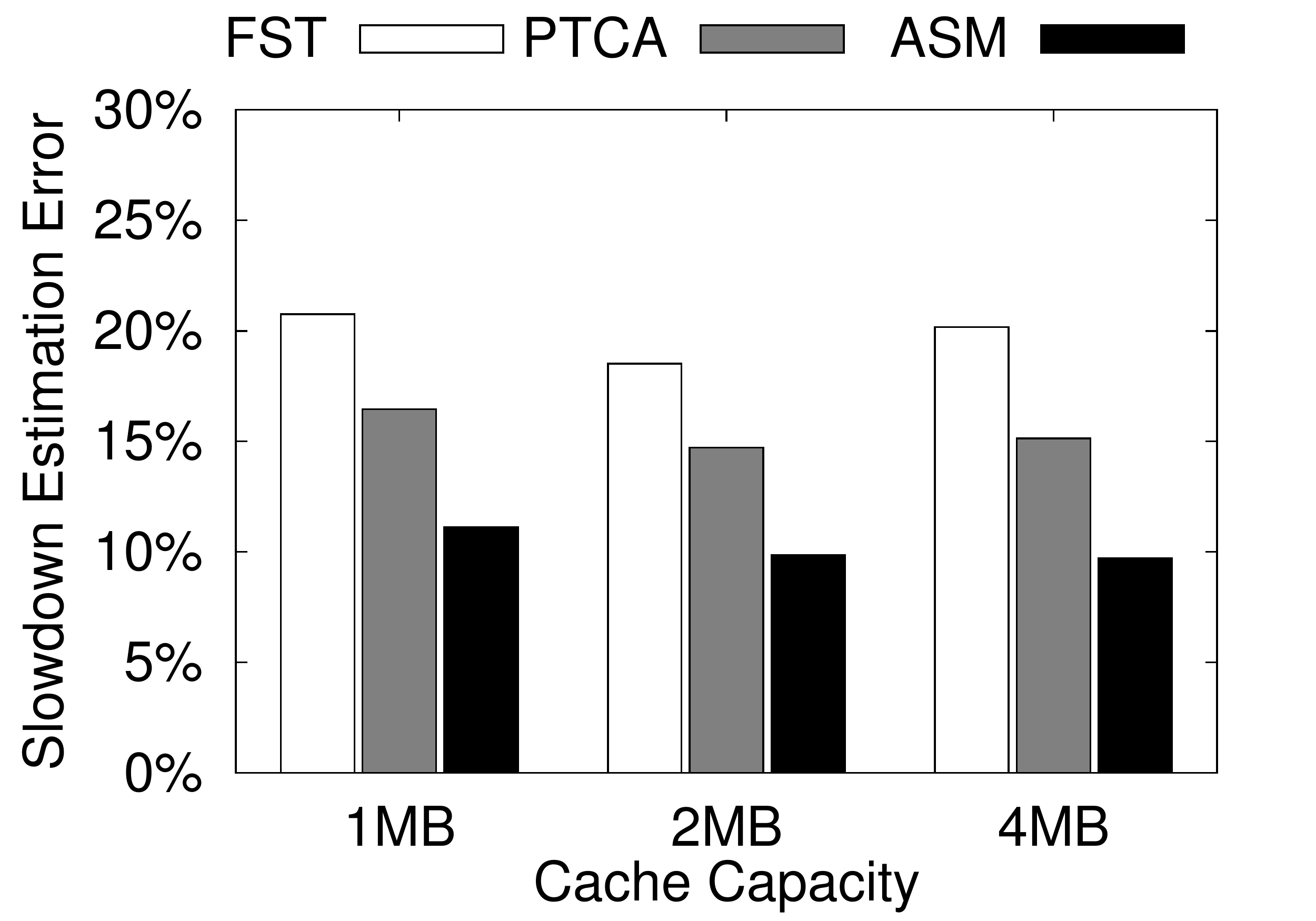}
  \caption{Sensitivity to cache capacity}
  \label{fig:sensitivity-cache-size}
  \end{minipage}
\end{figure*}

\noindent\textbf{Core Count.}
Figure~\ref{fig:sensitivity-core-count} presents the sensitivity
of slowdown estimates from FST, PTCA and \asm to the number of
cores. Since PTCA and FST's slowdown estimation error degrades
significantly with sampling, for all studies from now on, we
present results for prior works with no sampling.
However, for \asm, we still present results with a sampled
auxiliary tag store. We evaluate 100 workloads for each core
count. We draw two conclusions. First, \asm's slowdown estimates
are significantly more accurate than slowdown estimates from FST
and PTCA across all core counts. Second, \asm's accuracy gains over
FST and PTCA increase with increasing core count. As core count
increases, interference at the shared cache and main memory
increases and consequently, request behavior at the shared cache
and main memory is even more different from alone run behavior.
%This makes it more difficult for FST and PTCA to estimate alone
%run behavior accurately by tracking per-request interference,
%resulting in inaccurate slowdown estimates. 
\asm tackles this
problem by tracking aggregate request behavior, thereby scaling more
effectively to larger systems with more number of cores.

\noindent\textbf{Cache Capacity.}
Figure~\ref{fig:sensitivity-cache-size} shows the sensitivity of
slowdown estimates from \asm and previous schemes to cache
capacity across all our 4-core workloads. \asm's slowdown estimates
are significantly more accurate than slowdown estimates from FST
and PTCA, across all cache capacities.

%Two conclusions are in
%order. First, \asm's slowdown estimates are significantly more
%accurate than slowdown estimates from FST and PTCA, across all
%cache capacities. Second, \asm's accuracy benefits increase as
%cache capacity decreases. As cache size decreases, contention for
%shared cache capacity increases, resulting in more main memory
%accesses and consequently, more memory contention. As described in
%Section~\ref{sec:sens-core-count}, such increased contention makes
%alone run behavior estimation more challenging for prior models. 

%\begin{figure}[h]
%  \vspace{-4mm}
%  \centering
%  \includegraphics[scale=0.20,angle=270]{asm/plots/sensitivity_core_count}
%  \caption{Sensitivity to core count}
%  \label{fig:sensitivity-core-count}
%  \vspace{-6mm}
%\end{figure}
%
%\begin{figure}[h]
%  \vspace{-3mm}
%  \centering
%  \includegraphics[scale=0.20,angle=270]{asm/plots/sensitivity_cache_size}
%  \caption{Sensitivity to cache capacity}
%  \label{fig:sensitivity-cache-size}
%  \vspace{-6mm}
%\end{figure}

\subsection{Sensitivity to Algorithm Parameters}
\label{sec:sensitivity-algo-parameters}

\begin{figure}
  \centering
  \includegraphics[scale=0.4,angle=270]{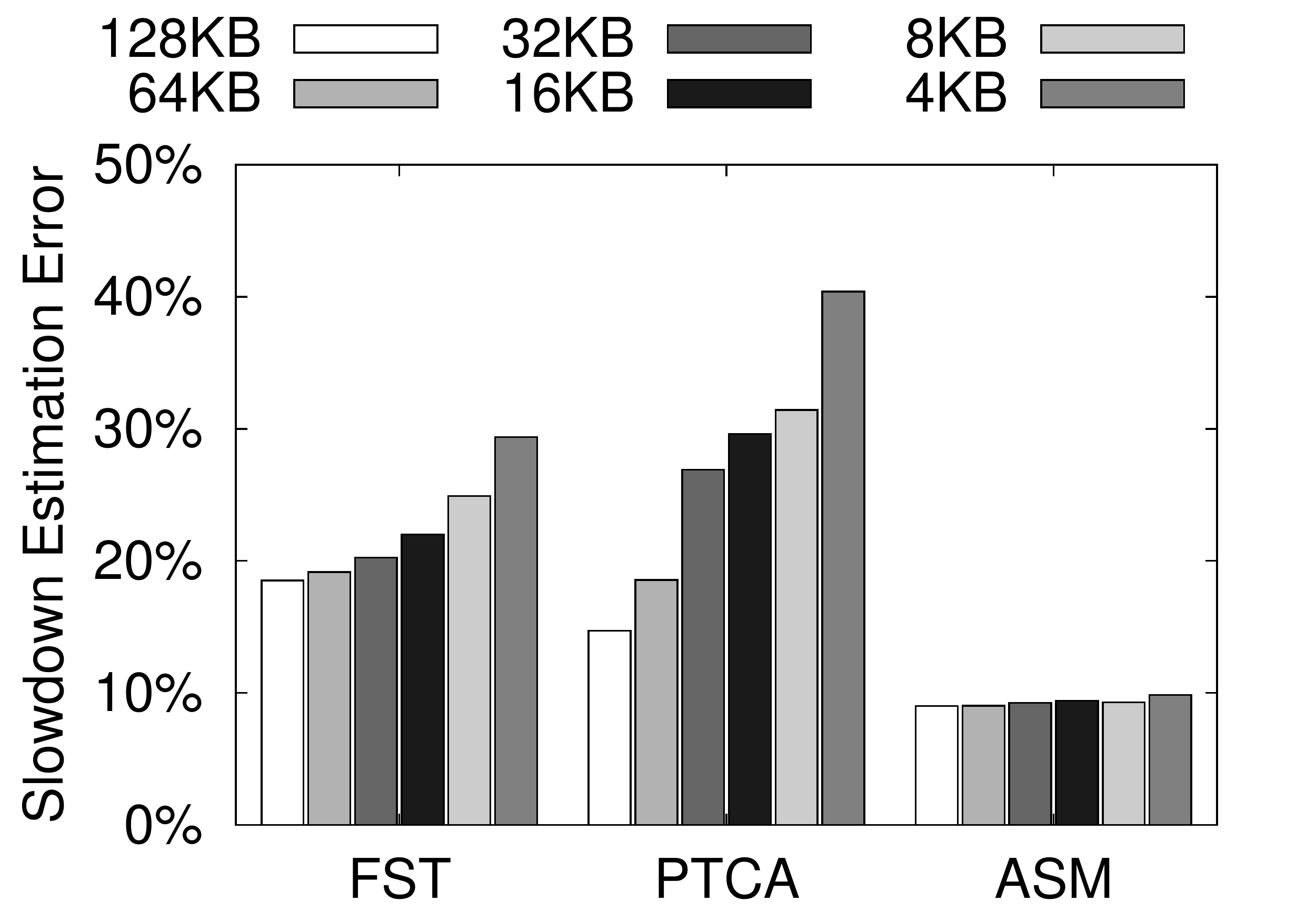}
  \caption{Sensitivity to ATS size}
  \label{fig:sensitivity-ats-size}
\end{figure}

\noindent\textbf{ATS/Pollution Filter Size.}  As
we already explained in Section~\ref{sec:main-model-results}, \asm
is robust to reduction in the size of the auxiliary tag store
(ATS). In contrast, FST and PTCA are significantly affected when
the size of the pollution filter and ATS respectively are reduced. 
Figure~\ref{fig:sensitivity-ats-size} illustrates this
further by plotting the sensitivity of slowdown estimates from
\asm, FST and PTCA to size of the auxiliary tag store
(ATS)/pollution filter. The left most bar for PTCA and \asm
corresponds to no sampling (128KB ATS) and as we move to the right,
we decrease the number of sampled sets. The right most bar
corresponds to 64 sampled sets (4KB ATS). For FST, we set the size
of the pollution filter to be the same as the corresponding
ATS. As expected, varying the size of the ATS has
no visible impact on the estimation error of \asm, whereas
the estimation error of FST and PTCA increase with more
aggressive sampling.

\noindent\textbf{Epoch and Quantum Lengths.}
Table~\ref{tab:alg-sensitivity} shows the average slowdown
estimation error, across all our workloads, for different values
of the quantum ($Q$) and epoch lengths ($E$). As the table shows,
the estimation error increases with decreasing quantum length and
increasing epoch length. This is because the number of epochs
($\sfrac{Q}{E}$) decreases as quantum length ($Q$) decreases
and/or epoch length ($E$) increases. With fewer epochs, certain
applications may not be assigned enough epochs to enable \asm to
reliably estimate their \acarabb. For our main evaluations, we use
a quantum length of 5,000,000 cycles and epoch length of 10,000
cycles.

\begin{table}[ht!]\footnotesize
  \centering
  \input{asm/tables/sensitivity-epoch-quantum}

  \caption{Sensitivity to epoch and quantum lengths}
  \label{tab:alg-sensitivity}
\end{table}

%\begin{table}[ht]\footnotesize
%%  \vspace{-1mm}
%  \centering
%  \input{asm/tables/sensitivity-epoch-quantum}
%  \caption{Sensitivity to epoch and quantum lengths}
%  \label{tab:alg-sensitivity}
%  \vspace{-1mm}
%\end{table}

%% file: asm/tables/sensitivity-epoch-quantum.tex
\begin{tabular}{cccccc}
  \toprule
  \backslashbox{Quantum Length}{Epoch Length} & 10000 & 50000 & 100000\\
  \toprule
  1000000 & 12\% & 14\% & 16.6\% \\
  \midrule
  5000000 & 9.9\% & 10.6\%  & 11.5\%\\
  \midrule
  10000000 & 9.2\% & 9.9\% & 10.5\%\\
  \bottomrule
\end{tabular}

%% file: asm/summary.tex
\section{Summary}
We present the \asmfull (\asm) to estimate the slowdowns of
applications running concurrently on a multicore system due to
\emph{both} shared cache and main memory interference. We observe
that the performance of each application is proportional to the
rate at which the application accesses the shared cache. \asm
exploits this observation to quantify interference using the
aggregate request behavior of each application, by minimizing
interference at the main memory and quantifying interference at
the shared cache. As a result, \asm estimates slowdown more
accurately than prior works, which rely on quantifying
interference at a much finer per-request granularity.

%% file: chapters/applications-asm.tex
\chapter{Applications of ASM}
\label{chap:asm-applications}

\asm's ability to estimate application slowdowns due to both
shared cache and main memory interference can be leveraged to
build various hardware, software and hardware-software-cooperative
slowdown-aware resource management mechanisms to improve
performance, fairness, and provide soft slowdown guarantees.
Furthermore, accurate slowdown estimates can be used to drive fair
pricing schemes based on slowdowns, rather than just resource
allocation or virtual machine migration, in a cloud computing
setting~\cite{azure,ec2}. We explore five such use cases of \asm,
in this chapter.

\input{applications-asm/application-cache.tex}

\input{applications-asm/application-memory.tex}
\input{applications-asm/application-performance-guarantees.tex}
\input{applications-asm/application-fair-billing.tex}
\input{applications-asm/application-migration-admission-control.tex}

\input{applications-asm/summary.tex}

%% file: applications-asm/application-cache.tex
\section{\asm Cache Partitioning (\asmcache)}
\label{sec:asm-cache}

\asmcache partitions the shared cache capacity among applications
with the goal of minimizing slowdown. The basic idea is to
allocate more cache ways to applications whose slowdowns reduce
the most when given additional cache capacity.

\subsection{Mechanism}

\asmcache consists of two main components. First, to partition the
cache in a slowdown-aware manner, we estimate the slowdown of each
application when the application is given different number of
cache ways. Next, we determine the cache way allocation for each
application based on the slowdown estimates using a mechanism
similar to Utility-based Cache Partitioning~\cite{ucp}.

%% \asm-Cache consists of two main components. First, in order to
%% estimate the slowdown reduction from giving additional ways to an
%% application, we estimate the slowdown of an application when it is
%% given different number of cache ways, using principles from the
%% \asm model. Next, we determine the slowdown reduction from giving
%% additional cache ways to each application and allocate ways using
%% a scheme similar to utility-based cache partitioning
%% (UCP)~\cite{ucp}.

\setlength{\belowdisplayskip}{1pt} \setlength{\belowdisplayshortskip}{1pt}
\setlength{\abovedisplayskip}{1pt} \setlength{\abovedisplayshortskip}{1pt}

\newcommand{\slowdown}{\textrm{\emph{slowdown}}\xspace}
\newcommand{\ncarabb}{\textrm{\emph{CAR}}$_n$\xspace}

\noindent\textbf{Slowdown Estimation.} Using the observation
described in Section~\ref{sec:observations}, we estimate slowdown
of an application when it is allocated $n$ ways as
\begin{eqnarray*}
  \slowdown_n = \frac{\textrm{\emph{CAR}}_{\textrm{alone}}}{\textrm{\emph{CAR}}_n}
\end{eqnarray*}
where, \ncarabb is the cache access rate of the application when
$n$ ways are allocated for the application. We estimate \acarabb
using the mechanism described in Section~\ref{sec:asm-model}. While
\ncarabb can be estimated by measuring it while giving all
possible way allocations to each application, such an approach is
expensive and detrimental to performance as the search space is
huge. Therefore, we propose to estimate \ncarabb using a mechanism similar to
estimating \acarabb.

\newcommand{\qhits}{\textrm{\emph{quantum-hits}}\xspace}
\newcommand{\qmisses}{\textrm{\emph{quantum-misses}}\xspace}
\newcommand{\qhitsn}{\textrm{\emph{quantum-hits}$_n$}\xspace}
\newcommand{\qhittime}{\textrm{\emph{quantum-hit-time}}\xspace}
\newcommand{\qmisstime}{\textrm{\emph{quantum-miss-time}}\xspace}

Let \qhits and \qmisses be the number of shared cache hits and
misses for the application during a quantum. At the end of the
quantum,
\begin{eqnarray*}
  \textrm{\emph{CAR}}_n = \frac{\textrm{\qhits +
      \qmisses}}{\textrm{\# Cycles to serve above accesses with
      $n$ ways}}
\end{eqnarray*}
The challenge is in estimating the denominator, i.e., the number of
cycles taken to serve an application's shared cache accesses
during the quantum, if the application had been given $n$ ways. To
estimate this, we first determine the number of shared cache
accesses that would have hit in the cache had the application been
given $n$ ways (\qhitsn). This can be directly obtained from the
auxiliary tag store. (We use a sampling auxiliary tag store and
scale up the sampled \qhitsn value using the mechanism described
in Section~\ref{sec:aux-sampling}.)

There are three cases: 1)~\qhitsn = \qhits, 2)~\qhitsn >
\qhits, and 3)~\qhitsn < \qhits. In the first case, when the
number of hits with $n$ ways is the same as the number of hits during
the quantum, we expect the system to take the same number of
cycles to serve the requests even with $n$ ways, i.e., $Q$
cycles. In the second case, when there are more hits with $n$
ways, we expect the system to serve the requests in fewer than $Q$
cycles. Finally, in the third case, when there are fewer hits with
$n$ ways, we expect the system to take more than $Q$ cycles to
serve the requests. Let $\Delta\textrm{\emph{hits}}$ denote
\qhitsn $-$ \qhits. If \qhittime and \qmisstime are the average
cache hit service time and average cache miss service time for the
accesses of the application during the quantum, we estimate the
number of cycles to serve the requests with $n$ ways as,
\begin{eqnarray*}
  \textrm{\emph{cycles}}_n = Q - \Delta\textrm{\emph{hits}} (\qmisstime -
  \qhittime)
\end{eqnarray*}
wherein we remove/add the estimated excess cycles spent in serving the
additional hits/misses respectively for the application with $n$ ways.  Hence,
\ncarabb is,
\begin{eqnarray*}
  \frac{\qhits + \qmisses}{Q - \Delta\textrm{\emph{hits}} (\qmisstime -
  \qhittime)}
\end{eqnarray*}

It is important to note that extending \asm to estimate
application slowdowns for different possible cache allocations is
straightforward since we use aggregate cache access rates to
estimate slowdowns. Cache access rates for different cache
capacity allocations can be estimated in a straightforward manner.
In contrast, extending previous slowdown estimation techniques
such as FST and PTCA to estimate slowdowns for different possible
cache allocations would require estimating if every individual
request would have been a hit/miss for every possible cache
allocation, which is non-trivial.

\noindent\textbf{Cache Partitioning.} Once we have each
application's slowdown estimates for different way allocations, we
use the look-ahead algorithm used in Utility-based Cache
Partitioning (UCP)~\cite{ucp} to partition the cache ways such
that the overall slowdown is minimized. Similar to the marginal
miss utility (used by UCP), we define marginal slowdown utility as
the decrease in slowdown per extra allocated way. Specifically,
for an application with a current allocation of $n$ ways, the
marginal slowdown utility of allocating $k$ additional ways is,
\begin{eqnarray*}
  \textrm{Slowdown-Utility}^{n+k}_{n} = \frac{\slowdown_n - \slowdown_{n+k}}{k}
\end{eqnarray*}

%% \noindent\textbf{Cache Partitioning.} Once the slowdowns of an
%% application, for every possible cache way allocation, have been
%% estimated, we leverage them to perform slowdown utility-based
%% cache partitioning, similar to the miss utility based cache
%% partitioning scheme proposed by Qureshi et al. in~\cite{ucp}. The
%% basic idea is to allocate each way to the application that shows
%% maximum slowdown reduction from being given the way.  The marginal
%% slowdown utility of an application is calculated as the decrease
%% in slowdown per extra allocated way. Specifically, if an
%% application has a current allocation of M ways, its marginal
%% slowdown utility when it is given N ways (where N is larger than
%% M), is calculated as \begin{equation}
%% Slowdown Utility^{N}_{M} = \frac{(Slowdown_{N ways} - Slowdown_{M ways})}{(N-M)}
%% \end{equation}
Starting from zero ways for each application, the marginal slowdown utility
is computed for all possible way allocations for all
applications. The application that has maximum slowdown utility
for a certain allocation is given those number of ways. This
process is repeated until all ways are allocated. For more details
on the partitioning algorithm, please refer to the look-ahead
algorithm presented in~\cite{ucp}.

\subsection{Evaluation}
\label{sec:ccmm-cachepart-eval}

\begin{figure}[h!]
  \centering
  \begin{minipage}{0.48\textwidth}  
    \centering
    \includegraphics[scale=0.3, angle=270]{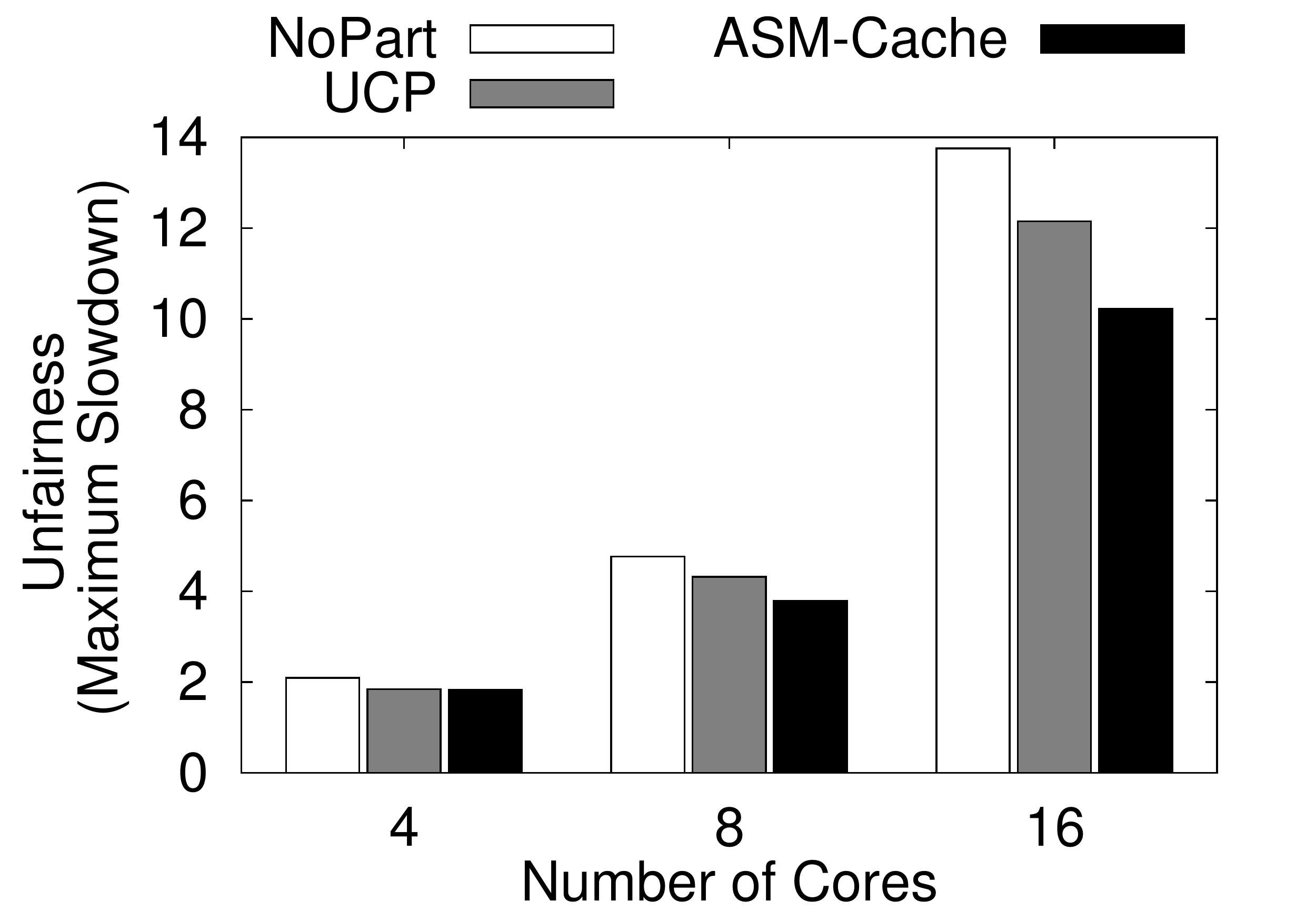}
  \end{minipage}
  \hspace{2mm}
  \begin{minipage}{0.48\textwidth}
    \centering
    \includegraphics[scale=0.3, angle=270]{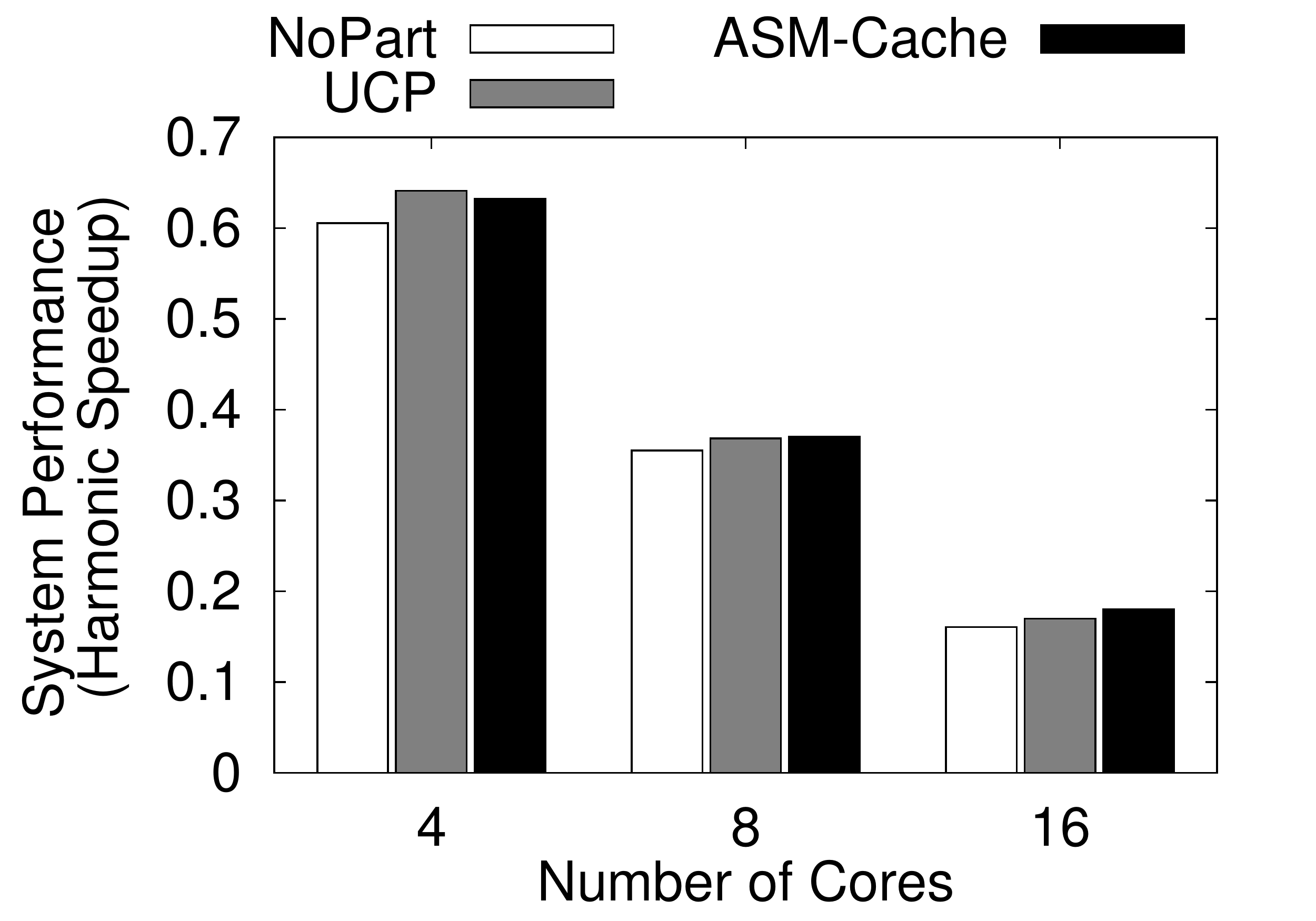}
  \end{minipage}
  \caption{\asm-Cache: Fairness and performance}
  \label{fig:cachepart-results}
\end{figure}

Figure~\ref{fig:cachepart-results} compares the system performance
and fairness of \asm-Cache against a baseline that employs no cache
partitioning (NoPart) and utility-based cache partitioning
(UCP)~\cite{ucp}, for different core counts. We simulate 100
workloads for each core count. We use the harmonic speedup metric
to measure system performance and the maximum slowdown metric to
measure unfairness. Three observations are in order. First,
\asm-Cache provides significantly better fairness and
comparable/better performance across all core counts, compared to
UCP. This is because \asm-Cache explicitly takes into account
application slowdowns in performing cache allocation, whereas UCP
uses miss counts as a proxy for performance. Second, \asm-Cache's
gains increase with increasing core count, reducing unfairness by
12.5\% on the 8-core system and reducing unfairness by 15.8\% and
improving performance by 5.8\% on the 16-core system. This is
because contention for cache capacity increases with increasing
core count, offering more opportunity for \asm-Cache to mitigate
unfair application slowdowns. Third, we see significant fairness
improvements of 12.5\% with a larger (4 MB) cache, on a 16-core
system (plots not presented due to space constraints). We conclude
that accurate slowdown estimates from \asm can enable effective
cache partitioning among contending applications, thereby
improving fairness and performance.

%% file: applications-asm/application-memory.tex
\section{\asm Memory Bandwidth Partitioning} 
\label{sec:asm-mem}

In this section, we present \asm Memory Bandwidth Partitioning
(\asm-Mem), a scheme to partition memory bandwidth among
applications, based on slowdown estimates from \asm, with
the goal of improving fairness. The basic
idea behind \asm-Mem is to allocate bandwidth to each
application proportional to its estimated slowdown, such that
applications that have higher slowdowns are given more bandwidth.

\subsection{Mechanism}

\asm is used to estimate all applications' slowdowns at
the end of every quantum, $Q$. These slowdown estimates are then
used to determine the bandwidth allocation of each application.
Specifically, the probability with which an epoch is assigned to
an application is proportional to its estimated slowdown. The
higher the slowdown of the application, the higher the probability
that each epoch is assigned to the application. For an application
$A_i$, probability that an epoch is assigned to the application is
given by,
\begin{eqnarray*}
\textrm{Probability of assigning an epoch to $A_i$} =
\frac{\slowdown(A_{i})}{\sum_{k} \slowdown(A_{k})}
\end{eqnarray*}
At the beginning of each epoch, the epoch is assigned to one of
the applications based on the above probability distribution and
requests of the corresponding application are prioritized over
other requests during that epoch, at the memory controller.
%\begin{itemize}
%\item
%The goal of this mechanism is to improve system performance and
%fairness by partitioning memory bandwidth based on application
%slowdowns.
%\item
%The basic idea is to allocate an application a fraction of the
%memory bandwidth proportional to its slowdown. This determines
%what fraction of an application's execution time it receives
%highest priority.
%
%$Highest Priority Fraction_{i} = \frac{Slowdown_{i}}{\sigma
%Slowdown_{i}}$
%\item
%The allocated bandwidth fraction is enforced by using a lottery
%scheduling-like scheme. A random number is drawn between 1 and
%100. Depending on which iapplication's range it falls in, the
%corresponding application is given highest priority.
%\item
%We observe that while this scheme provides fairness, in order to
%improve performance of low memory-intensity applications in the
%system, they need high priority. Hence, we identify low
%memory-intensity applications and prioritize them. 
%\end{itemize}

\subsection{Evaluation}

We compare \asm-Mem with three previously proposed memory
schedulers, FRFCFS, PARBS and TCM.
FRFCFS~\cite{frfcfs,frfcfs-patent} is an application-unaware
scheduler that prioritizes row-buffer hits (to maximize bandwidth
utilization) and older requests (for forward progress). FRFCFS
tends to unfairly slow down applications with low row-buffer
locality and low memory intensity. To tackle this
problem, application-aware schedulers such as
PARBS~\cite{parbs} and TCM~\cite{tcm} have been proposed that
reorder applications' requests at the memory
controller, based on their access characteristics.

\begin{figure}[ht!]
  \centering
  \begin{minipage}{0.48\textwidth}  
    \centering
    \includegraphics[scale=0.3, angle=270]{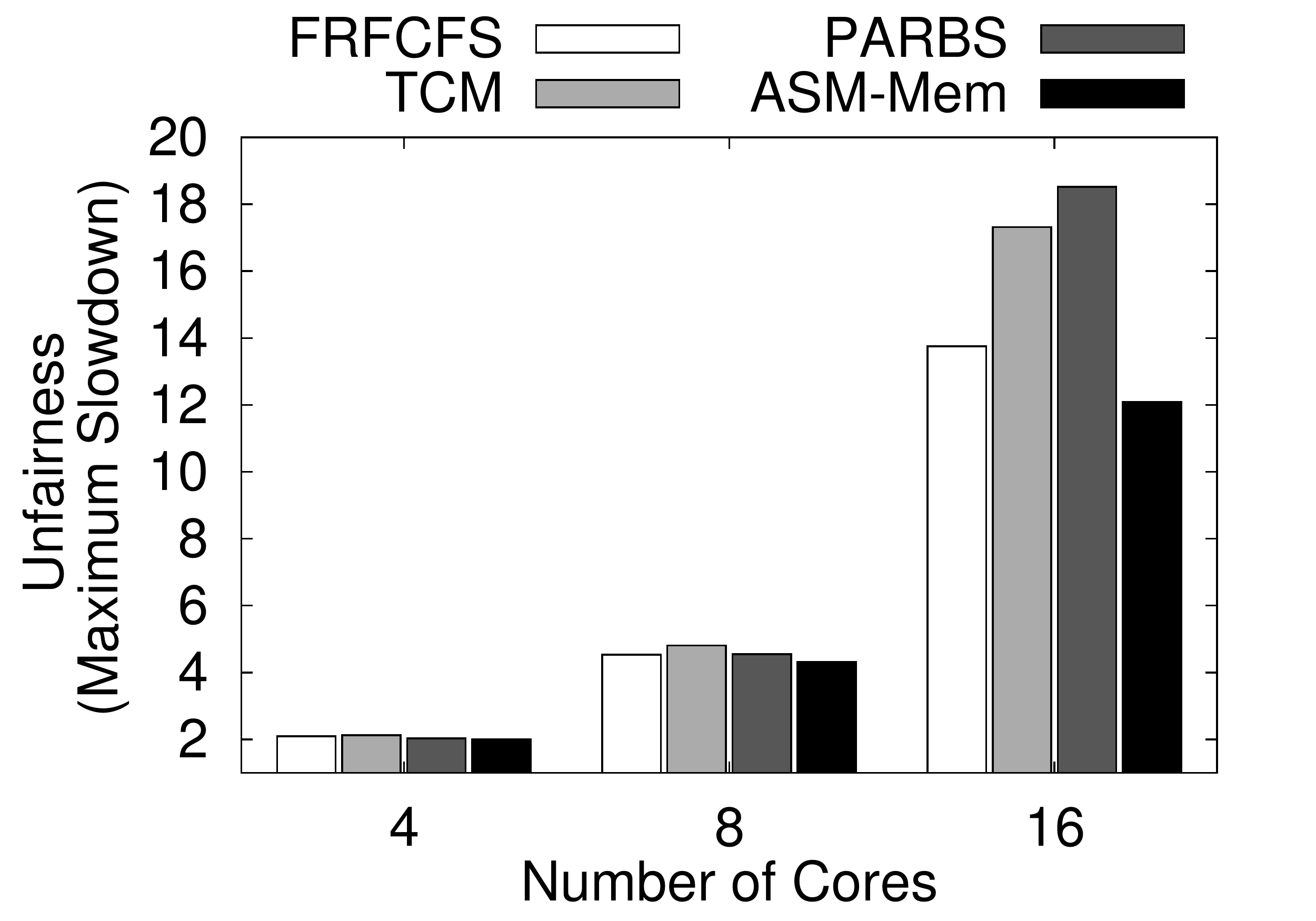}
  \end{minipage}
  \begin{minipage}{0.48\textwidth}
    \centering
    \includegraphics[scale=0.3, angle=270]{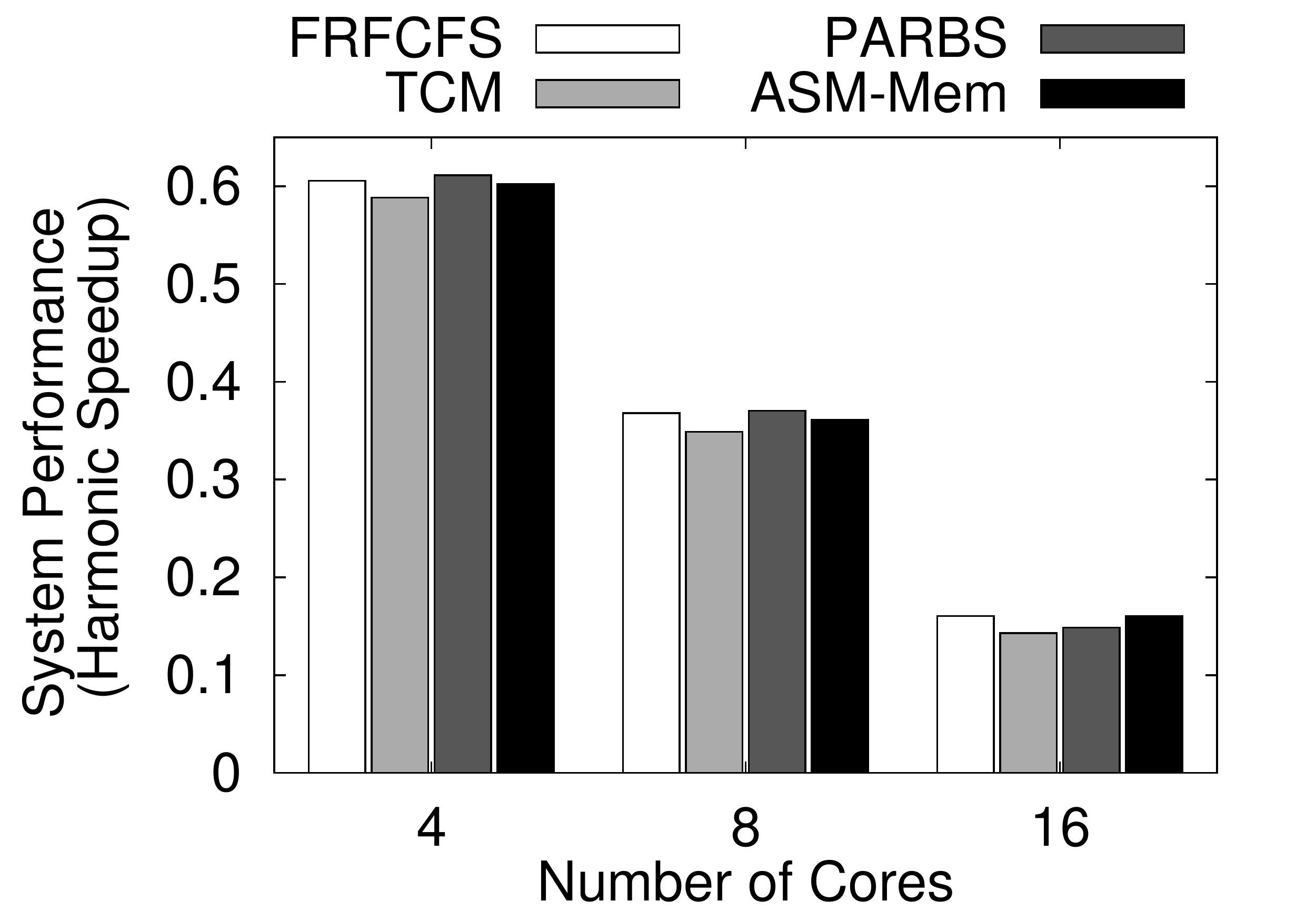}
  \end{minipage}
  \caption{\asm-Mem: Fairness and performance}
  \label{fig:bwpart-results}
\end{figure}

Figure~\ref{fig:bwpart-results} shows the fairness and performance
of \asm-Mem, FRFCFS, PARBS and TCM, for three core
counts, averaged over 100 workloads for each core count. We draw
three major observations. First, \asm-Mem achieves better fairness
than the three previously proposed scheduling policies, while
achieving comparable/better performance. This is because \asm-Mem
directly uses \asm's slowdown estimates to allocate more bandwidth
to highly slowed down applications, while previous works employ
metrics such as memory intensity and row-buffer locality that are
proxies for performance/slowdown. Second, \asm-Mem's gains increase
as the number of cores increases, achieving 5.5\% 
and 12\% improvement in fairness on the 8- and 16-core systems
respectively. Third, we see fairness gains on systems with larger
channel counts as well -- 6\% on a 16-core 2-channel system (do
not plots due to space constraints). We conclude that
\asm-Mem is effective in mitigating interference between
applications at the main memory, thereby improving fairness.

\subsection{Combining \asm-Cache and \asm-Mem} 
\begin{figure}[h!]
  \centering
  \begin{minipage}{0.48\textwidth}  
    \centering
    \includegraphics[scale=0.3, angle=270]{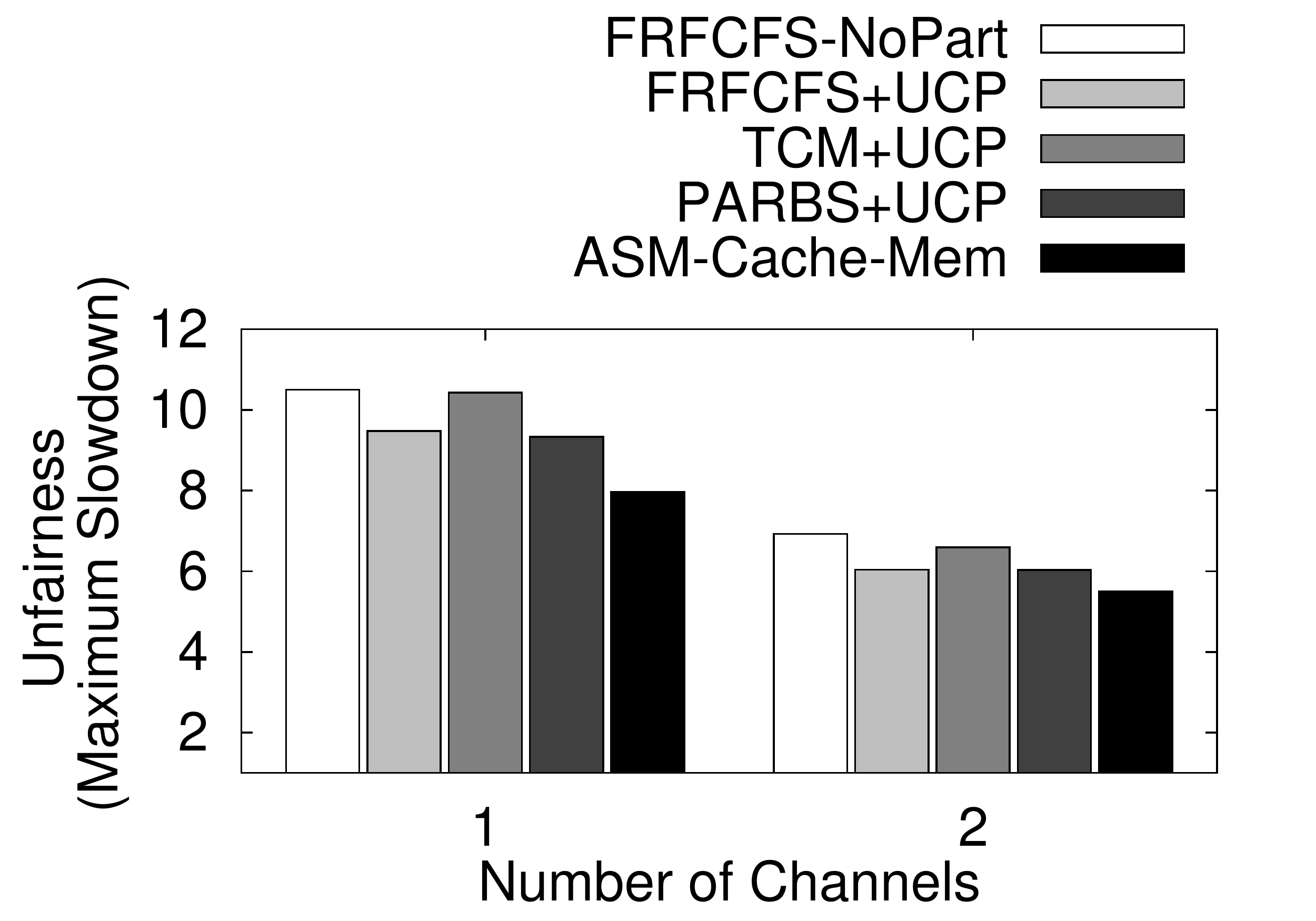}
  \end{minipage}
  \hspace{2mm}
  \begin{minipage}{0.48\textwidth}
    \centering
    \includegraphics[scale=0.3, angle=270]{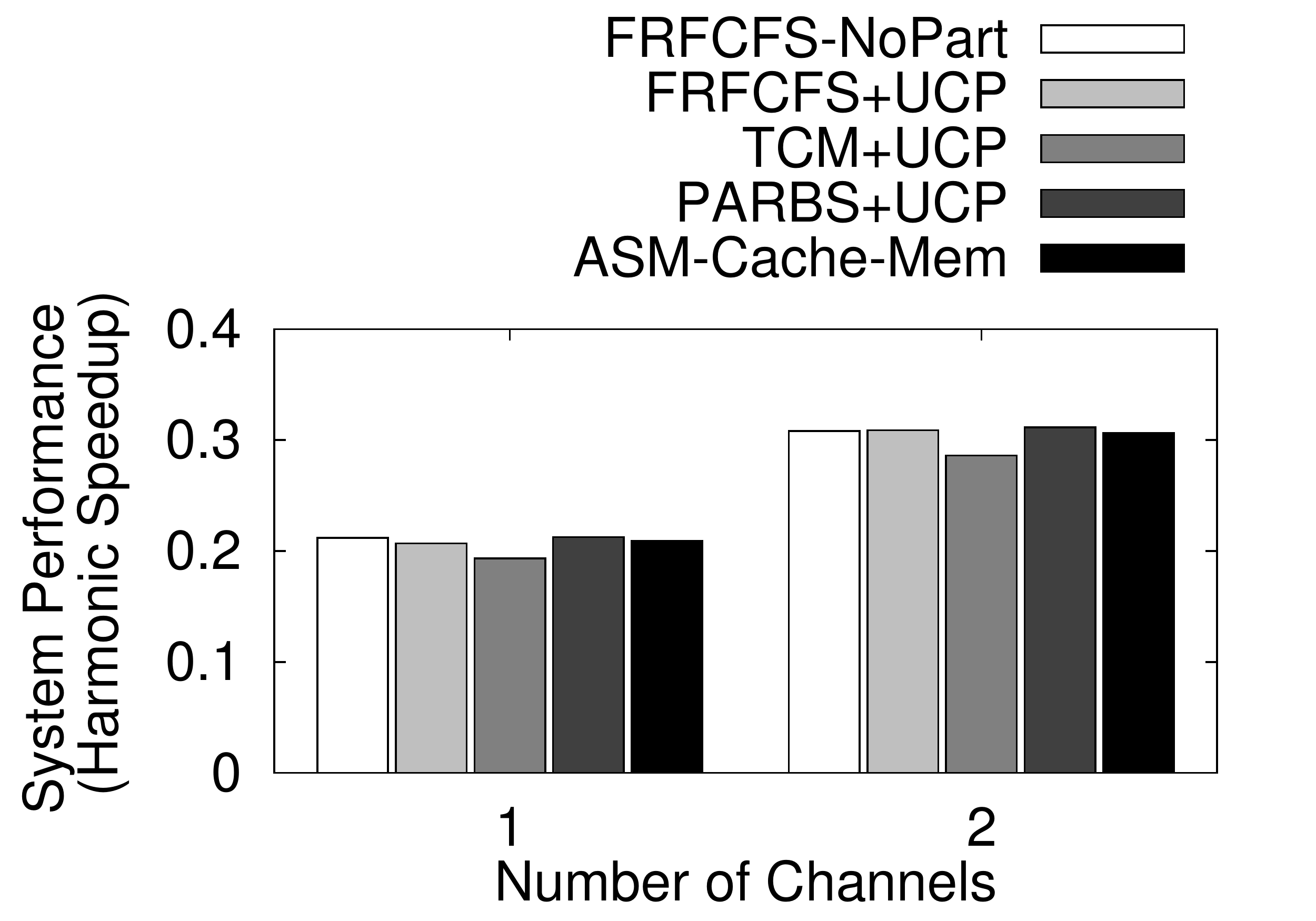}
  \end{minipage}
  \caption{Combining \asm-Cache and \asm-Mem}
  \label{fig:cachepart-bwpart-results}
\end{figure}

We combine
\asm-Cache and \asm-Mem to build a coordinated cache-memory
management scheme. \asm-Cache-Mem performs cache partitioning
using \asm-Cache and conveys the slowdown estimated by \asm-Cache
for each application (corresponding to its cache way allocation)
to the memory controller. The memory controller uses these
slowdowns to partition memory bandwidth across applications using
\asm-Mem.  Figure~\ref{fig:cachepart-bwpart-results} compares
\asm-Cache-Mem with combinations of FRFCFS, PARBS and TCM with
UCP, across 100 16-core workloads with 4MB shared cache and 1/2
memory channels. \asm-Cache-Mem improves fairness by 14.6\%/8.8\%
on the 1/2 channel systems respectively, compared to the fairest
previous mechanism (FRFCFS+UCP), while achieving performance
within 1\% of the highest performing previous combination
(PARBS+UCP). We conclude that \asm-Cache-Mem is effective in
mitigating interference at both the shared cache and main memory,
achieving significantly better fairness than combining previously
proposed cache partitioning/memory scheduling schemes. 

%\begin{figure}[h!]
%%  \vspace{-3mm}
%  \centering
%  \begin{minipage}{0.22\textwidth}  
%    \centering
%    \includegraphics[scale=0.17, angle=270]{applications-asm/plots/cachepart_bwpart_ms}
%  \end{minipage}
%  \hspace{2mm}
%  \begin{minipage}{0.22\textwidth}
%    \centering
%    \includegraphics[scale=0.17, angle=270]{applications-asm/plots/cachepart_bwpart_hs}
%  \end{minipage}
%  \vspace{-1mm}
%  \caption{Combining \asm-Cache and \asm-Mem}
%  \label{fig:cachepart-bwpart-results}
%%  \vspace{-1mm}
%\end{figure}

%% file: applications-asm/application-performance-guarantees.tex
\section{Providing Soft Slowdown Guarantees}
\label{sec:app-slowdown-guarantees}

%\begin{figure}[h!]
%  \vspace{-4mm}
%  \centering
%  \begin{minipage}{0.22\textwidth}  
%    \centering
%    \includegraphics[scale=0.17, angle=270]{applications-asm/plots/cachepart_bwpart_ms}
%  \end{minipage}
%  \hspace{2mm}
%  \begin{minipage}{0.22\textwidth}
%    \centering
%    \includegraphics[scale=0.17, angle=270]{applications-asm/plots/cachepart_bwpart_hs}
%  \end{minipage}
%  \vspace{-2mm}
%  \caption{Combining \asm-Cache and \asm-Mem}
%  \label{fig:cachepart-bwpart-results}
%  \vspace{-6mm}
%\end{figure}
%
In a multi core system, multiple applications are consolidated on
the same system. In such systems, \asm's slowdown estimates can
be leveraged to \emph{bound the application slowdowns}.

\begin{figure}[h!]
  \begin{minipage}{0.48\textwidth}
    \centering
    \includegraphics[scale=0.35, angle=270]{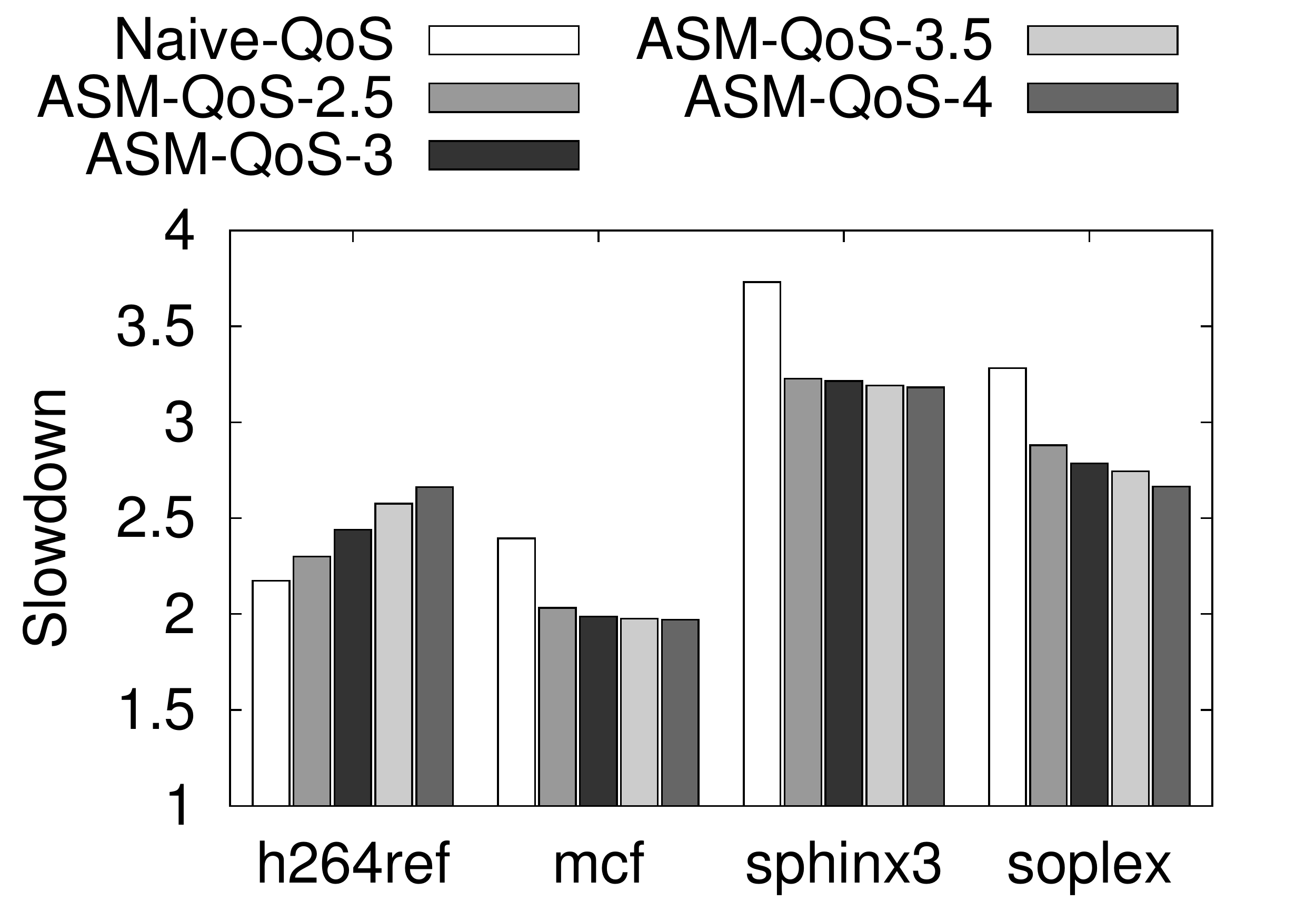}
%    \caption{\asm-QoS: Application Slowdowns}
    \label{fig:asm-qos-slowdown}
  \end{minipage}
  \begin{minipage}{0.48\textwidth}
    \centering
    \includegraphics[scale=0.35, angle=270]{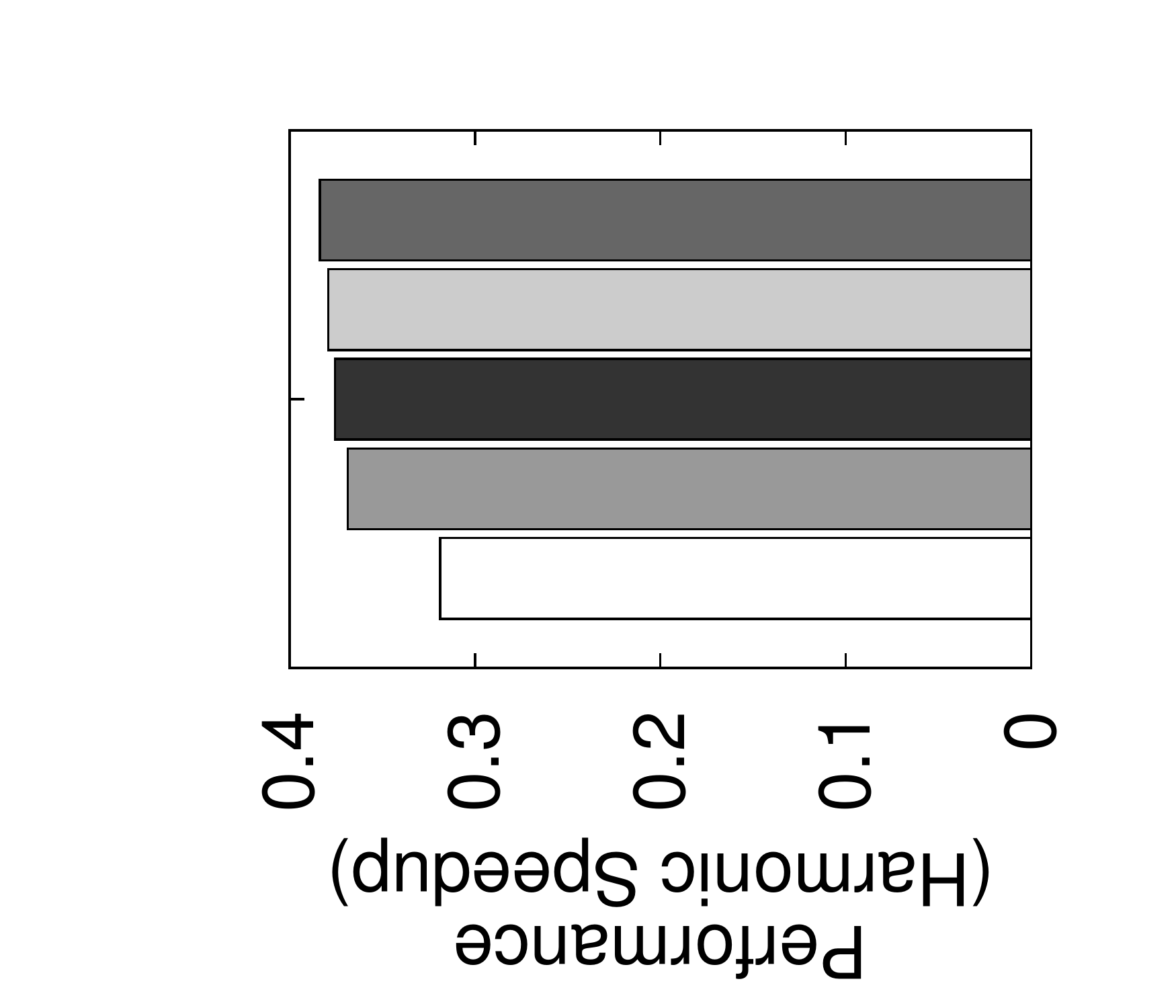}
%    \caption{\asm-QoS: Harmonic Speedup}
    \label{fig:asm-qos-hspeedup}
  \end{minipage}
  \caption{\asm-QoS: Slowdowns and performance}
  \label{fig:asm-qos-results}
\end{figure}

Figure~\ref{fig:asm-qos-results} shows the slowdowns of four
applications in a workload using a naive cache allocation scheme
and a slowdown-aware scheme based on \asm. The goal is to achieve a
specified slowdown bound for the first application,
\textit{h264ref}. The Naive-QoS scheme, which is unaware of
application slowdowns, allocates all caches ways to
\textit{h264ref}, the application of interest. \asm-QoS, on the
other hand, allocates just enough cache ways to the application of
interest, \textit{h264ref}, such that a specific slowdown bound
(indicated by X in \asm-QoS-X) is met. Such a scheme is enabled by
\asm's ability to estimate slowdowns for all possible
cache allocations (Section~\ref{sec:asm-cache}). Naive-QoS
minimizes \textit{h264ref}'s slowdown enabling it to meet any
slowdown bound greater than 2.17. However, it does so at the cost
of slowing down other applications significantly.
\asm-QoS, on the other hand, allocates just enough cache ways to
\textit{h264ref} such that it meets the specified bound, while
also reducing slowdowns for the other three applications,
\textit{mcf}, \textit{sphinx3} and \textit{soplex}, compared to
Naive-QoS, thereby improving overall performance significantly
(15\%/20\% for \asm-QoS-2.5/\asm-QoS-4 over Naive-QoS).

This is one example policy that leverages \asm's
slowdown estimates to partition the shared cache capacity to
achieve a specific slowdown bound. More sophisticated schemes can
be built on top of \asm's slowdown estimates that control the
allocation of \emph{both} memory bandwidth and cache capacity such
that different applications' slowdown bounds are met, while still
achieving high overall system performance. We propose to explore
such schemes as part of future work.

%Figure~\ref{fig:asm-qos-results} shows the results of a simple scheme that
%attempts to bound the slowdown of one application (XXX) to \emph{Nx} by allocating
%just the required number of caches ways to the application, based
%on our slowdown estimates for different cache allocation (computed
%using the scheme described in Section~\ref{sec:asm-cache}).
%
%Employing slowdown estimates from \asm enables allocating just
%enough ways to the application, XXX, such that its slowdown is
%bounded, while also improving other applications' performance.
%A naive scheme that allocates all ways to the application of
%interest, XXX, in the absence of slowdown estimates for different
%cache allocations, would slowdown other applications
%significantly, as shown in Figure~\ref{}.

%% file: applications-asm/application-fair-billing.tex
\section{Fair Pricing in Cloud Systems}
\label{sec:app-fair-billing}

Applications from different users could be consolidated onto the
same machine in a cloud server cluster. Pricing schemes in
cloud systems bill users based on CPU core, memory/storage
capacity allocation and run length of a job
~\cite{azure-billing,ec2-billing}. However, they do not account for
interference at the cache and main memory. For instance,
when two jobs A and B are run together on the same system, job A
runs for three hours due to cache/memory interference from job B,
but would have run for only an hour, had it been run alone. In
this scenario, accurate slowdown estimates from \asm can enable
pricing based on how much an application is slowed down due to
interference (especially since profiling every application to get
alone run times is not feasible). In the example above, \asm would
estimate job A's slowdown to be 3x, enabling the
user to be billed for only one hour, as against three hours with a
scheme that bills based only on resource allocation and run time.

%% file: applications-asm/application-migration-admission-control.tex
\section{Migration and Admission Control}
\label{sec:app-migration-admission-control}
\asm's slowdown estimates could be leveraged by the
system software to make migration and admission control decisions.
Previous works monitor different metrics such as cache misses,
memory bandwidth utilization across machines in a cluster and and
migrate applications across machines based on these
metrics~\cite{tang-thread-scheduling,Liu-ISCA14,Rao-HPCA13,adrm}. While such
metrics serve as proxies for interference, accurate slowdown
estimates are a direct measure of the impact of interference on
performance. Hence, periodically communicating slowdown estimates
from \asm to the system software could enable better migration
decisions. For instance, the system software could migrate
applications away from machines on which slowdowns are very high
or it could perform admission control and prevent new applications
from being scheduled on machines where currently running
applications are experiencing significant slowdowns.

%% file: applications-asm/summary.tex
\section{Summary}
We present several use cases/mechanisms that can leverage accurate slowdown
estimates from ASM, towards different goals such as achieving high
fairness, system performance and providing soft slowdown guarantees. Our
evaluations show that several of these mechanisms improve
fairness/performance over state-of-the-art schemes, thereby
demonstrating the effectiveness of \asm in enabling higher and
more controllable performance. We conclude that \asm is a promising substrate that
can enable the design of effective mechanisms to estimate and
control application slowdowns in modern and future multicore
systems.

%% file: chapters/conclusions.tex
\chapter{Conclusions and Future Directions}
\label{chap:conclusions}

\section{Conclusions}

In a multicore system, interference between applications at shared
resources is a major challenge and degrades both overall system
performance and fairness and individual application performance.
Furthermore, as we showed in Chapter~\ref{chap:introduction}, an
application's performance varies depending on the co-running
applications and the amount of available shared resources in a
system. 

Several previous works have tackled the problem of
memory interference mitigation, with the goal of achieving high
performance, with the prevalent direction being memory request
scheduling. State-of-the-art memory schedulers rank individual
applications with a total order based on their memory access
characteristics. Such a total order based ranking scheme increases
hardware complexity significantly, to the point that the scheduler
cannot always meet the fast command scheduling requirements of
state-of-the-art DDR protocols. Furthermore, employing a total
order ranking across individual applications also causes unfair
application slowdowns, as we demonstrated in Chapter~\ref{chap:blacklisting}.

We presented the Blacklisting memory scheduler (BLISS) in
Chapter~\ref{chap:blacklisting}, that tackles these shortcomings
of previous schedulers and achieves high performance and fairness,
while incurring low hardware complexity. BLISS does so based on
two new observations. First, it is sufficient to i) separate
applications into \emph{only two} groups, one containing
applications that are vulnerable to interference and another
containing applications that cause interference, and ii)
prioritize requests of the {\em vulnerable-to-interference}
group over the requests of the {\em interference-causing} group.
Second, we observe the applications can be classified as {\em
interference-causing} or {\em vulnerable} by simply monitoring the
number of consecutive requests served from an application in a
short time interval.

While BLISS is able to achieve high performance, it does not
tackle the problem of providing performance guarantees in the
presence of shared resource interference. Specifically, while
BLISS mitigates application slowdowns, it cannot precisely
quantify and control application slowdowns. Towards achieving the
goal of quantifying and controlling slowdowns, we presented a model to
accurately estimate application slowdowns in the presence of
memory interference in Chapter~\ref{chap:mise}. The Memory
Interference induced Slowdown Estimation (MISE) model estimates
application slowdowns based on the observation that a memory-bound
application's performance is roughly proportional to the rate at
which its memory requests are served. This enables estimating
slowdown as a ratio of request service rates. The
alone-request-service-rate of an application can be estimated by
giving its requests highest priority in accessing main memory. 

Accurate slowdown estimates from the MISE model can be leveraged
to drive various hardware, software and hardware-software
cooperative resource management techniques that strive to achieve
high performance, fairness and provide performance guarantees. We
demonstrate two such use cases of the MISE model in
Chapter~\ref{chap:mise-applications}, one that bounds the slowdown
of critical applications in a workload, while also optimizing for
overall system performance and another that minimizes slowdowns
across all applications in a workload.

The MISE model estimates slowdown accurately in the presence of
main memory interference, but does not take into account
contention for shared cache capacity. The Application Slowdown
Model (ASM) that we presented in Chapter\ref{chap:asm} takes into
account shared cache capacity interference, in addition to memory
bandwidth interference. ASM observes that an application's
performance is roughly proportional to the rate at which it
accesses the shared cache. This observation is more general than
MISE's observation that holds only for memory-bound applications.
ASM exploits this observation to estimate slowdown as a ratio of
cache access rates and estimates alone-cache-access-rate by
minimizing interference at the main memory and quantifying
interference at the shared cache. Slowdown estimates from ASM can
enable various resource management techniques to manage both the
shared cache and main memory. We discuss and evaluate several such
techniques in Chapter~\ref{chap:asm-applications}, demonstrating
the effectiveness of ASM in estimating and controlling application
slowdowns.

We believe the mechanisms and models we proposed in this thesis
could have wide applicability both in terms of being effective in
achieving high and controllable performance and also in terms of
inspiring future research. Furthermore, beyond the models and
mechanisms, the key principles and ideas that we conceive and
employ in building our mechanisms could have applicability and
impact in several different contexts. Specifically,

\begin{itemize}
\item
The principle of using request service/access rate as a proxy for
performance is a general observation that would hold in any closed
loop system. This principle can be applied to manage contention
and estimate interference-induced slowdowns in the context of
other resources such as storage and network too, besides at the
shared cache and main memory.
\item
The notion of achieving interference mitigation by simply
classifying applications into two groups, rather than employing a
full ordered ranking across all applications can be applied in the
context of managing contention at other resources too.
\end{itemize}

\section{Future Research Directions}

Our models, mechanisms and principles can inspire future research
in multiple different directions. We describe some potential
research directions in the next sections.

\subsection{Leveraging Slowdown Estimates for Cluster Management}

Slowdown estimates from our models can be leveraged to drive
various resource management policies. The hardware resource
management policies that we presented and evaluated in
Chapters~\ref{chap:mise-applications}
and~\ref{chap:asm-applications} partition resources such as caches
and main memory bandwidth on a single node. Especially in the
context of providing soft slowdown guarantees, such a node-level
management policy might not be able to meet the required slowdown
bounds/performance requirements. In this scenario, communicating
the slowdown estimates to the system software/hypervisor can
enable various application/virtual machine migration and admission
control policies. 

We have built one such policy that employs a simple linear model
that relates performance of an application to its memory bandwidth
consumption to detect contention and drive virtual machine
migration decisions in our VEE 2015 paper~\cite{adrm} (part of Hui
Wang's PhD thesis). This policy strives to achieve high
performance. 

We believe there is ample scope to explore and build several more
virtual cluster management policies that exploit slowdown
estimates to achieve various different goals such as meeting
performance guarantees, improving system fairness etc. both in
real systems and in the simulation realm. For instance, we were
relatively constrained by what counters are available in existing
systems when designing our model and virtual machine migration
policy. The new Haswell machines provide more counters and support
for monitoring and managing the shared cache capacity. This would
enable more effective cluster management policies and would also
enable combining cluster management policies with resource
allocation policies.

\subsection{Performance Guarantees in Heterogeneous Systems}

While the focus of this thesis was on providing soft performance
guarantees in the context of homogeneous multicore systems,
meeting different kinds of performance requirements in heterogeneous systems with
different kinds of agents is an important research problem. We
have explored this problem in the context of SoC systems with
different agents such as CPU cores, GPUs and hardware
accelerators~\cite{squash-tech-report}. Our goal in this work is
to meet the deadlines/frame rate requirements of hardware
accelerators and GPUs, while still achieving high CPU performance. 

We believe there are several interesting and unsolved problems and
challenges in this space. For instance, different agents could
have different kinds of performance requirements in a
heterogeneous system. Some agents might need to meet deadlines,
whereas other agents might have requirements on resources such as
memory bandwidth, while agents such as CPU cores might have
slowdown/latency requirements. 

The design of a memory system and a memory controller that is able
to take into account the different and often, conflicting
requirements of different agents and applications is a significant
challenge. Furthermore, building a memory system that can take in
such requirements and strive to meet them in a general manner and
not be limited by the specific configuration of the agents and the
system is an even bigger challenge. We believe this is a rich area
with ample scope for future exploration.

\subsection{Integration of Memory Interference Mitigation
Techniques}

The main focus of this thesis was on mitigating and quantifying
memory interference with the goal of building better resource
management and allocation techniques. This thesis focused heavily
on memory request scheduling to perform slowdown estimation and
resource management (e.g., BLISS, MISE and the mechanisms built on
top of MISE). However, as mentioned in
Section~\ref{sec:other-interf-mitigation}, there are several other
approaches to mitigate memory interference. For instance, we have
explored memory channel partitioning~\cite{mcp}. Other previous
works have proposed bank
partitioning~\cite{bank-part,pact-bank-part,bank-part-hpca14},
interleaving~\cite{mop}, source
throttling~\cite{fst,hat,nychis,cc-hotnets10}.

These different approaches could be effectively combined together
rather than relying on one approach, to address the memory
interference problem, as we discuss in our papers describing the
challenges in the main memory system~\cite{kiise,superfri}. For instance, channel partitioning maps
applications' data onto different channels depending on their
access patterns. In this context, the memory request scheduling
policy could be tailored to better match the access
characteristics of the specific applications that are mapped to
the channel. We briefly explored this idea in our work on memory
channel partitioning~\cite{mcp} (part of Sai Prashanth
Muralidhara's thesis). However, there is plenty of scope to
explore this further. In fact, this could lead to the notion of
programmable memory controllers, where the memory request
scheduling policy can be tuned at run time, depending on the
workload. 

The address interleaving policy heavily influences the row-buffer
locality and bank-level parallelism of different applications'
accesses. Hence, co-design of the address interleaving policy
along with the memory scheduling policy can enable a memory
controller design that is more amenable to the access
characteristics of different applications, given a specific
address interleaving policy.

We expand more on some of these challenges
in~\cite{superfri,kiise}. We classify resource management
techniques into dumb vs. smart resource techniques. Dumb resources
do not have the intelligence to manage themselves and rely on a
centralized agent to manage and allocate them. Smart resources
have the intelligence to manage and control their own allocation.
We discuss the trade-offs involved in the design and effectiveness
of these different kinds of techniques and the challenges in
combining them effectively.

The interactions between these different memory interference
mitigation techniques offer a wide range of different choices and
opportunities to design a memory system that leverages these
different degrees of freedom in a synergistic manner. Hence, we
believe this is an important and promising direction for future
exploration.

\subsection{Resource Management for Multithreaded Applications}

This thesis focused predominantly on estimating and managing
contention between multiple single-threaded applications when they
are run together on a system. The problem of managing contention
between different threads in a multithreaded application and 
between multiple multithreaded applications is an important
challenge. 

Multiple threads in a multithreaded application work towards
achieving a common goal. This is a different scenario than when
multiple competing single threaded applications, each with a
different goal, are run together on a system. Although the
different threads work towards the same goal, it is still
important to apportion resources accordingly among the different
threads such that the threads that lag the most are given more
resources. 

There are several different unaddressed research challenges in
this space. For instance, accurately estimating how much each
thread is slowed down is an important aspect of determining the
amount of progress made by each thread. Once accurate metrics are
developed to capture the progress of multithreaded applications,
they can be leveraged to build resource allocation policies that
partition resource among different threads of a multithreaded
application. Furthermore, managing resources between multiple
multithreaded applications is yet another important and promising
research area that has not been explored much.

\subsection{Coordinated Management of Main Memory and Storage}

The primary focus of this thesis was on managing main memory and
shared caches. However, the storage system is an important
component that needs to be taken into account in order to build a
comprehensive resource management substrate. While there has been
a large body of work on storage QoS, as we describe in
Section~\ref{sec:storage-qos}, the interactions between memory
bandwidth, memory capacity and storage bandwidth have not been
explored much. Our ideas on coordinated shared cache capacity and
memory bandwidth management could potentially be leveraged in the
context of main memory capacity and storage bandwidth.
Furthermore, our observations on request service rate correlating
linearly with performance could be leveraged to estimate progress
and performance in the context of other resources such as the
storage bandwidth.

In the past decade, there has been a proliferation of storage
class non-volatile memory technologies such as flash and phase
change memory. In this context, the notion of how long storage
accesses take changes. An application might potentially not need
to be context switched on a page fault. Furthermore, such fast
non-volatile memory technologies provide the opportunity to manage
the DRAM main memory and the NVM storage system, as a single
address space, as described by Meza et al. in~\cite{meza-weed13}.
Coordinated management of main memory and such fast storage
technologies, in light of these advancements, presents new and
rich opportunities. Furthermore, the coordinated management of
main memory and storage also opens up opportunities for more
hardware-software cooperative solutions, since both the hardware
and software layers need to be involved for effective management
of the main memory and storage in a coordinated manner. We
believe there are several very important and intriguing challenges
in this space.

\subsection{Comprehensive Slowdown Estimation}

Estimating slowdown due to contention at all resources in a system
enables understanding the impact of contention at all resources in
a comprehensive manner and consequently, the management of these
different resources. The previous section on expanding our work to
include the storage system was in this spirit. However, other
resources such as the on-chip interconnect, the off-chip network
should be taken into account to build a comprehensive slowdown
estimation model.

Our principles on request service rate correlating with
performance can potentially be used to estimate slowdown due to
different shared resources. However, access characteristics and
bottleneck behavior at these different resources could be
different, providing ample scope for new insights and ideas in
this space. Furthermore, once slowdown estimates are available,
managing a large set of resources and specifically, doing do in a
coordinated manner are important and challenging problems.